\documentclass[twocolumn,showpacs,amsmath,amssymb,prd]{revtex4-1}

\usepackage{graphicx}% Include figure files
\usepackage{dcolumn}% Align table columns on decimal point
\usepackage{bm}% bold math
\usepackage{epsfig}
\usepackage{amsmath}
\usepackage{color}
\usepackage{lipsum}
\usepackage{hyperref}
        
\newcommand{\be}{\begin{equation}}
\newcommand{\ee}{\end{equation}}
\newcommand{\ba}{\begin{eqnarray}}
\newcommand{\ea}{\end{eqnarray}}
\def\eps{\epsilon}

\def\g{\gamma}

\begin{document}
\input{epsf}

\title{Ultra-High Energy Neutrino Afterglows of nearby Long Duration Gamma-Ray Bursts}

\author{Jessymol K.\ Thomas}
\email{jessymolkt@uj.ac.za}

\author{Reetanjali Moharana}
\altaffiliation{Currently at the Racah Institute of Physics, Jerusalem, Israel}
\email{moharana.reetanjali@mail.huji.ac.il}

\author{Soebur Razzaque}
\email{srazzaque@uj.ac.za}

\affiliation{Department of Physics, University of Johannesburg, PO Box  524, Auckland Park 2006, South Africa}

\begin{abstract} 
Detection of ultra-high energy (UHE, $\gtrsim 1$~PeV) neutrinos from astrophysical sources will be a major advancement in identifying and understanding the sources of UHE cosmic rays (CRs) in nature. Long duration gamma-ray burst (GRB) blast waves have been considered as potential acceleration sites of UHECRs.  These CRs are expected to interact with GRB afterglow photons, which is synchrotron radiation from relativistic electrons co-accelerated with CRs in the blast wave, and naturally produce UHE neutrinos. Fluxes of these neutrinos are uncertain, however, and crucially depend on the observed afterglow modeling.  We have selected a sample of 23 long duration GRBs within redshift 0.5 for which adequate electromagnetic afterglow data are available and which could produce high flux of UHE afterglow neutrinos, being nearby.  We fit optical, X-ray and $\gamma$-ray afterglow data with an adiabatic blast wave model in a constant density interstellar medium and in a wind environment where the density of the wind decreases as inverse square of the radius from the center of the GRB.  The blast wave model parameters extracted from these fits are then used for calculating UHECR acceleration and $p\gamma$ interactions to produce UHE neutrino fluxes from these GRBs.  We have also explored the detectability of these neutrinos by currently running and upcoming large area neutrino detectors, such as the Pierre Auger Observatory, IceCube Gen-2 and KM3NeT observatories.  We find that our realistic flux models from nearby GRBs will be unconstrained in foreseeable future.      
\end{abstract}


\date{\today}
\maketitle

\section{Introduction}

\label{intro}

The origin of UHECRs is still unknown despite recent detection of astrophysical neutrinos, which are produced by interactions of CRs, with the IceCube neutrino observatory \cite{aartsen2014observation,Aartsen:2015zva}.  Detection of UHE neutrinos in principle can identify the sources of UHECRs \cite{Stanev:1996qj}, as the latter is deflected by the Galactic and intergalactic magnetic fields from the line-of-sight to their origin.  At energies $\gtrsim 4\times 10^{19}$~eV the deflection of UHECRs, if protons, can be small and a correlation between the UHECR and neutrino arrival directions is expected.  Preliminary studies with IceCube astrophysical neutrinos in the energy range between $\sim 20$~TeV and $\sim 2$~PeV and UHECRs in the energy range $\gtrsim 10^{20}$~eV show hints of correlation~\cite{Moharana:2015nxa,Aartsen:2015dml}, but no evidence of a source population producing those \cite{Moharana:2015nxa}.  Moreover the source population for relatively low energy IceCube neutrinos and UHECRs may be different. Detection of $\gtrsim 1$~PeV neutrinos can therefore probe the UHE neutrino and CR sources at similar energy range. 

Long-duration GRBs, possibly originating from core-collapse of massive stars \cite{MacFadyen:1998vz, Woosley:2005gy}, have long been hypothesized to accelerate UHECRs.  Protons and/or ions can be accelerated during the prompt $\gamma$-ray emission phase, in the internal shocks between clumpy material in the GRB jet  \cite{Waxman:1995vg}, and during the afterglow emission, in the external shocks of the GRB jet ejecta and blast wave \cite{Vietri:1995hs}.  High energy neutrinos can be produced by interactions of these UHECRs with prompt $\gamma$-ray photons in case of internal shocks \cite{Waxman:1997ti} and with afterglow photons in case of external shocks \cite{Waxman:1999ai, Dai+01}.  Detection of these neutrinos from a GRB can provide a telltale signature of UHECR acceleration in this powerful astrophysical source, as well as probe highly interesting nature of relativistic jets.  Searches of TeV-PeV neutrinos from the prompt GRB phase by the IceCube Neutrino Observatory \cite{Abbasi:2012zw} and the ANTARES neutrino telescope \cite{Adrian-Martinez:2013dsk} have resulted in null detection. Separate analysis of short and long GRBs done for four year IceCube data search has put more stringent constrains on neutrino flux from GRBs \cite{Casier:2015boc}. The GRB triggers used in this analysis consists of 82 short and 491 long bursts and the IceCube data search window was fixed from $-1$ to 4~s and one hour for short and Long bursts, respectively. These results constrain significant UHECR production in the GRB internal shocks \cite{Ahlers:2011jj, Aartsen:2016qcr}. 

Recently UHECR acceleration in the GRB blast wave (external shock with circumburst environment) and subsequent UHE neutrino emission, arising from UHECR interactions with afterglow photons, scenario has gained attention \cite{Razzaque:2013dsa, Xiao:2014vga, Razzaque:2014ola, Tamborra:2015qza, Asano:2015kma} because of a possibility to detect these neutrinos with very large detectors such as the proposed IceCube Gen-2 \cite{Aartsen:2015dkp}, KM3NeT \cite{Adrian-Martinez:2016fdl} and the currently operating Pierre Auger Observatory (PAO) for cosmic-rays \cite{Aab:2015kma}. This neutrino afterglow is contemporaneous with the $\gamma$-ray, X-ray and optical afterglow, thus can be searched for within well-localized time and position windows with very little or no background.  Estimated neutrino fluence for IceCube corresponding to 468 long GRBs detected within 4 years (2011-2015), reported in Ref.~\cite{Brayeur2015}, gives an upper limit $1 \times 10^{-4}$~GeV~cm$^{-2}$~sr$^{-1}$.  This calculation considers both prompt and afterglow neutrino production in GRBs.  The afterglow neutrino flux in this case, however, does not rely on the observed radiations and an $E^{-2}$ energy spectrum was used for the flux. 

Availability of multi-wavelength data on GRBs at an unprecedented level now allow modeling of broadband spectra and long-duration light curves to constrain the afterglow models \cite{Toma:2006iu,Laskar:2013uza,DePasquale:2015wga,Zhang:2016sve}. Earlier we have done this analysis for GRB 130427A \cite{jesH,Thomas:2016vna}. This in turn allow realistic calculation of UHECR acceleration and neutrino flux from the GRB blast wave than was previously possible.  It was also pointed out sometime ago that detection of neutrinos from individual GRBs may be possible if they are nearby and have high gamma-ray fluence (see, e.g., \cite{Razzaque:2003uw, Guetta:2003wi}).  Indeed a few nearby bright GRBs can dominate stacked neutrino fluence from all GRBs (see, e.g., \cite{Razzaque:2006qa}).  These serve as motivations for our current study. 
 
In this work  we have modeled UHE neutrino afterglows, following Refs.~\cite{Razzaque:2013dsa, Razzaque:2014ola}, for a well defined set of 23 nearby long-duration GRBs within redshift $z = 0.5$, which are the most promising sources for very large neutrino detectors.  We have used data from the {\it Fermi-}Large Area Telescope (LAT), {\it Fermi-}Gamma-ray Burst Monitor (GBM), {\it Swift-}Burst Alert Telescope (BAT),  {\it Swift-}X-ray Telescope (XRT), {\it Swift-}Ultraviolet/Optical Telescope (UVOT) and by ground-based optical and radio telescopes to fit afterglow synchrotron radiation models \cite{Sari:1997qe,Piran:2010ew}.  The two most popular models, evolutions of an adiabatic GRB blast wave in a constant density interstellar medium (ISM) and in a wind-type medium in which the density decreases as $1/R^{2}$, where $R$ is the radial distance from the center of the GRB, are then compared for each GRB in our sample.  Using the parameters from the afterglow model fits we have calculated individual neutrino fluxes from all 23 GRBs for different time intervals, after the duration ($T_{90}$) of the prompt phase.  The neutrino flux calculations are also done in both the ISM and wind medium. Next, we have calculated neutrino events for the IceCube Gen-2, KM3NeT and PAO observatories, using neutrino fluence, both from individual GRBs and by stacking them.  We have also calculated possible upper limits for the stacked fluence, in case of no detection by the relevant neutrino observatories.

The plan of this paper is the following.  We provide details of our GRB sample in Sec.~\ref{selection} including sources of electromagnetic data. In Sec.~\ref{modelling} we discuss afterglow modeling of electromagnetic data and obtain the model parameters both in a constant density interstellar medium and in a wind medium.  We calculate neutrino flux from the GRB sample in Sec.~\ref{neutrino-flux} using the same afterglow model parameters obtained in Sec.~\ref{modelling}.   In Sec.~\ref{detection} we compute neutrino fluence of individual GRBs and stacked GRBs for the ones, falling in the observable sky of the corresponding detectors and calculate upper limits on those for different observatories.  We discuss our results in Sec.~\ref{discussion}.

%%%%%%% Section II %%%%%%%%%%%%%%%%%%
\section{Long-duration GRB sample within redshift $z=0.5$}
%%%%%%%%%%%%%%%%%%%%%%%%%%%%%%
\label{selection}
We have done synchrotron afterglow modeling of spectral energy distribution (SED) at different time intervals and the light curves at different frequencies for all 23 long-duration GRBs within redshift $z=0.5$ detected by {\it Swift} until March 2017 \cite{mpi_grb}. The lowest redshift is $z = 0.03$ for GRB 060218 \cite{4792} and the highest redshift is $z = 0.49$ for GRB 091127 \cite{10400}. In our afterglow modeling we have used the {\it Fermi-}LAT and {\it Swift-}XRT/UVOT data as well as optical and radio data from different ground-based telescopes.  

Table~\ref{tab1} lists {\it Swift-}BAT/XRT/UVOT and ground-based optical telescope flux data points for {the 21} GRBs.  Tables~\ref{tab2} and \ref{tab3} list two more GRBs for which radio flux data points (GRB 130702A) and GeV data points (GRB 130427A) are also available.  We have collected {\it Swift-}XRT data for all GRBs from the UK Swift Science Data Centre~\cite{www}.  {\it Swift-}UVOT and optical data are collected from different GRB Circular Notices (GCNs) and other published papers: 
GRB 050803 (GCN 3759 \cite{3759}), 
GRB 050826 (GCN 3887 \cite{3887}), 
GRB 051109B (GCN 4233 \cite{4233} and GCN 4259 \cite{4259}), 
GRB 051117B (GCN 4303 \cite{4303}), 
GRB 060218 (Ref.~\cite{SCamp}), 
GRB 060512 (GCN 5130 \cite{5130}),  
GRB 060614 (GCN 5255 \cite{5255}), 
GRB 061021 (GCN 5745 \cite{5745}), 
GRB 090417B (GCN 9174 \cite{9174, 9156}),
GRB 091127B (GCN 10199 \cite{10199}), 
GRB 100316D (Ref.~\cite{Cano}), 
GRB 101225A (GCN 11499, \cite{11499}),
GRB 111225A (GCN 12735 \cite{12735} and GCN 12740 \cite{12740}), 
GRB 120422A (GCN 13263 \cite{13263}), 
GRB 120714B (GCN 13478 \cite{13478} and GCN 13484 \cite{13484}), 
GRB 130427A (Ref.~\cite{maseli}), 
GRB 130702A (Ref.~\cite{Leo1}), 
GRB 130831A (GCN 15168 \cite{15168}), 
GRB 150727A (GCN 18084, \cite{18084}),
GRB 150818A (GCN 18161 \cite{18161}),
GRB 151027A (GCN 18478, \cite{18478}),
GRB 160623A (GCN 19572, \cite{19572}),
GRB 161219B (GCN 20300, \cite{20300}).
The radio data points at 4-8 GHz range from the Very Large Array (VLA) for GRB 130702A are taken from Ref.~ \cite{Leo1} and the {\it Fermi-}LAT data for GRB 130427A  are taken from Ref.~\cite{maseli}.

The XRT data for all GRBs are already given in the $\nu F_\nu$ flux format with unit erg~cm$^{-2}$~s$^{-1}$ at the UK Swift Science Data Centre \cite{www} and we have downloaded corresponding Flux light curves (ascii data files).  UVOT and optical data in GCNs are often given in magnitudes and we have converted those to flux units of erg~cm$^{-2}$~s$^{-1}$ using magnitude to flux converter on the Gemini Observatory website \cite{gemini}.  {\it Fermi-}LAT data points for GRB 130427A are already given in the erg~cm$^{-2}$~s$^{-1}$ flux units in Ref.~\cite{maseli}.  We have converted the radio data points from Jansky as given in Refs.~\cite{Leo1}. For afterglow synchrotron modeling of the spectral energy distribution (SED) and light curves, we have directly used these flux data.

%%%%%%%%%%%%%%%%%%%%%%%%%Table1%%%%%%%%%%%%%%%%%%%%%%%%%
\begin{table*}[th!]
\setlength\tabcolsep{0.4cm}
\centering
\caption{Afterglow flux data at different times for 21 long-duration GRBs within redshift $z=0.5$}
\label{tab1}
\begin{tabular}{llll l l l}
\hline
\hline
GRB & $z$ & Time  &  XRT (0.3-10 keV) & UVOT/optical & Filter & References \\ 
  &  &  $T-T_0$~s    &     flux (erg~cm$^{-2}$~s$^{-1}$)  & Magnitude (mag)\footnote{We have averaged over the magnitudes in time interval around the reported times. }  &      \\ \hline

050803 & 0.422 & 1150 & (7.08 $\pm$ 1.60)$\times 10^{-11}$ & $>$ 19.7  &  B & \cite{www,3759,3758} \\ 
    
             &           & 17650 & (7.57 $\pm$ 1.70)$\times 10^{-12}$ & $>$ 20.7 &  U         \\ 

050826 & 0.297 & 10000 & (4.67 $\pm$ 1.21)$\times 10^{-13}$ & $>$21.2 &  B & \cite{www,3887,5982} \\

             &           & 16000 & (4.00 $\pm$ 0.96)$\times 10^{-13}$ & 21.0 $\pm$0.2  &  R & \\

051109B & 0.08 & 5800 & (1.71 $\pm$ 0.44)$\times 10^{-12}$  &$>$20.8 & v & \cite{www,4233,4259,5441} \\

               &         & 43000 & (1.45 $\pm$ 0.37)$\times 10^{-13}$  & 22.04$\pm$ 0.09 &  B & \\

051117B  &0.481  &     135        &   (3.15 $\pm$ 0.69)$\times 10^{-11}$   &      $>$20.53  &    v      &\cite{www,4303,grbhost}    \\

         &  & 450        &       (5.49  $\pm$ 1.21)$\times 10^{-12}$  &       $>$19.66    &   u         \\ 

060218 &0.033   &    50000   &        (4.01 $\pm$ 0.88)$\times 10^{-12}$   &    $>$18.22    &   v &\cite{www,SCamp,4819} \\         

          & & 450000    &        (4.00 $\pm$ 0.79)$\times 10^{-13}$  &        $>$22.4   &   uvm1         \\

060512 &0.443  &       4275        &     (1.63 $\pm$ 0.42)$\times 10^{-12}$   &        19.50 $\pm$ 0.21    &  B  &\cite{www,5130}         \\

         & &  22500       &          (3.02 $\pm$  0.69)$\times 10^{-13}$   &      20.40 $\pm$ 0.46 &   u          \\   

 060614 &0.125  &      4900    &          (7.43 $\pm$  2.01)$\times 10^{-12}$    &     18.81 $\pm$ 0.14 &    u  &\cite{www,5255,5276}        \\ 

           &  &   5750    &         (6.15 $\pm$  1.40)$\times 10^{-12}$   &       19.54 $\pm$ 0.30  &     v   \\ 

061021 &0.346    &  185         &        (3.20 $\pm$ 0.53)$\times 10^{-10}$   &      16.76 $\pm$ 0.06   &   v   &\cite{www,5745}       \\ 

         & & 4580         &        (2.23 $\pm$ 0.50)$\times 10^{-11}$    &       18.18 $\pm$ 0.12  &    u          \\ 

         & & 6220          &        (1.20 $\pm$ 0.26)$\times 10^{-11}$    &      19.33 $\pm$ 0.19   &   B         \\

090417B & 0.345  & 6850  & (6.27 $\pm$ 1.42)$\times 10^{-11}$ & $>$ 21.3  &  V & \cite{www, 9174} \\ 
    
       &        & 13290 & (2.19 $\pm$ 0.49)$\times 10^{-11}$ & $>$ 22 &  U         \\ 

091127B &0.49   & 3500         &        (4.41 $\pm$ 0.60)$\times 10^{-10}$    &     17.0 $\pm$ 0.1     &    v   &\cite{www,10199, 10400}  \\ 

          &   &          &          -      &       16.42 $\pm$ 0.15    &    i'    \\ 

          &    &       &        -     &    16.53 $\pm$ 0.15   &   r'    \\

           &  &11000    &      (6.12 $\pm$  1.90)$\times 10^{-11}$      &     17.02 $\pm$ 0.2   &   R  \\ 

          &  &78000    &       (8.68  $\pm$  1.76)$\times 10^{-12}$    &      18.90  $\pm$ 0.15   &   R \\

100316D & 0.059   &  60000    &       (5.074  $\pm$ 1.29)$\times 10^{-13}$     &        $20.5 \pm 0.65$   &    B &\cite{www,Cano, 10513} \\ 
        &         & 80670     &       (2.28  $\pm$ 0.58)$\times 10^{-13}$  &  $20.1  \pm  0.43  $ & R\\
101225A & 0.40  & 29880 & (1.51 $\pm$ 0.24)$\times 10^{-11}$ & $>$  21.40  &  V & \cite{www,11499,11522} \\

  111225A &0.297   &  10500    &        (2.88 $\pm$ 0.60) $\times 10^{-13}$       &      $>$20.66     &   u &\cite{www,12735,12740}\\

         &  &  62500    &        (5.94 $\pm$ 1.67) $\times 10^{-14}$       &         $>$ 22.5 &   R\\                

 120422A & 0.28 &  23000    &        (2.69 $\pm$  0.65) $\times 10^{-13}$      &      $>$ 21.5 &    g'  &\cite{www,13263} \\

 120714B & 0.398 & 10000   &         (3.34  $\pm$  0.88) $\times 10^{-13}$      &     $>$21.5  &    b   &\cite{www,13478,13484}  \\

           & & 22000   &       (2.20 $\pm$ 0.01) $\times 10^{-13}$       &       $>$23.0 $\pm$ 0.2   &   g'\\

  130831A &0.479  &   600    &         (2.33  $\pm$  0.24)$\times 10^{-10}$       &    14.9 $\pm$ 0.1   &    b &\cite{www,15168,15144} \\

           & & 23725    &       (5.41  $\pm$  0.92)$\times 10^{-12}$      &        19.2 $\pm$ 0.2        &   b\\

           & & 40650    &        (2.50  $\pm$  0.57)$\times 10^{-12}$      &      20.1 $\pm$ 0.2          &  b\\ 
150727A &  0.313  & 4150 & (1.97 $\pm$ 0.46)$\times 10^{-12}$ & $>$  18.7 &  V& \cite{www,18084,18080} \\ 
    
        &         & 24600 & (5.01 $\pm$ 1.10)$\times 10^{-13}$ & $>$ 18.9 &  r        \\

150818A &0.282  & 4350      &        (2.24  $\pm$ 0.55)$\times 10^{-12}$       &   20.56  $\pm$ 0.33      &    b & \cite{www,18161,18177} \\

          & & 8000      &        (1.91  $\pm$ 0.54)$\times 10^{-12}$      &     $>$19.36  &    v \\  
151027A &  0.38   & 1034  & (4.22 $\pm$ 0.63)$\times 10^{-10}$ & 15.37 $\pm$ 0.03  &  B & \cite{www,18478,18482} \\ 
    
        &         & 53300 & (7.57 $\pm$ 1.70)$\times 10^{-12}$ & 17.49 $\pm$ 0.10&  V  \\ 
160623A &   0.367  &  46164 & (2.23 $\pm$ 0.50)$\times 10^{-11}$ & $>$  20.1  &  V & \cite{www,19572,19708} \\ 
    
             &           & 100000 & (7.55 $\pm$ 1.73)$\times 10^{-12}$ & $>$ 20.4 &  g        \\ 
161219B &  0.148  & 6948& (8.47 $\pm$ 1.85)$\times 10^{-11}$ &  17.38 $\pm$ 0.02  &  R & \cite{www,20300,20306} \\ 
    
       &         & 45415 & (1.46 $\pm$ 0.33)$\times 10^{-11}$ &  18.0 $\pm$ 0.01 &  U \\ 
       \hline
\hline
\end{tabular}
\end{table*}

%%%%%%%%%%%%%%%%%%%%%%%%%Table2%%%%%%%%%%%%%%%%%%%%%%%%%
\begin{table*}
\setlength\tabcolsep{0.2cm}
\centering
\caption{Afterglow flux data at different times for GRB 130702A}
\label{tab2}
\begin{tabular}{clllllll}
\hline
\hline
GRB & $z$ & Time  &  XRT (0.3-10 keV) & Radio/VLA & UVOT/optical & Filter & References \\ 
  &  &  $T-T_0$~s    &     flux (erg~cm$^{-2}$~s$^{-1}$)  & flux (erg~cm$^{-2}$~s$^{-1}$)  & Magnitude (mag)\footnote{We have averaged over the magnitudes in time interval around the reported times. }  &      \\ \hline

130702A  & 0.145 & 1.165$\times 10^{-5}$ &   (8.13 $\pm$ 1.76)$\times 10^{-12}$  & - &  18.42 $\pm$ 0.04  & i' & \cite{www,14985,Leo1} \\   
                &           &                                     & -                                                           & - & 18.80 $\pm$ 0.04 & g'  \\   
                &           & 2.0 $\times 10^{5}$ & (5.44 $\pm$ 1.22)$\times 10^{-12}$ & (8.9  $\pm$ 0.45)$\times 10^{-17}$ & - &   \\          
                &           &                                 &  - &  (9.6 $\pm$ 0.49)$\times 10^{-18}$ & - &   \\ 
                
\hline
\end{tabular}
\end{table*}

%%%%%%%%%%%%%%%%%%%%%%%%%Table3%%%%%%%%%%%%%%%%%%%%%%%%%                
\begin{table*}
\setlength\tabcolsep{0.2cm}
\centering
\caption{Afterglow flux data at different times for GRB 130427A}
\label{tab3}
\begin{tabular}{clllllll}
\hline
\hline        
GRB & $z$ & Time  &  LAT & XRT (0.3-10 keV) & UVOT/optical & Filter & References \\ 
  &  &  $T-T_0$~s    &     flux (erg~cm$^{-2}$~s$^{-1}$)  & flux (erg~cm$^{-2}$~s$^{-1}$)  & Magnitude (mag)\footnote{We have averaged over the magnitudes in time interval around the reported times. } &      \\ \hline

130427A & 0.34& 193       &      (3.01  $\pm$ 1.38) $\times 10^{-7}$  &    (1.42 $\pm$ 0.24)$\times 10^{-7}$  &     - &    &\cite{www,maseli,14455}  \\   

        & & 505 &     (6.81  $\pm$ 2.95) $\times 10^{-8}$  &    (1.99 $\pm$ 0.39 $\times 10^{-8}$    &        11.37 $\pm$ 0.03 &  $w1$  \\
         
        &  &    &      -  &    - &           11.19 $\pm$ 0.35  &   $u$   \\
        &   &   &      -  &    - &          11.48 $\pm$ 0.56  &   $b$   \\

        & &4500 &     (6.03 $\pm$ 2.84) $\times 10^{-9}$    &    - &          14.66 $\pm$ 0.01   &    $i'$  \\ 

         & &    &                             &                          &    14.72 $\pm$ 0.02   &   $r'$  \\

        &  & 23000 &  (4.93 $\pm$ 2.24) $\times 10^{-10}$    &   (2.01 $\pm$ 0.42) $\times 10^{-10}$  &  15.47 $\pm$ 0.01  &  $w1$ \\ \hline
     
\hline
\end{tabular}
\end{table*}

%%%%%%% Section III %%%%%%%%%%%%%%%%%%%%
\section{Synchrotron afterglow model fit to multiwavelength data}
%%%%%%%%%%%%%%%%%%%%%%%%%%%%%%%%%
\label{modelling}
Synchrotron radiation by shock-accelerated electrons in the GRB ejecta (reverse shock) and blast wave (forward shock) is widely believed to be responsible for GRB afterglow emission \cite{Sari:1997qe,Piran:2010ew}.  The dynamics of the GRB blast wave crucially depends on the surrounding environment.  A constant density interstellar medium (ISM) and a wind blown by the GRB progenitor star are typically discussed in literature.  We have modeled afterglow data of 23 GRBs in our sample with synchrotron radiation from the forward shock of an adiabatic GRB blast wave  in the ISM and wind environment, where the density of the wind decreases as $\propto R^{-2}$ from the GRB centre,  $R$ being the radial distance.  We refer readers to Refs.~\cite{Sari:1997qe, Chevalier+00, Granot:2001ge, Panaitescu:2001fv, Ghisellini+10,Razzaque:2013dsa} for details of the blast wave evolution and synchrotron radiation model.  Here we briefly discuss the main formulae used for modeling. 

The instantaneous spectrum of low energy photons from the GRBs can be modeled by the synchrotron emission of a single power law distribution of electrons, $\propto \gamma_e^{-p}$, with a minimum Lorentz factor $\gamma_{e, m}$ and an e-folding Lorentz factor $\gamma_{e, s}$.  However the photon spectrum will have breaks at different frequencies due to absorption and cooling effects of electrons.  These break frequencies are $\nu_m$, the photon frequency corresponding to  $\gamma_{e,m}$ of electrons; $\nu_c$, the photon frequency corresponding to the cooling Lorentz factor of electrons $\gamma_{e,c}$; $\nu_a$, the photon frequency due to self absorption; and $\nu_s$, the photon frequency corresponding to the e-folding electron Lorentz factor $\gamma_{e,s}$.  The cooling Lorentz factor $\gamma_{e,c}$ is determined from balancing the synchrotron cooling timescale in a random magnetic field generated in the shock  and the dynamic timescale of the blast wave evolution.  The spectrum is different for the case when all electrons radiate away their energy within the dynamic time scale (fast cooling) and the case when electrons only above a certain energy ($\gamma_{e.c}m_ec^2$) can radiate away their energy within the dynamic time scale (slow cooling).  The fast-cooing ($\nu_m > \nu_c$) spectrum is given by \cite{Piran:2004ba},
\begin{eqnarray}
F_\nu &=& F_{\nu, \rm max} \nonumber \\
&\times & 
\left\{
\begin{array}{ll}
\left(\frac{\nu}{\nu_a}\right)^2\left(\frac{\nu_a}{\nu_c}\right)^\frac{1}{3}; &  \nu < \nu_a \\
\left(\frac{\nu}{\nu_c}\right)^\frac{1}{3}; & \nu_a \le \nu < \nu_c \\
\left(\frac{\nu}{\nu_c}\right)^{-\frac{1}{2}}; &  \nu_c \le \nu < \nu_m \\ 
\left(\frac{\nu_m}{\nu_c}\right)^{-\frac{1}{2}} \left(\frac{\nu}{\nu_m}\right)^{-\frac{p}{2}} e^{{-\frac{\nu}{\nu_s}}}; & \nu \ge \nu_m,
\end{array} 
\right . 
\label{fastcool}
\end{eqnarray}
similarly the slow-cooling ($\nu_m < \nu_c$) spectrum is given by,
\begin{eqnarray}
F_\nu &=& F_{\nu, \rm max} \nonumber \\
&\times & 
\left\{
\begin{array}{ll}
\left(\frac{\nu}{\nu_a}\right)^2 \left(\frac{\nu_a}{\nu_m}\right)^\frac{1}{3}; &  \nu < \nu_a \\
\left(\frac{\nu}{\nu_m}\right)^\frac{1}{3}; & \nu_a \le \nu < \nu_m \\
\left(\frac{\nu}{\nu_m}\right)^{-\frac{(p-1)}{2}}; &  \nu_m \le \nu < \nu_c \\ 
\left(\frac{\nu_c}{\nu_m}\right)^{-\frac{(p-1)}{2}} \left(\frac{\nu}{\nu_c}\right)^{-\frac{p}{2}} e^{-\frac{\nu}{\nu_{s}}}; & \nu \ge \nu_c.
\end{array} 
\right . 
\label{slowcool}
\end{eqnarray}
Here the break frequencies evolve with time and so do other blast wave parameters, as discussed below.  The initial forward-shock afterglow spectrum is in the fast-cooling regime as given by equation~(\ref{fastcool}) and it evolves into the slow-cooling spectrum in equation~(\ref{slowcool}) after a time $t_0$, such that $\nu_m (t_0) = \nu_c (t_0)$.

Depending on whether the blast wave is evolving in a ISM or wind medium, the evolution of the spectra are different.  The kinetic energy of the blast wave and several microphysical parameters also affect synchrotron flux.  For our calculation of the break frequencies and the maximum flux in equations~(\ref{fastcool}) and (\ref{slowcool}) we use $E_{55}$ as the nominal initial kinetic energy of blast wave in units of $10^{55}$ erg, $\epsilon_{e, 0.1} = \epsilon_e/0.1$ as the fraction of shock energy going to the relativistic electrons and $\epsilon_{b,0.1} = \epsilon_b/0.1$ as fraction of the shock energy going to the magnetic fields.  We use a reference luminosity distance $d_{l,28}$ in $10^{28}$ cm unit and $t_d$ as the time after prompt emission in days. The parameter $\phi_1$ = $\phi/10$ is the number of gyro-radius needed for the electron acceleration in the magnetic field.  

%%%%%%%%%%%%%%%%%%
\subsection{Flux parameters in ISM}
%%%%%%%%%%%%%%%%%%
\label{parameters-ISM}
We provide below equations for all the synchrotron frequencies in an ISM environment following Ref.~\cite{Razzaque:2013dsa}, that are used to fit flux data, as

\begin{eqnarray}
\nu_a &=& 5.5\times10^{9}\, \epsilon_{b,0.1}^{1/5}\epsilon_{e,0.1}^{-1}E_{55}^{1/5}(1+z)^{-1}n_0^{3/5} \,    \rm {Hz}
\\
\nu_m &=& 1.7\times10^{14}\, \epsilon_{b,0.1}^{1/2}\epsilon_{e,0.1}^2E_{55}^{1/2}(1+z)^{1/2}t_d^{-3/2} \,  \rm  {Hz}
\\
\nu_c &=& 2.0\times10^{13}\, \epsilon_{b,0.1}^{-3/2}E_{55}^{-1/2}(1+z)^{-1/2}t_d^{-1/2}n_0^{-1}  \,  \rm {Hz} ~~~~
\\
\nu_s  &=& 5.4\times10^{22}\, n_0^{-1/8}E_{55}^{1/8}(1+z)^{-5/8}t_d^{-3/8}\phi_1^{-1} \,     \rm {Hz}.
\end{eqnarray}
Here $n_0 = 1$~cm$^{-3}$ is the particle number density in the ISM.  The maximum flux in equations~(\ref{fastcool}) and (\ref{slowcool}) is given by
\begin{eqnarray}
F_{\nu, \rm max} &=& 8\times10^{-23} (1+z)^{-1}n_0^{1/2}d_{l,28}^{-2}E_{55} \epsilon_b^{1/2} \nonumber \\
&& ~{\rm erg~cm}^{-2}~{\rm s}^{-1}~{\rm Hz}^{-1} ,
\end{eqnarray}
and the transition time from the fast- to slow cooling spectrum is
\begin{equation}
t_0 = 8.5~ \epsilon_{b,0.1}^{2}\epsilon_{e,0.1}^2E_{55}(1+z)n_{0} \, \rm{days} .
\end{equation}

%%%%%%%%%%%%%%%%%%%
\subsection{Flux parameters in wind}
%%%%%%%%%%%%%%%%%%%
\label{parameters-wind}

Similar to ISM, equations for all the synchrotron frequencies in a wind environment can be written as~\cite{Razzaque:2013dsa}, 
\begin{eqnarray}
\nu_a &=& 8.3\times10^{9}\, \epsilon_{b,0.1}^{1/5}\epsilon_{e,0.1}^{-1}E_{55}^{-2/5}(1+z)^{-2/5}t_d^{-3/5}A_*^{6/5}\, \rm{Hz} ~
\\
\nu_m &=& 9.5\times10^{13}\, \epsilon_{b,0.1}^{1/2}\epsilon_{e,0.1}^2E_{55}^{1/2}(1+z)^{1/2}t_d^{-3/2}\, \rm{Hz}
\\
\nu_c &=& 2.1\times10^{15}\, \epsilon_{b,0.1}^{-3/2}E_{55}^{1/2}(1+z)^{-3/2}t_d^{1/2}A_*^{-2} \, \rm{Hz}
\\
\nu_s  &=& 8.2\times10^{22}\, A_*^{-1/4}E_{55}^{1/4}(1+z)^{-3/4}t_d^{-1/4}\phi_1^{-1}\, \rm{Hz}.
\end{eqnarray}
Here $A_* \equiv \dot{M}_{-5}/v_8$ corresponds to a mass-loss rate of $\dot{M}_w = 10^{-5}\dot{M}_{-5}M_{\odot}$~yr$^{-1}$ in the wind of the progenitor star, with velocity $v_w = 10^8v_8$ cm/s.  The maximum flux to be used in equations~(\ref{fastcool}) and (\ref{slowcool}) for the wind case is
\begin{eqnarray}
F_{\nu, \rm max} &=& 3.53\times10^{-24}\, \epsilon_{b,0.1}^{1/2}E_{55}^{1/2}(1+z)^{-1/2}t_d^{-1/2}d_{l,28}^{-2} A_{*}\nonumber \\
&& ~{\rm erg~cm}^{-2}~{\rm s}^{-1}~{\rm Hz}^{-1} ,
\end{eqnarray}
and the transition time from the fast- to slow-cooling spectrum is
\begin{equation}
t_0 = 0.2~ \epsilon_{b,0.1}\epsilon_{e,0.1}(1+z) A_* \, \rm{days}.
\end{equation}

%%%%%%%%%%%%%%%%%%%%%%%%%%%%%%%%%%%%
\subsection{Results from synchrotron model fit to SED and light curves}
\begin{figure*}[th!]
\includegraphics[trim =  0 21 0 10 , width= 0.85\columnwidth]{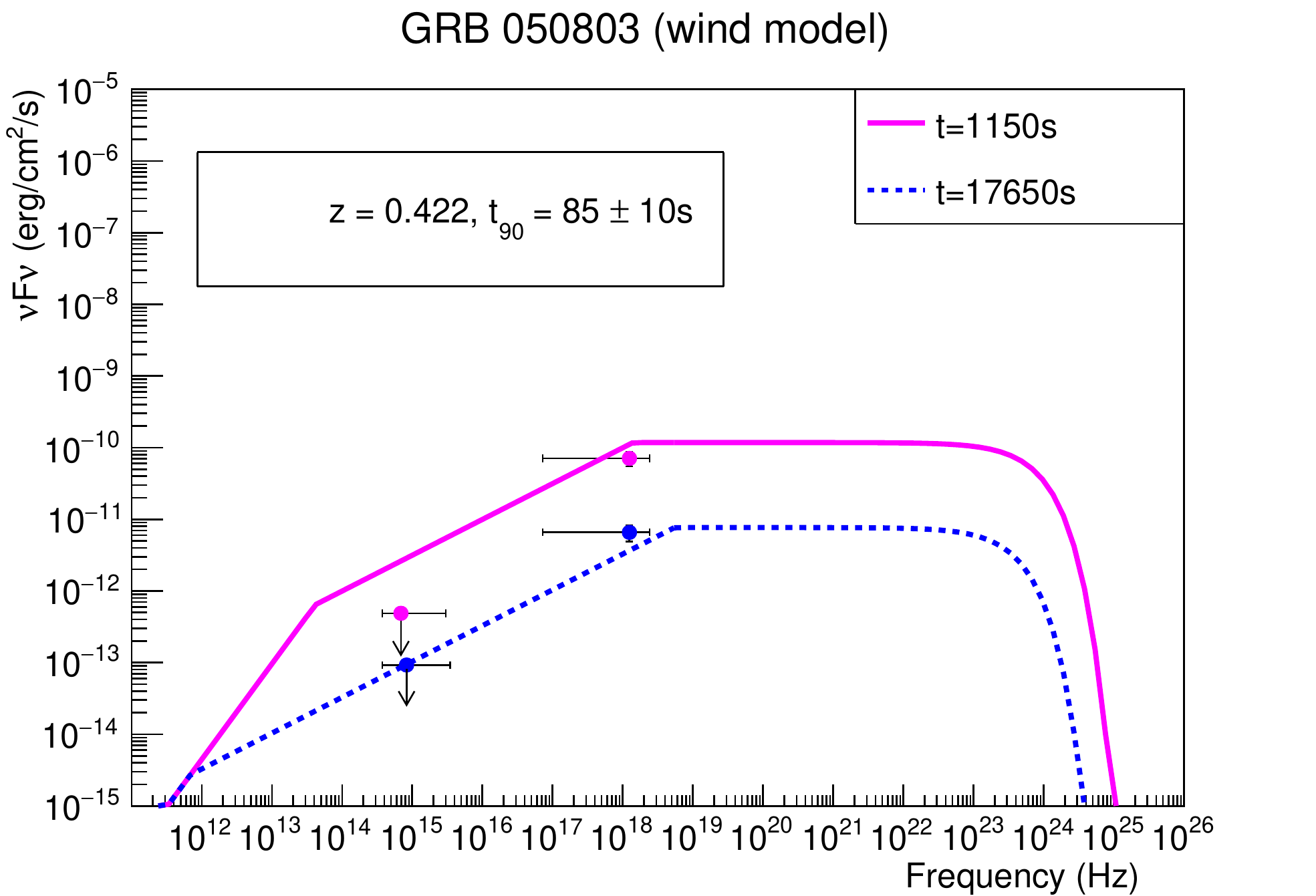}
\vspace{0.5 cm} 
\includegraphics[trim =  0 21 0 10 , width= 0.85\columnwidth]{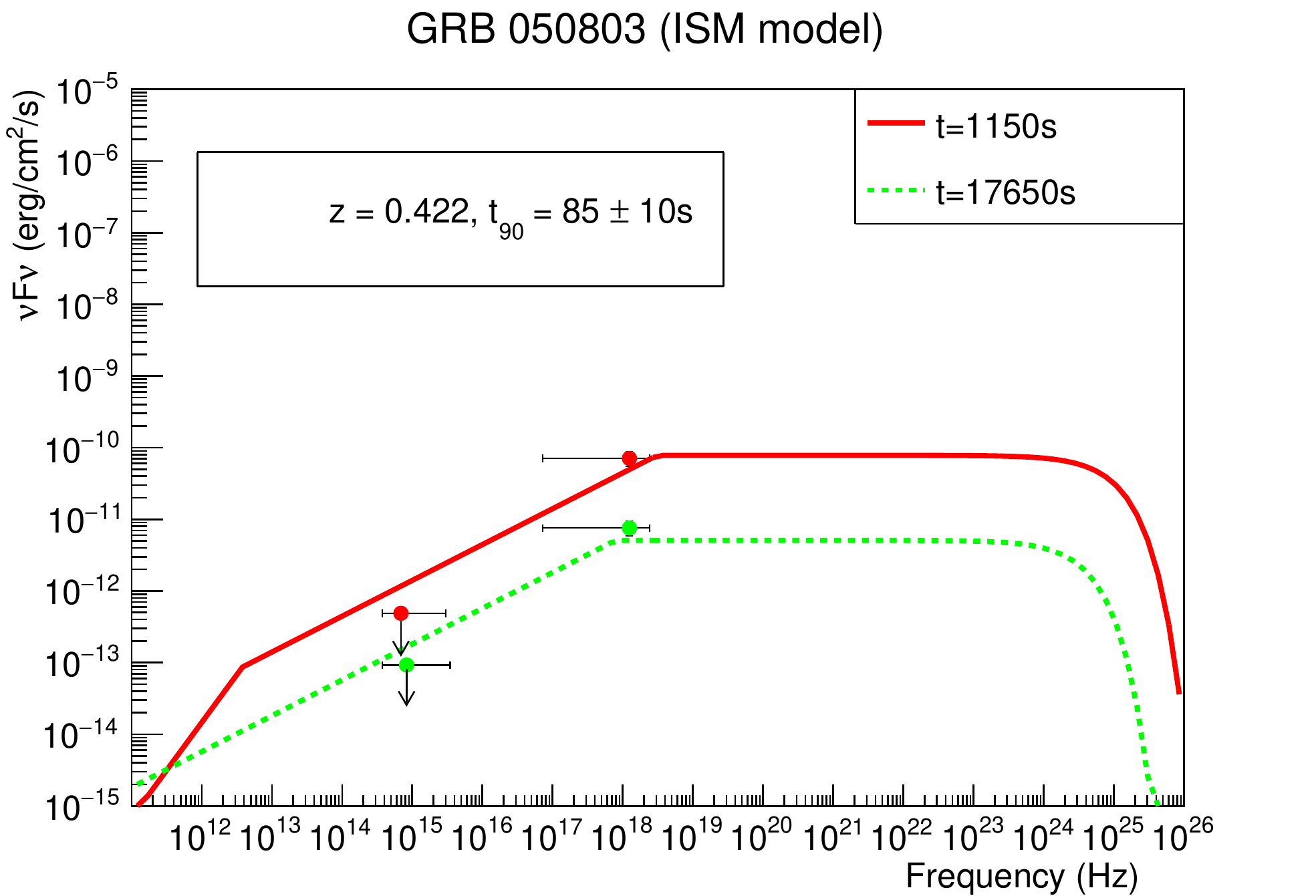}
\includegraphics[trim =  0 21 0 10, width= 0.85\columnwidth]{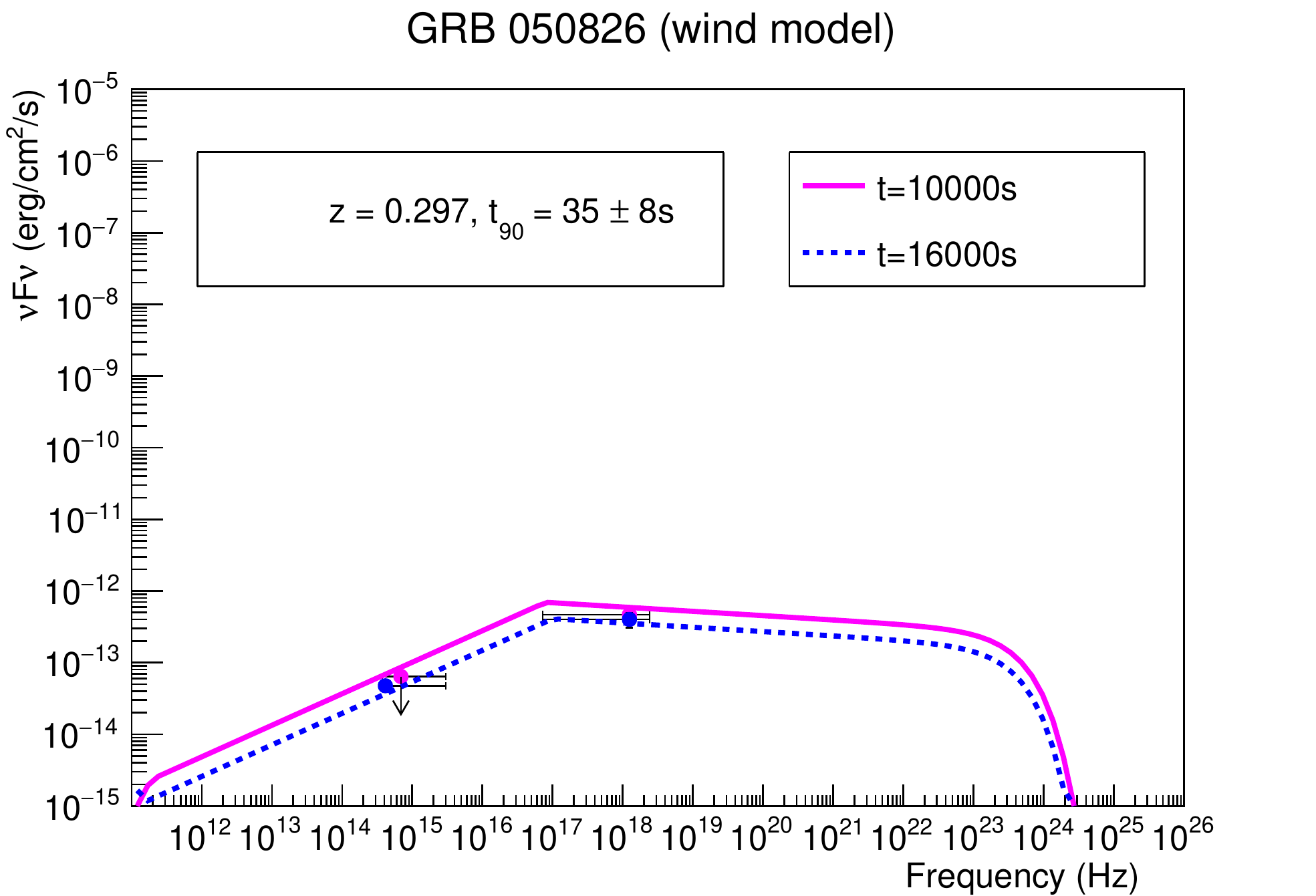}
\vspace{0.5 cm}  
\includegraphics[trim =  0 21 0 10, width= 0.85\columnwidth]{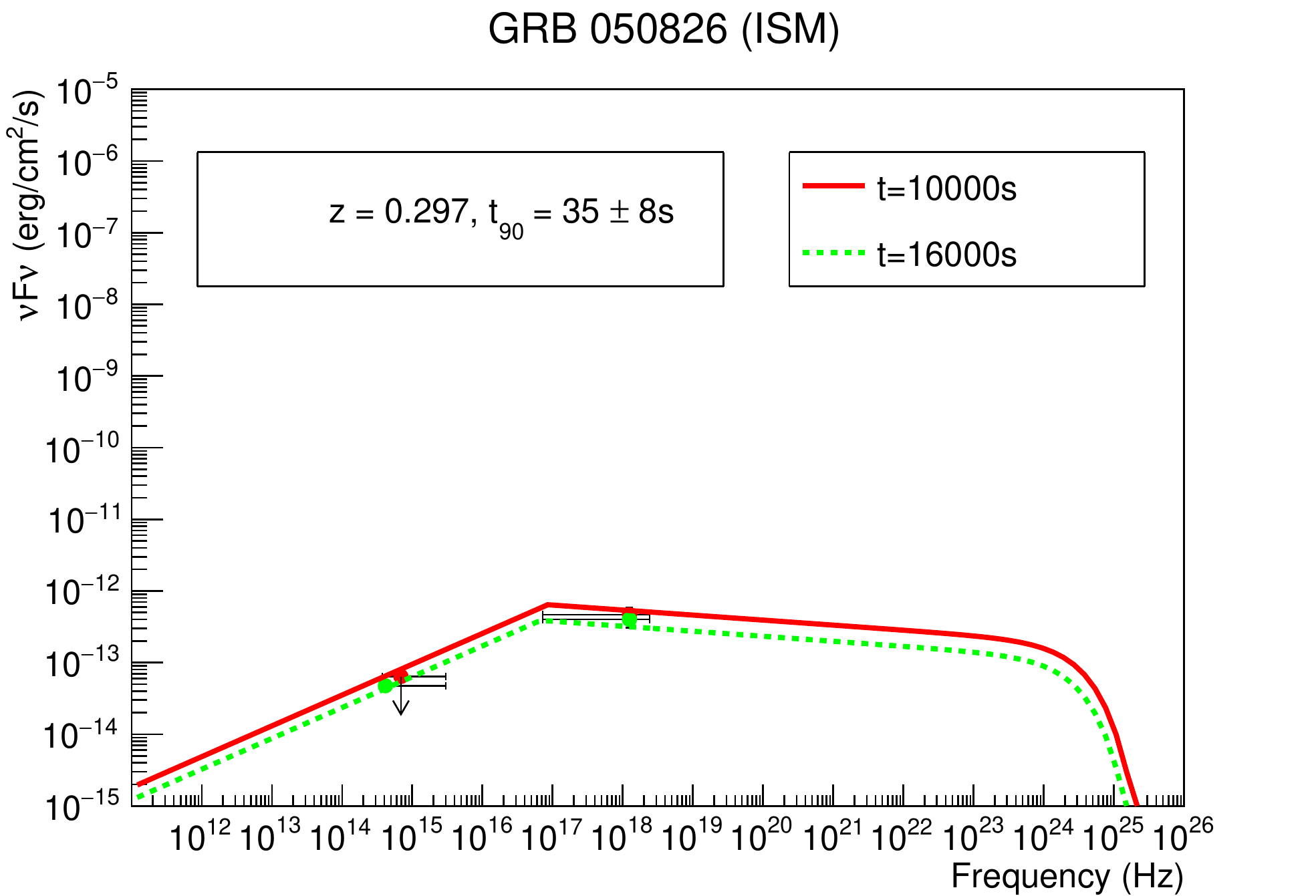}
\includegraphics[trim =  0 21 0 10, width= 0.85\columnwidth]{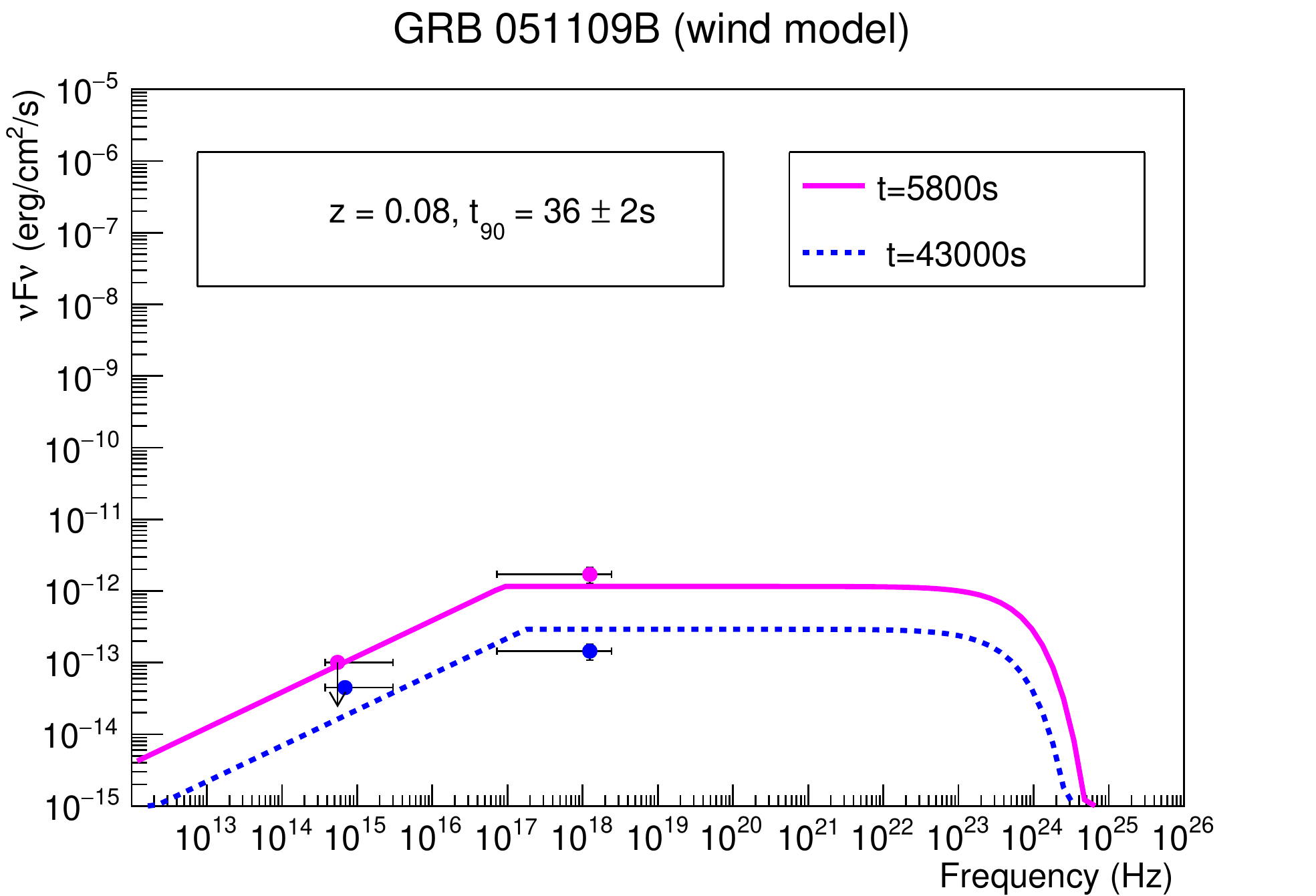}
\vspace{0.5 cm} 
\includegraphics[trim =  0 21 0 10, width= 0.85\columnwidth]{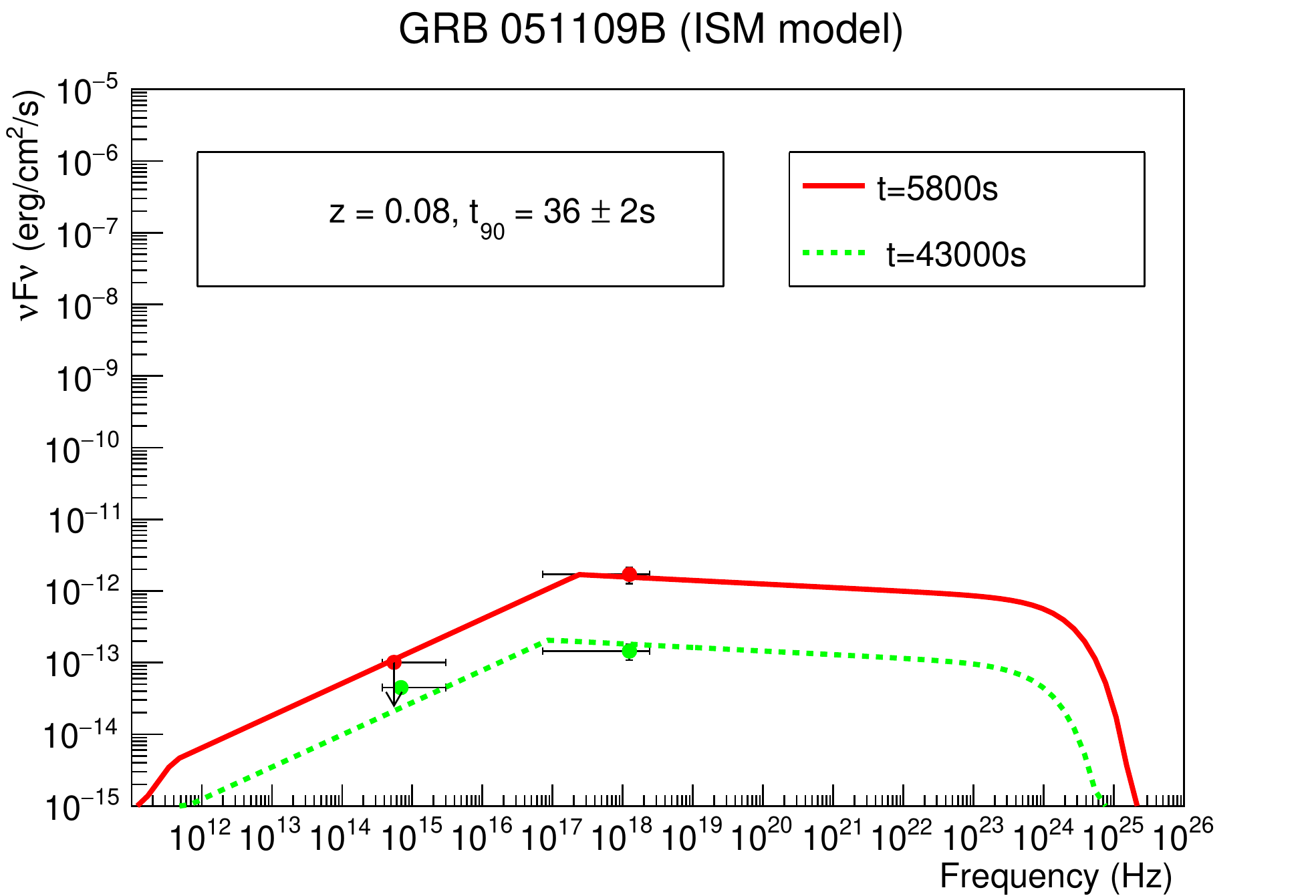}
\includegraphics[trim =  0 21 0 10 , width=0.85\columnwidth]{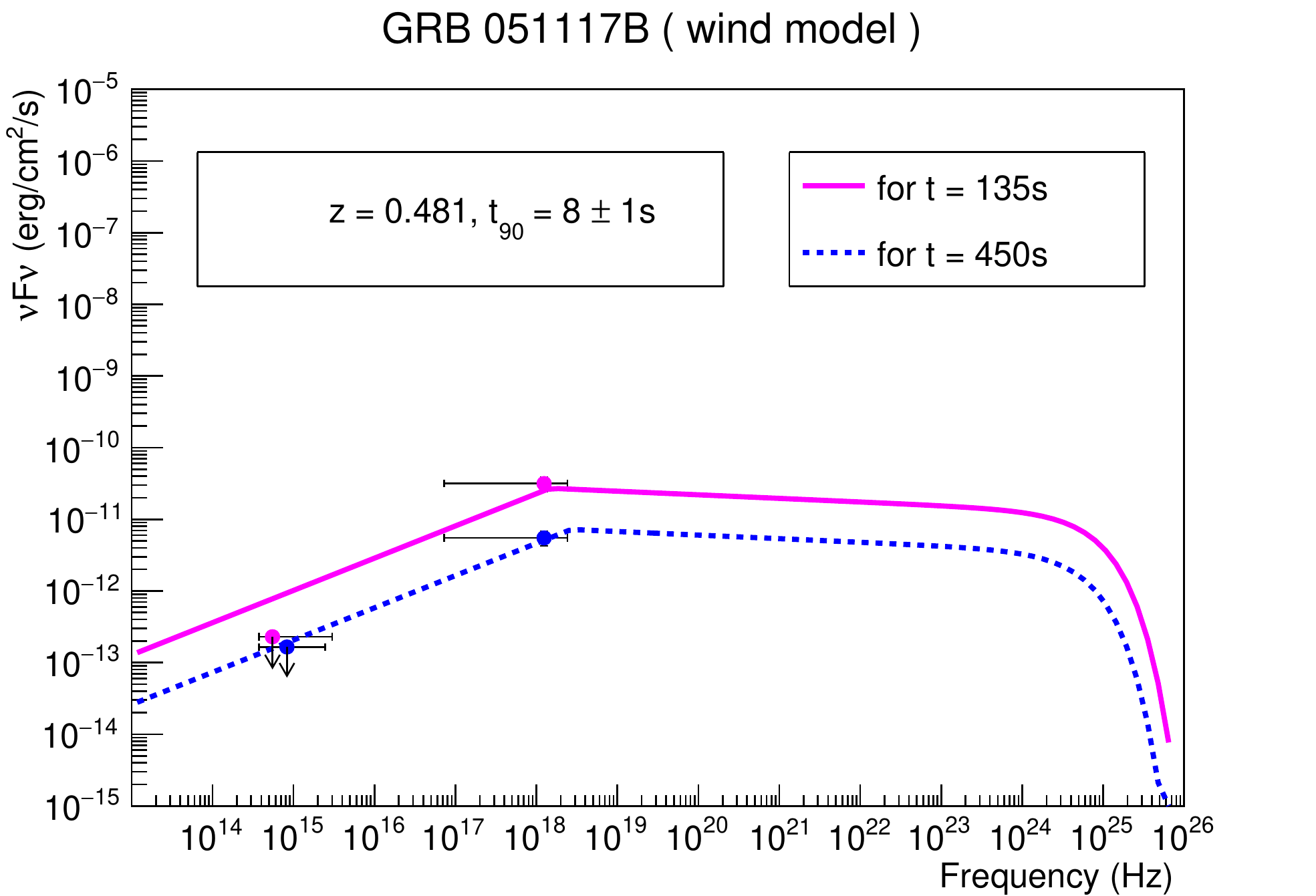}
\vspace{0.5 cm} 
\includegraphics[trim =  0 21 0 10 , width=0.85\columnwidth]{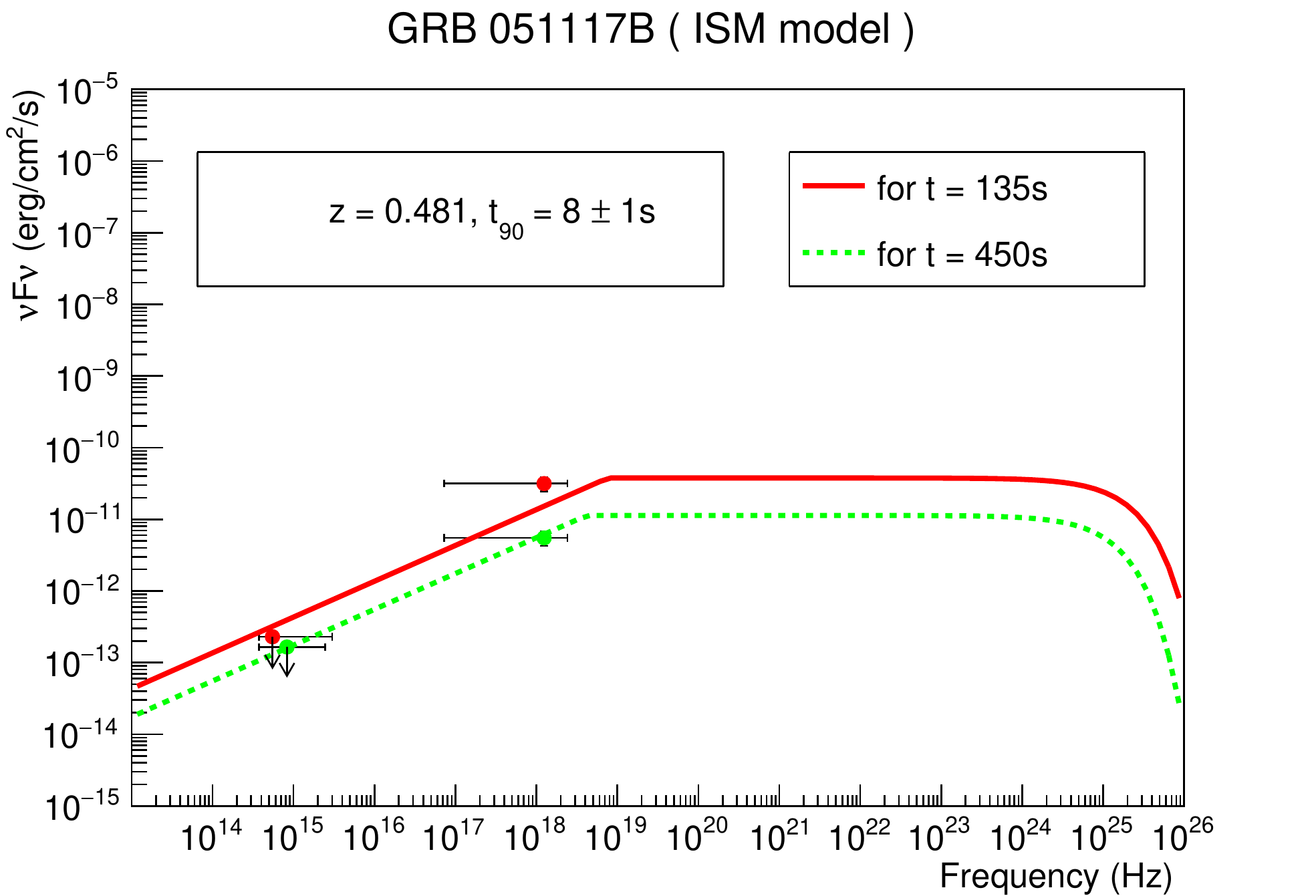}
 \caption{\label{fig1sed} SEDs for GRB~050803, GRB~050826, GRB~051109B and GRB~051117B, and synchrotron afterglow model fits.  Wind models are on the left panels and the ISM models are on the right.  The lines in different colours are the synchrotron model output plotted to fit the data at different time intervals.  The dots around $\nu = 10^{18}$ Hz represent the XRT data points and the dots which fall between  $10^{14}$ Hz to $10^{16}$ Hz represent the UVOT/optical data points.}
\end{figure*}

\begin{figure*}[th!]
\includegraphics[trim =  0 21 0 10 , width=0.85\columnwidth]{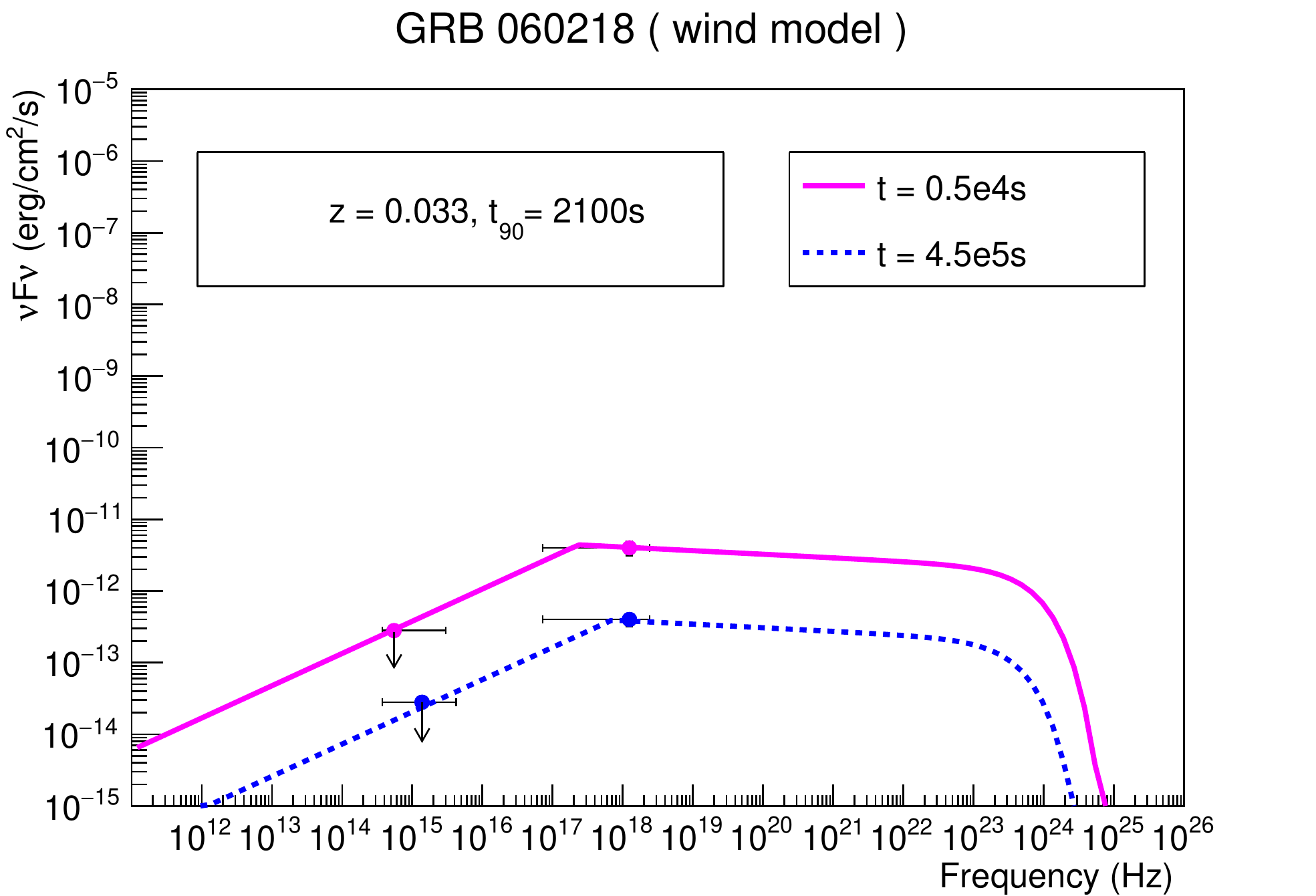}
\vspace{0.5 cm} 
\includegraphics[trim =  0 21 0 10 , width=0.85\columnwidth]{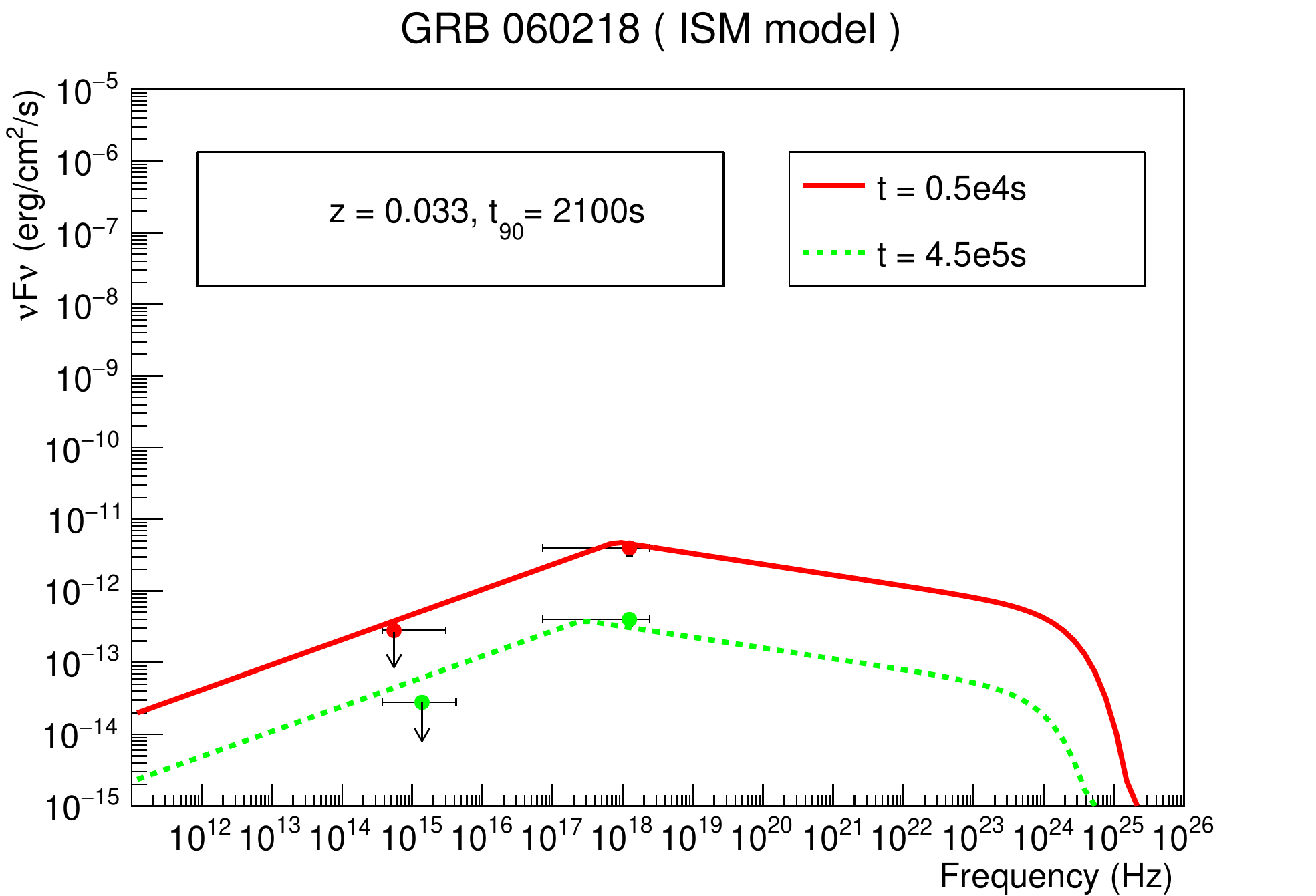}
\includegraphics[trim =  0 21 0 10, width=0.85\columnwidth]{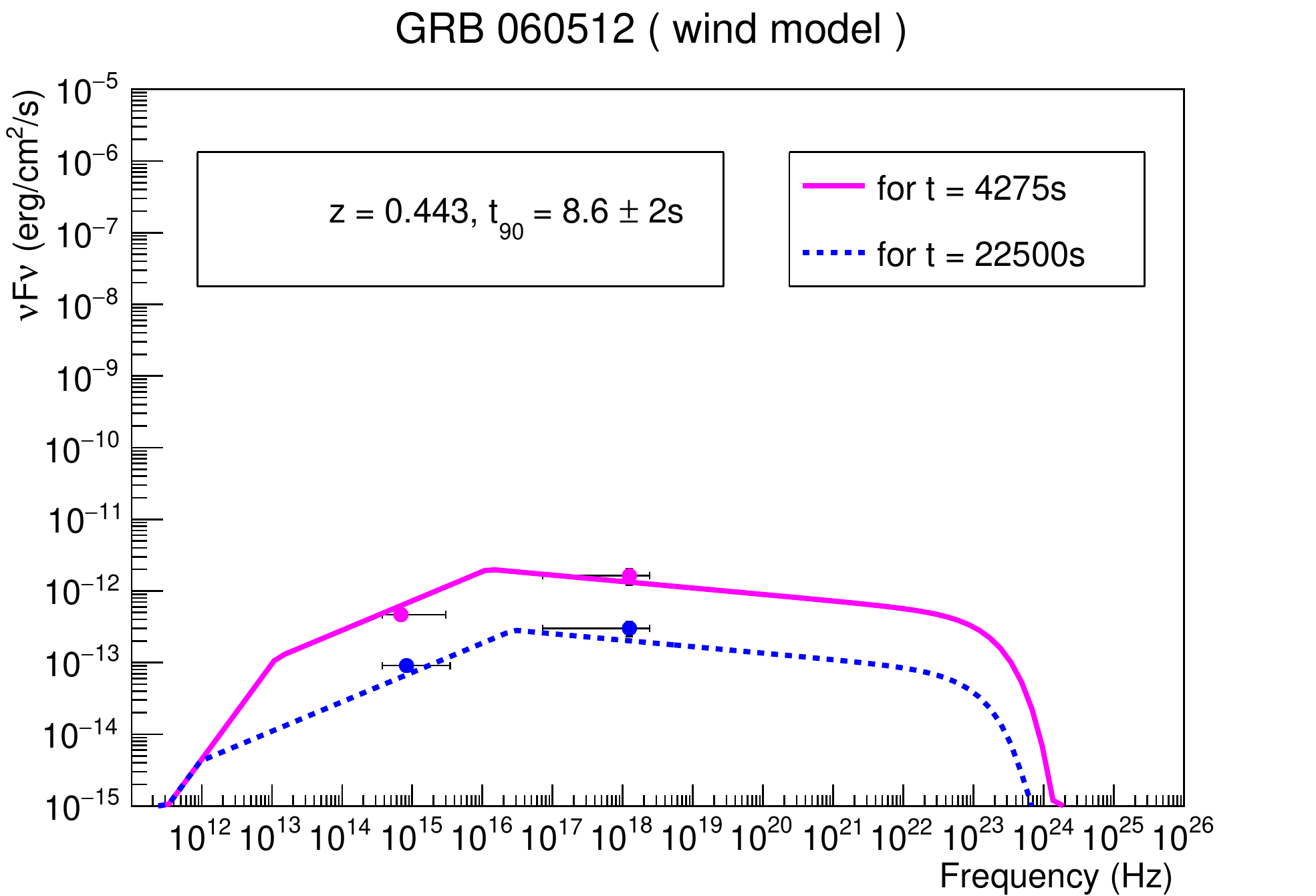}
\vspace{0.5 cm} 
\includegraphics[trim =  0 21 0 10, width=0.85\columnwidth]{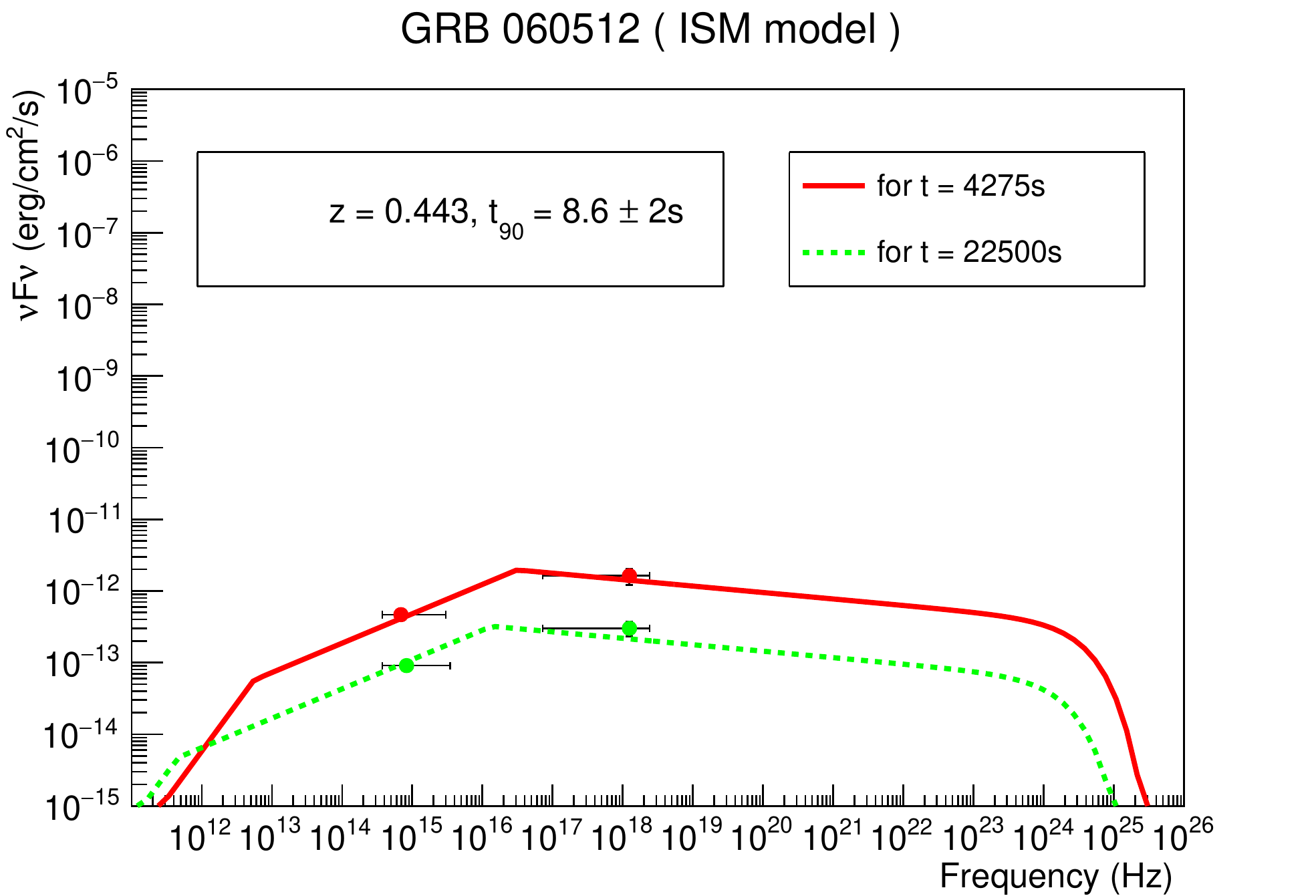}
\includegraphics[trim =  0 21 0 10, width=0.85\columnwidth]{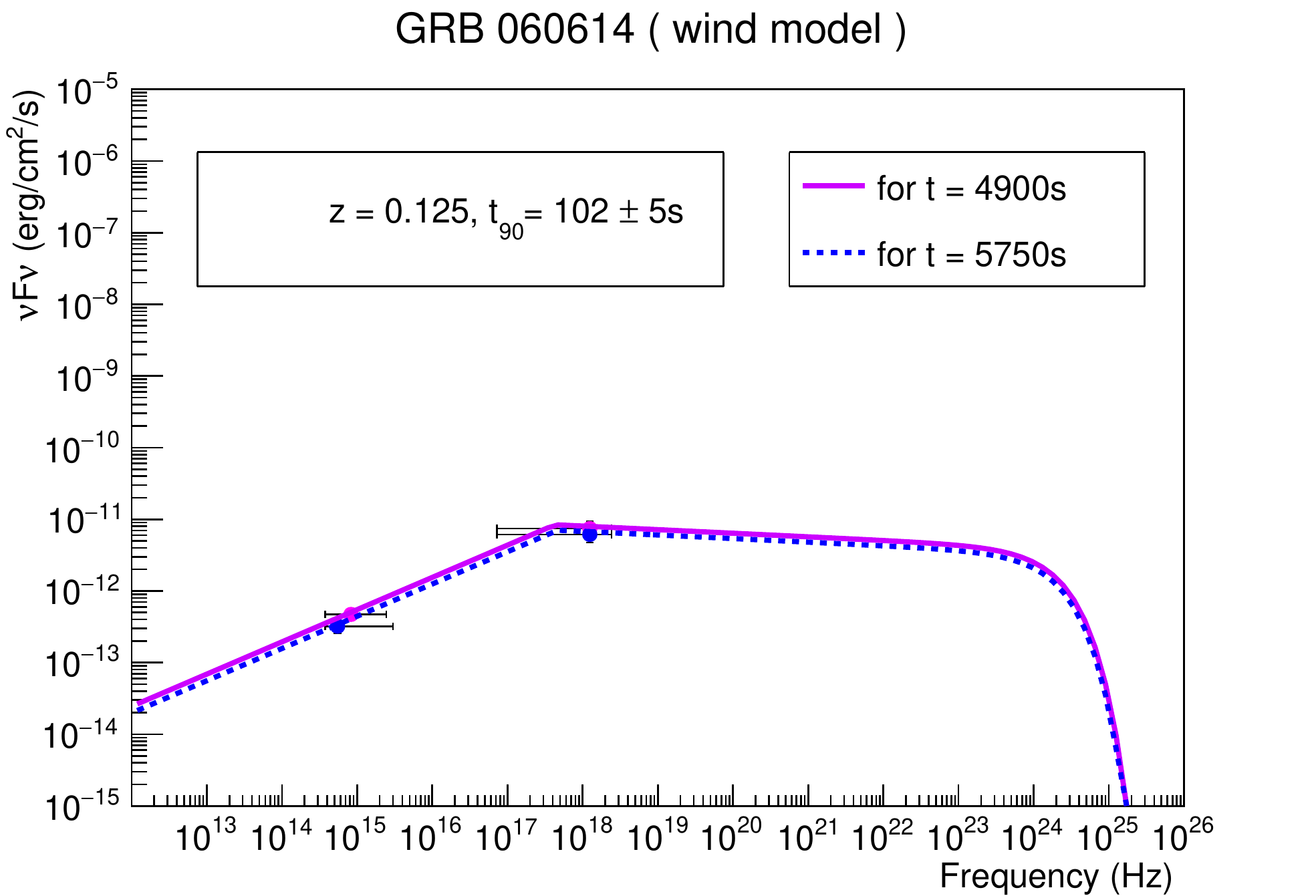}
\vspace{0.5 cm} 
\includegraphics[trim =  0 21 0 10, width=0.85\columnwidth]{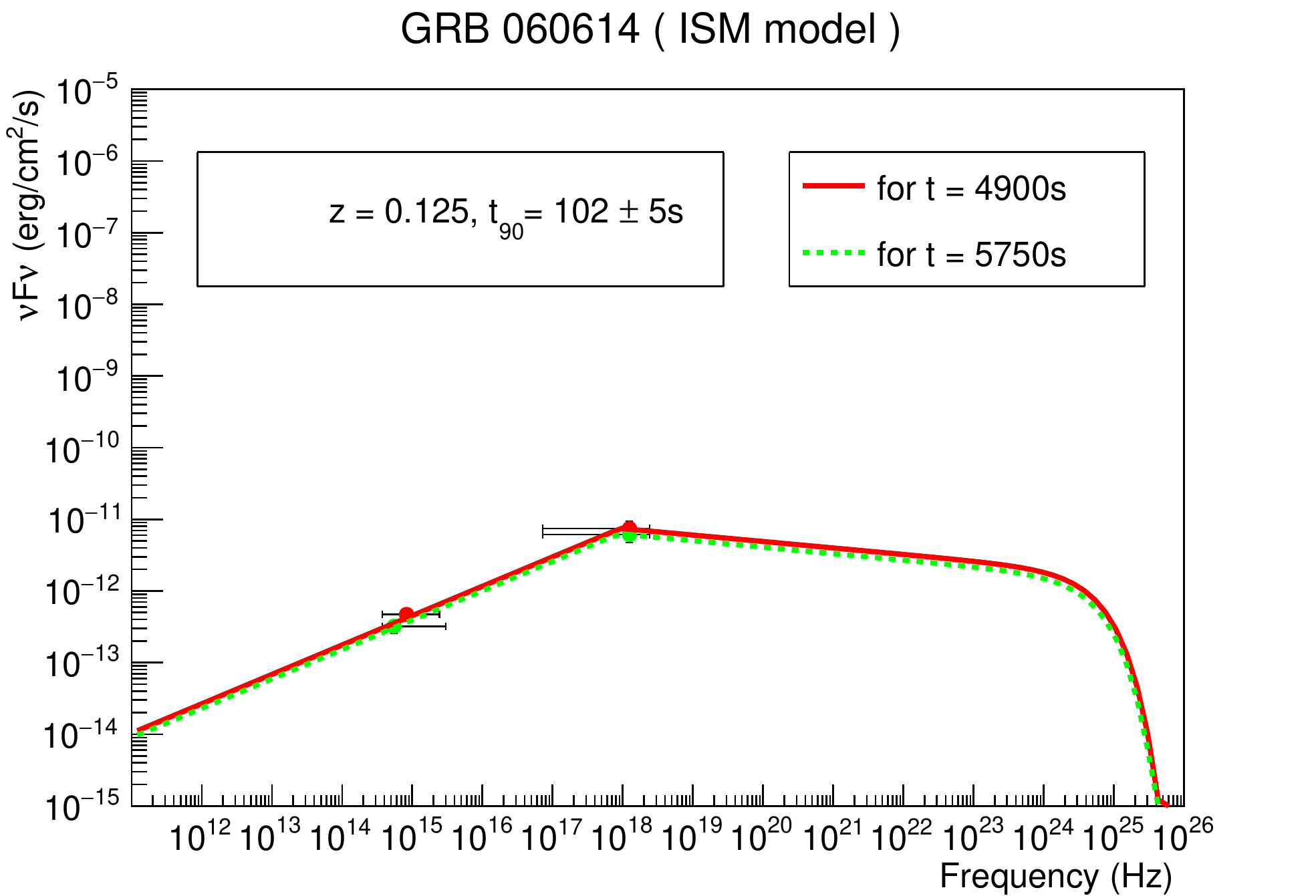}
\includegraphics[trim =  0 21 0 10, width=0.85\columnwidth]{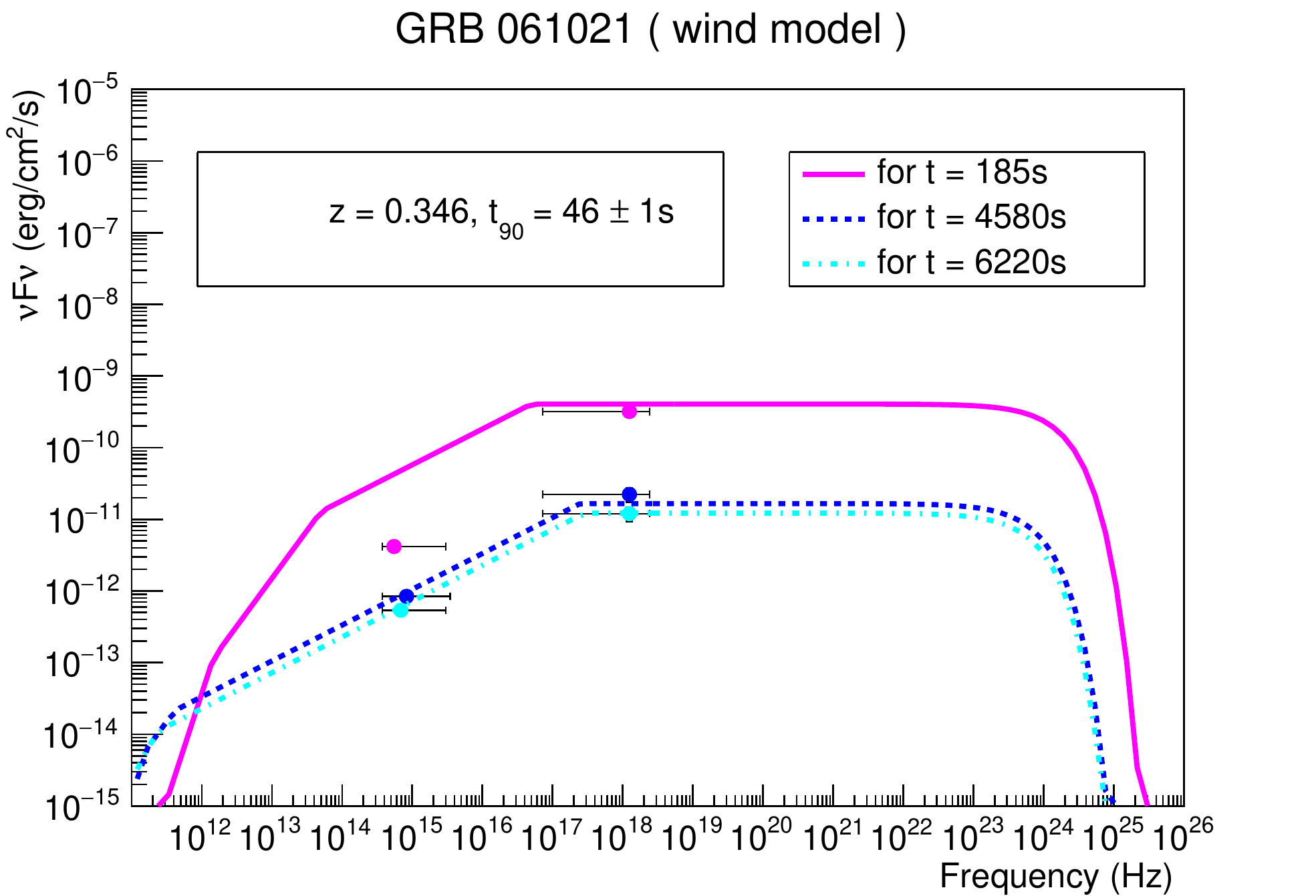}
\vspace{0.5 cm} 
\includegraphics[trim =  0 21 0 10, width=0.85\columnwidth]{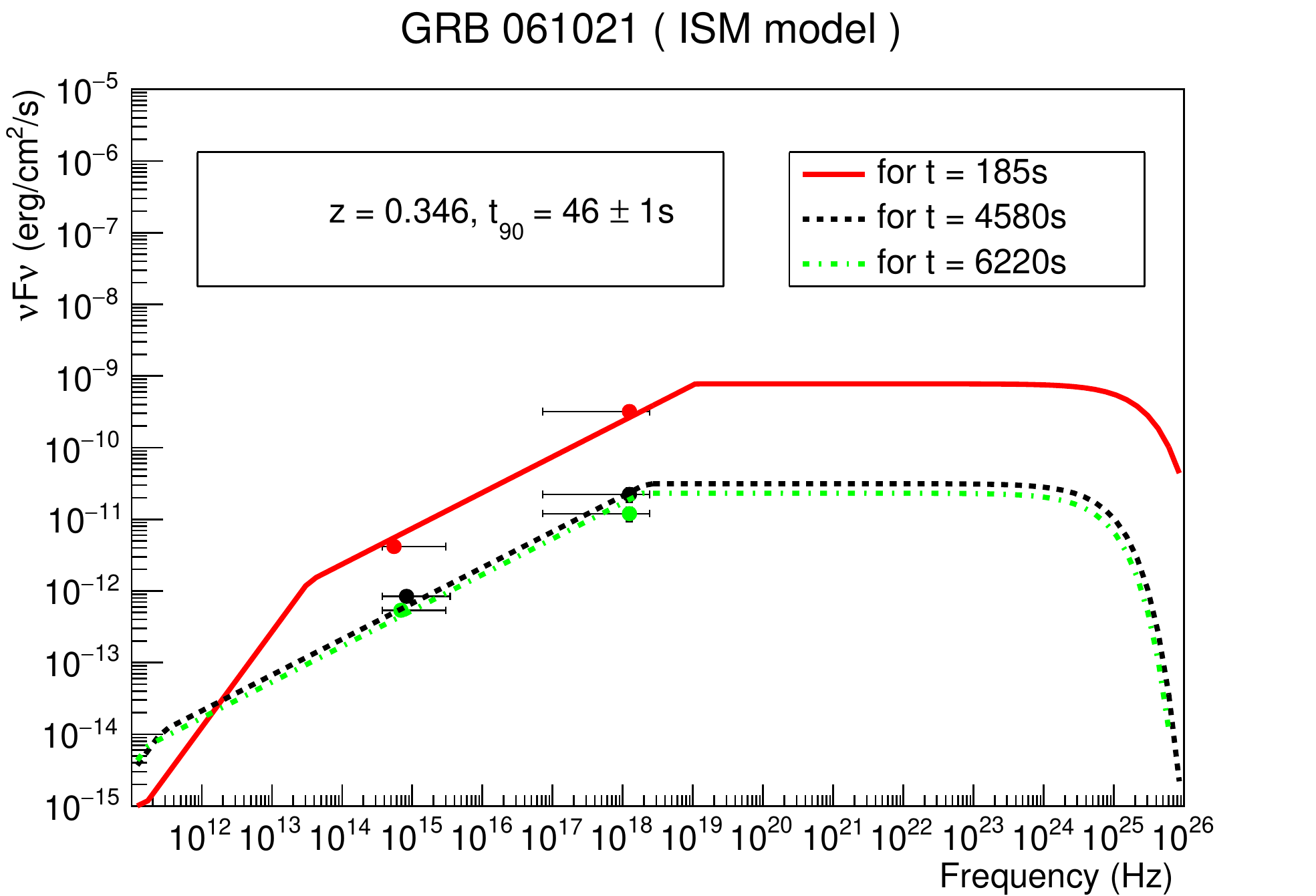}
\caption{\label{fig2sed} Same as Fig.~\ref{fig1sed} but for GRB~060218, GRB~060512, GRB~060614 and GRB~061021.} 
\end{figure*}

\begin{figure*}[th!]
\includegraphics[trim =  0 21 0 10, width=0.85\columnwidth]{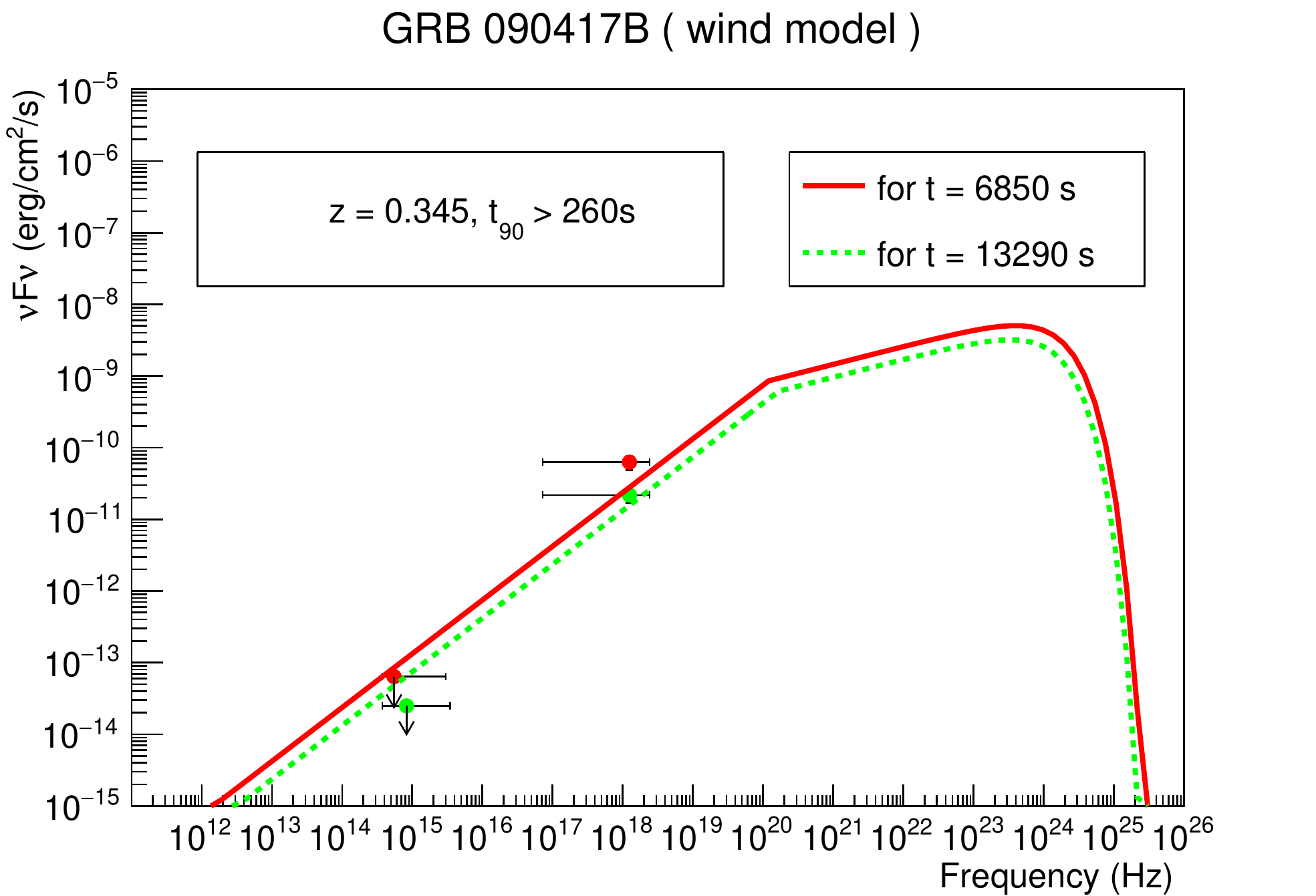}
\vspace{0.5 cm} 
\includegraphics[trim =  0 21 0 10, width=0.85\columnwidth]{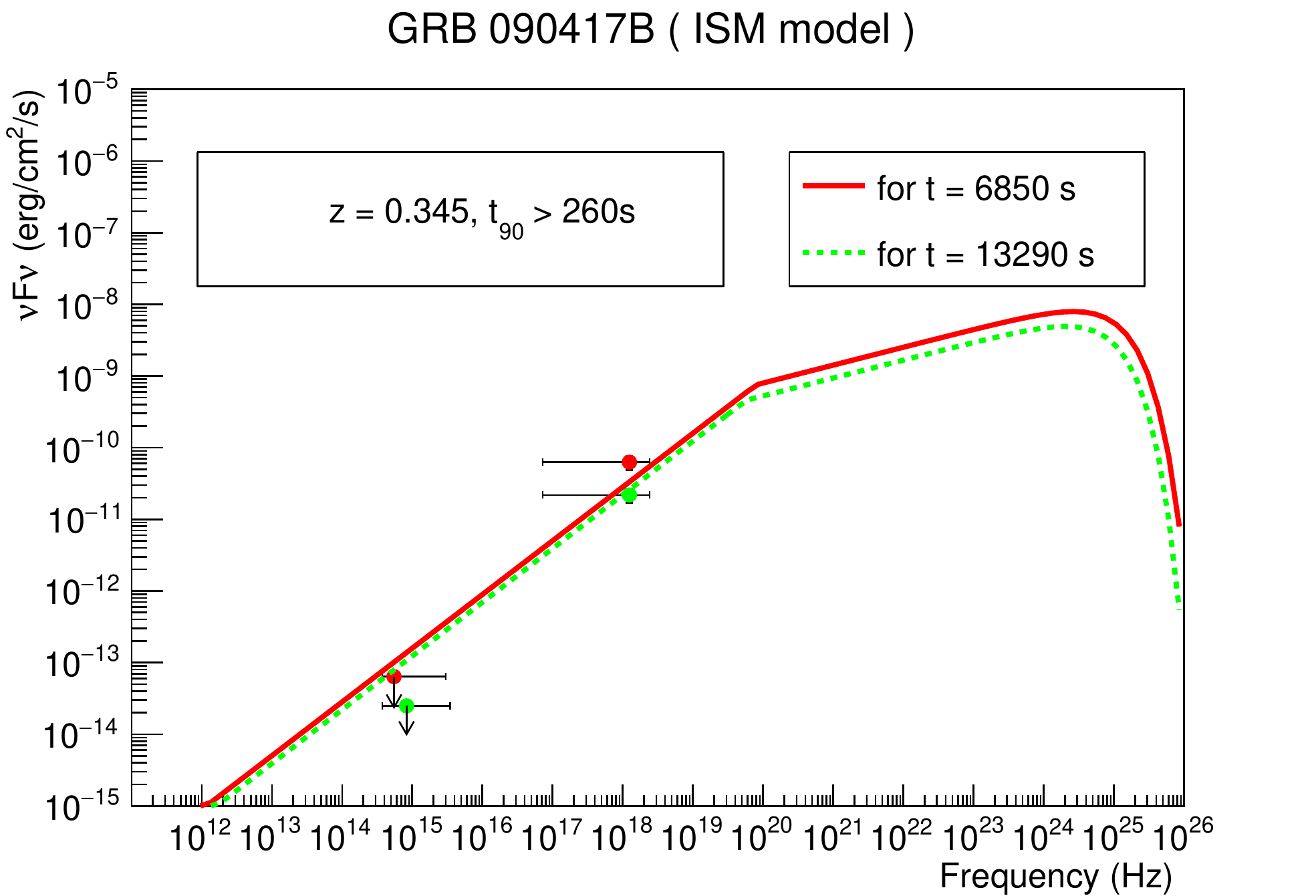}
\includegraphics[trim =  0 21 0 10, width=0.85\columnwidth]{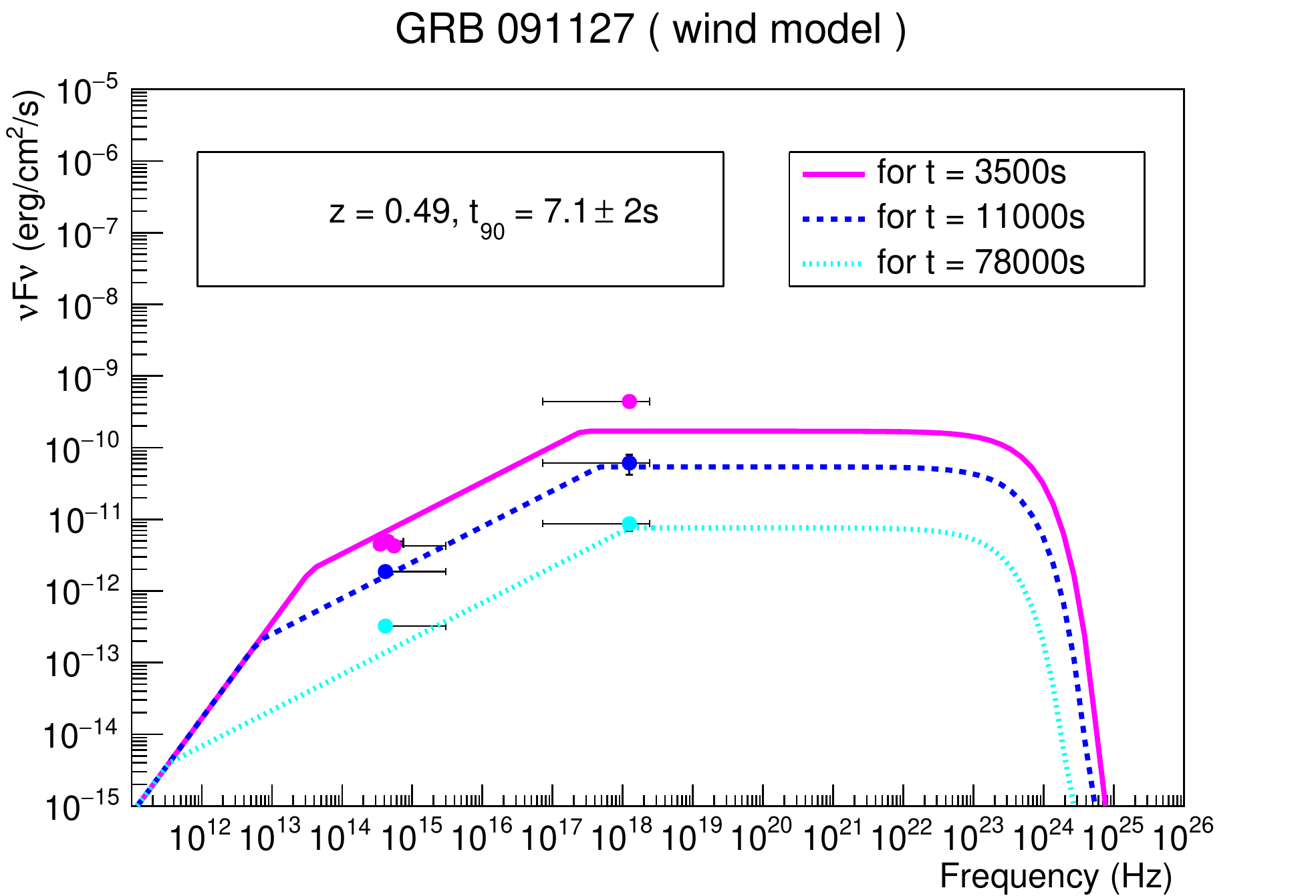}
\vspace{0.5 cm} 
\includegraphics[trim =  0 21 0 10, width=0.85\columnwidth]{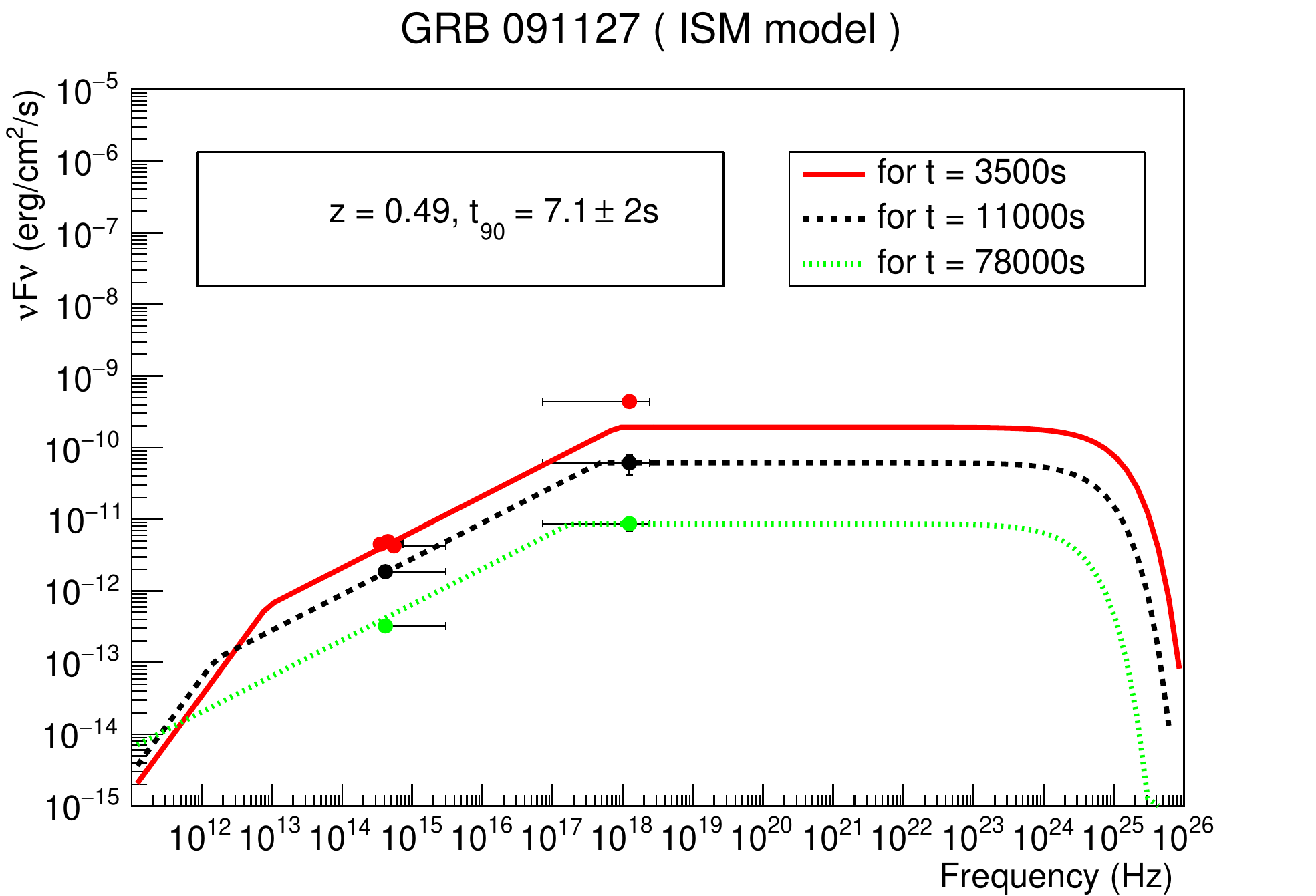}
\includegraphics[trim =  0 21 0 10, width=0.85\columnwidth]{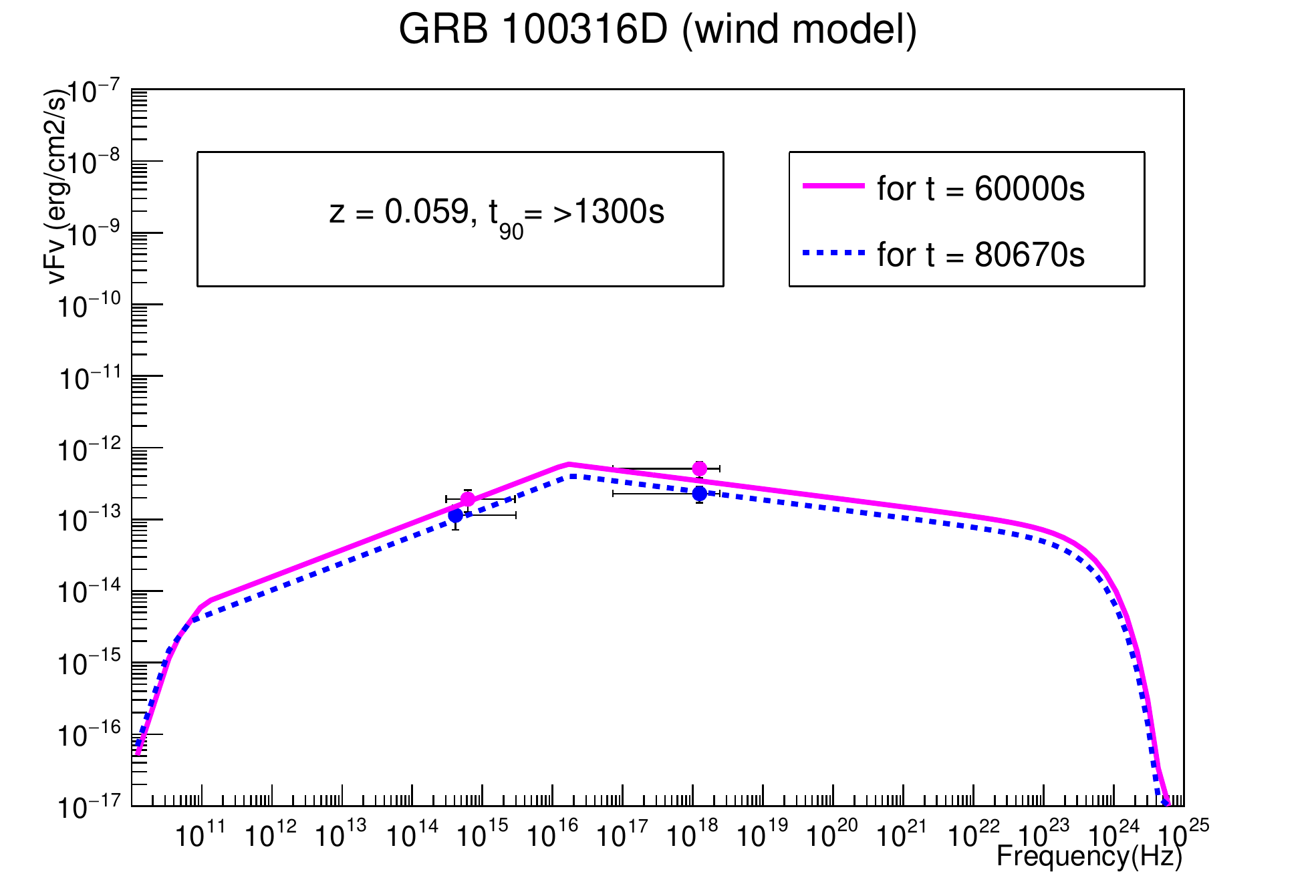}
\vspace{0.5 cm} 
\includegraphics[trim =  0 21 0 10, width=0.85\columnwidth]{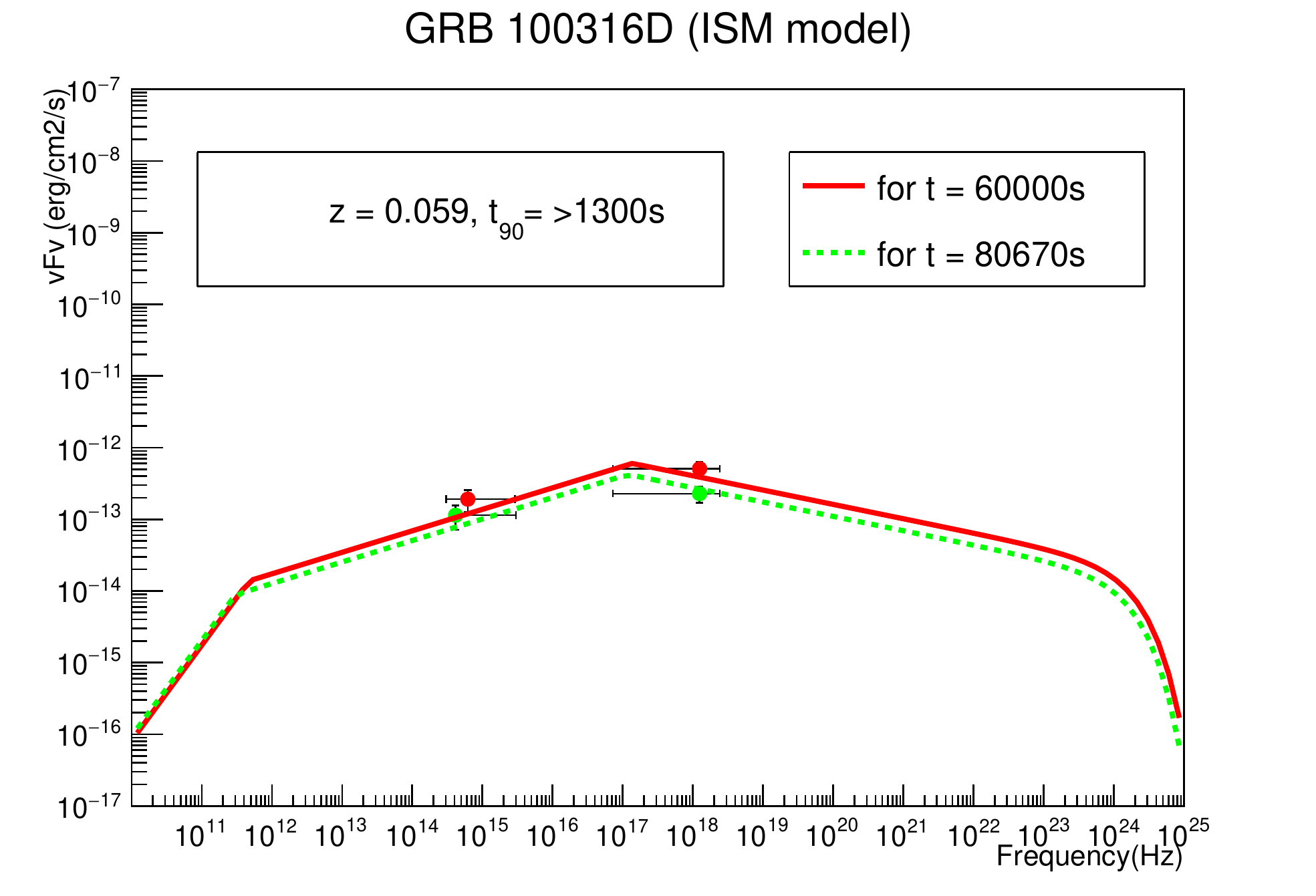}
\includegraphics[trim =  0 21 0 10, width=0.85\columnwidth]{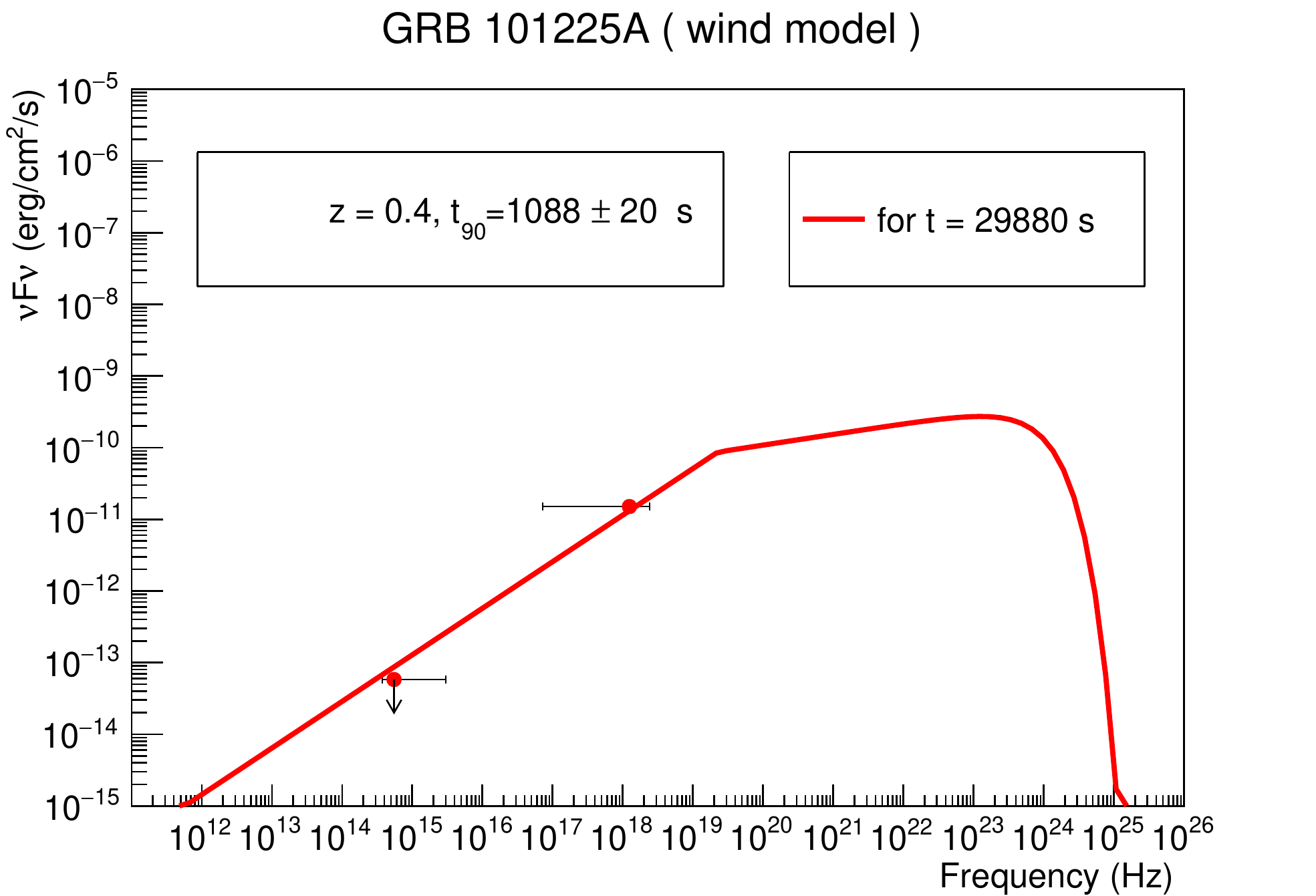}
\vspace{0.5 cm} 
\includegraphics[trim =  0 21 0 10, width=0.85\columnwidth]{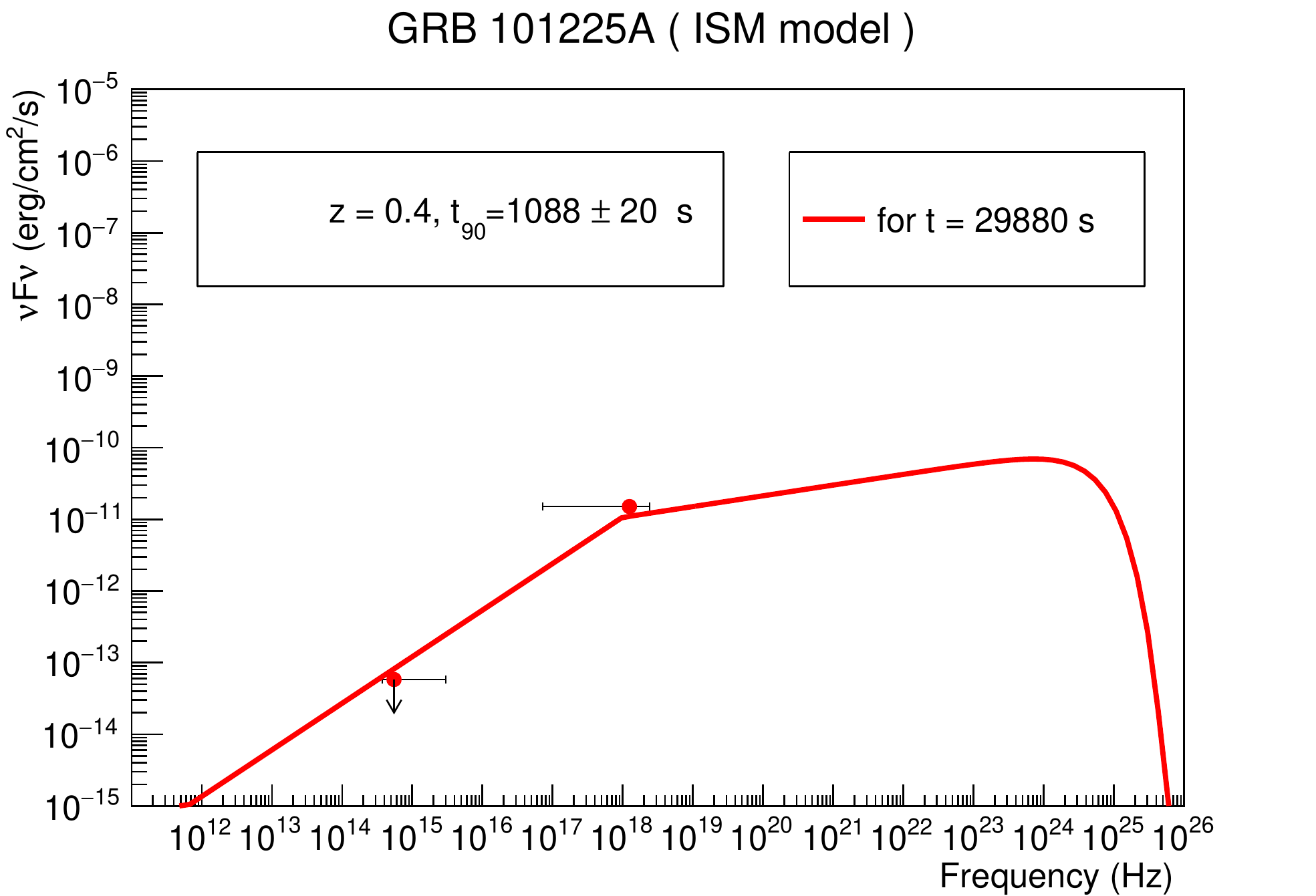}
\caption{\label{fig3sed} Same as Fig.~\ref{fig1sed} but for GRB~090417B, GRB~091127B, GRB~100316D and GRB~101225A.}
\end{figure*}

\begin{figure*}[th!]
\includegraphics[trim =  0 21 0 10, width=0.85\columnwidth]{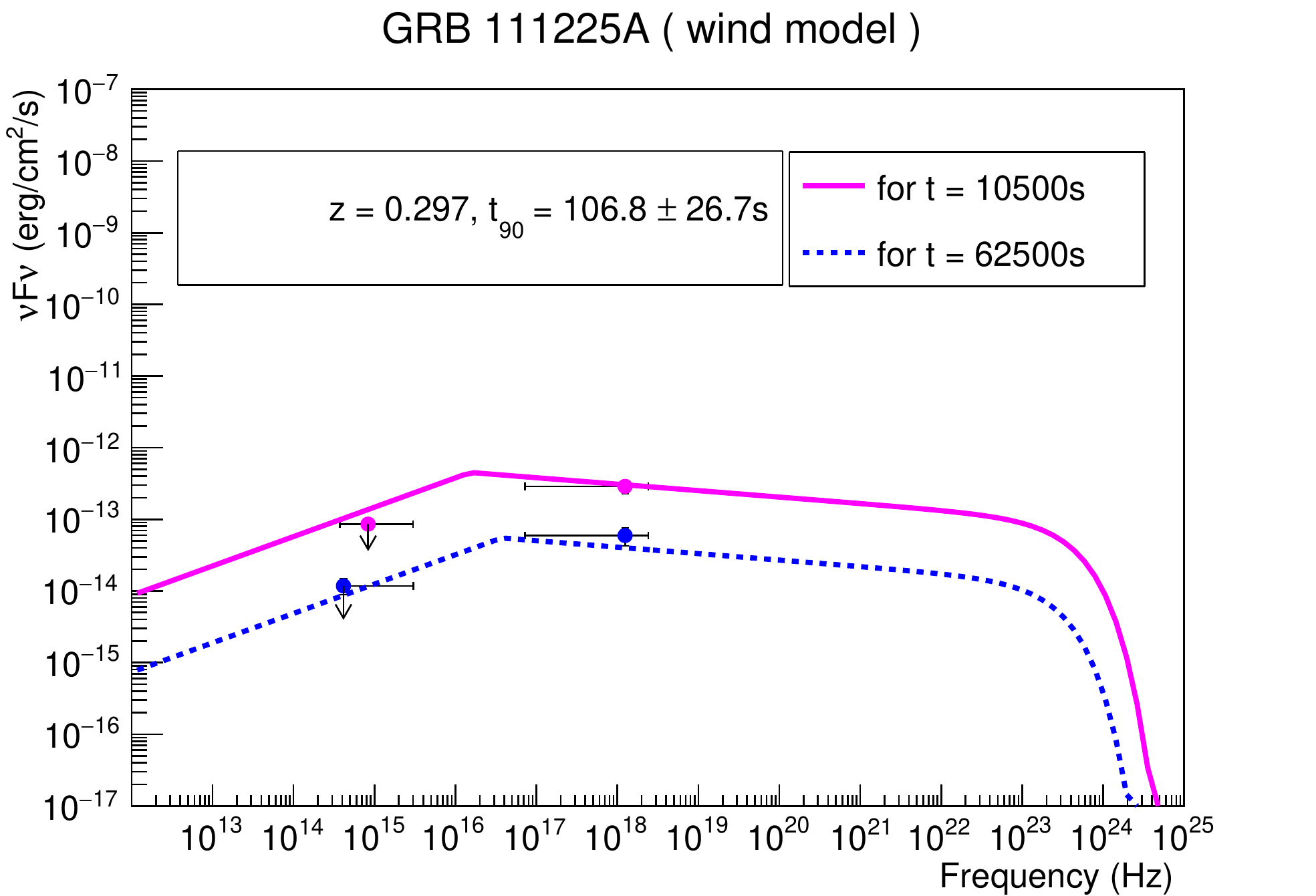}
\vspace{0.5 cm} 
\includegraphics[trim =  0 21 0 10, width=0.85\columnwidth]{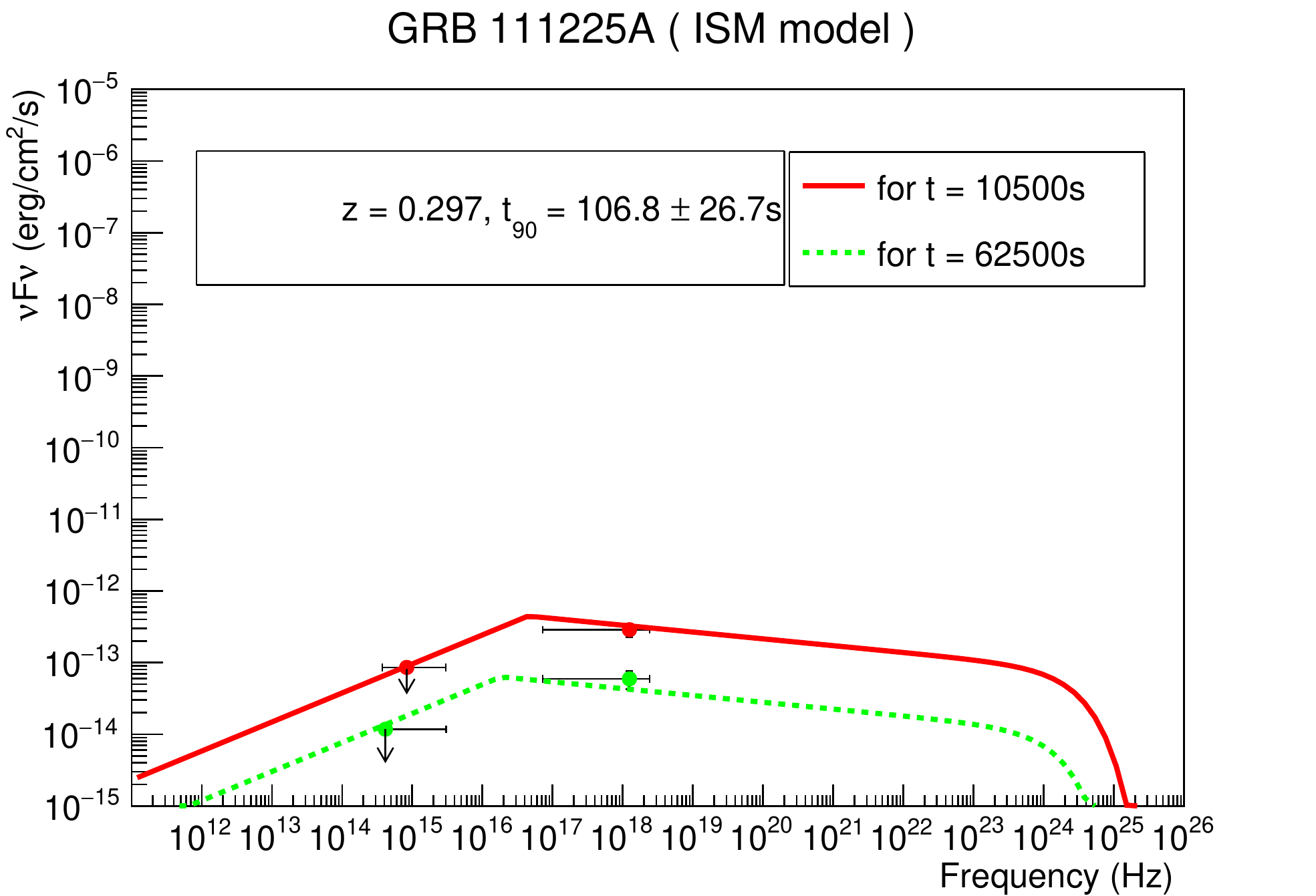}
\includegraphics[trim =  0 21 0 10, width=0.85\columnwidth]{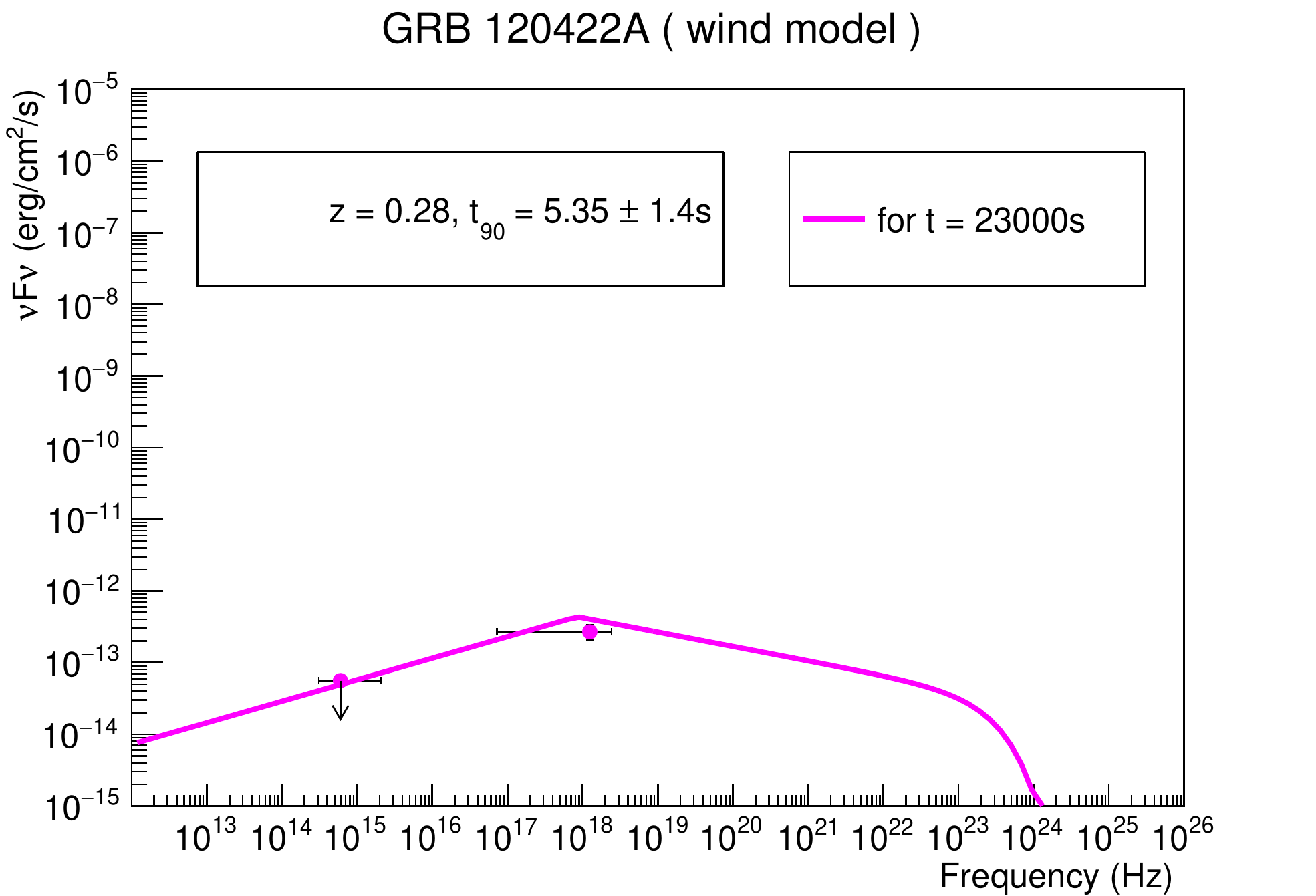}
\vspace{0.5 cm} 
\includegraphics[trim =  0 21 0 10, width=0.85\columnwidth]{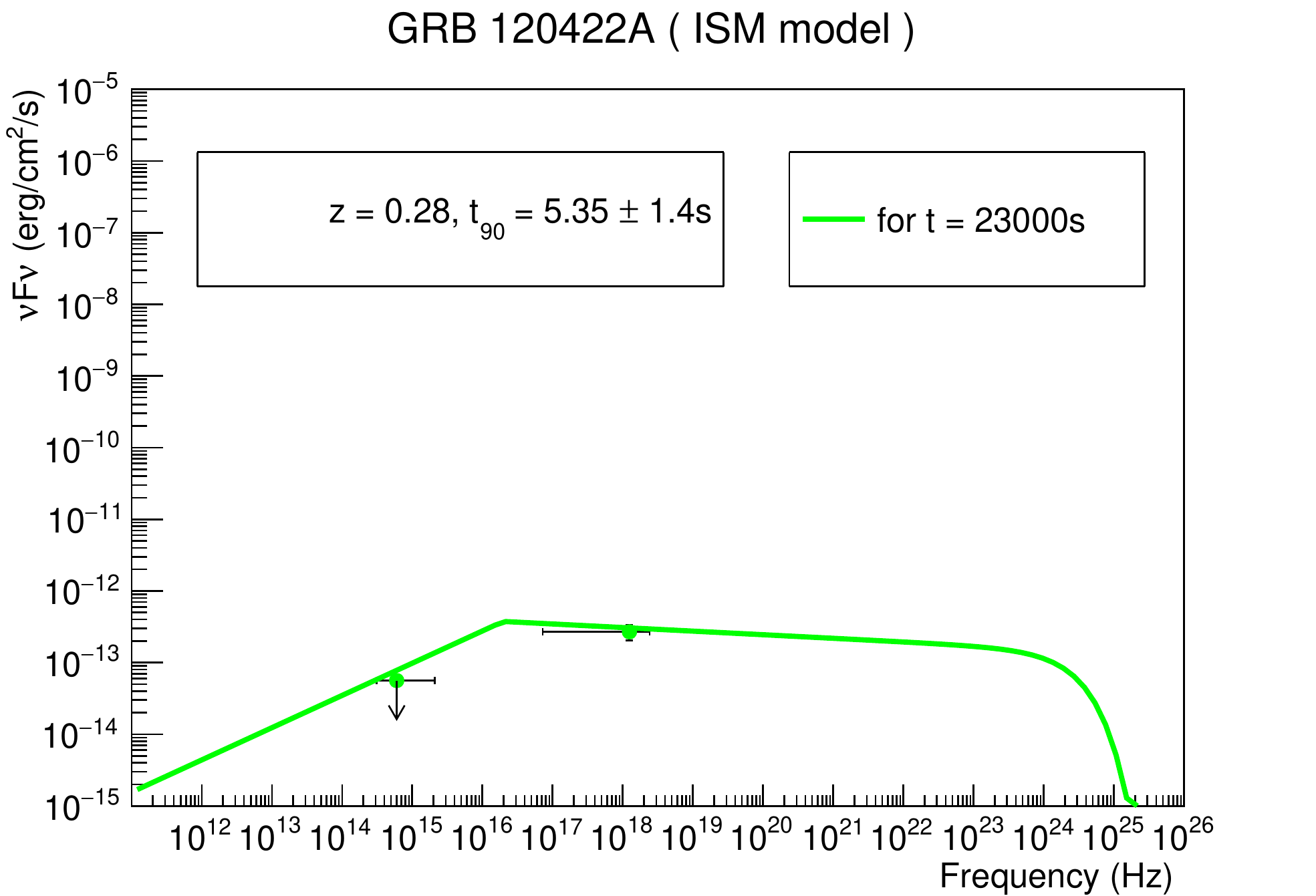}
\includegraphics[trim =  0 21 0 10, width=0.85\columnwidth]{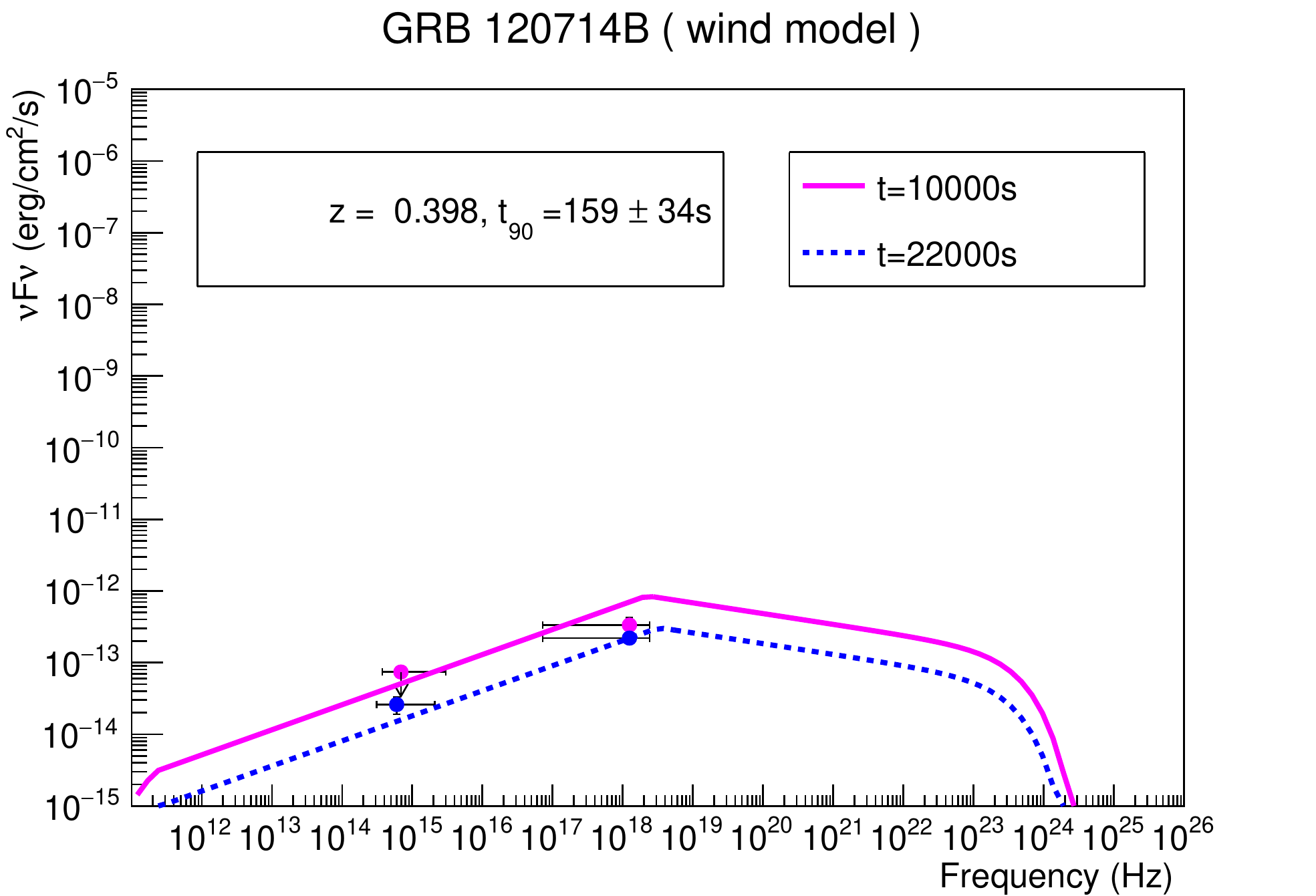}
\vspace{0.5 cm} 
\includegraphics[trim =  0 21 0 10, width=0.85\columnwidth]{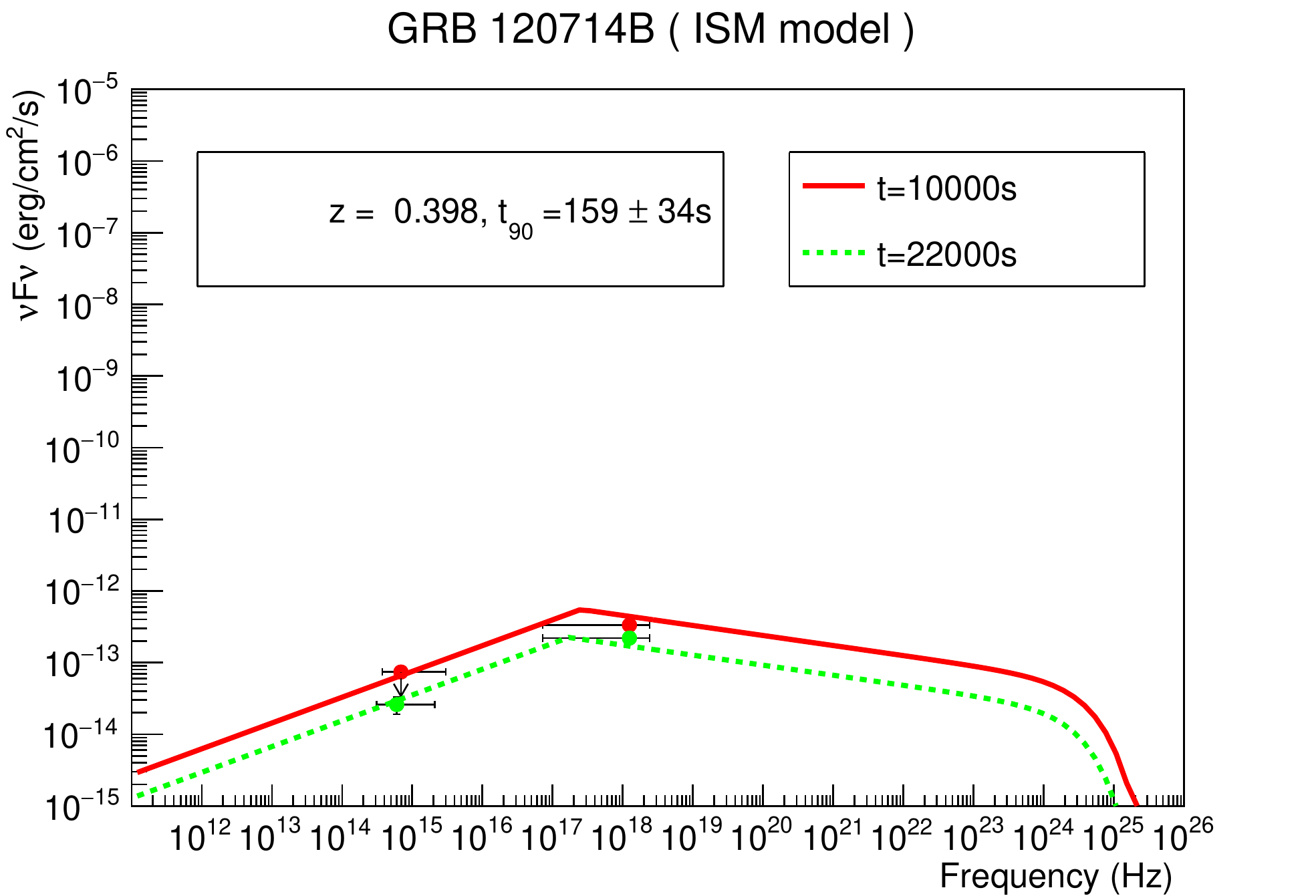}
\includegraphics[trim =  0 21 0 10, width=0.85\columnwidth]{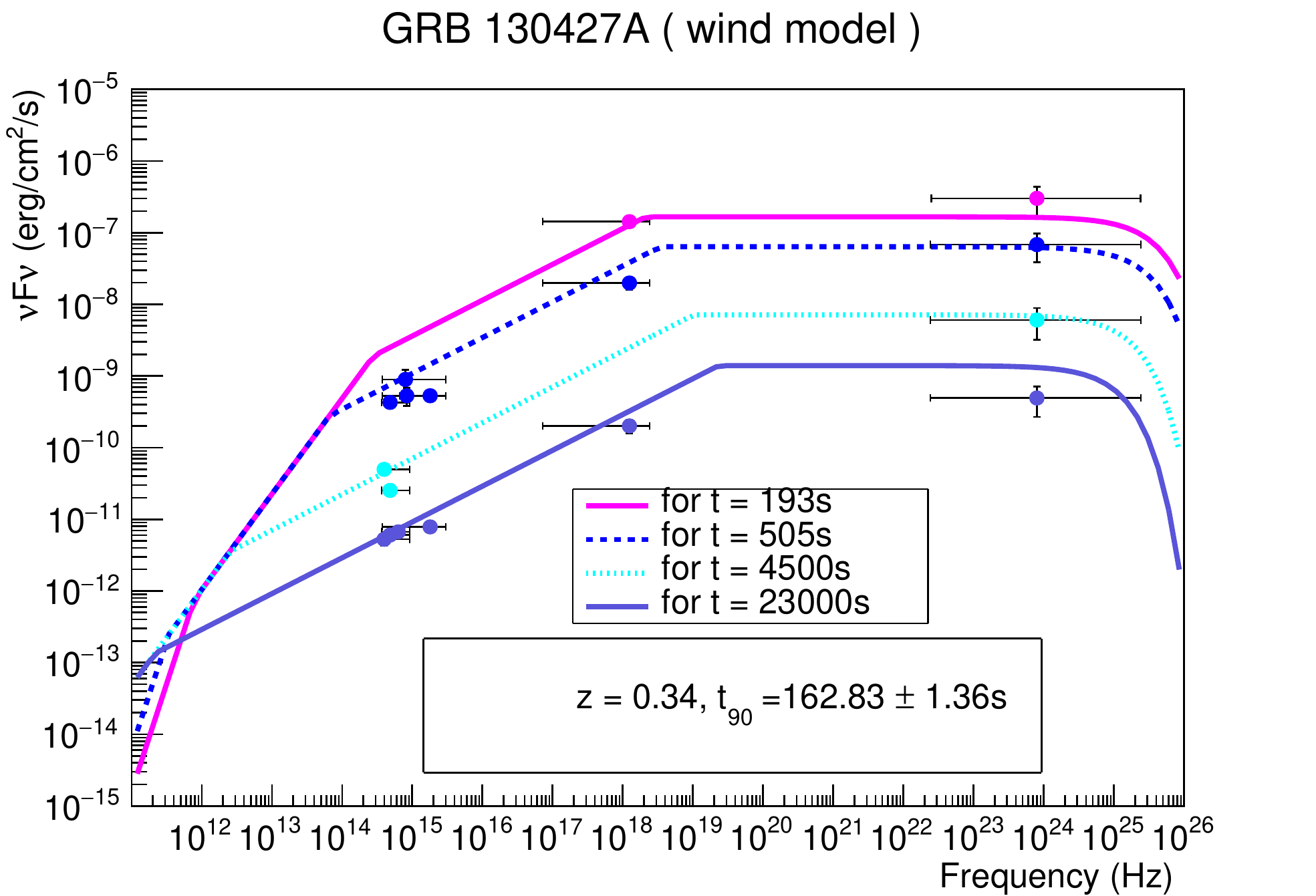}
\vspace{0.5 cm} 
\includegraphics[trim =  0 21 0 10, width=0.85\columnwidth]{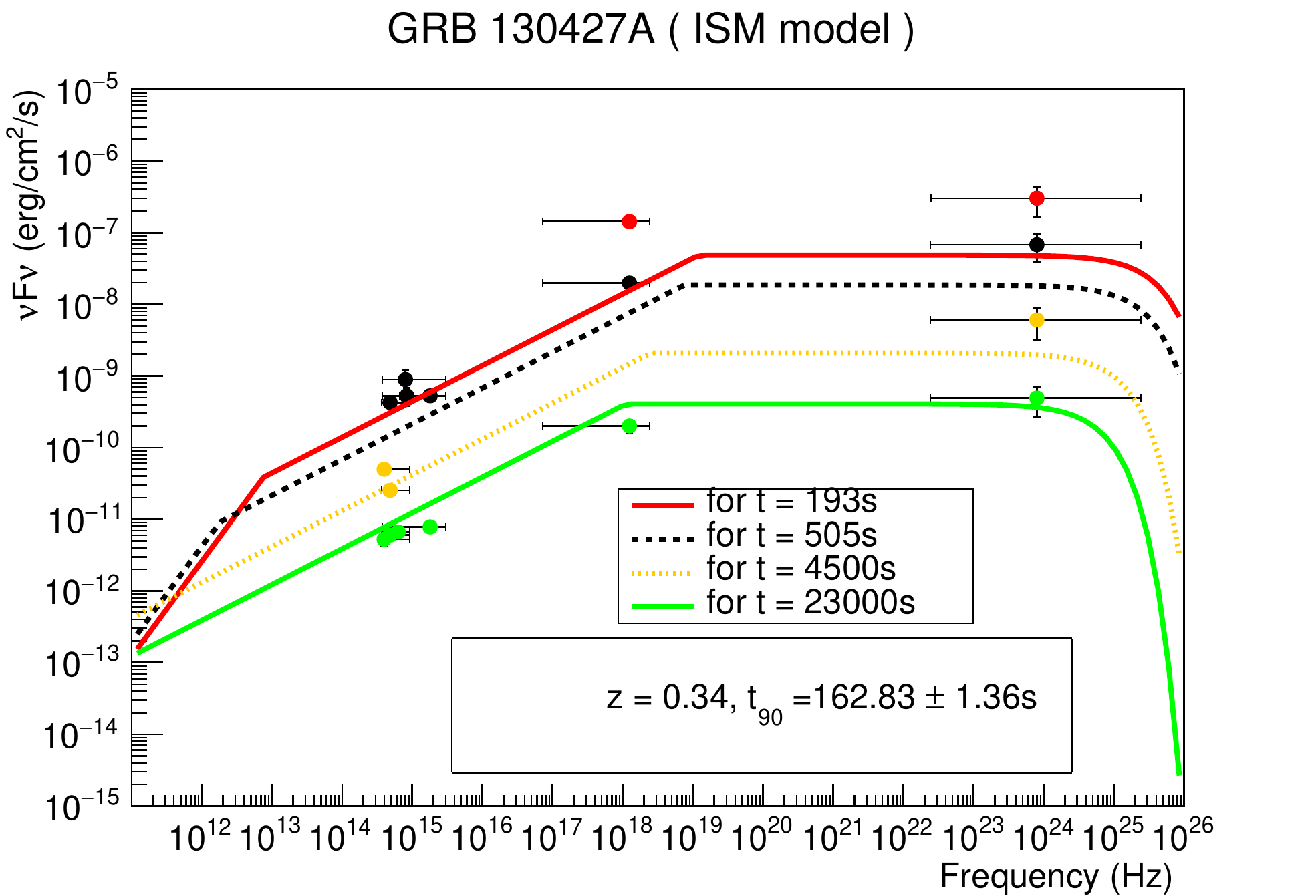}
\caption{\label{fig4sed} Same as Fig.~\ref{fig1sed} but for GRB~111225A, GRB~120422A, GRB~120714B and GRB~130427A.} 
\end{figure*}

\begin{figure*}[th!]
\includegraphics[trim =  0 21 0 10, width=0.85\columnwidth]{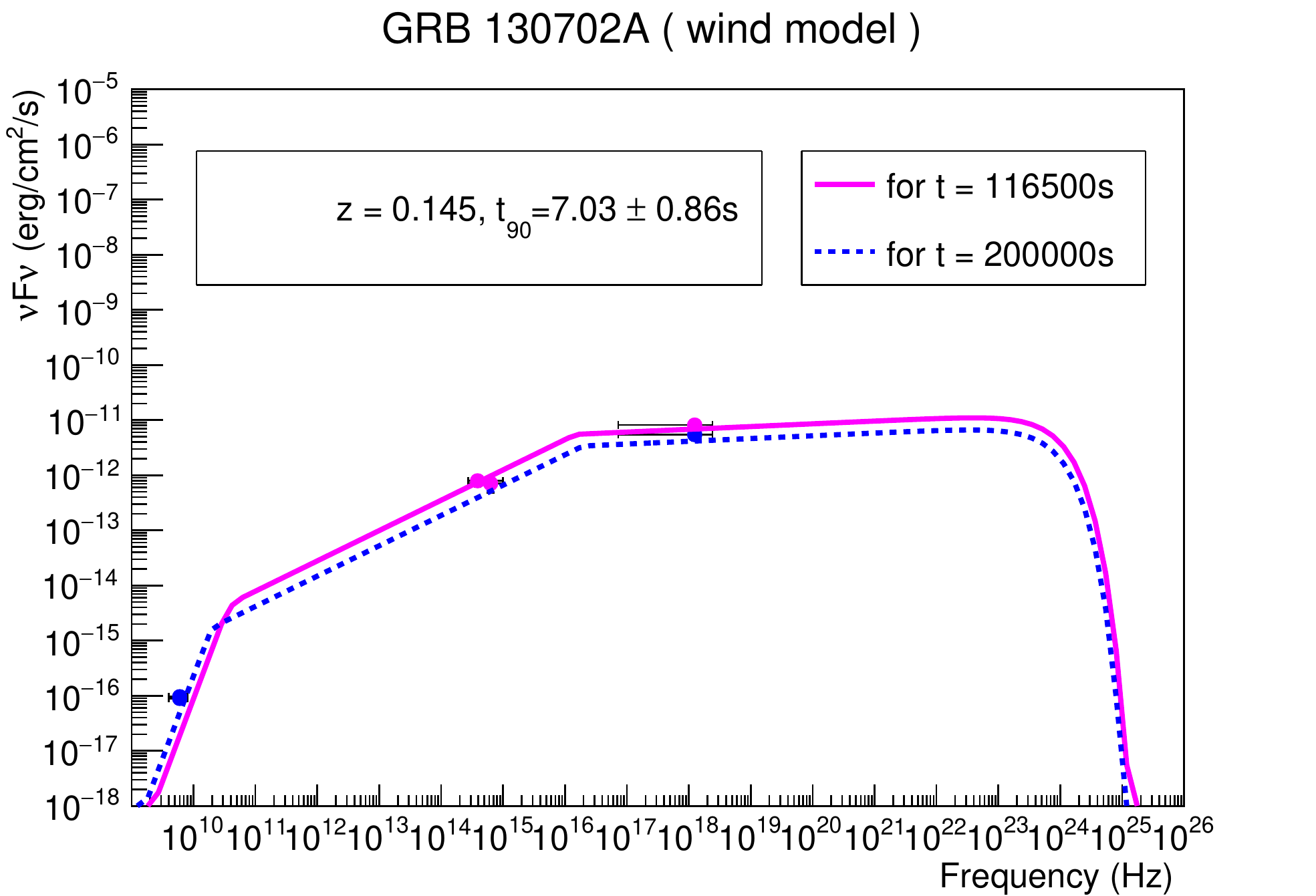}
\vspace{0.5 cm} 
\includegraphics[trim =  0 21 0 10, width=0.85\columnwidth]{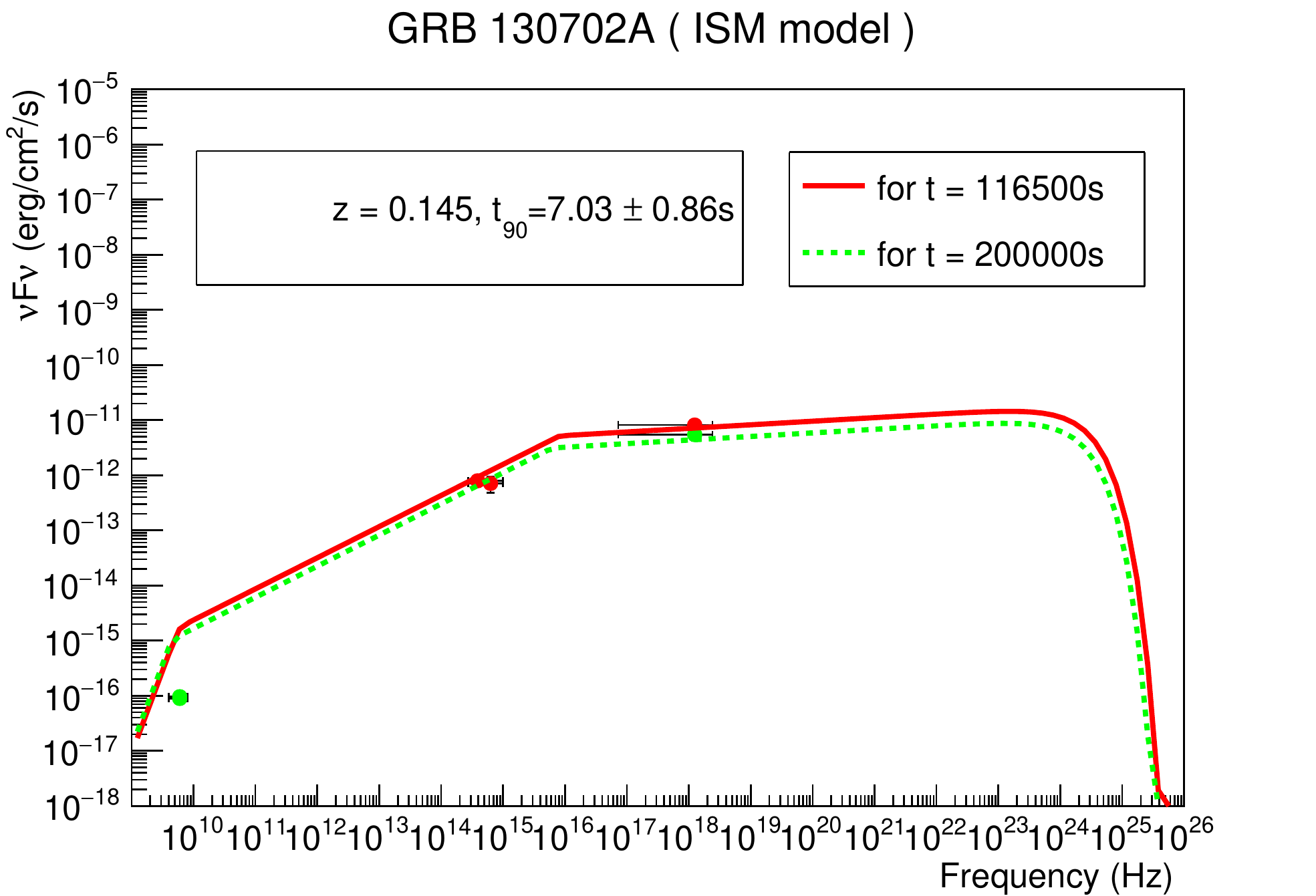}
\includegraphics[trim =  0 21 0 10, width=0.85\columnwidth]{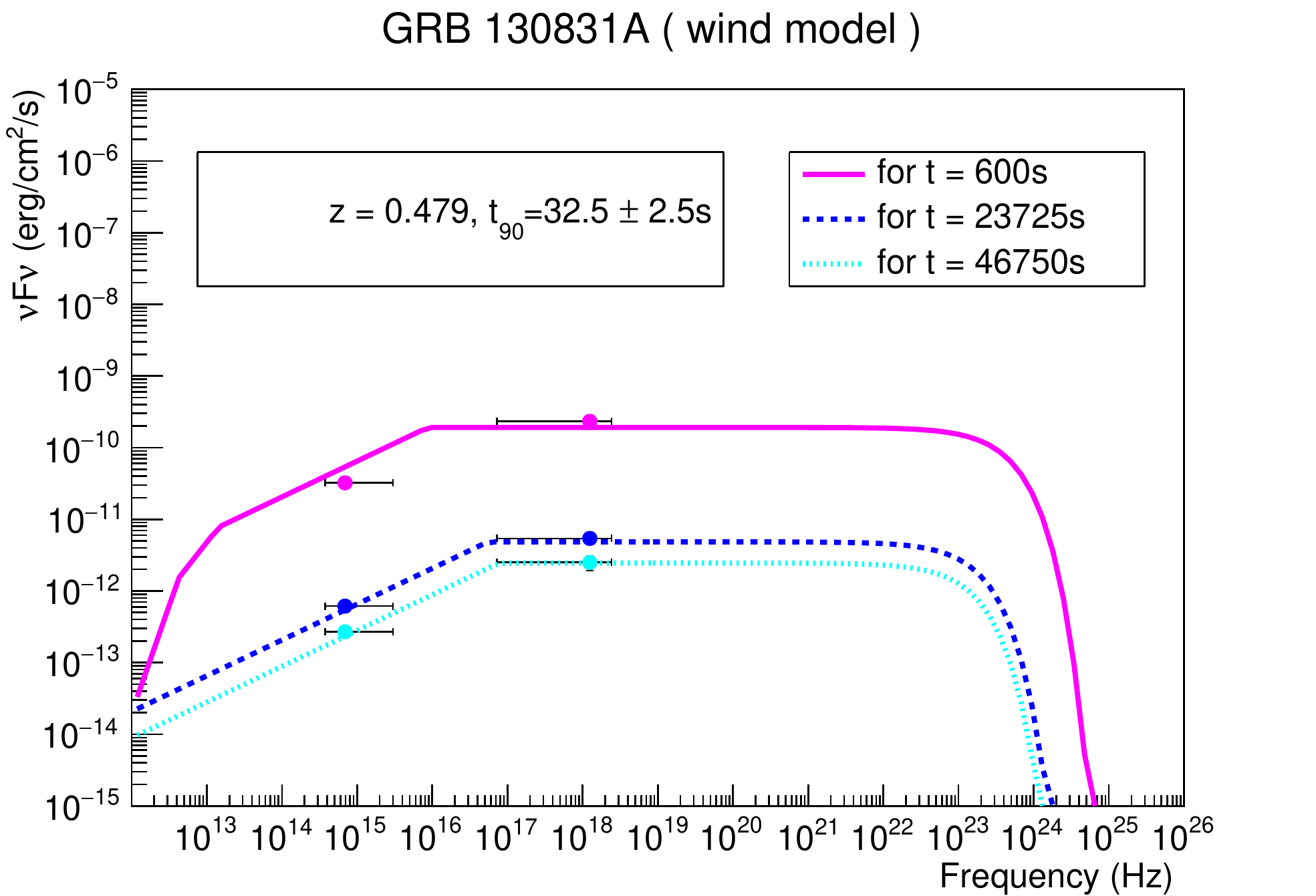}
\vspace{0.5 cm} 
\includegraphics[trim =  0 21 0 10, width=0.85\columnwidth]{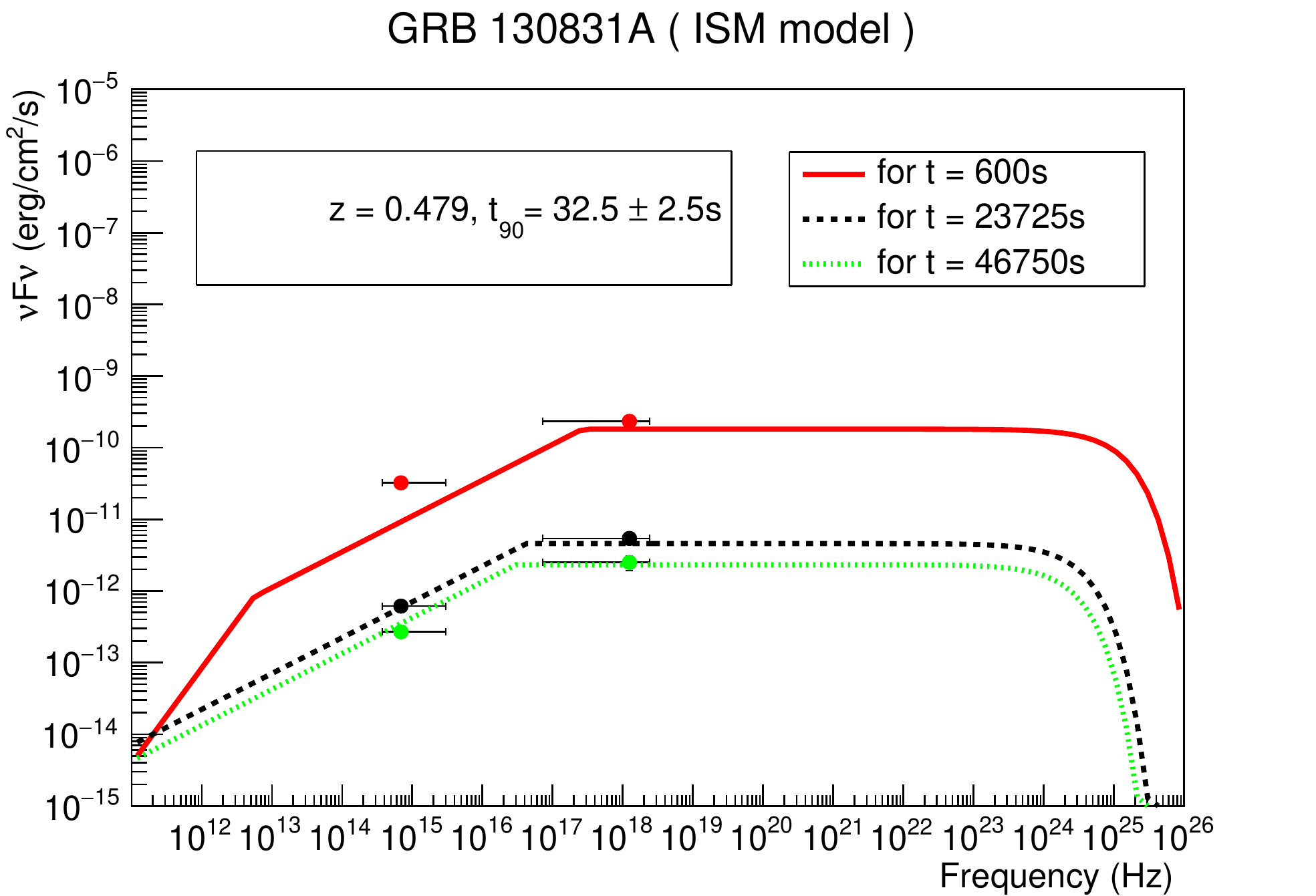}
\includegraphics[trim =  0 21 0 10, width=0.85\columnwidth]{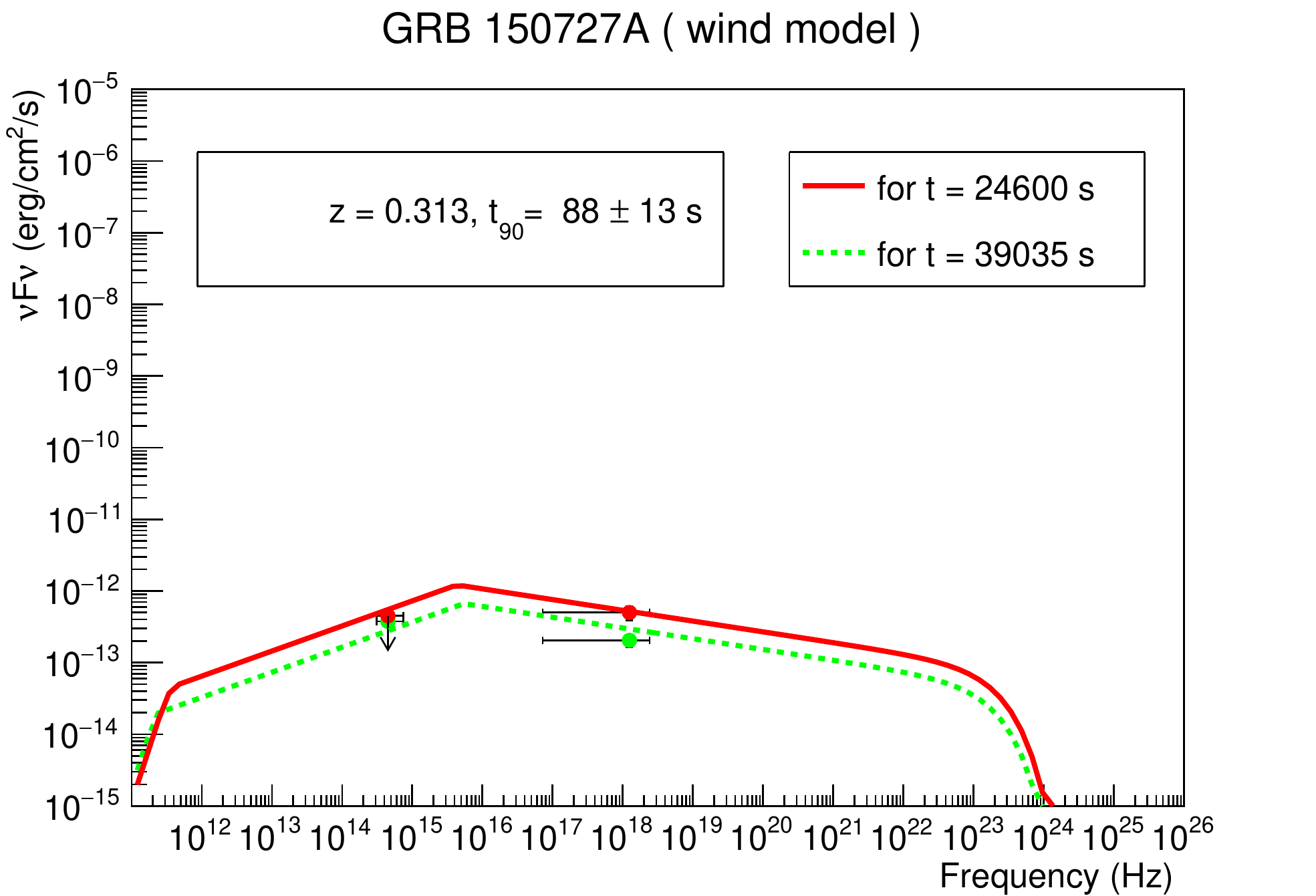}
\vspace{0.5cm}
\includegraphics[trim =  0 21 0 10, width=0.85\columnwidth]{150727wind.pdf}
\includegraphics[trim =  0 21 0 10, width=0.85\columnwidth]{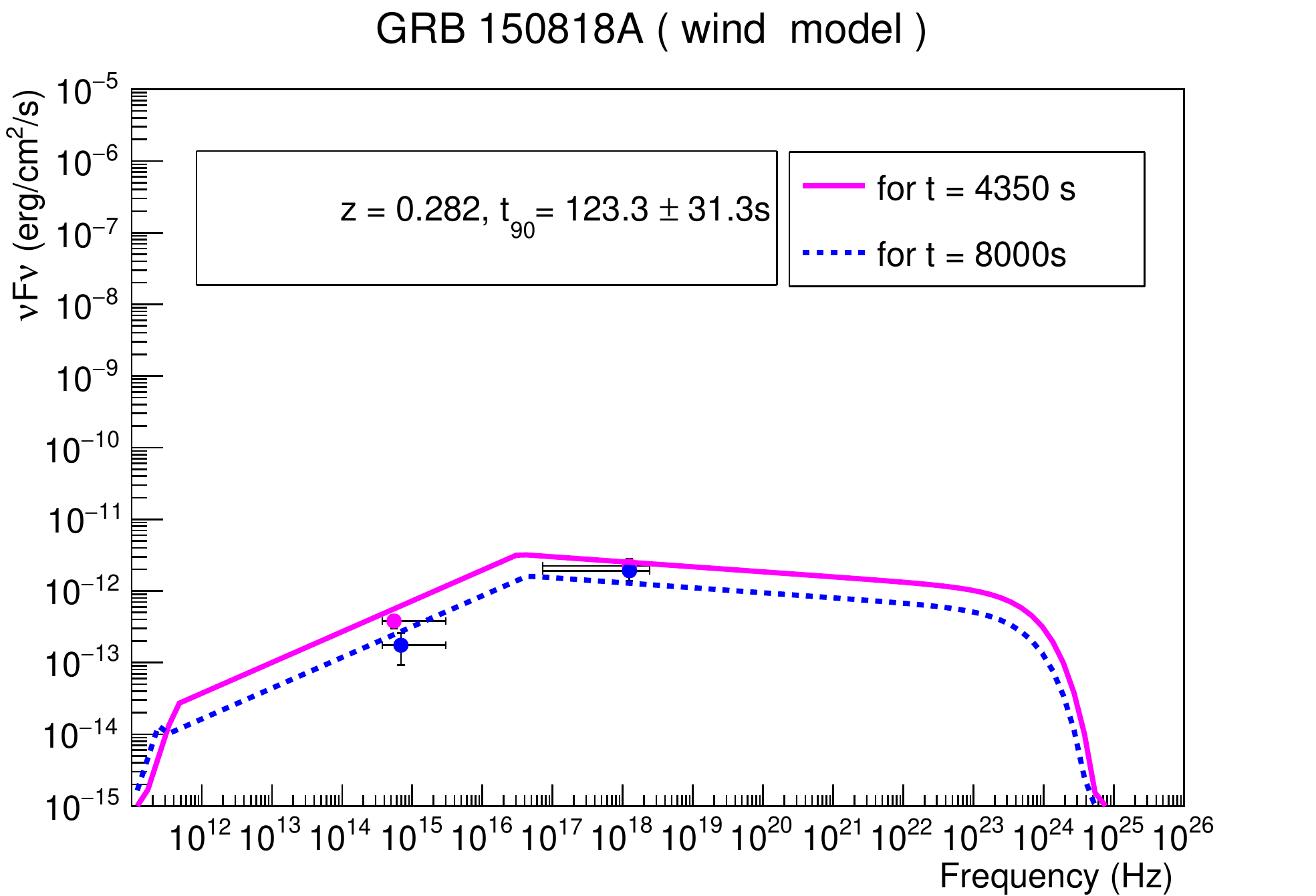}
\vspace{0.5 cm}
\includegraphics[trim =  0 21 0 10, width=0.85\columnwidth]{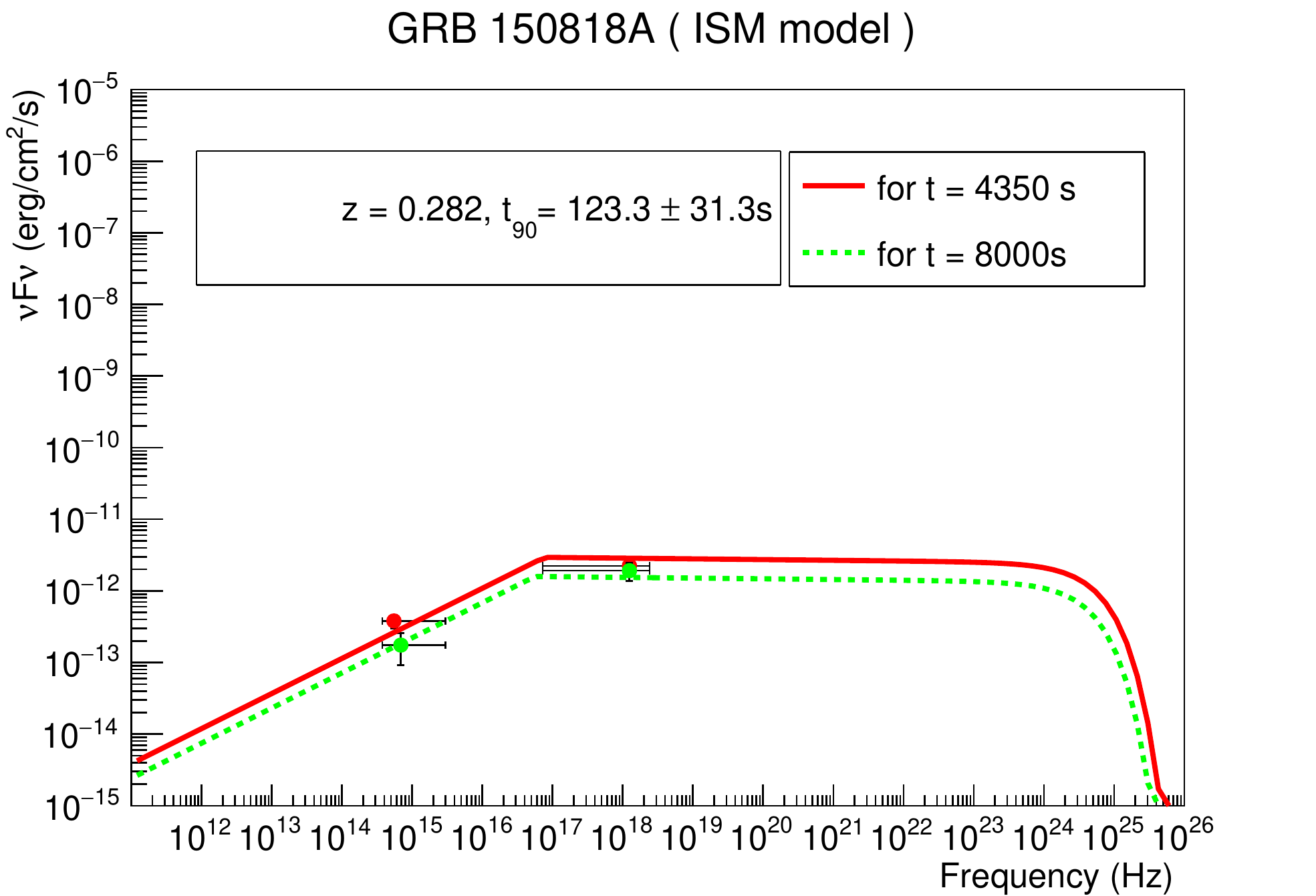}
\caption{\label{fig6sed} Same as Fig.~\ref{fig1sed} but for GRB~130702A, GRB~130831A, GRB~150727A and GRB~150818A.}
\end{figure*}

\begin{figure*}[th!]
\includegraphics[trim =  0 21 0 10, width=0.85\columnwidth]{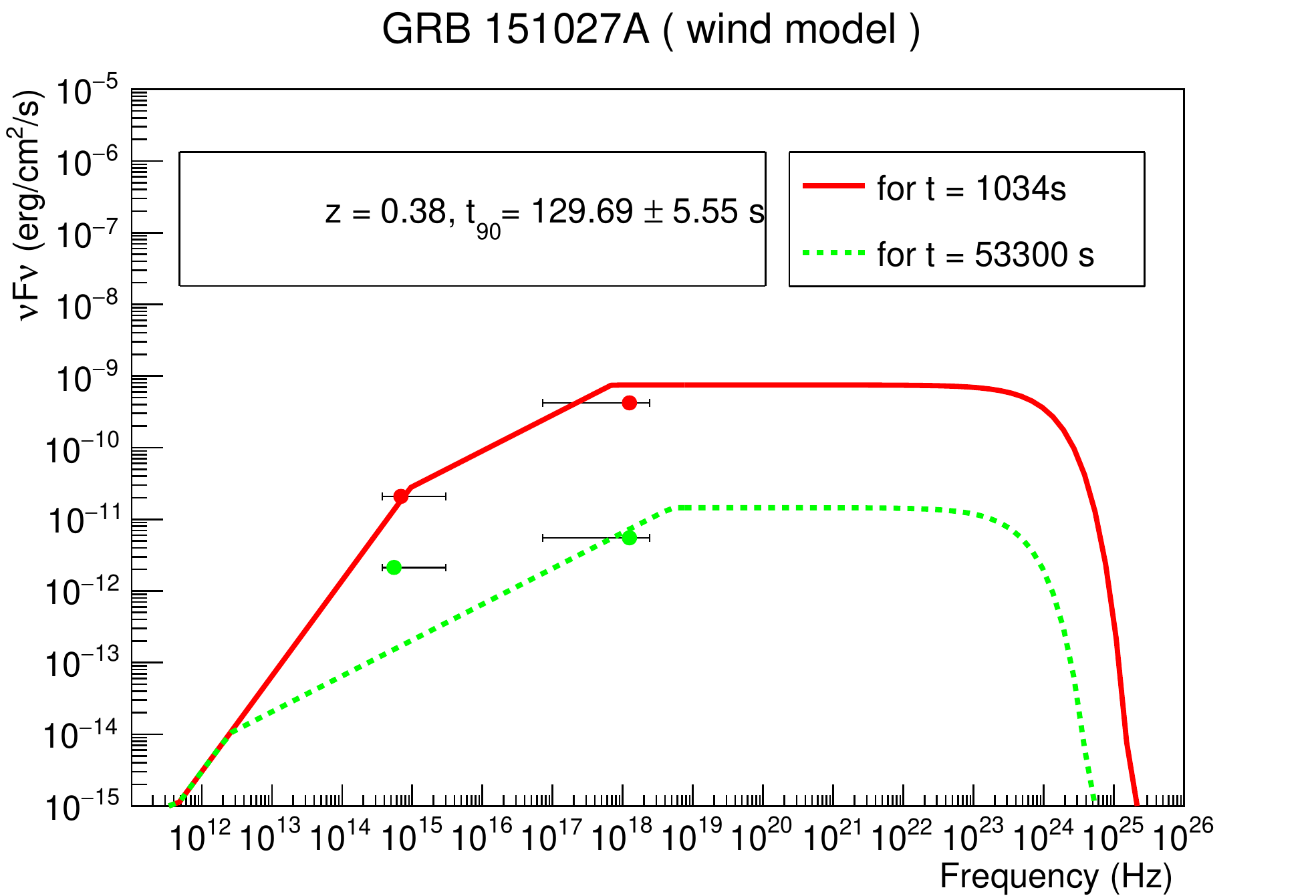}
\vspace{0.5 cm}
\includegraphics[trim =  0 21 0 10, width=0.85\columnwidth]{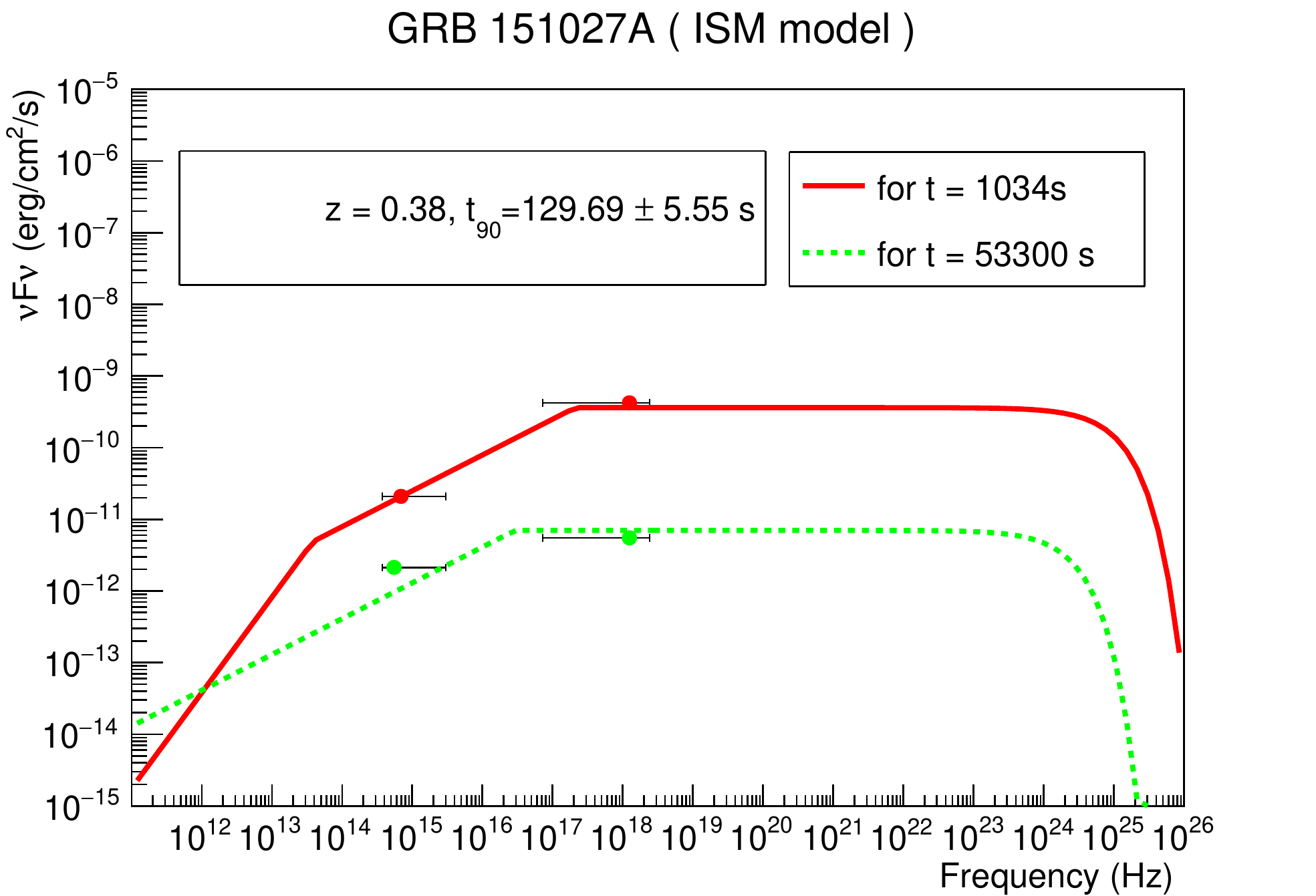}
\includegraphics[trim =  0 21 0 10, width=0.85\columnwidth]{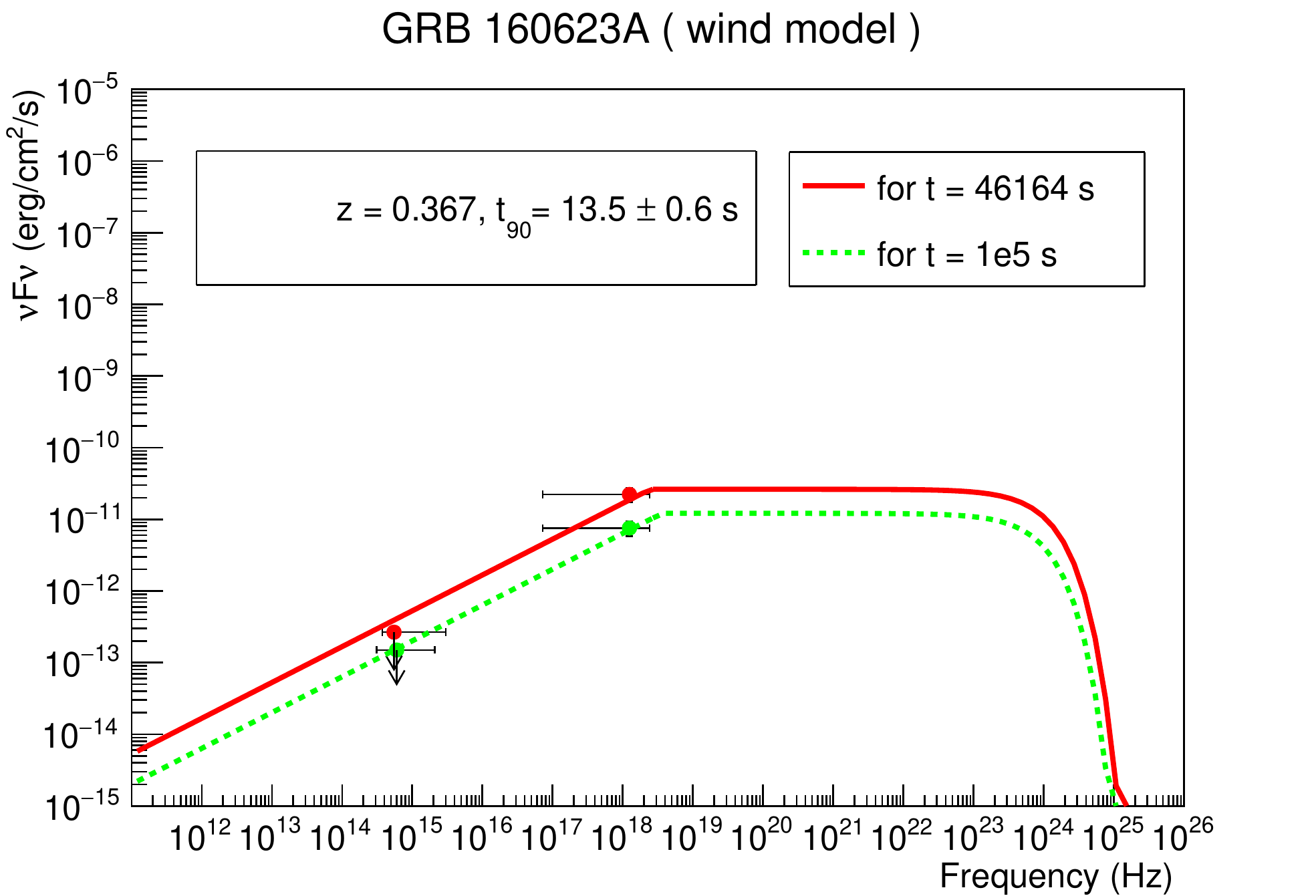}
\vspace{0.5 cm} 
\includegraphics[trim =  0 21 0 10, width=0.85\columnwidth]{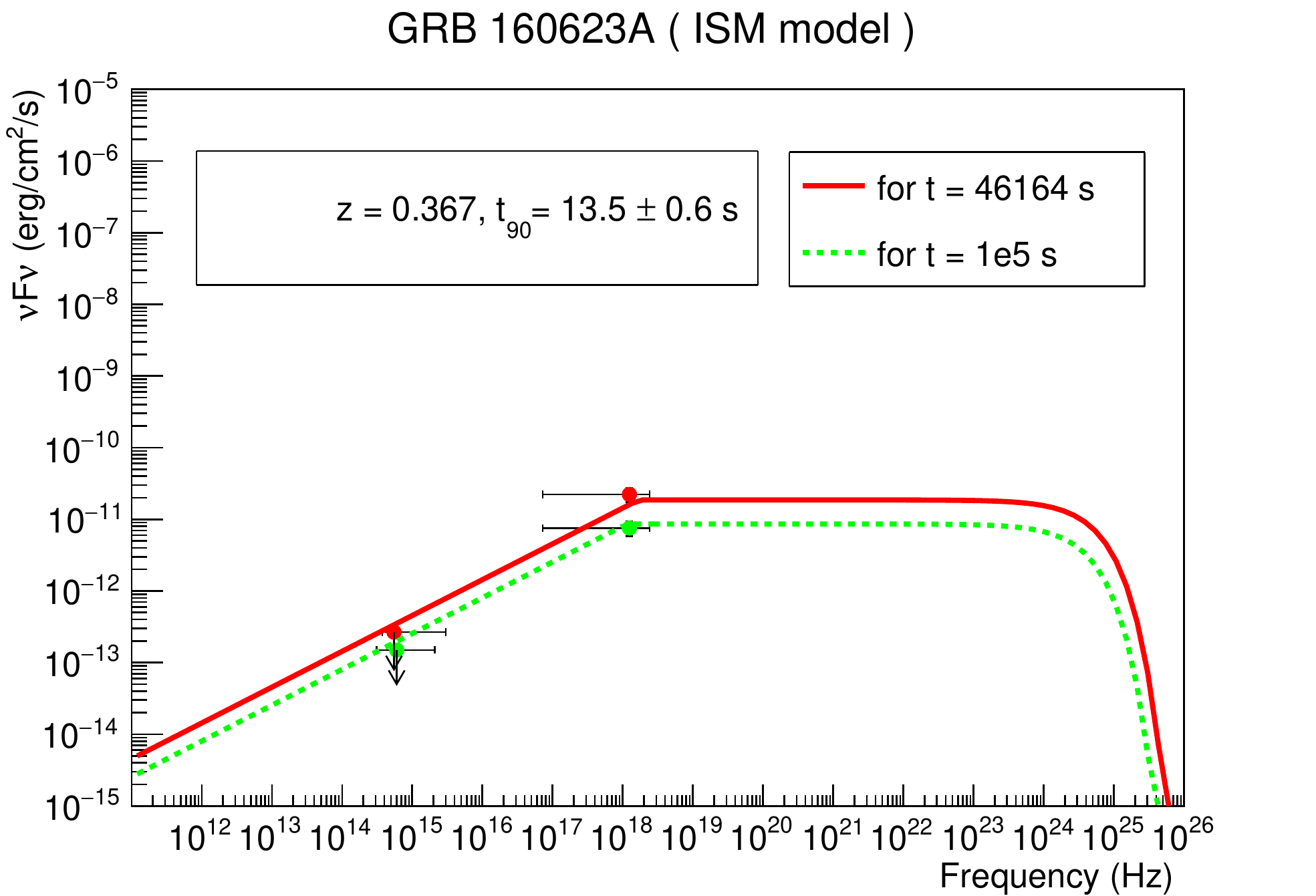}
\includegraphics[trim =  0 21 0 10, width=0.85\columnwidth]{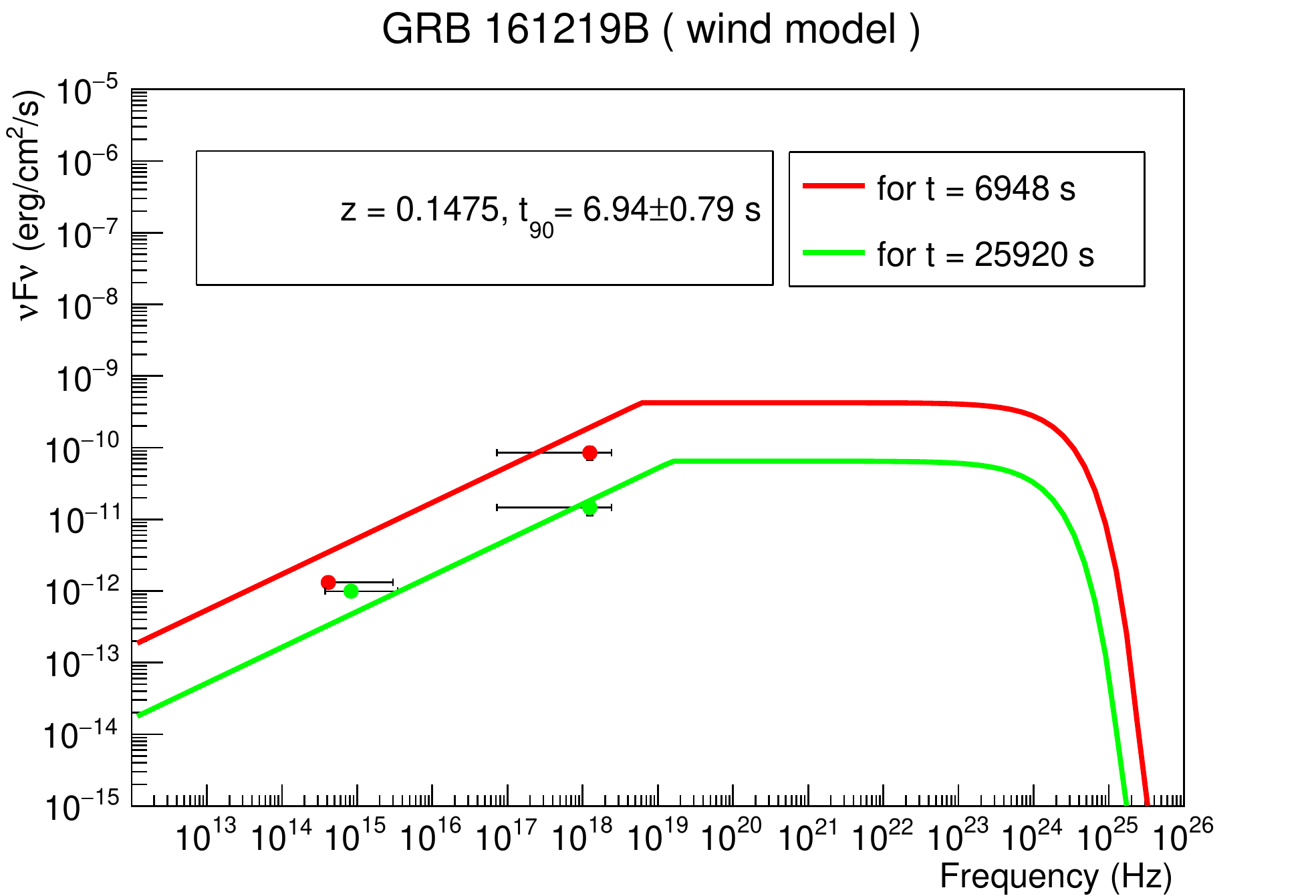}
\vspace{0.5 cm} 
\includegraphics[trim =  0 21 0 10, width=0.85\columnwidth]{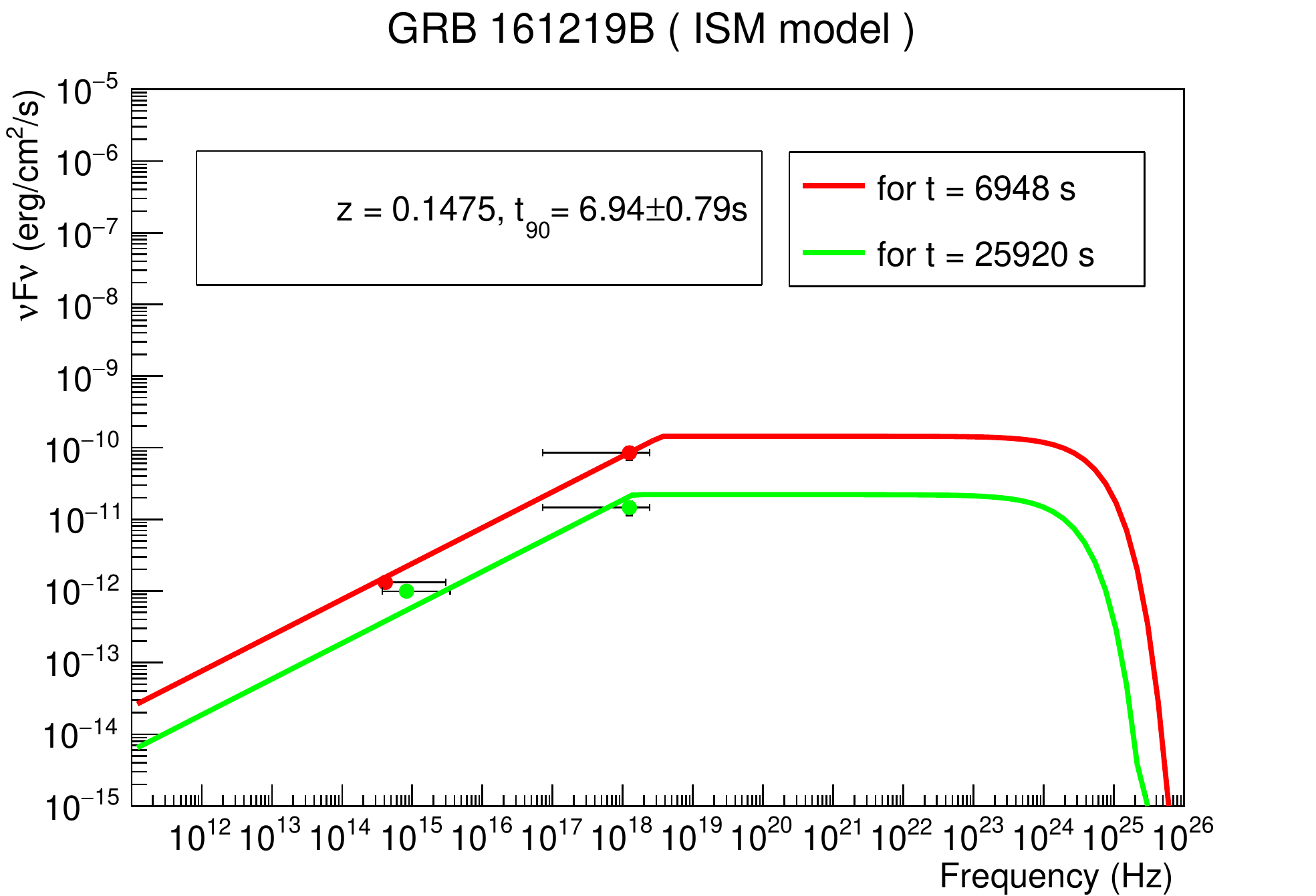}
\caption{\label{fig5sed} Same as Fig.~\ref{fig1sed} but for GRB~151027A, GRB~160623A, GRB~161219B.}
\end{figure*}

\begin{figure*}[th!]
\includegraphics[trim =  0 21 0 10 , width= 0.85\columnwidth]{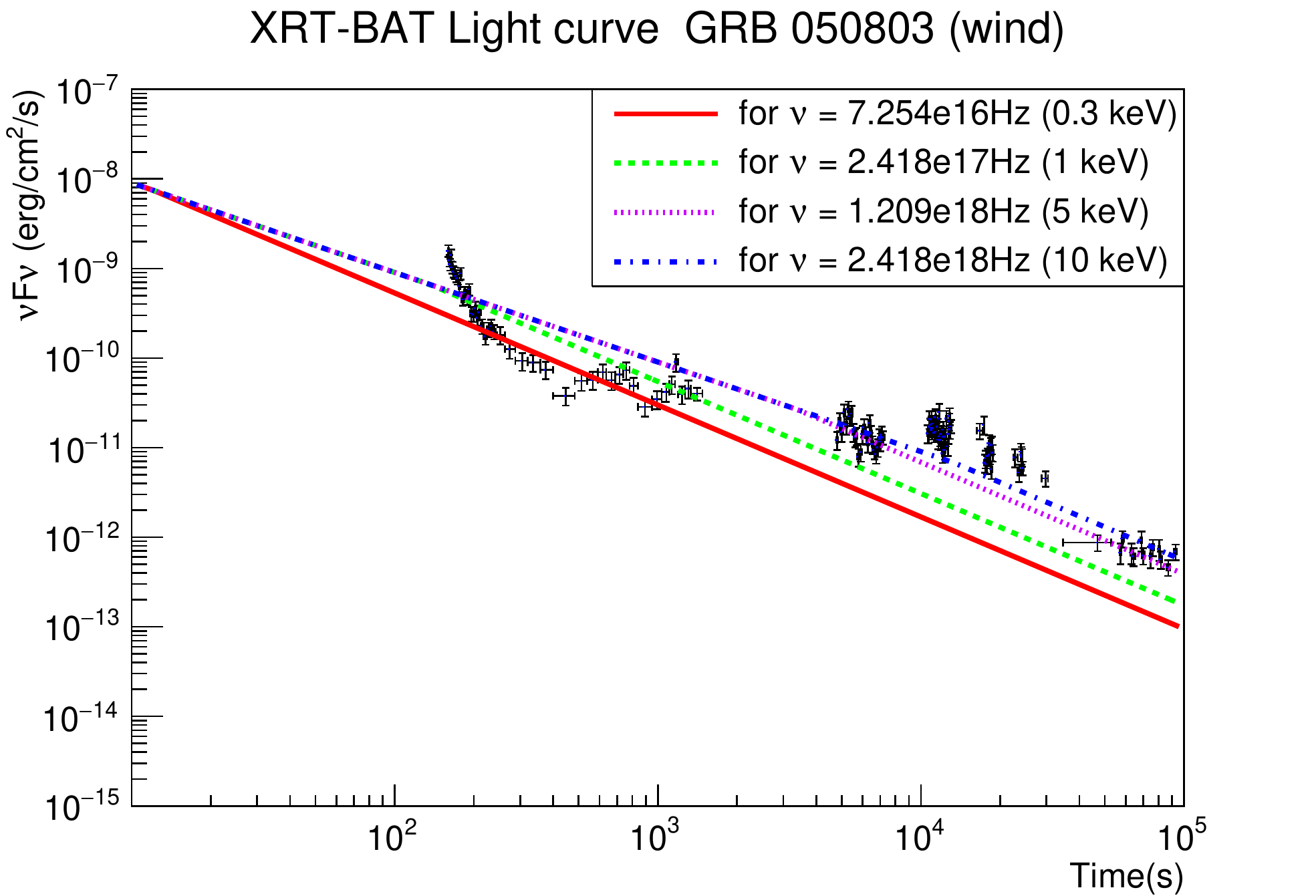}
\vspace{0.5 cm}
\includegraphics[trim =  0 21 0 10 , width= 0.85\columnwidth]{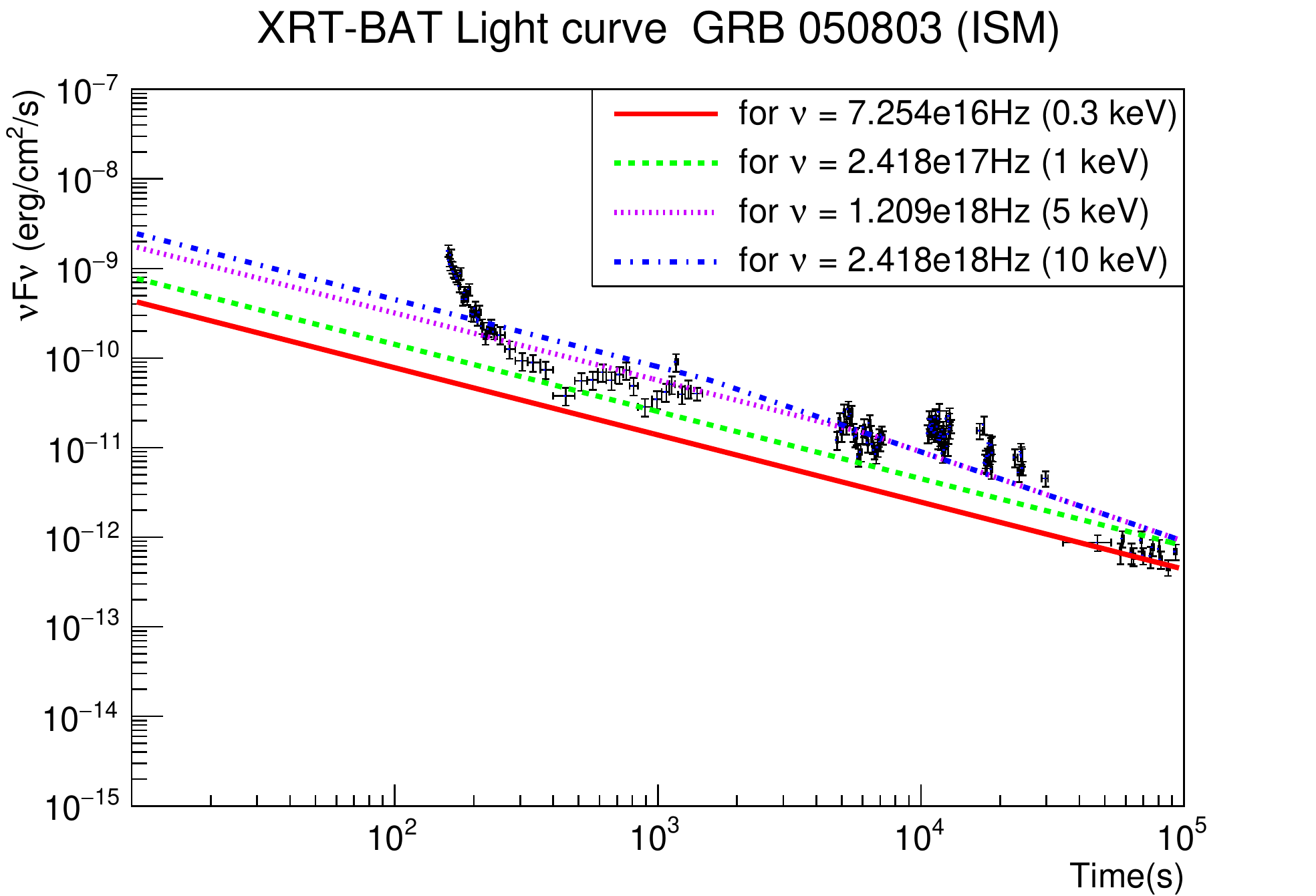}
\includegraphics[trim =  0 21 0 10, width= 0.85\columnwidth]{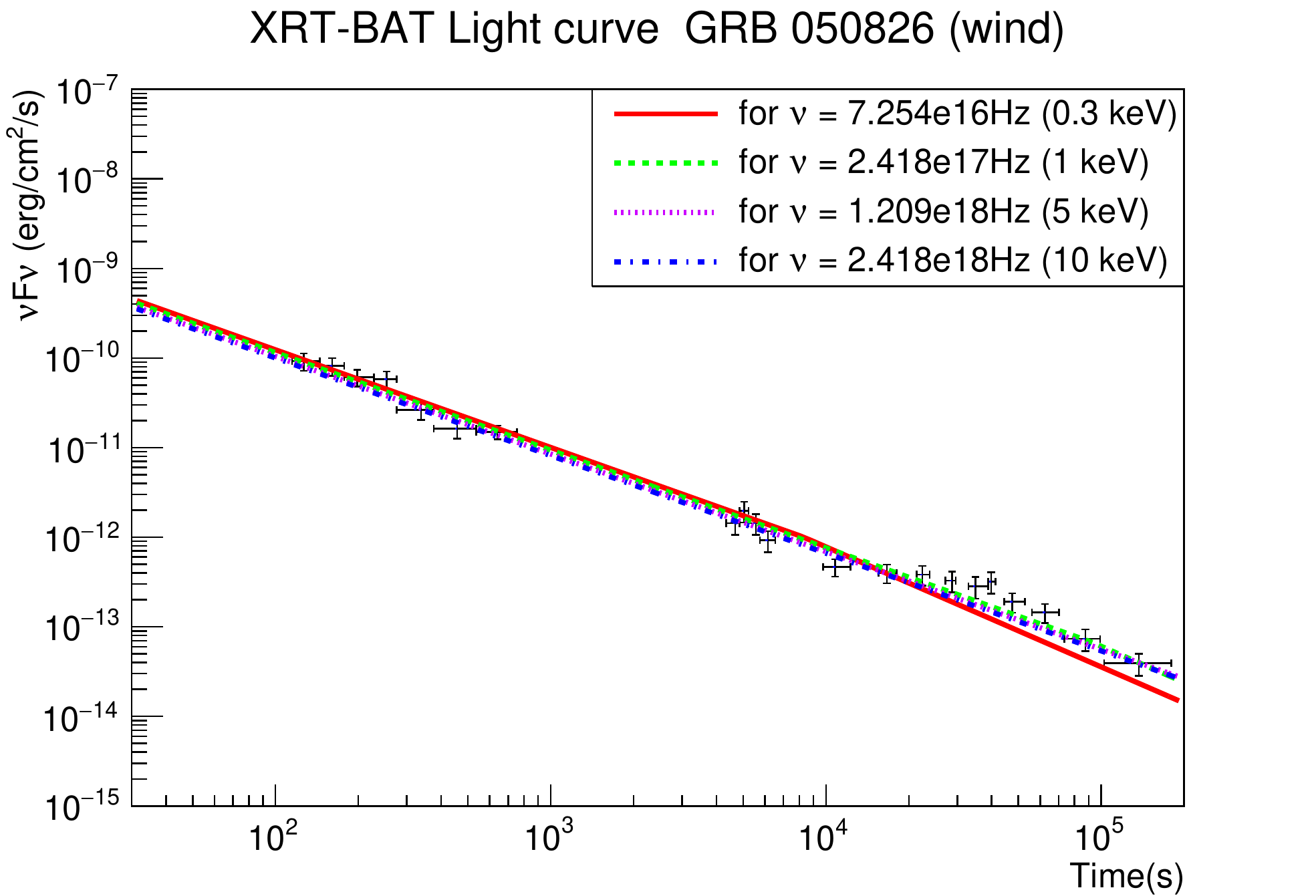}
\vspace{0.5 cm}
\includegraphics[trim =  0 21 0 10, width= 0.85\columnwidth]{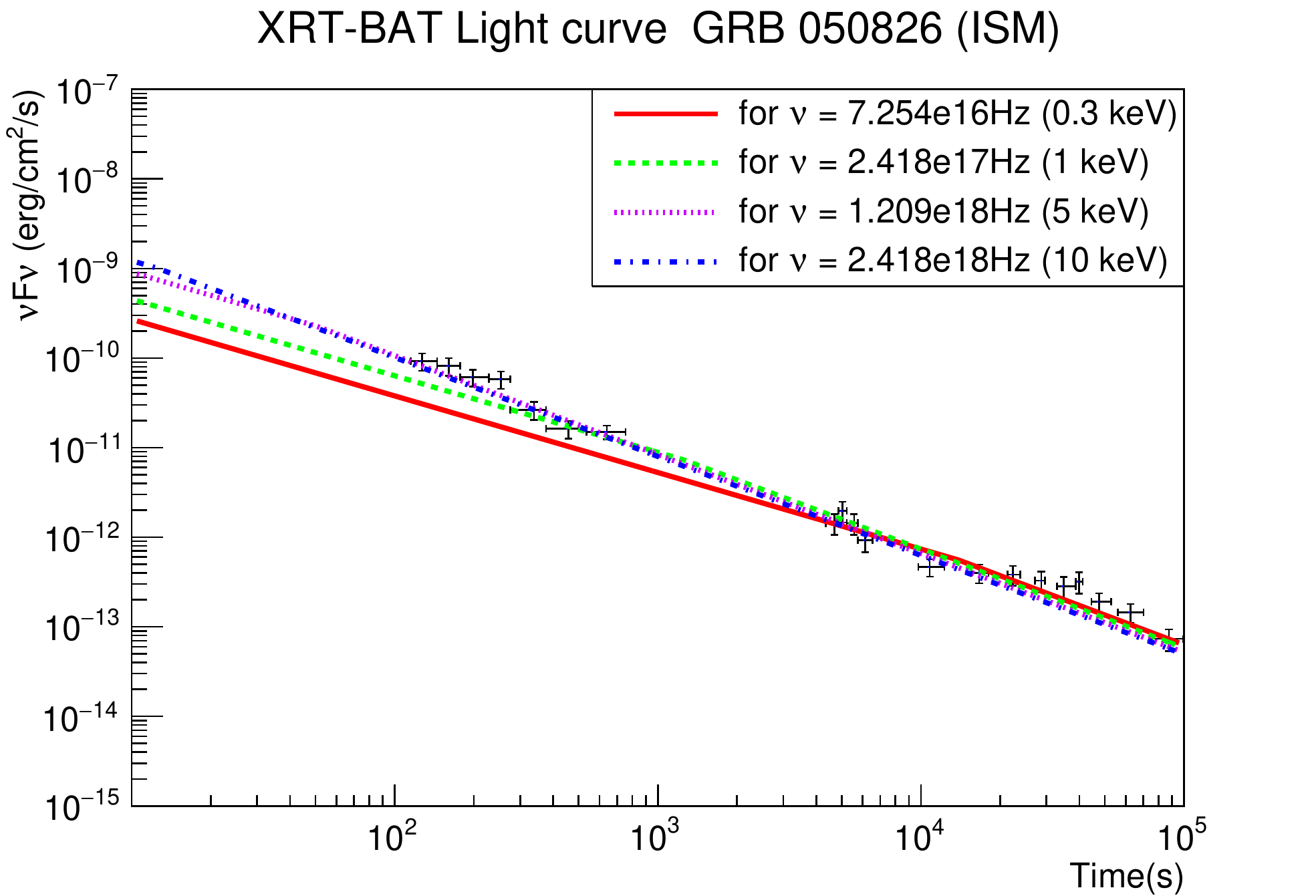}
\includegraphics[trim =  0 21 0 10, width= 0.85\columnwidth]{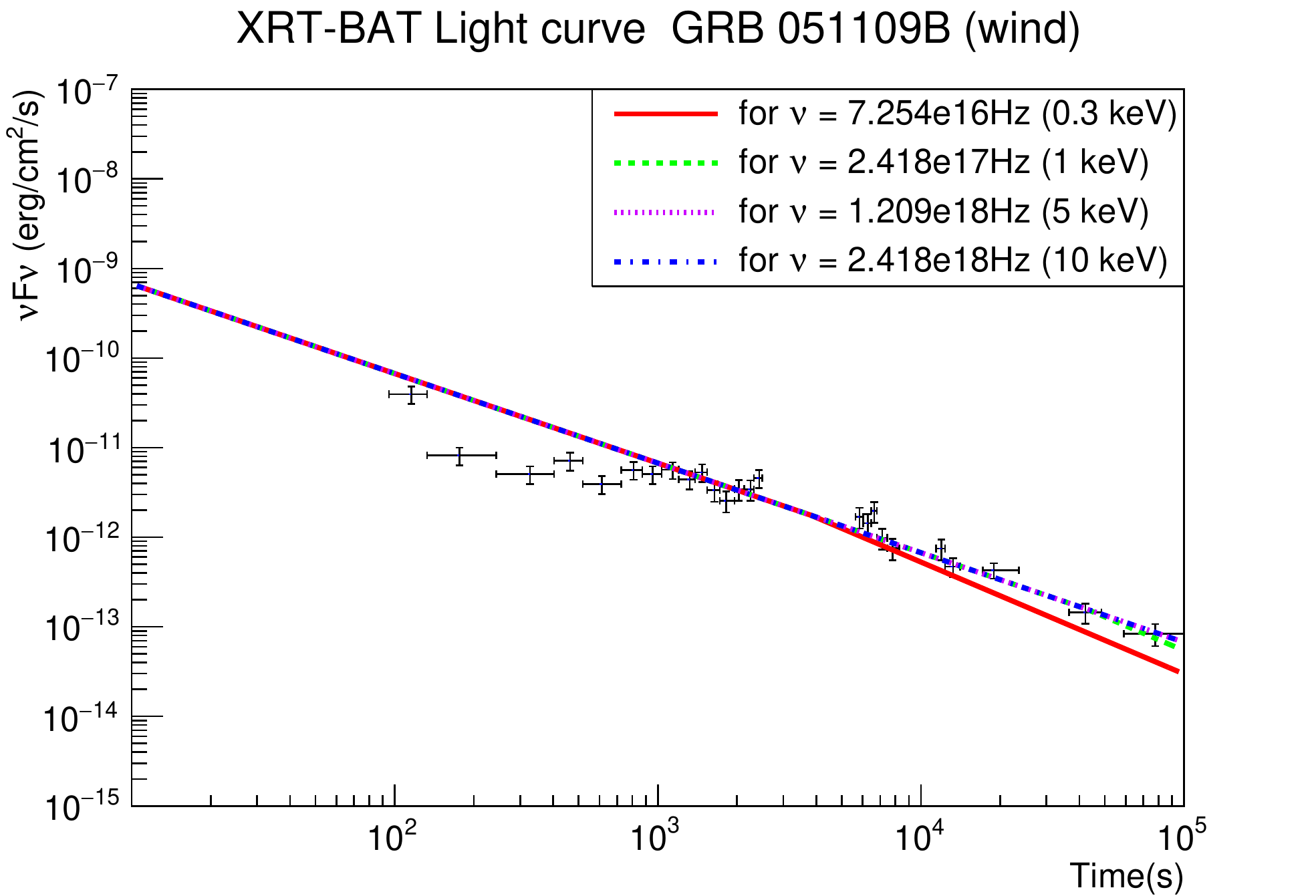}
\vspace{0.5 cm}
\includegraphics[trim =  0 21 0 10, width= 0.85\columnwidth]{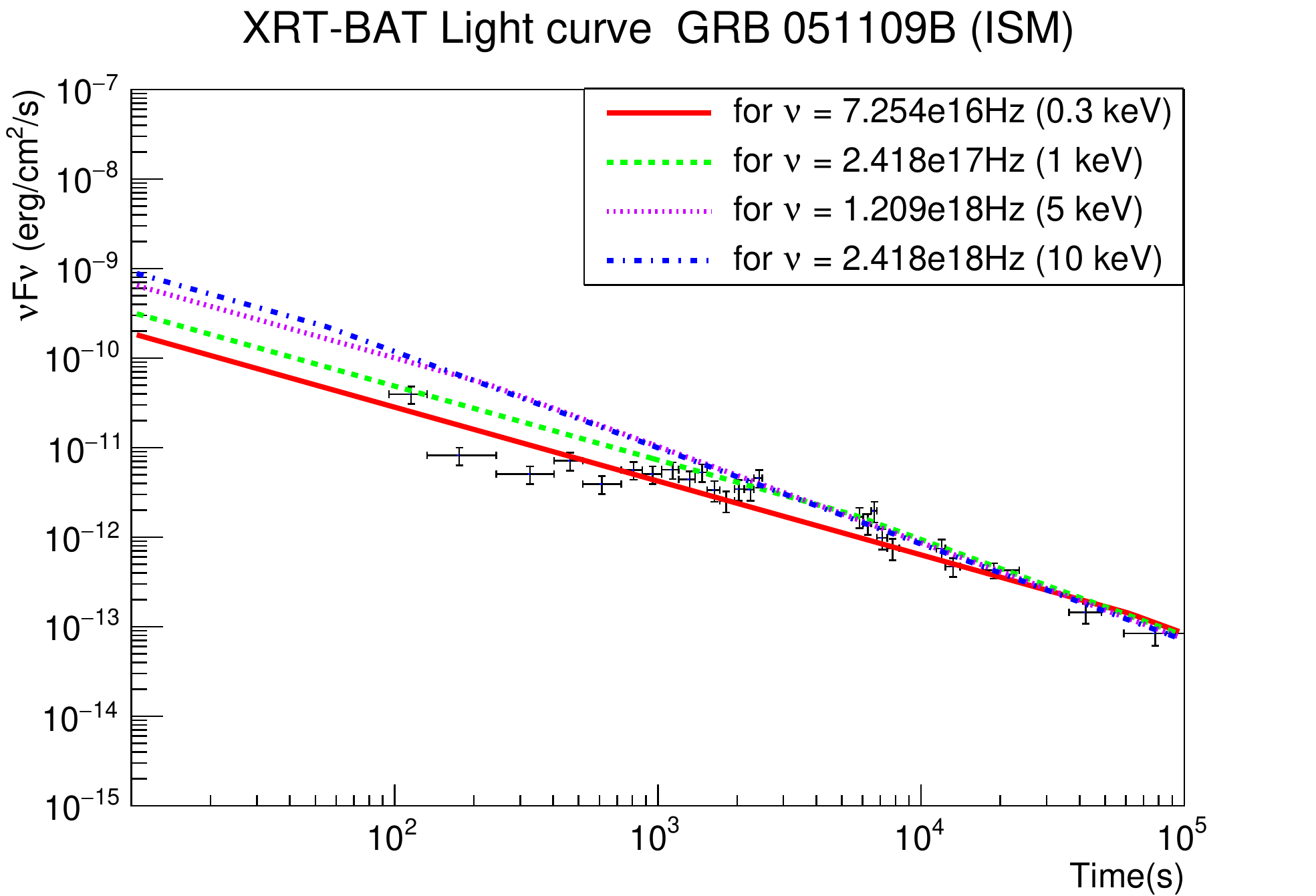}
\includegraphics[trim =  0 21 0 10 , width=0.85\columnwidth]{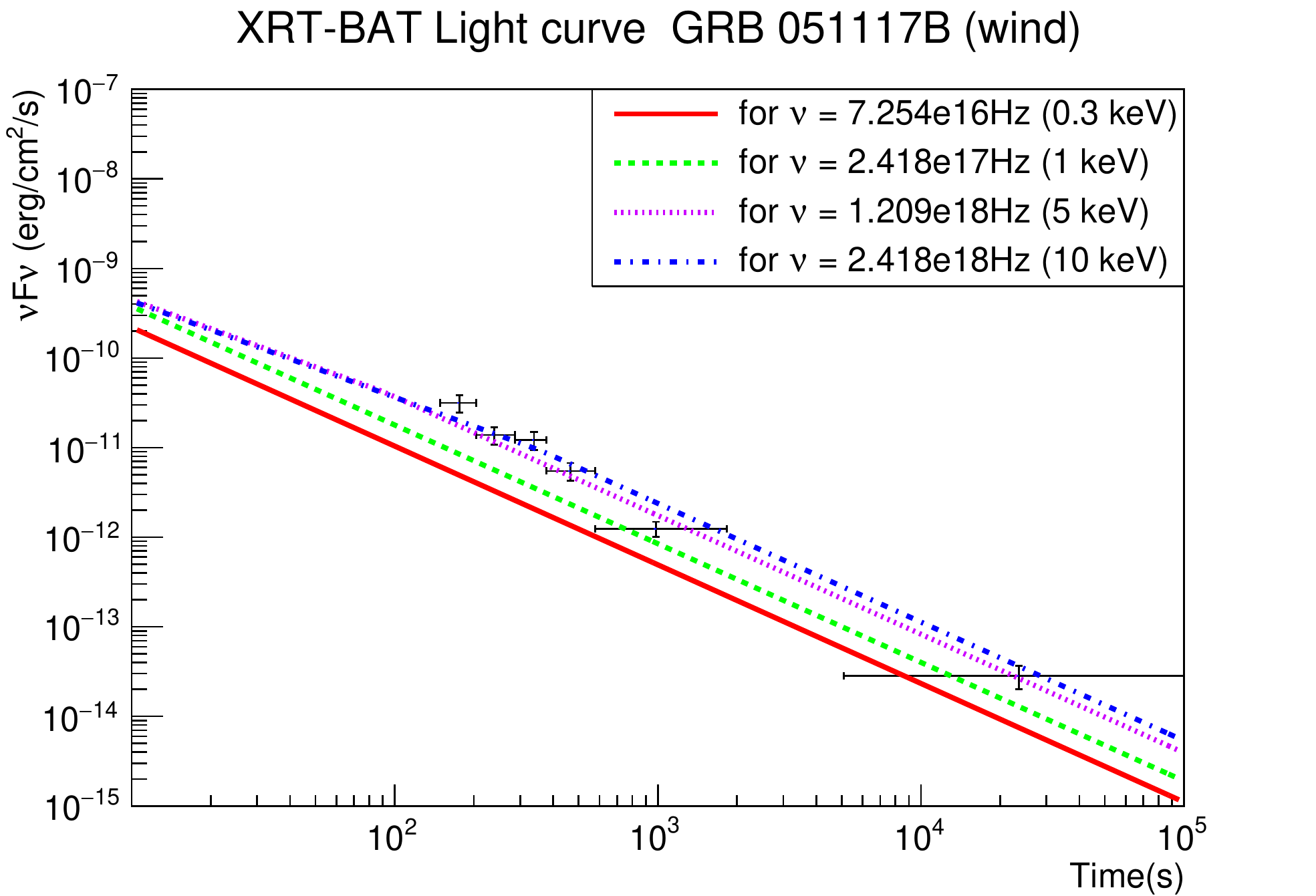}
\vspace{0.5 cm}
\includegraphics[trim =  0 21 0 10 , width=0.8\columnwidth]{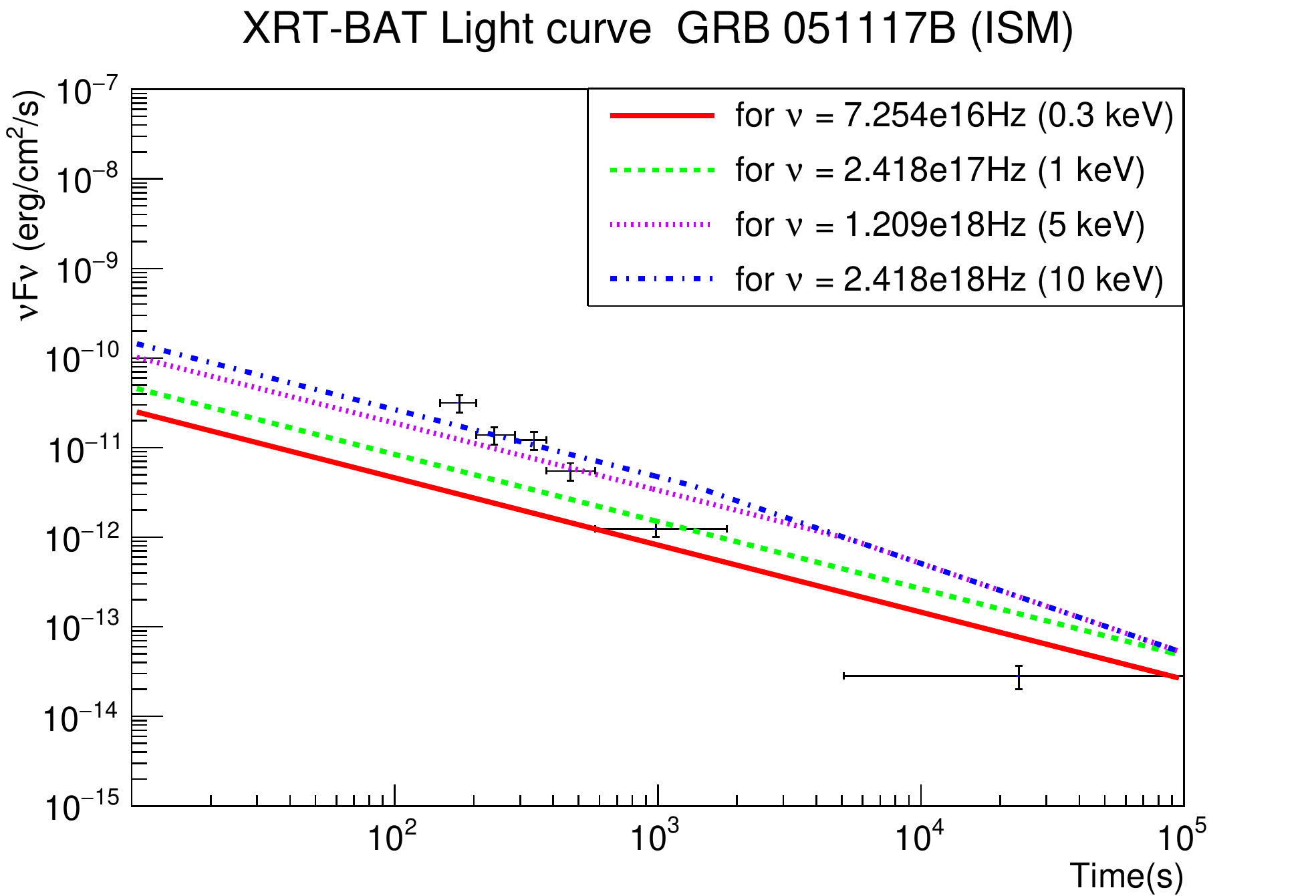}
\caption{\label{fig1lc} Light curves of GRB~050803, GRB~050826 GRB~051109B and GRB~051117B for different frequencies as well as synchrotron afterglow model fits in Wind (left panels) and ISM (right panels) environments.  The light curves are fitted after $T_{90}$ with the same model parameters as in SEDs.  The breaks apparent in some cases correspond to a change from fast-cooling to slow-cooling spectra.}
\end{figure*}

\begin{figure*}[th!]
\includegraphics[trim =  0 21 0 10 , width=0.85\columnwidth]{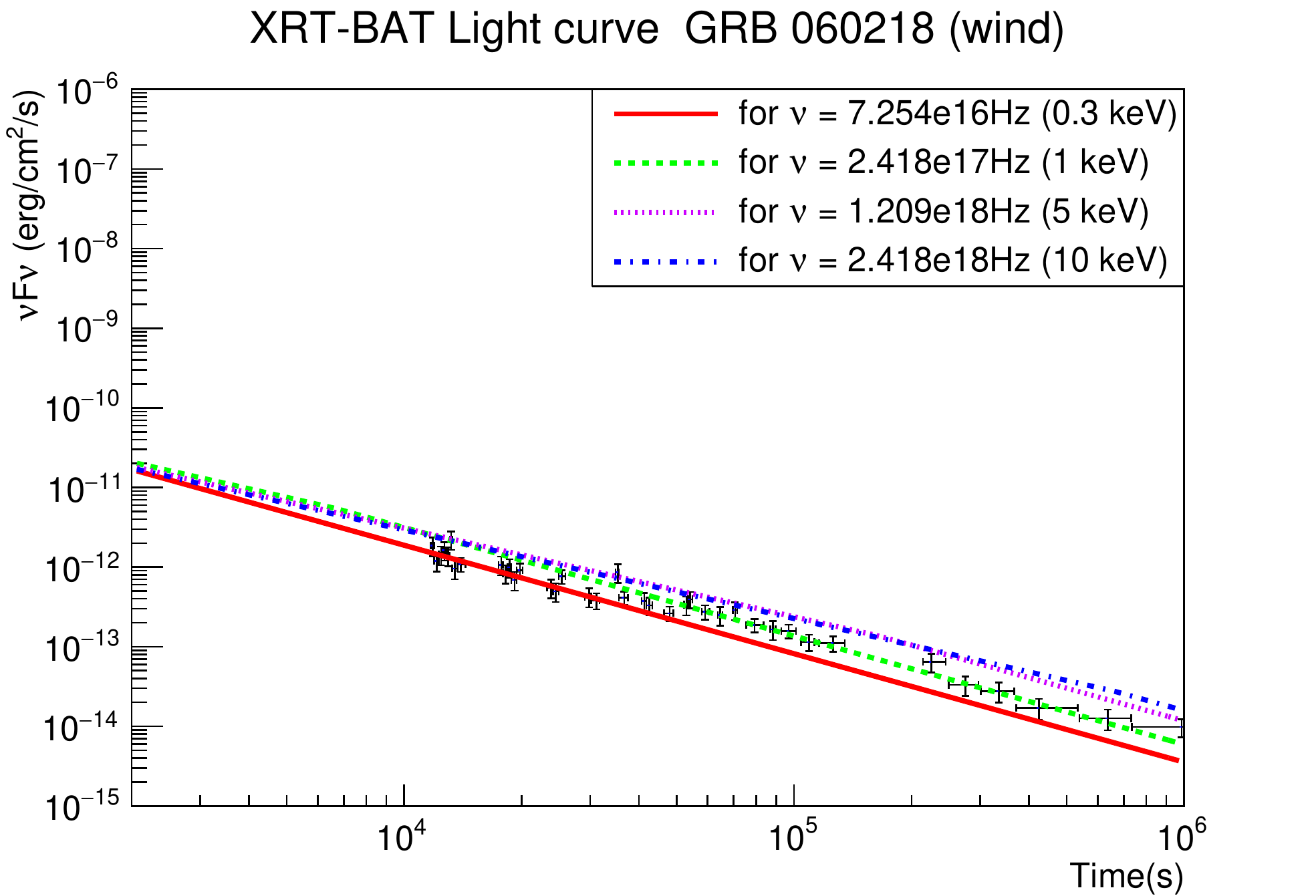}
\vspace{0.5 cm}
\includegraphics[trim =  0 21 0 10 , width=0.85\columnwidth]{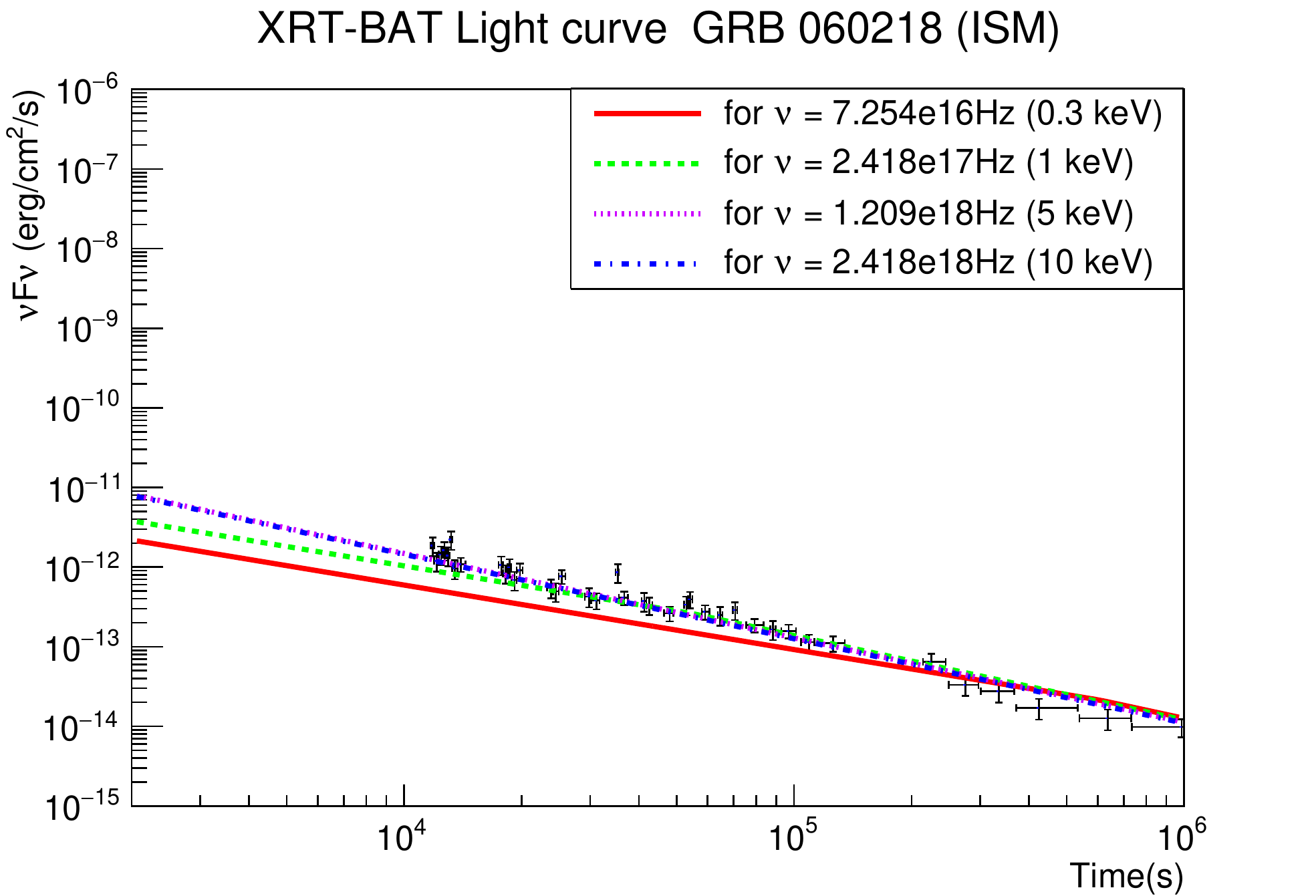}
\includegraphics[trim =  0 21 0 10, width=0.85\columnwidth]{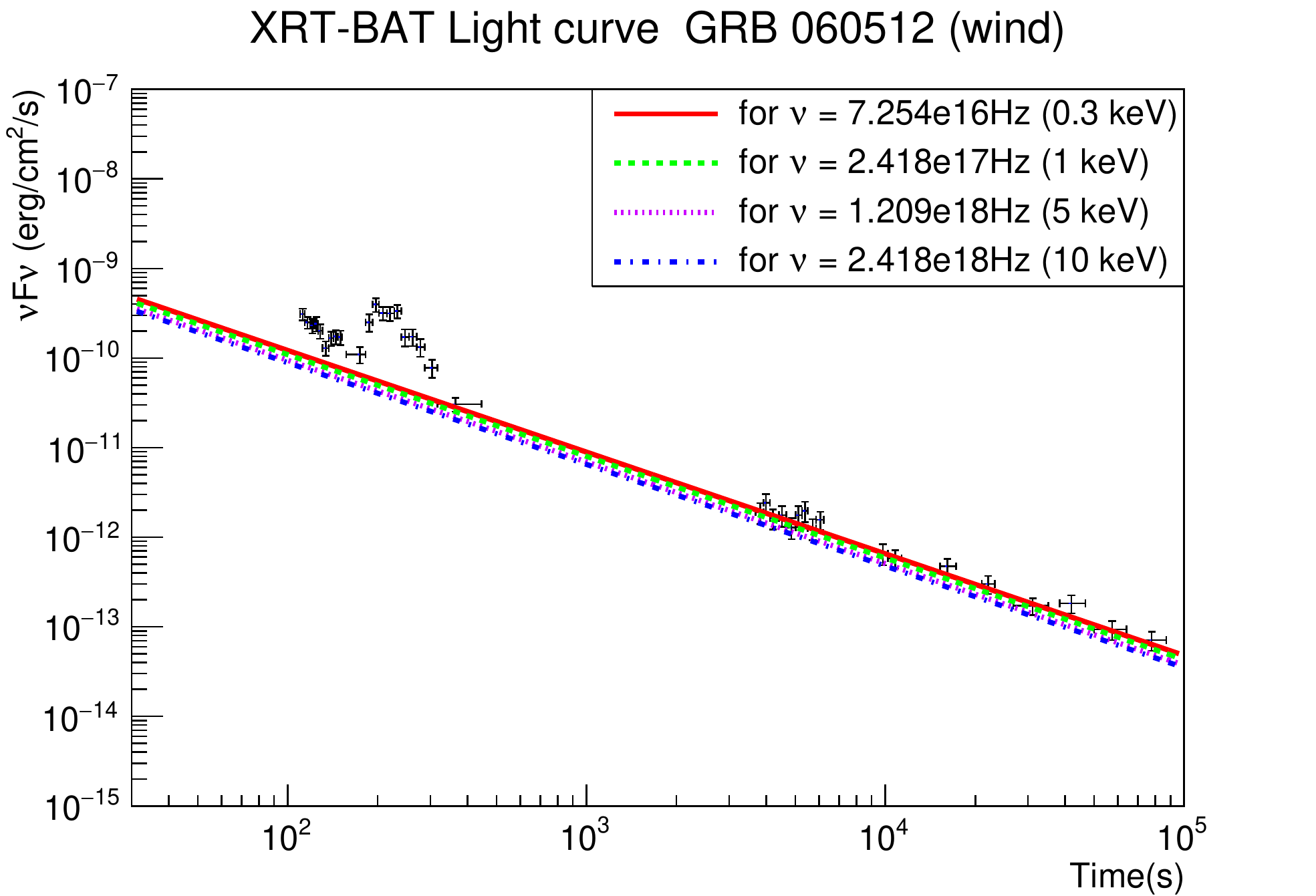}
\vspace{0.5 cm}
\includegraphics[trim =  0 21 0 10, width=0.85\columnwidth]{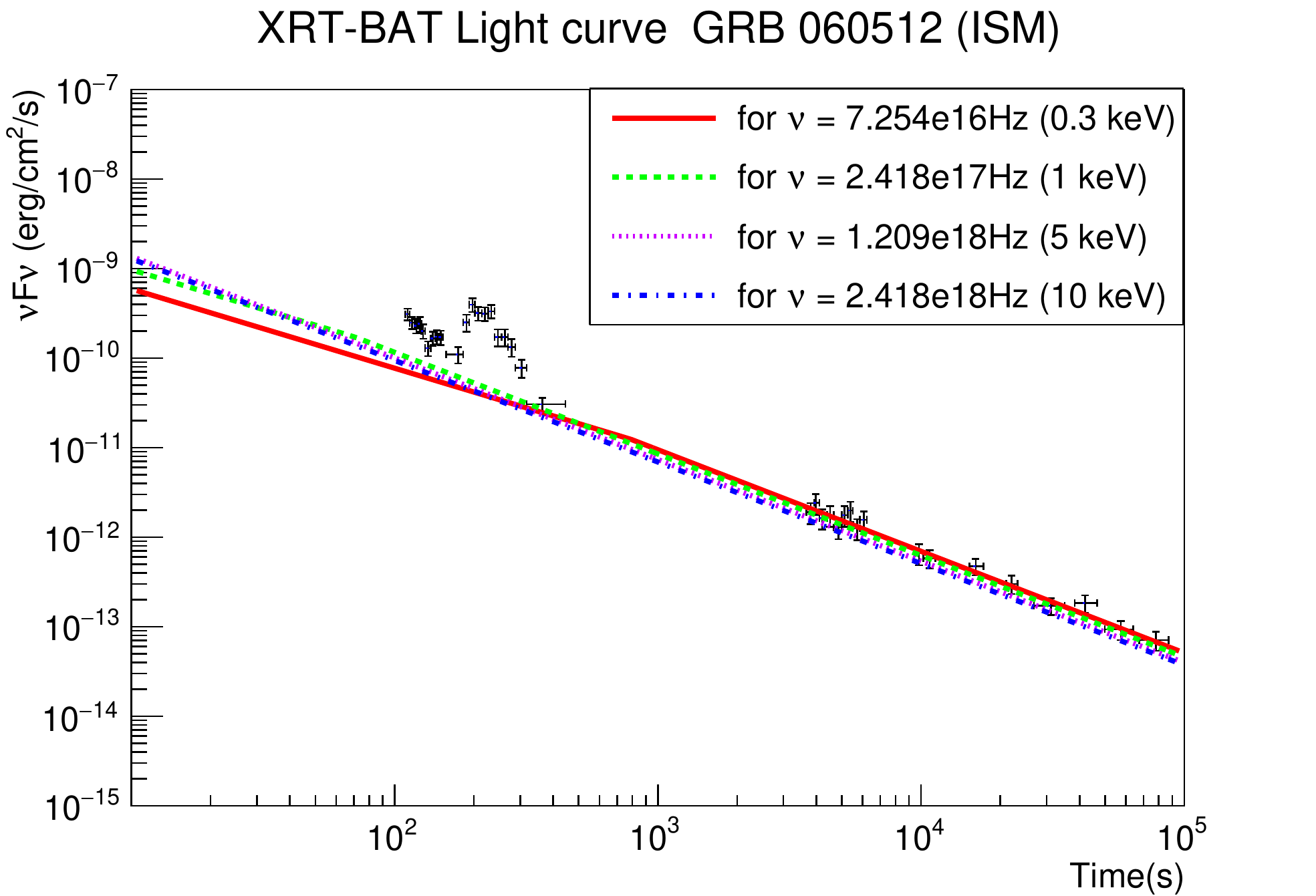}
\includegraphics[trim =  0 21 0 10, width=0.85\columnwidth]{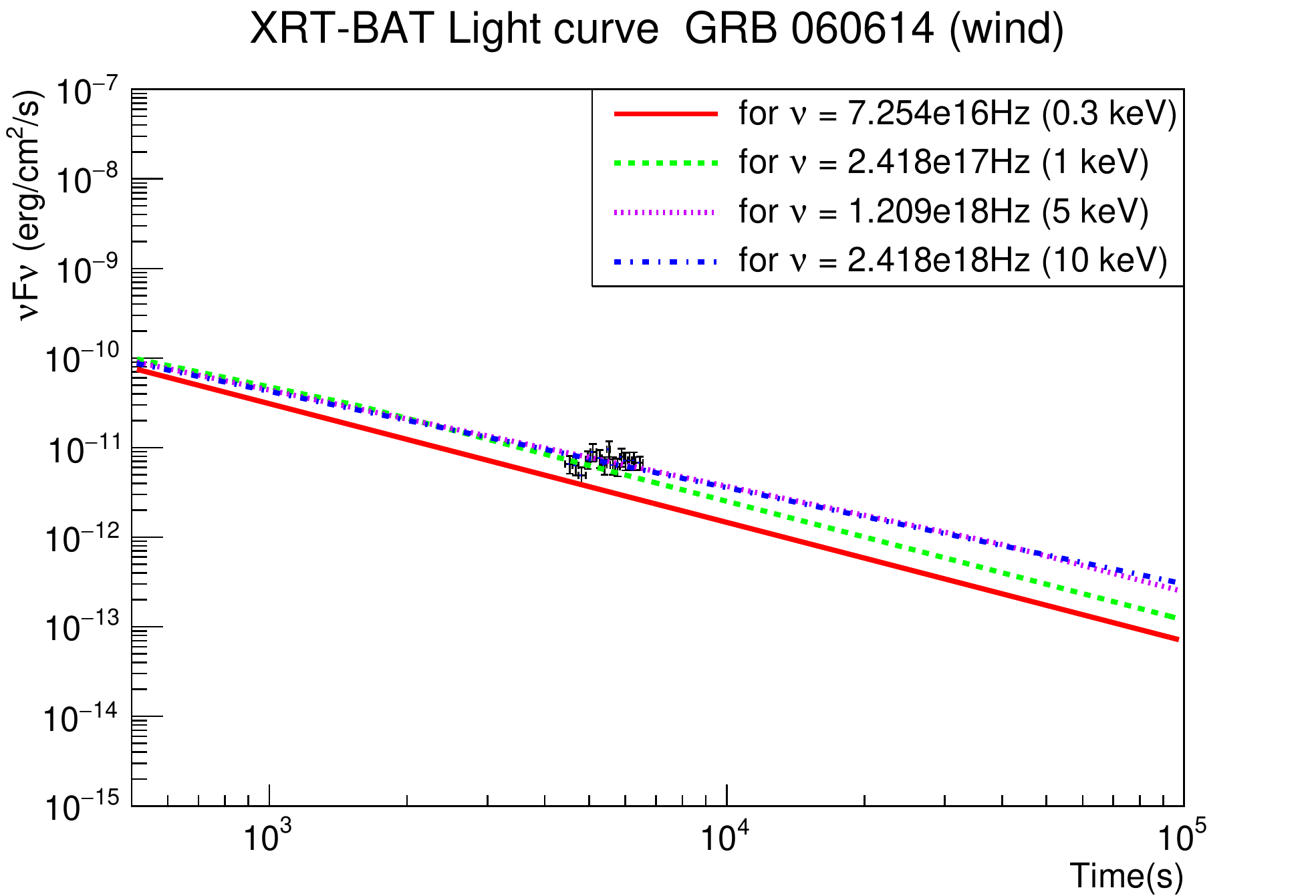}
\vspace{0.5 cm}
\includegraphics[trim =  0 21 0 10, width=0.85\columnwidth]{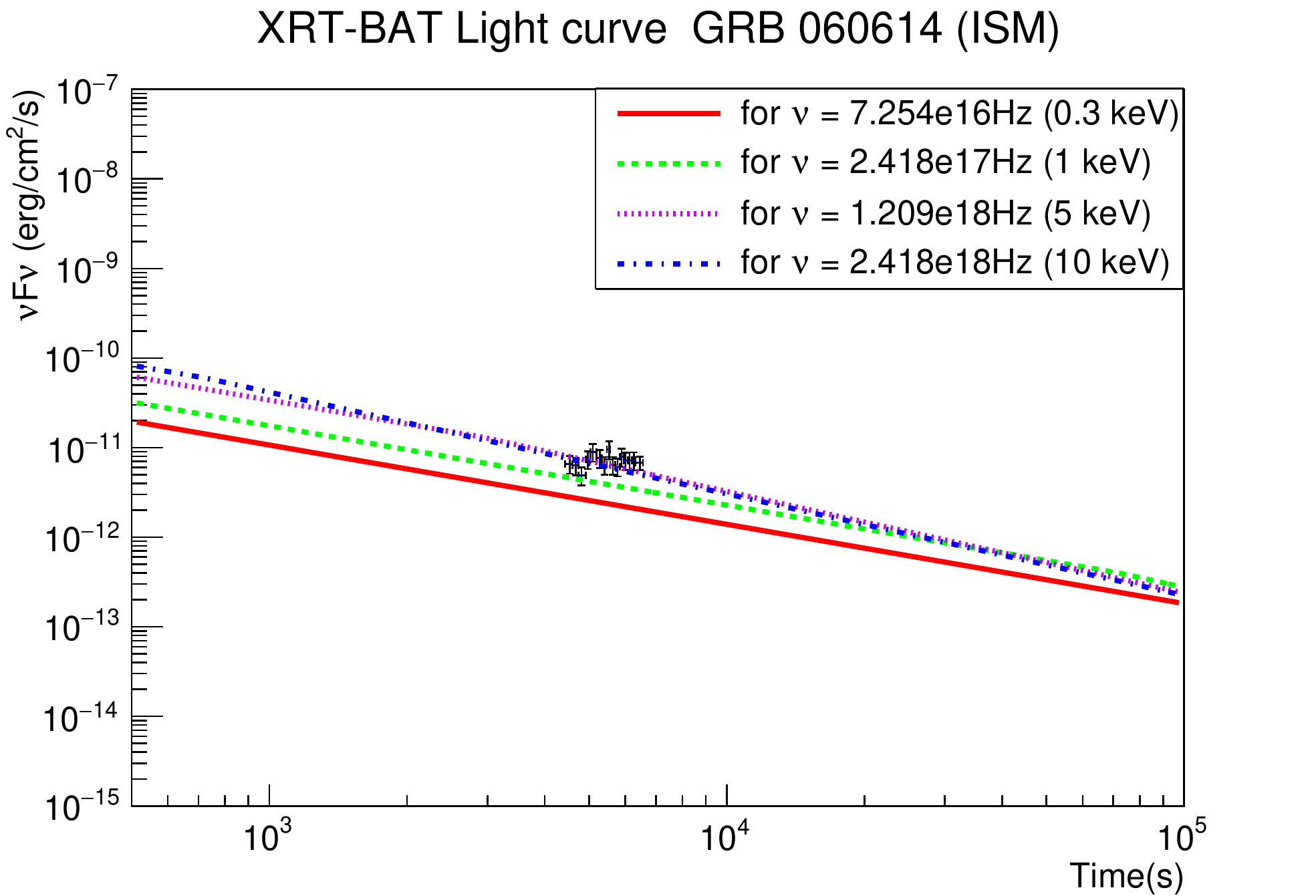}
\includegraphics[trim =  0 21 0 10, width=0.85\columnwidth]{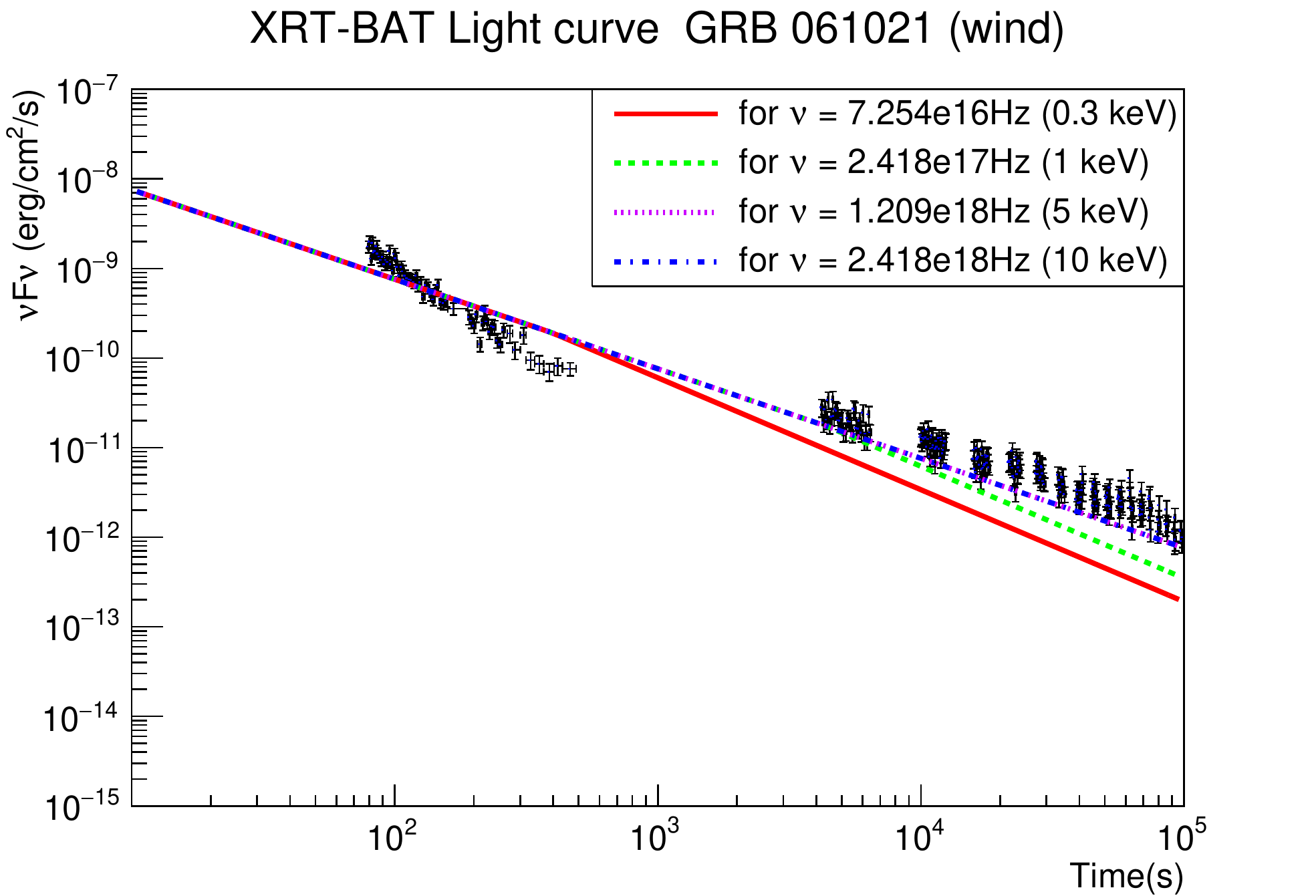}
\vspace{0.5 cm}
\includegraphics[trim =  0 21 0 10, width=0.85\columnwidth]{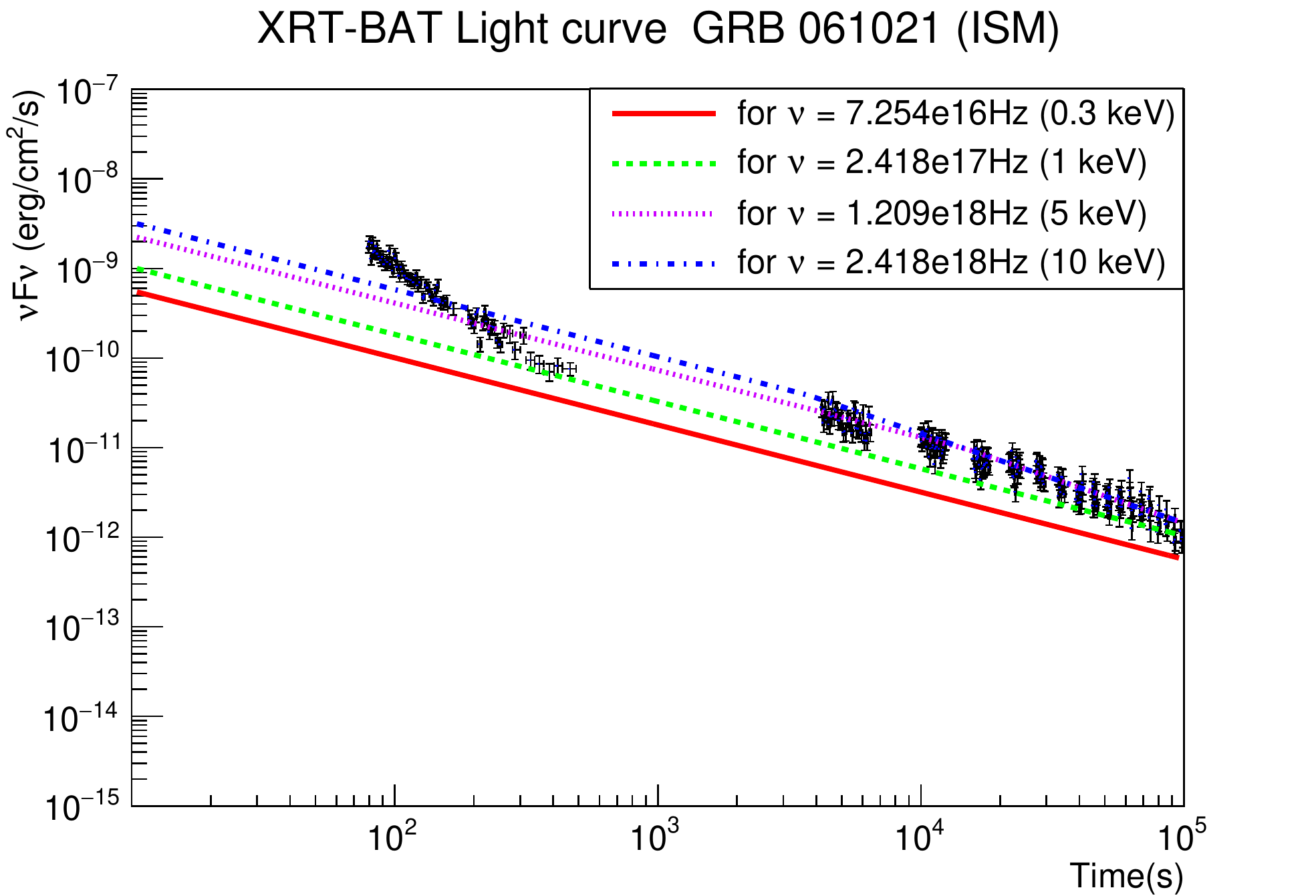}
\caption{\label{fig2lc} Same as Fig.~\ref{fig1lc} but for GRB~060218, GRB~060512, GRB~060614 and GRB~061021.}
\end{figure*}

\begin{figure*}[th!]
\includegraphics[trim =  0 21 0 10, width=0.85\columnwidth]{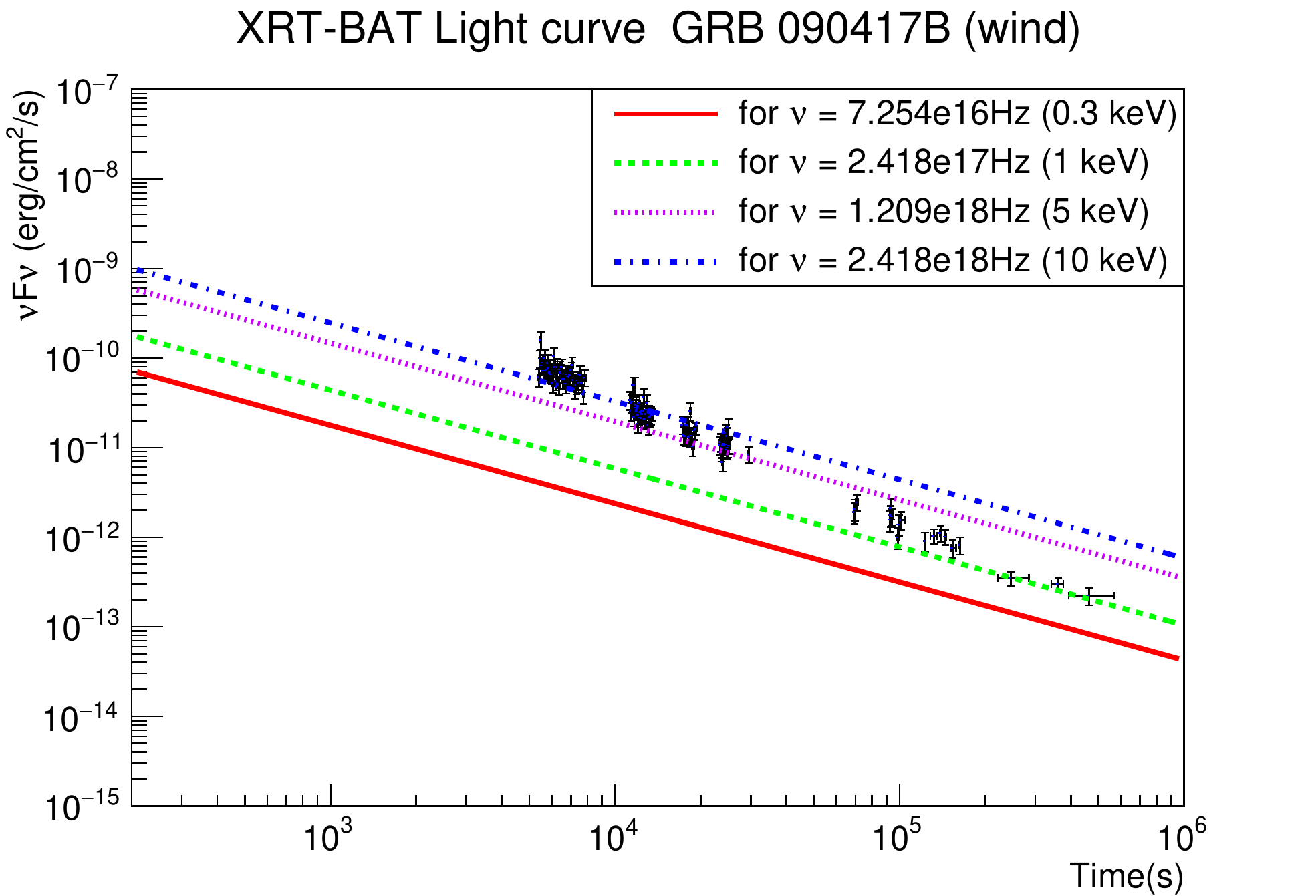}
\vspace{0.5 cm} 
\includegraphics[trim =  0 21 0 10, width=0.85\columnwidth]{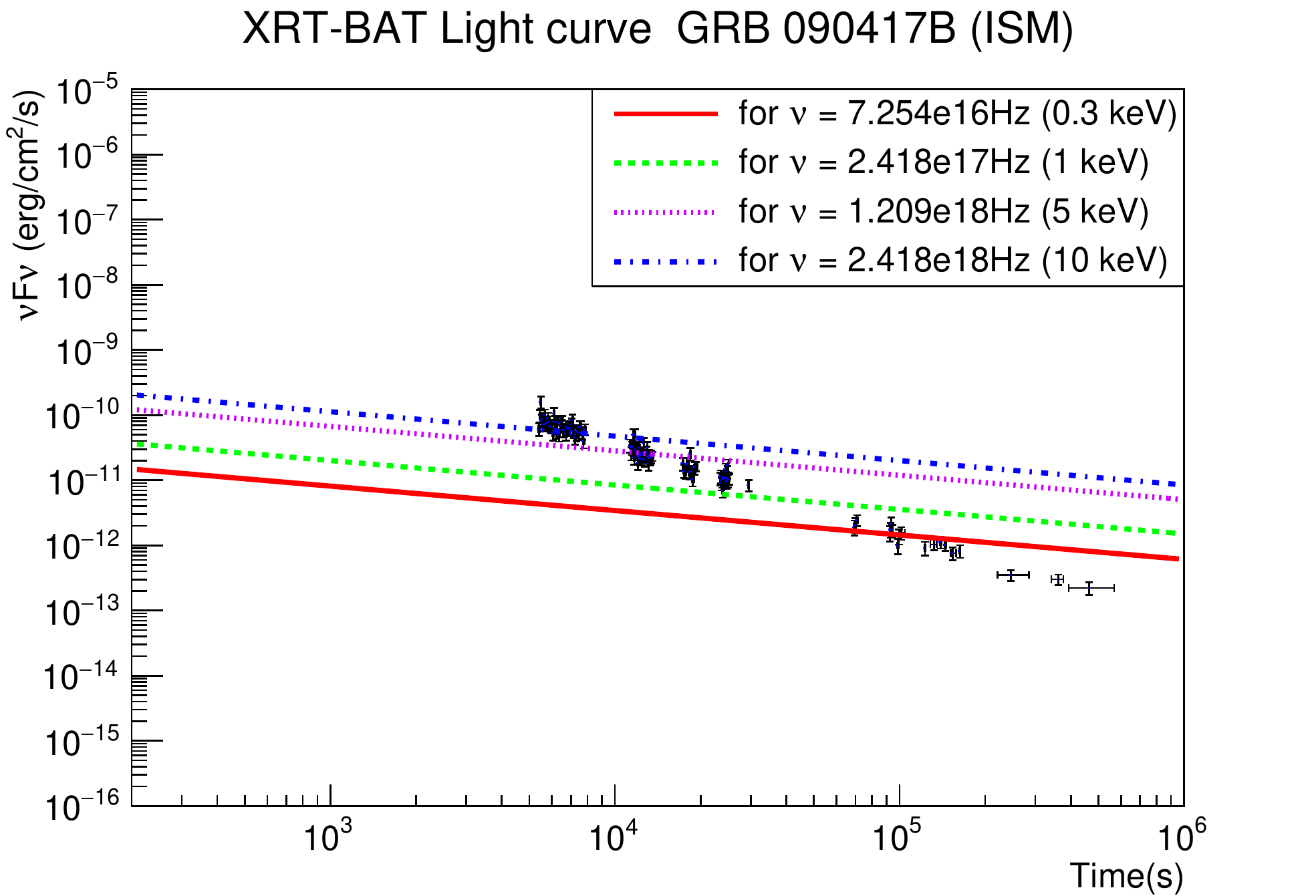}
\includegraphics[trim =  0 21 0 10, width=0.85\columnwidth]{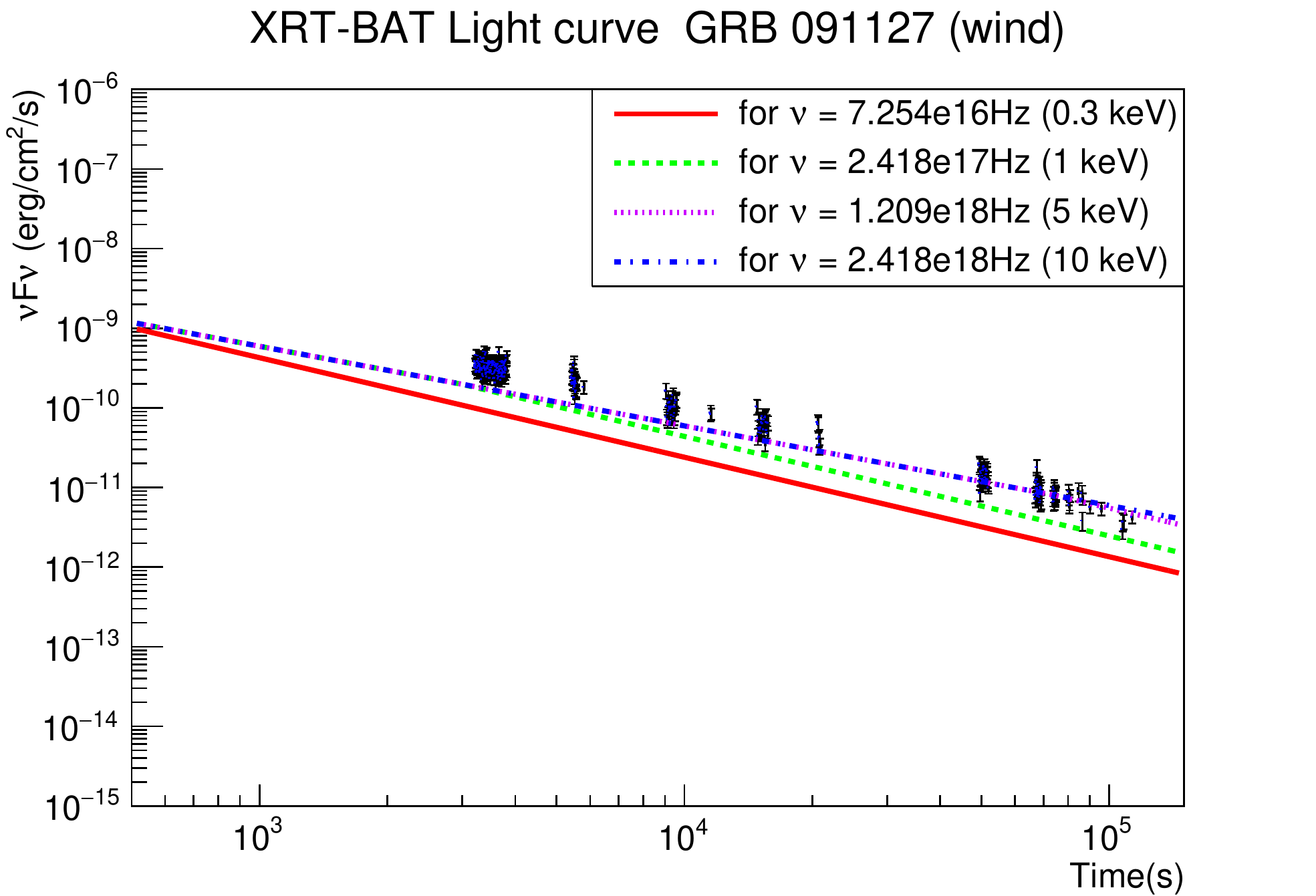}
\vspace{0.5 cm} 
\includegraphics[trim =  0 21 0 10, width=0.85\columnwidth]{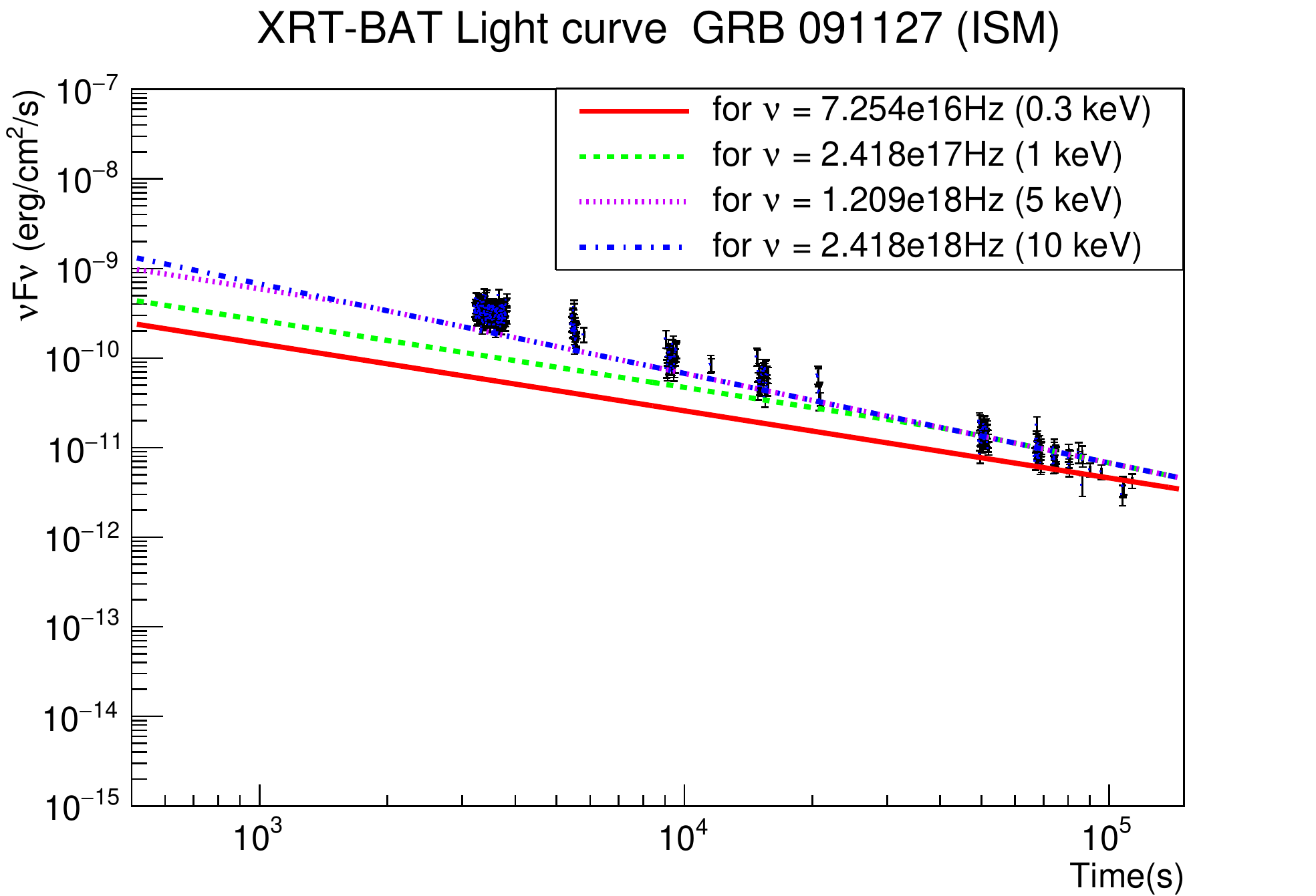}
\includegraphics[trim =  0 21 0 10, width=0.85\columnwidth]{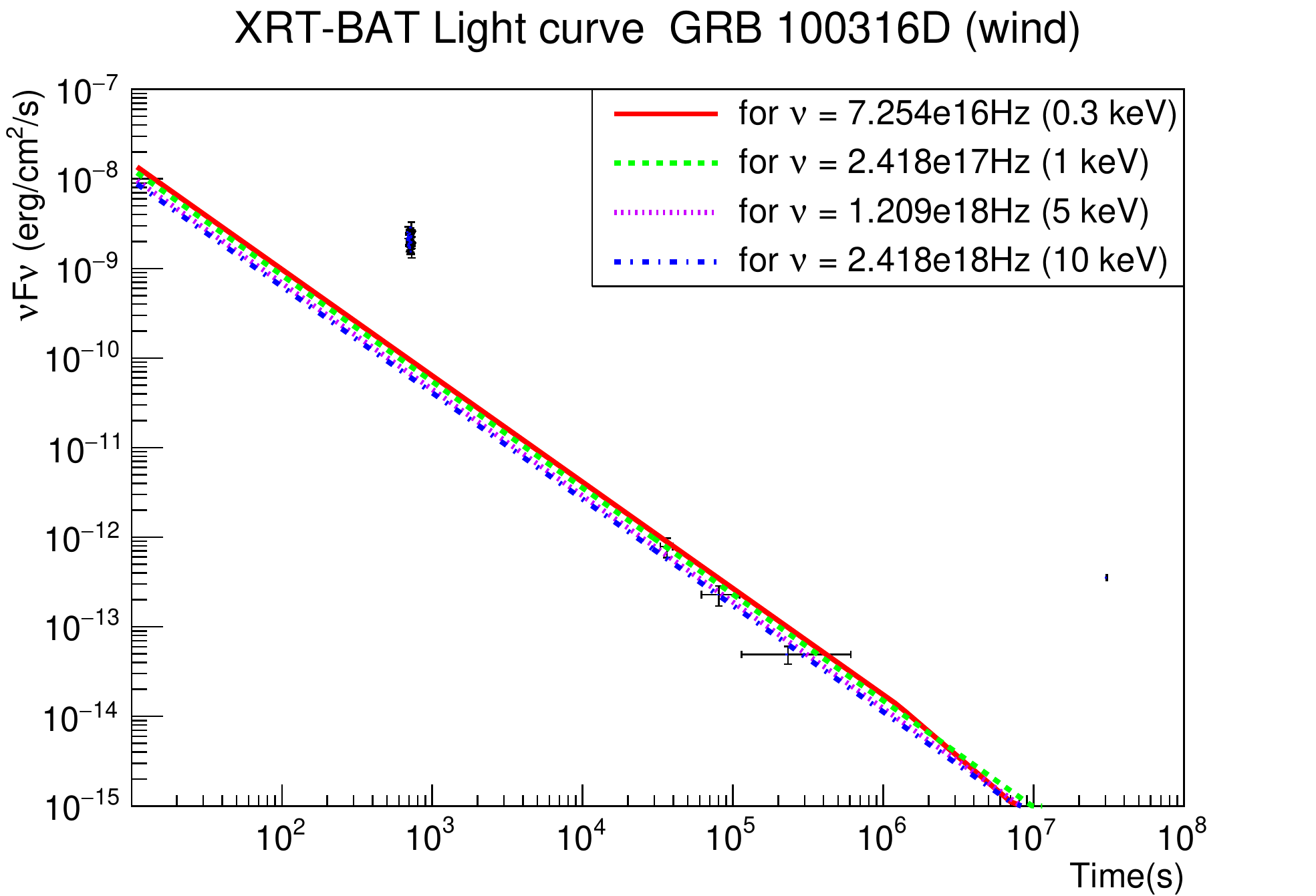}
\vspace{0.5 cm} 
\includegraphics[trim =  0 21 0 10, width=0.85\columnwidth]{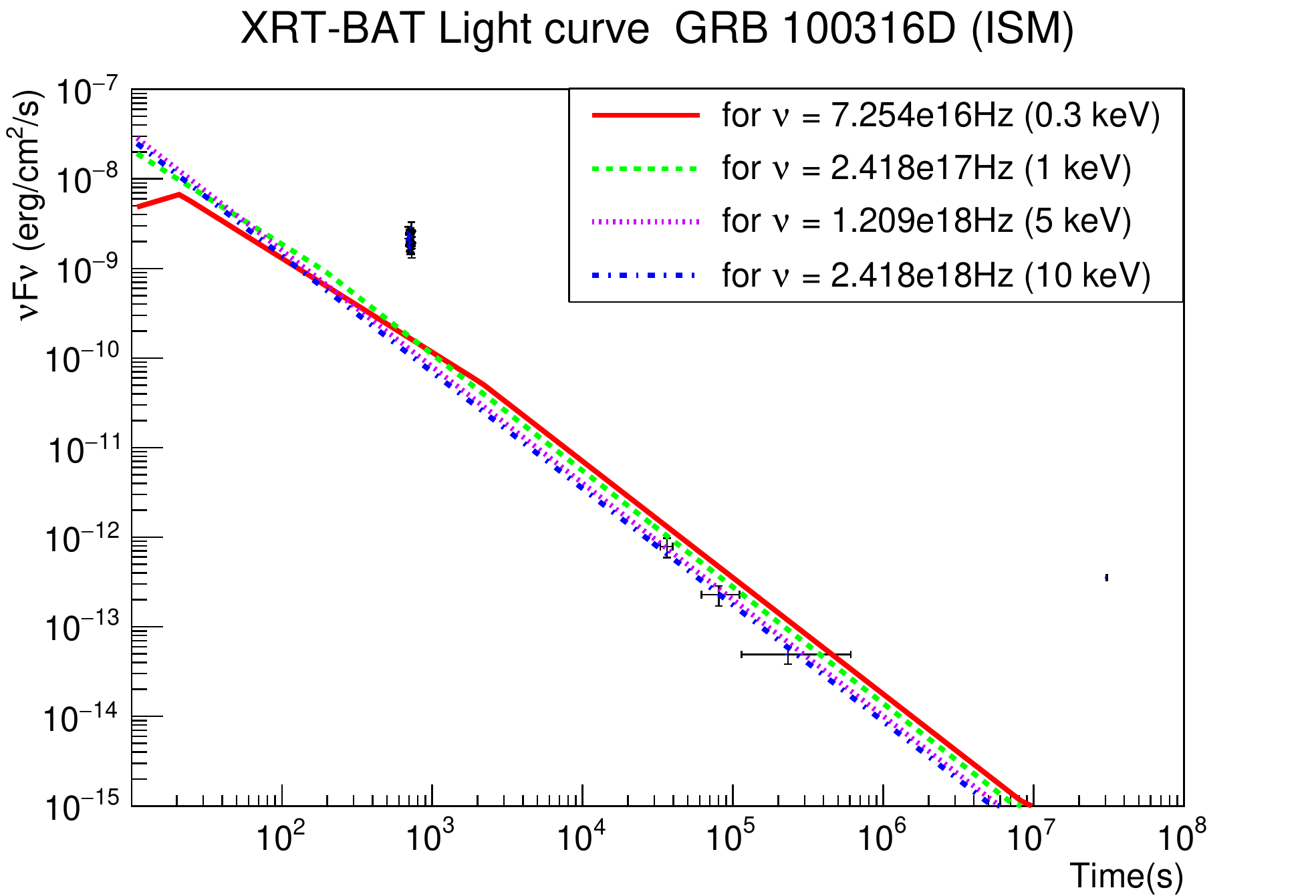}
\includegraphics[trim =  0 21 0 10, width=0.85\columnwidth]{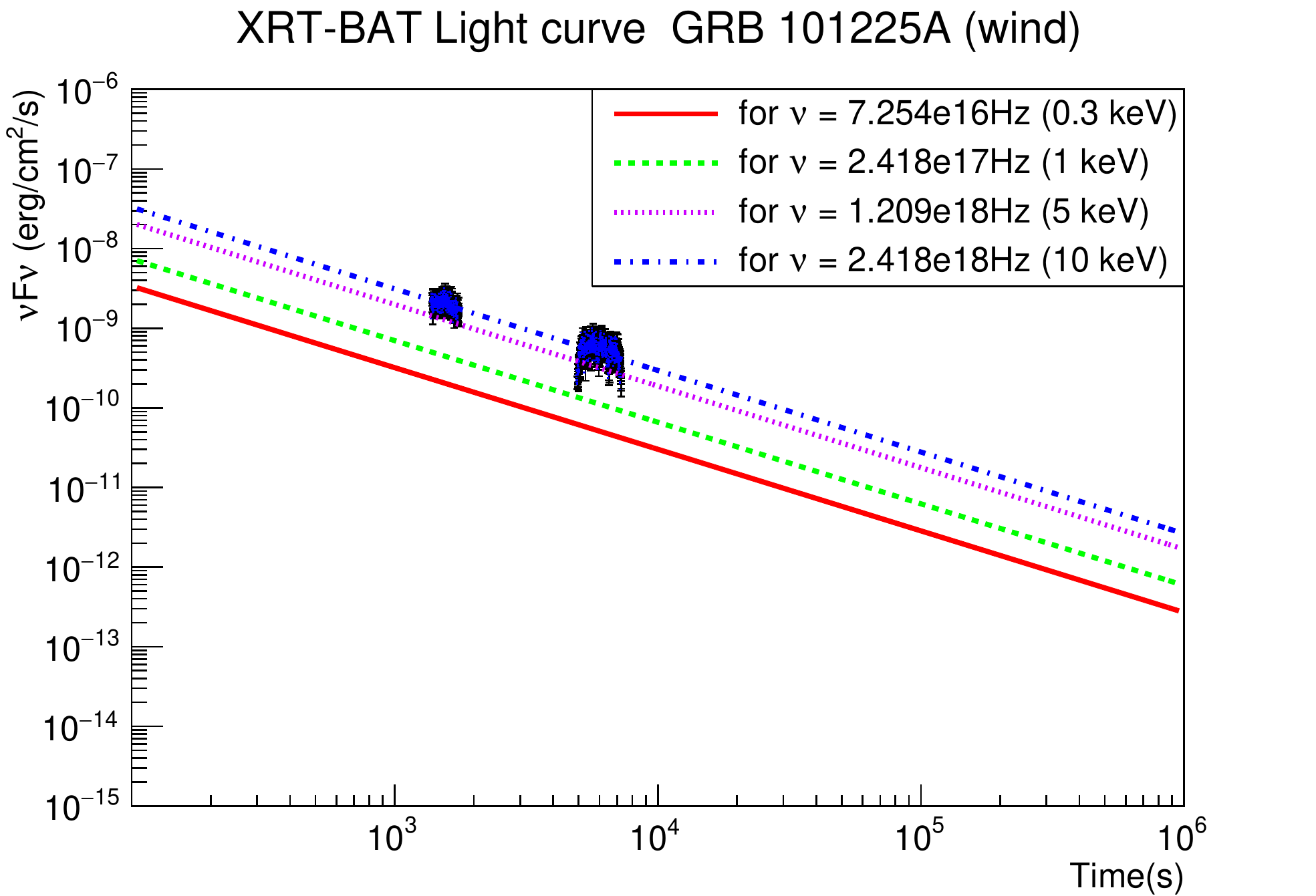}
\vspace{0.5 cm} 
\includegraphics[trim =  0 21 0 10, width=0.85\columnwidth]{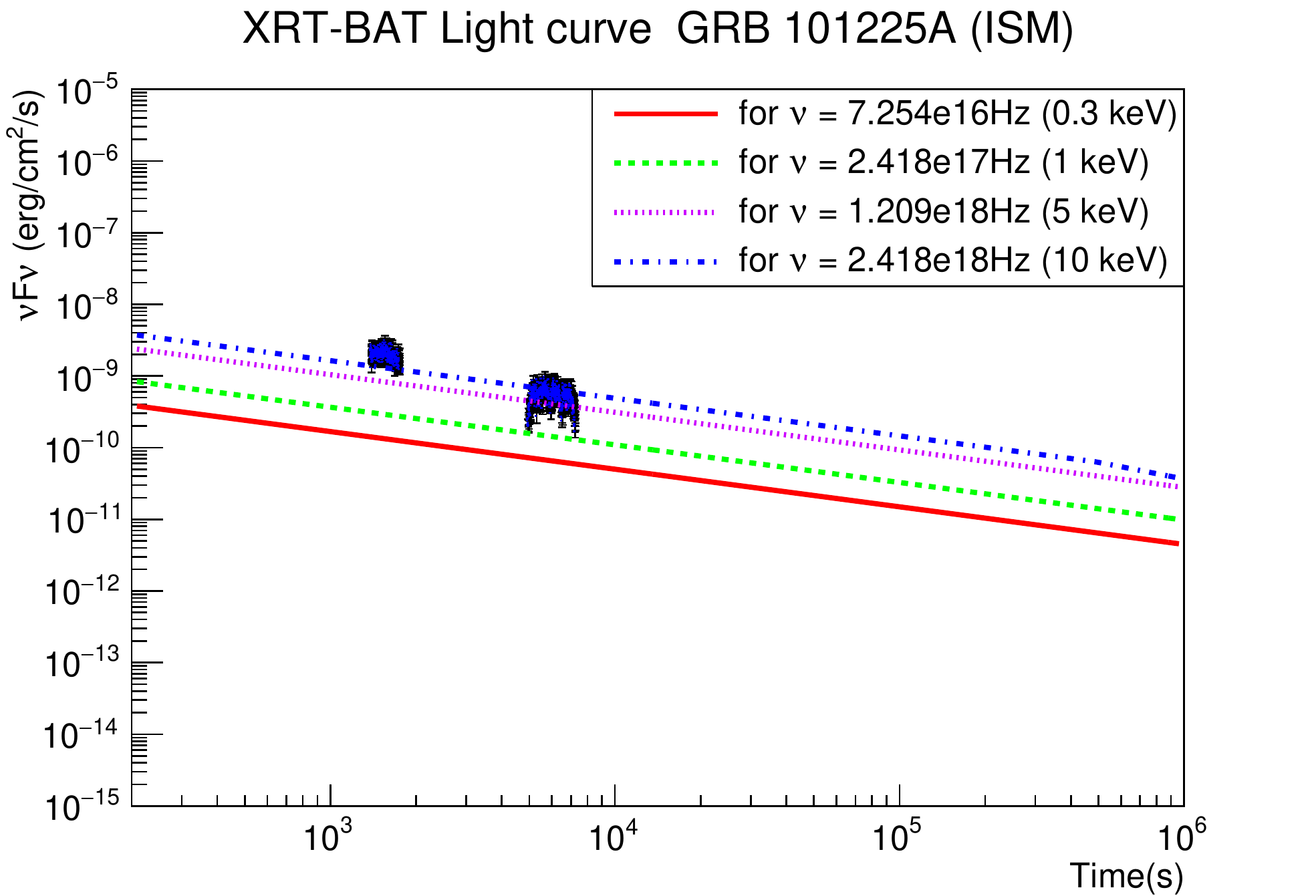}
\caption{\label{fig3lc} Same as Fig.~\ref{fig1lc} but for GRB~090417B, GRB~091127B, GRB~100316D and GRB~101225A.}
\end{figure*}

\begin{figure*}[th!]
\includegraphics[trim =  0 21 0 10, width=0.85\columnwidth]{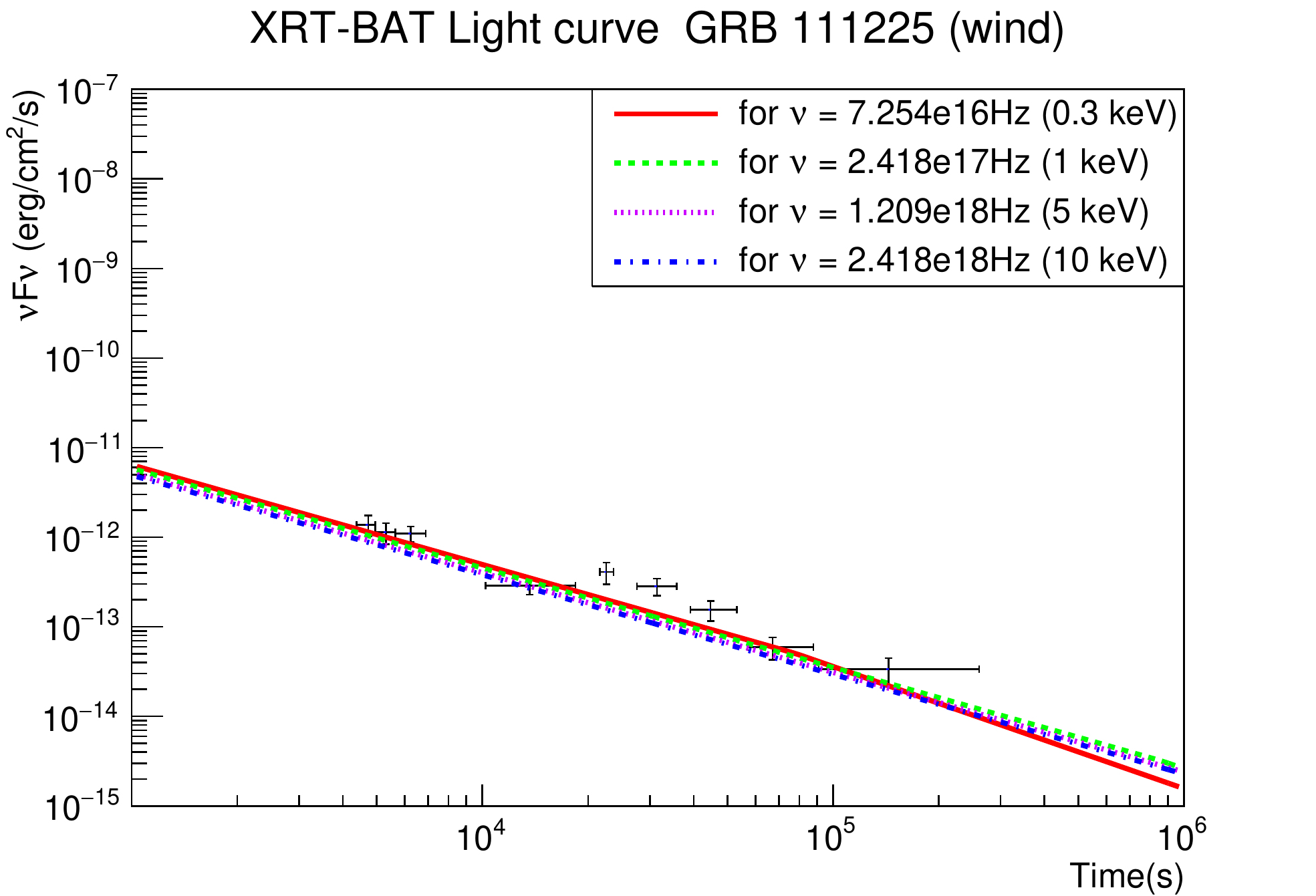}
\vspace{0.5 cm} 
\includegraphics[trim =  0 21 0 10, width=0.85\columnwidth]{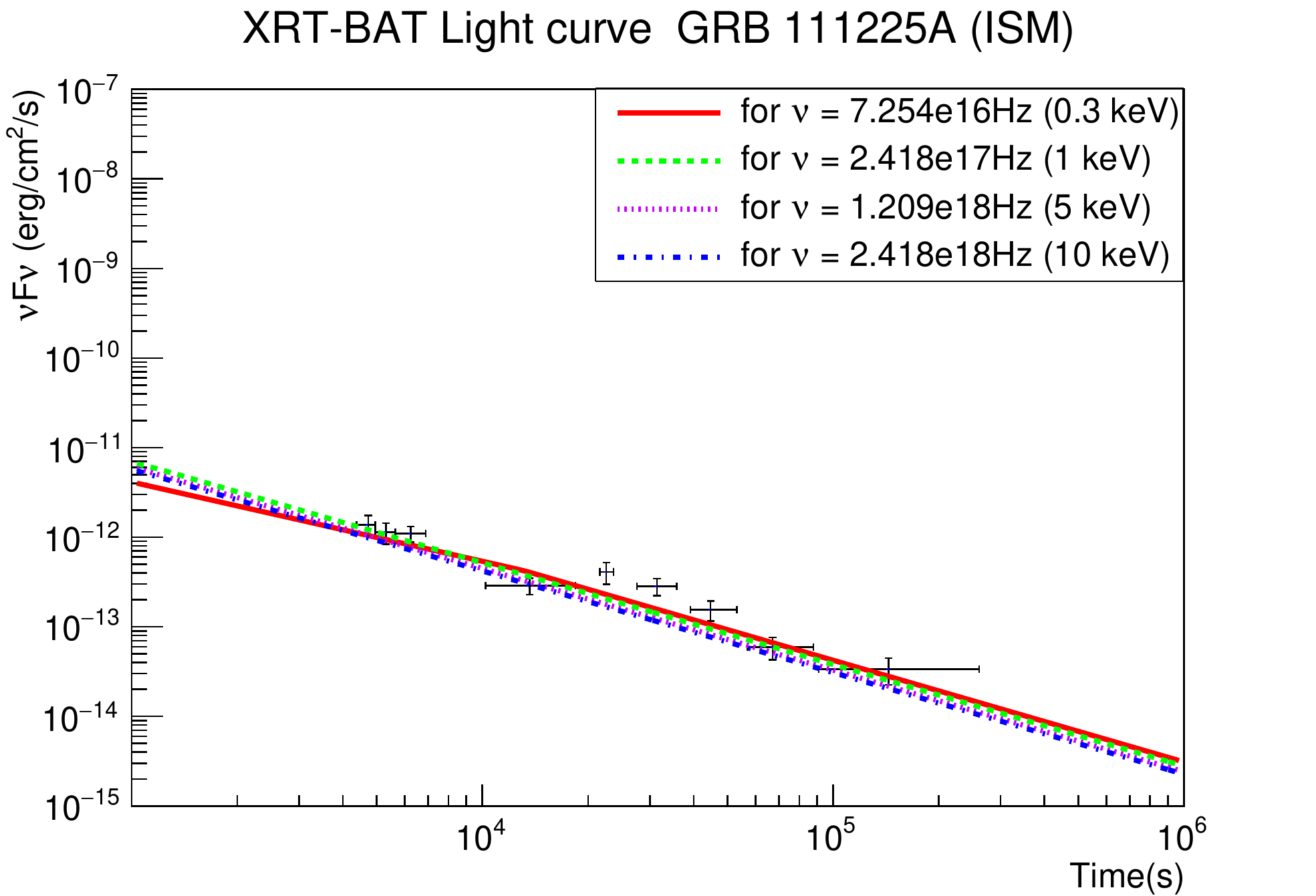}
\includegraphics[trim =  0 21 0 10, width=0.85\columnwidth]{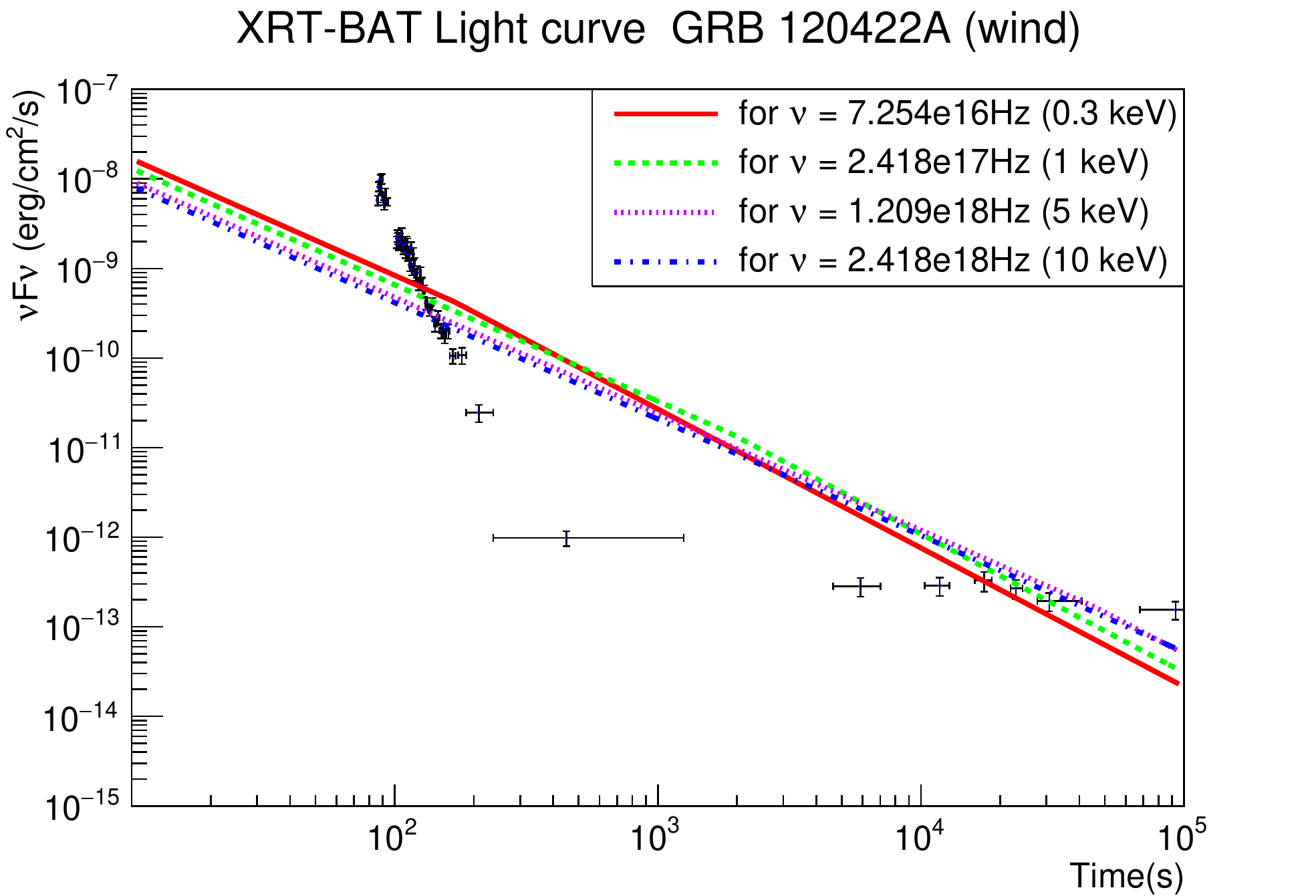}
\vspace{0.5 cm} 
\includegraphics[trim =  0 21 0 10, width=0.85\columnwidth]{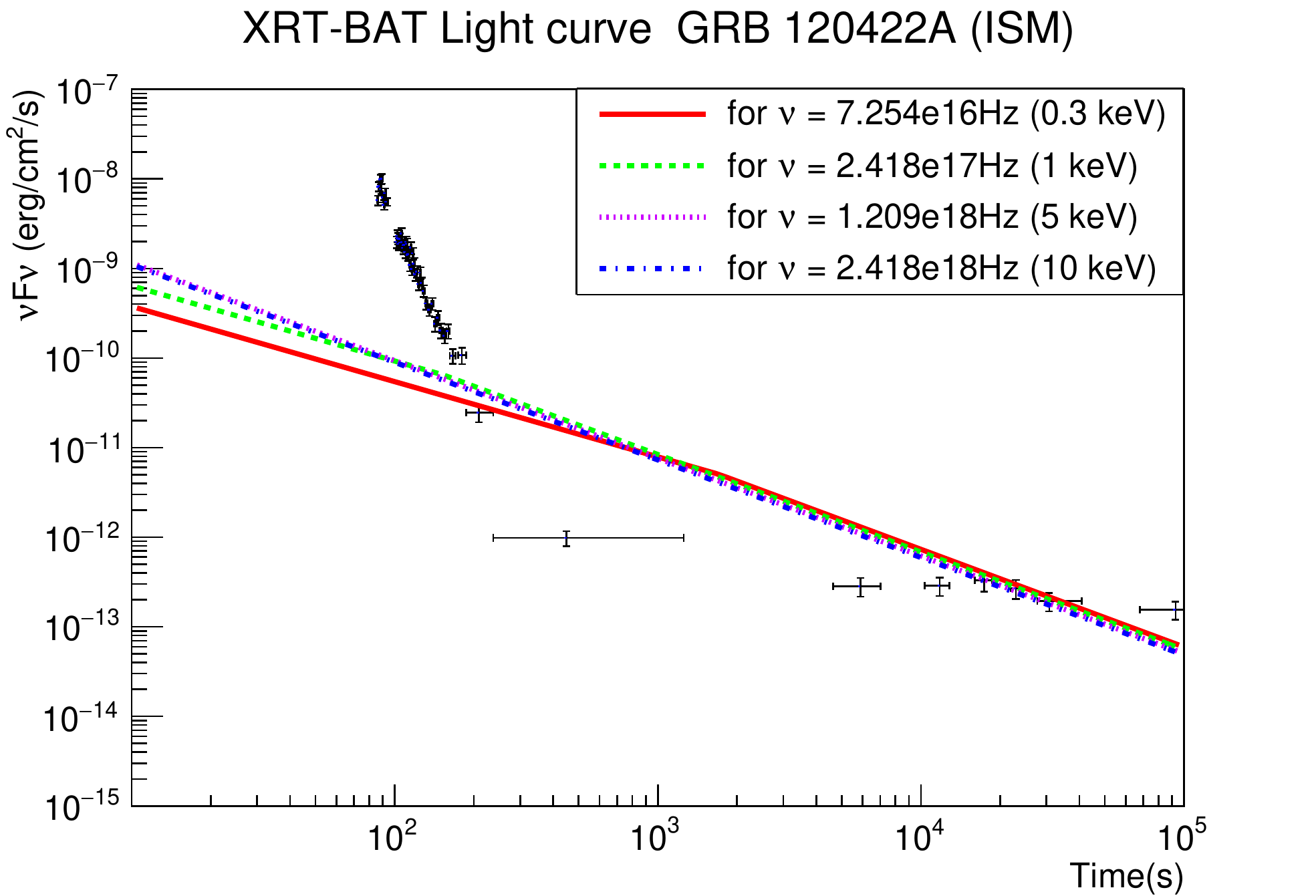}
\includegraphics[trim =  0 21 0 10, width=0.85\columnwidth]{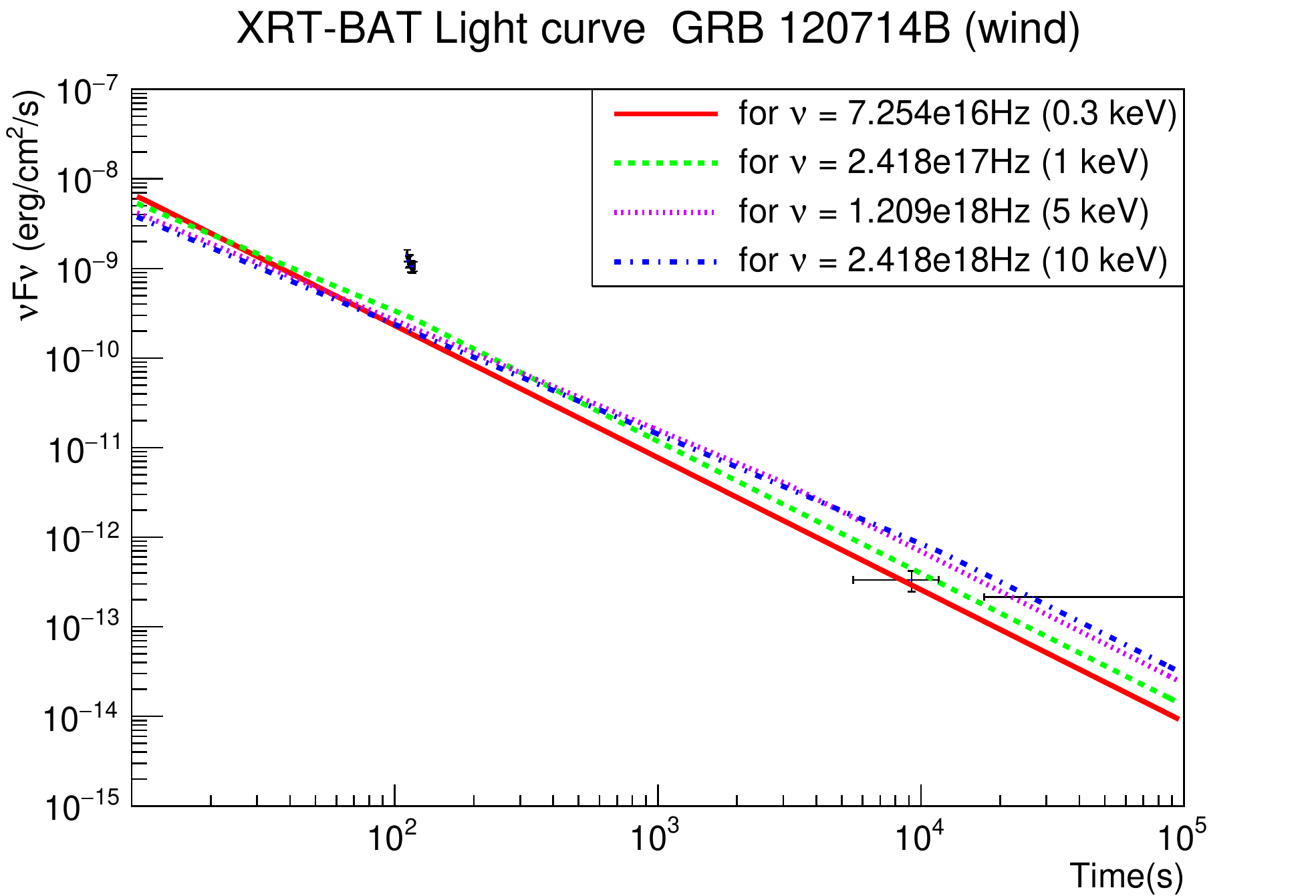}
\vspace{0.5 cm} 
\includegraphics[trim =  0 21 0 10, width=0.85\columnwidth]{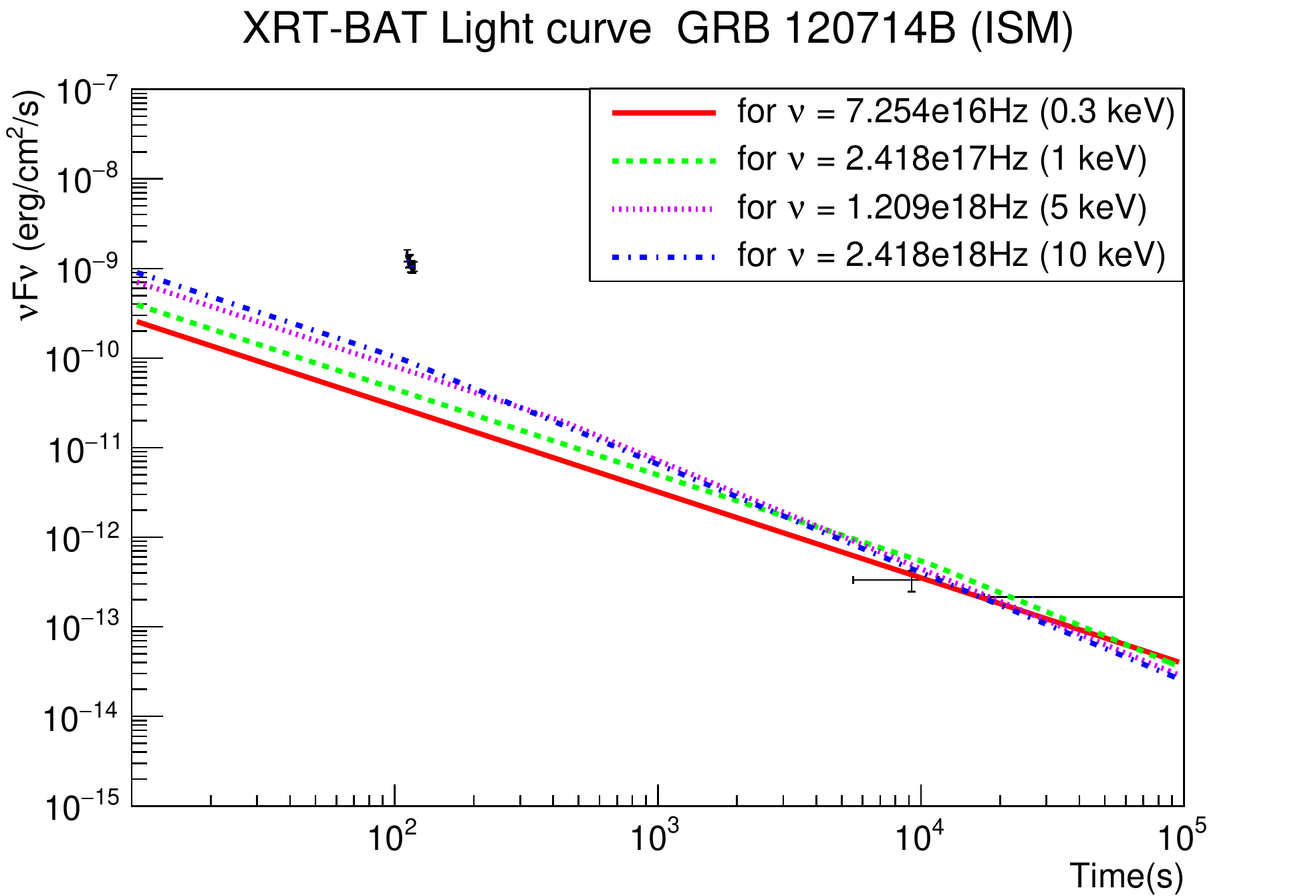}
\includegraphics[trim =  0 21 0 10, width=0.85\columnwidth]{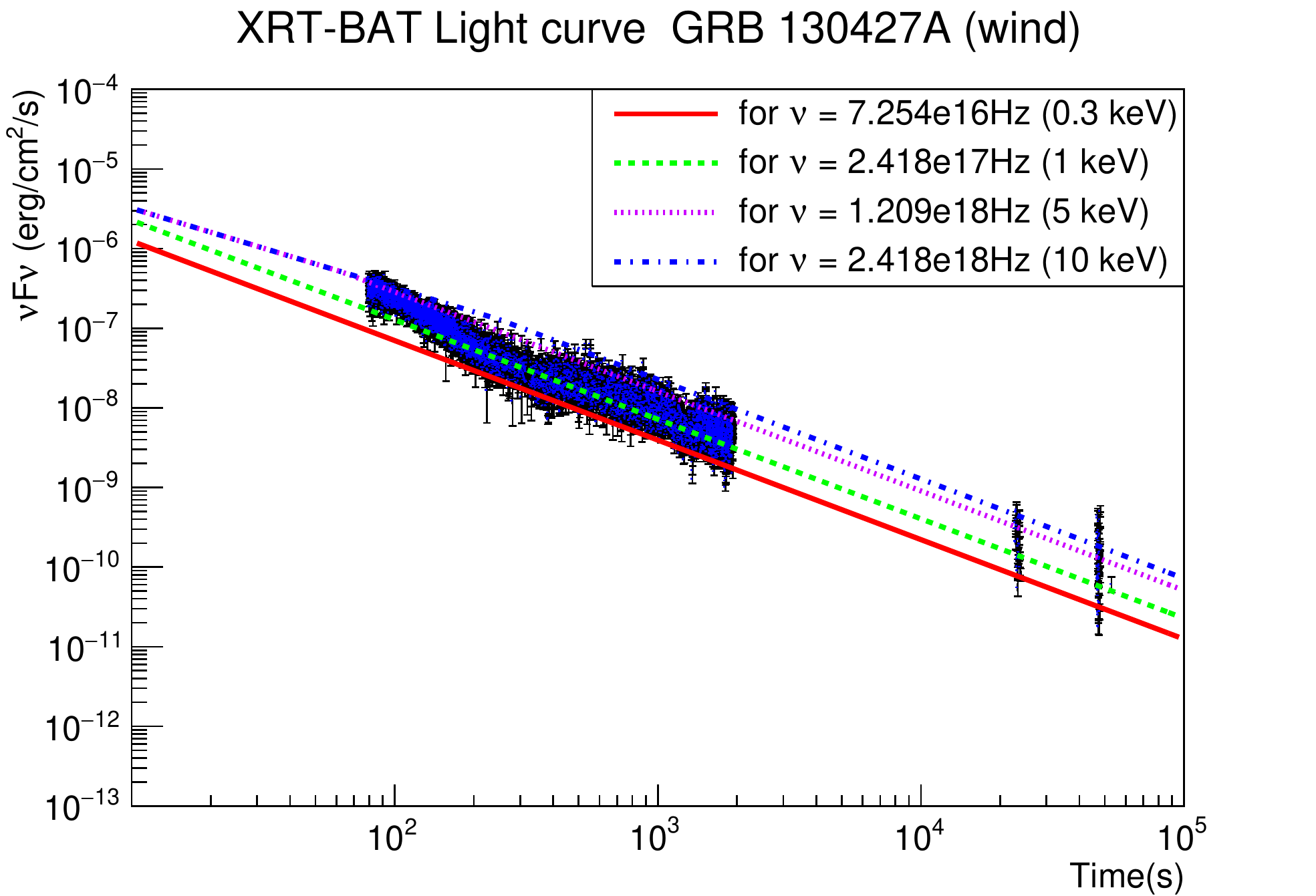}
\vspace{0.5 cm} 
\includegraphics[trim =  0 21 0 10, width=0.85\columnwidth]{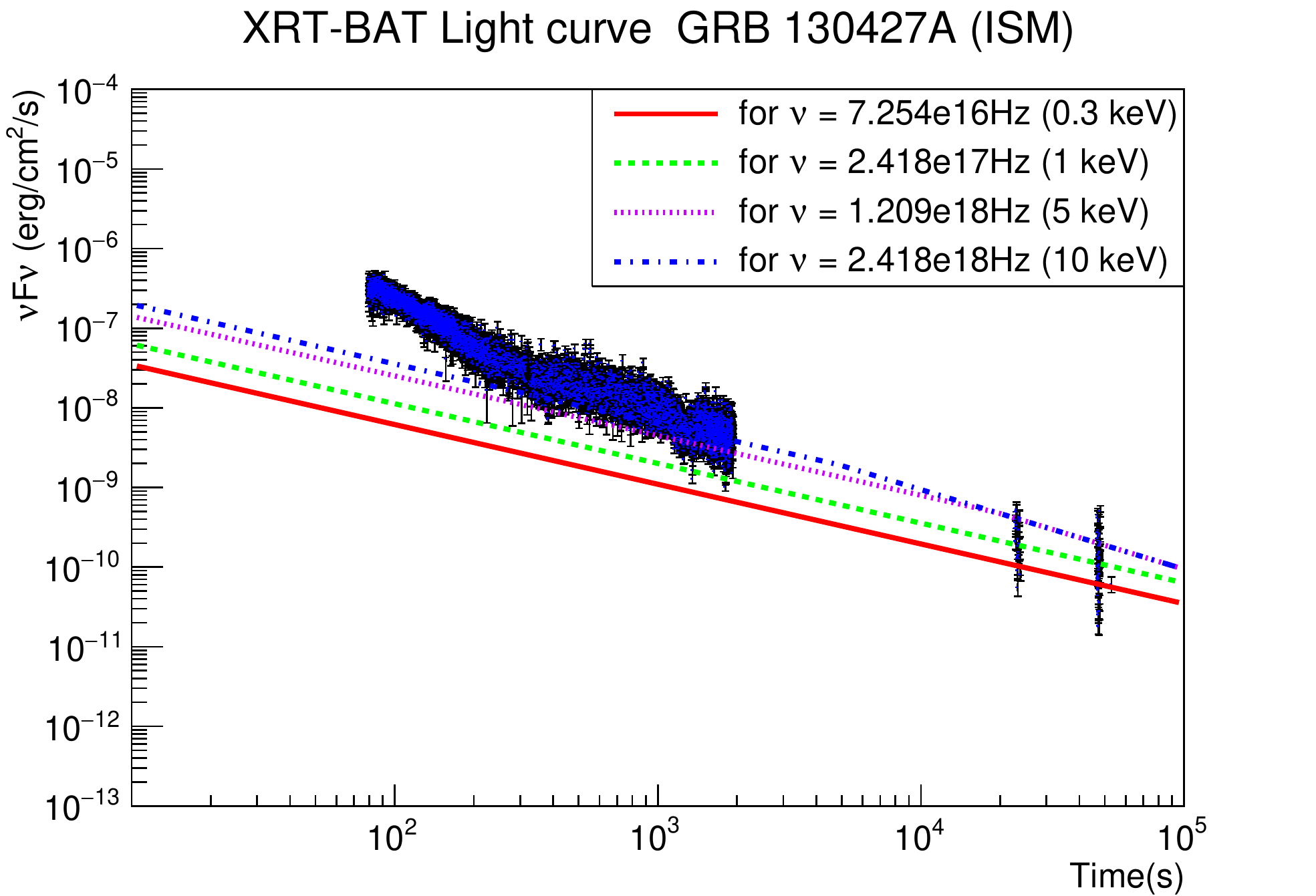}
\caption{\label{fig4lc} Same as Fig.~\ref{fig1lc} but for GRB~111225A, GRB~120422A, GRB~120714B and GRB~130427A.}
\end{figure*}

\begin{figure*}[th!]
\includegraphics[trim =  0 21 0 10, width=0.85\columnwidth]{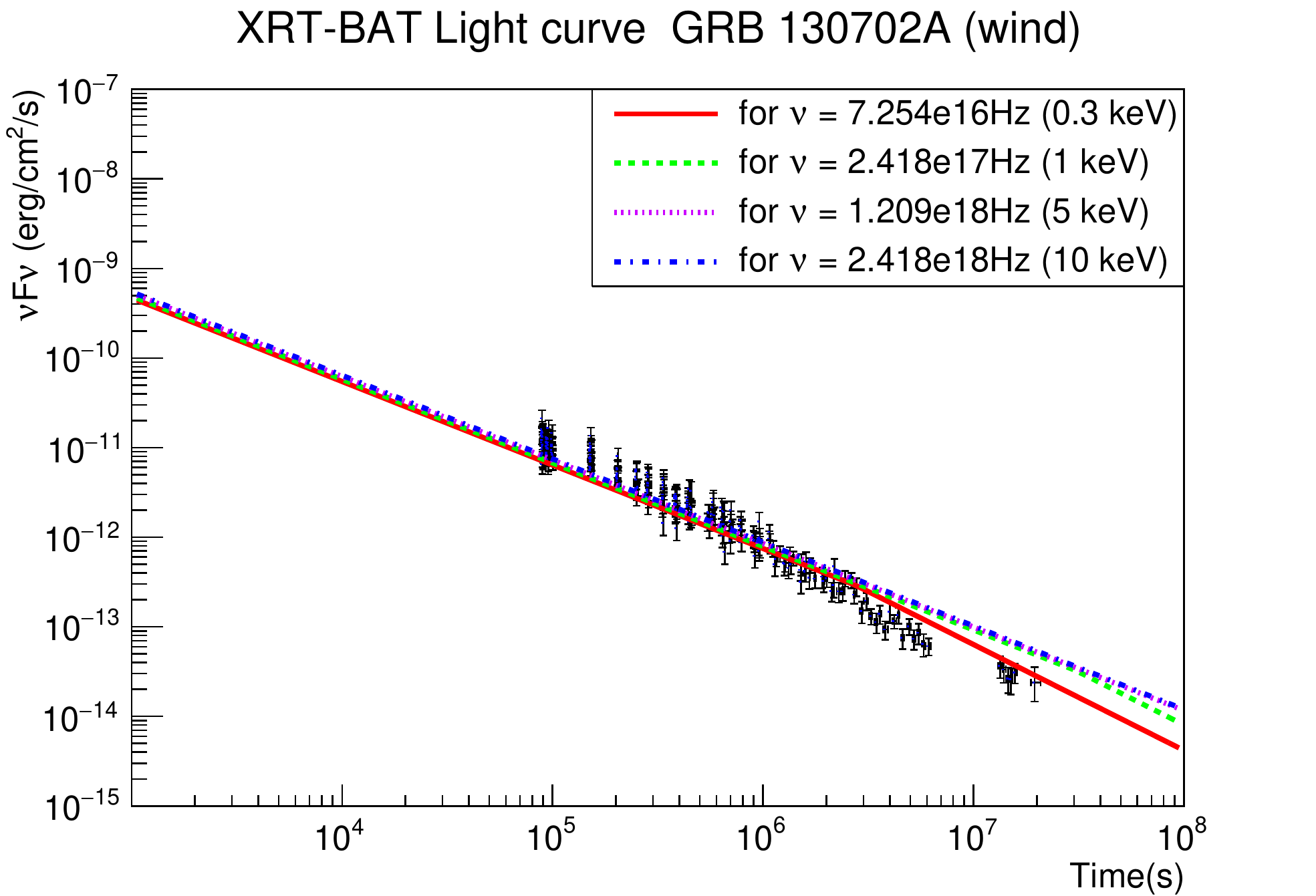}
\vspace{0.5 cm} 
\includegraphics[trim =  0 21 0 10, width=0.85\columnwidth]{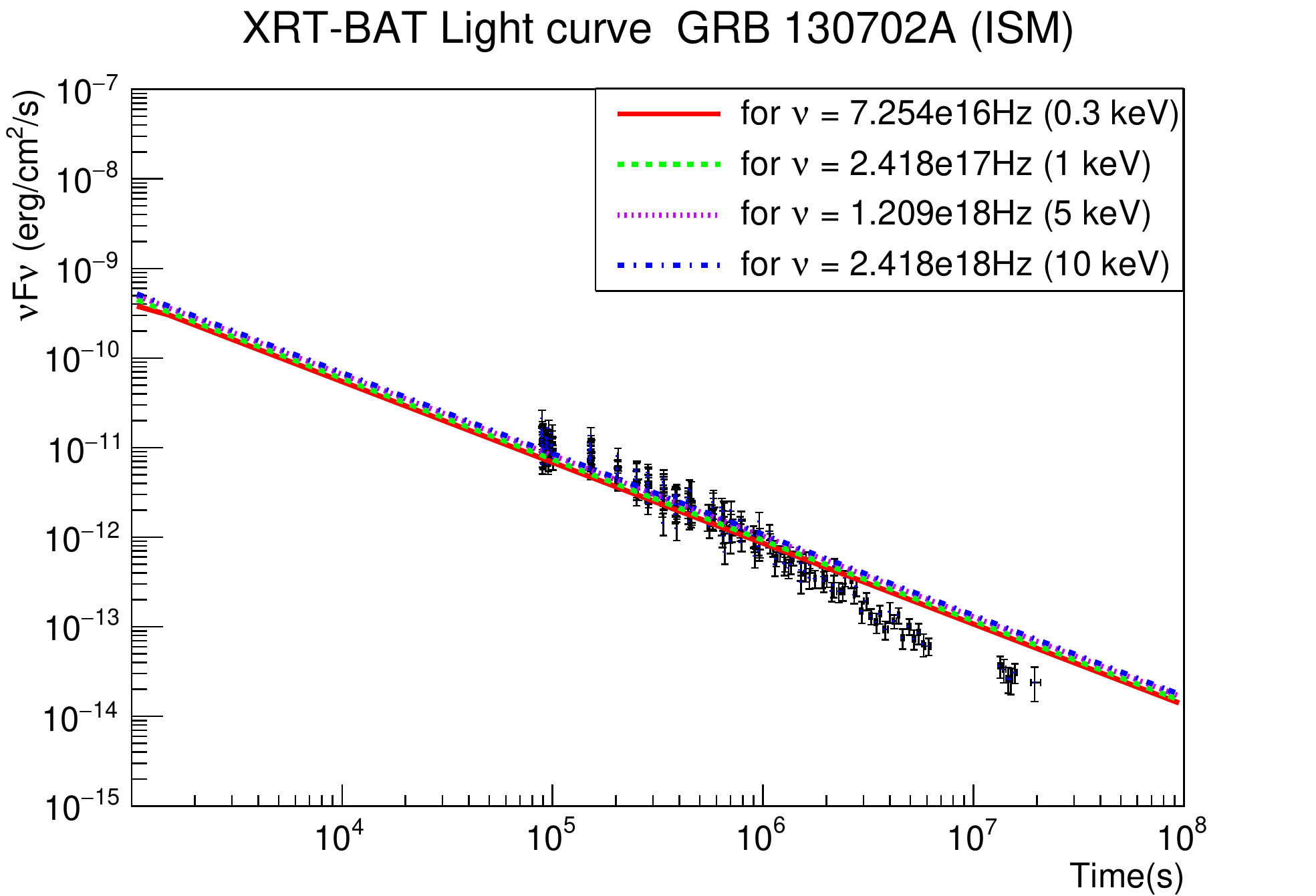}
\includegraphics[trim =  0 21 0 10, width=0.85\columnwidth]{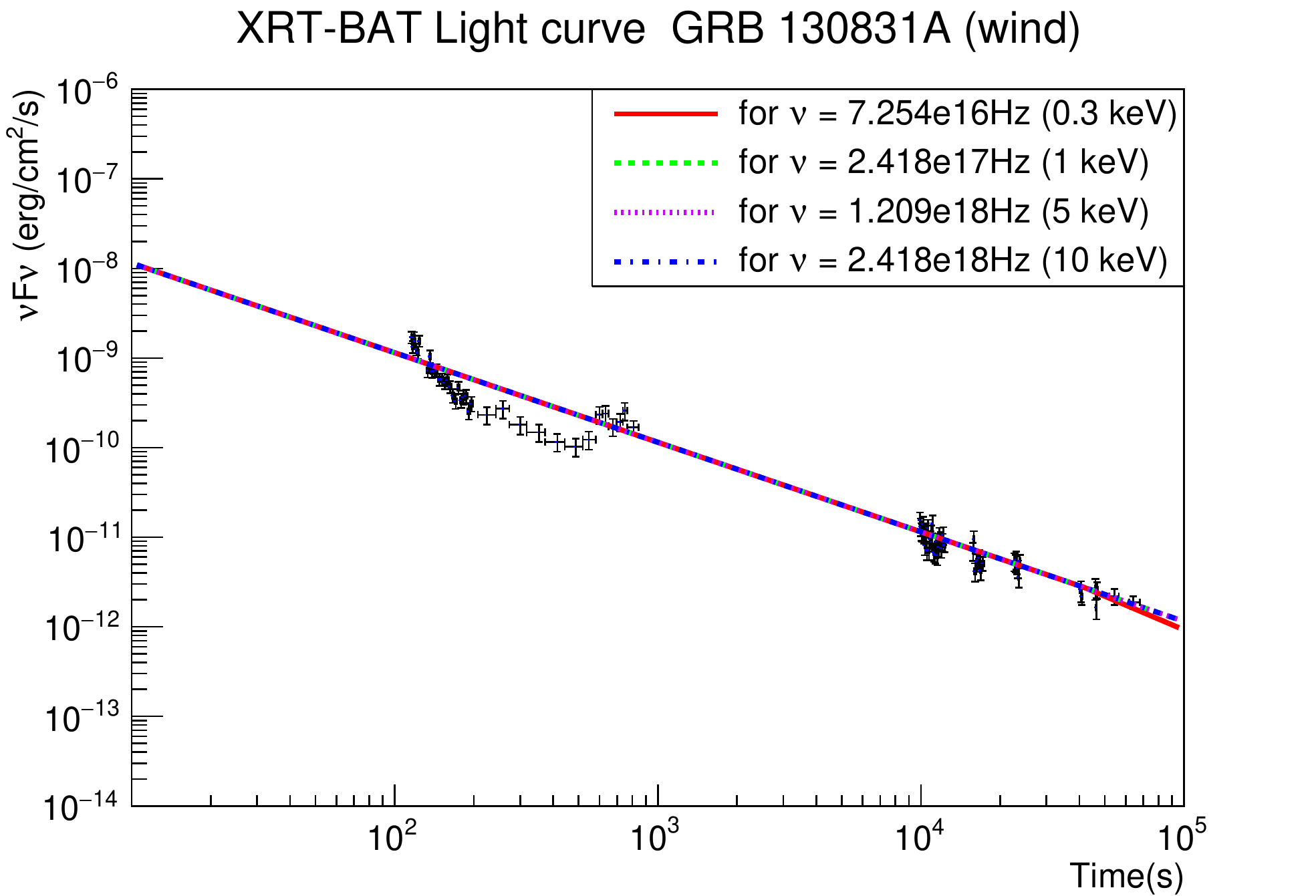}
\vspace{0.5 cm} 
\includegraphics[trim =  0 21 0 10, width=0.85\columnwidth]{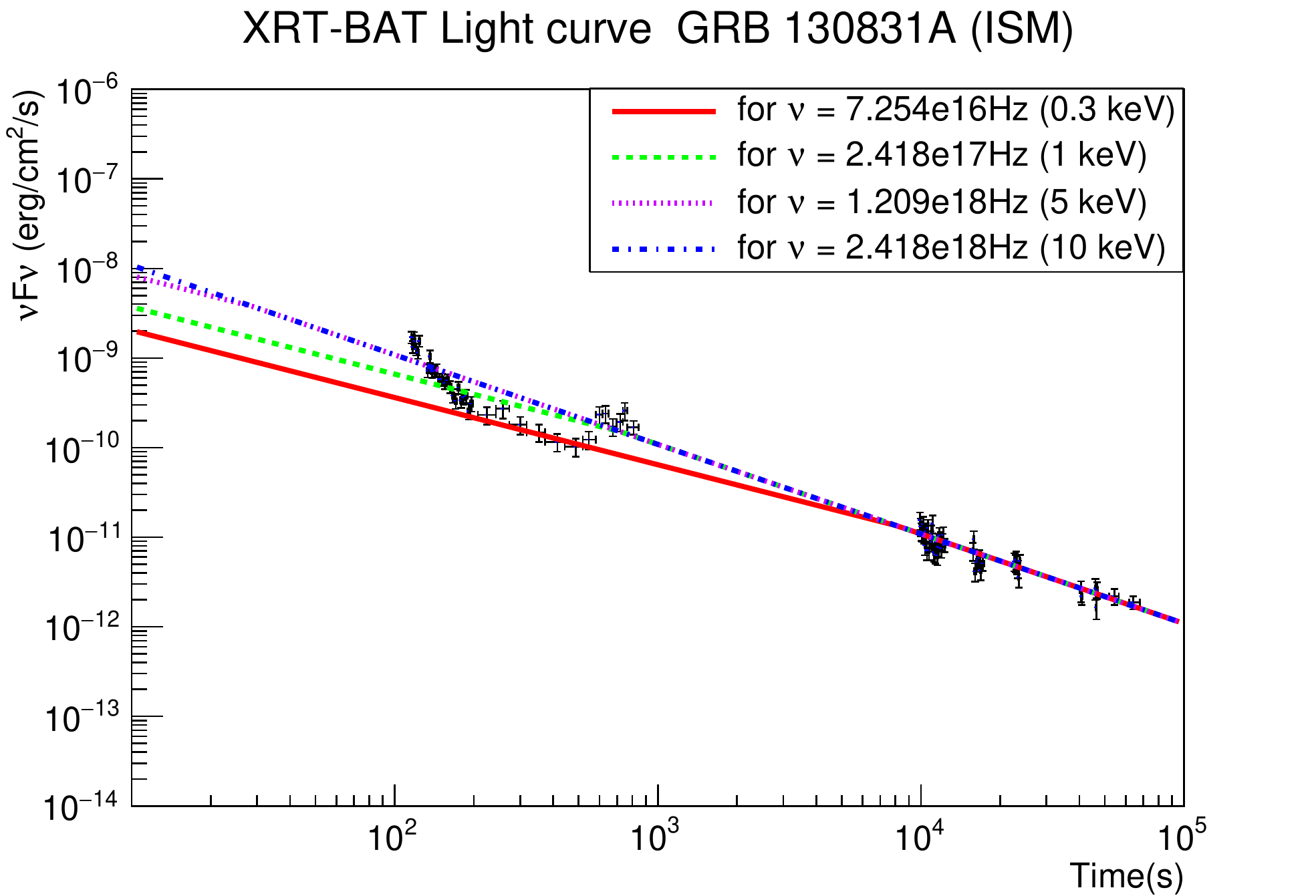}
\includegraphics[trim =  0 21 0 10, width=0.85\columnwidth]{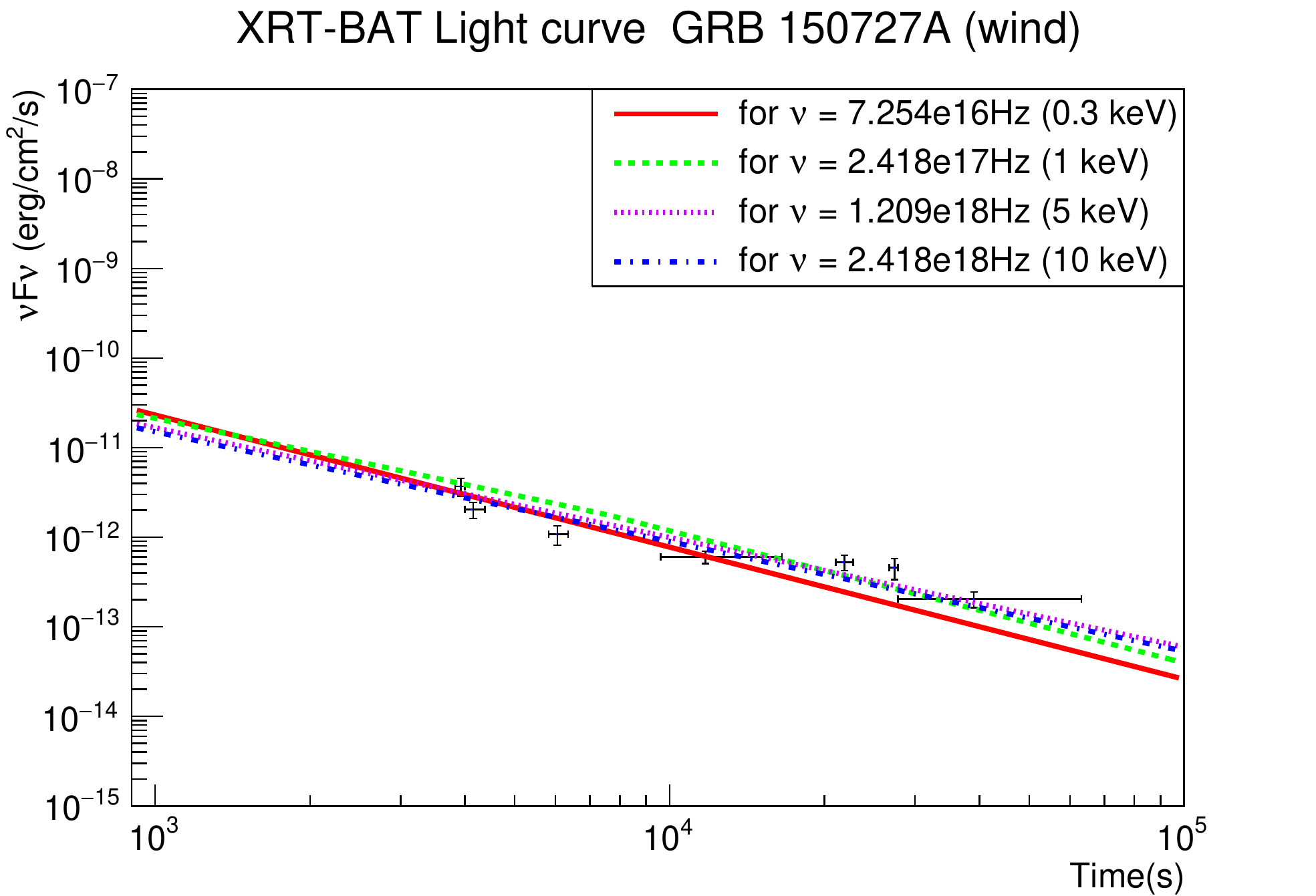}
\vspace{0.5cm}
\includegraphics[trim =  0 21 0 10, width=0.85\columnwidth]{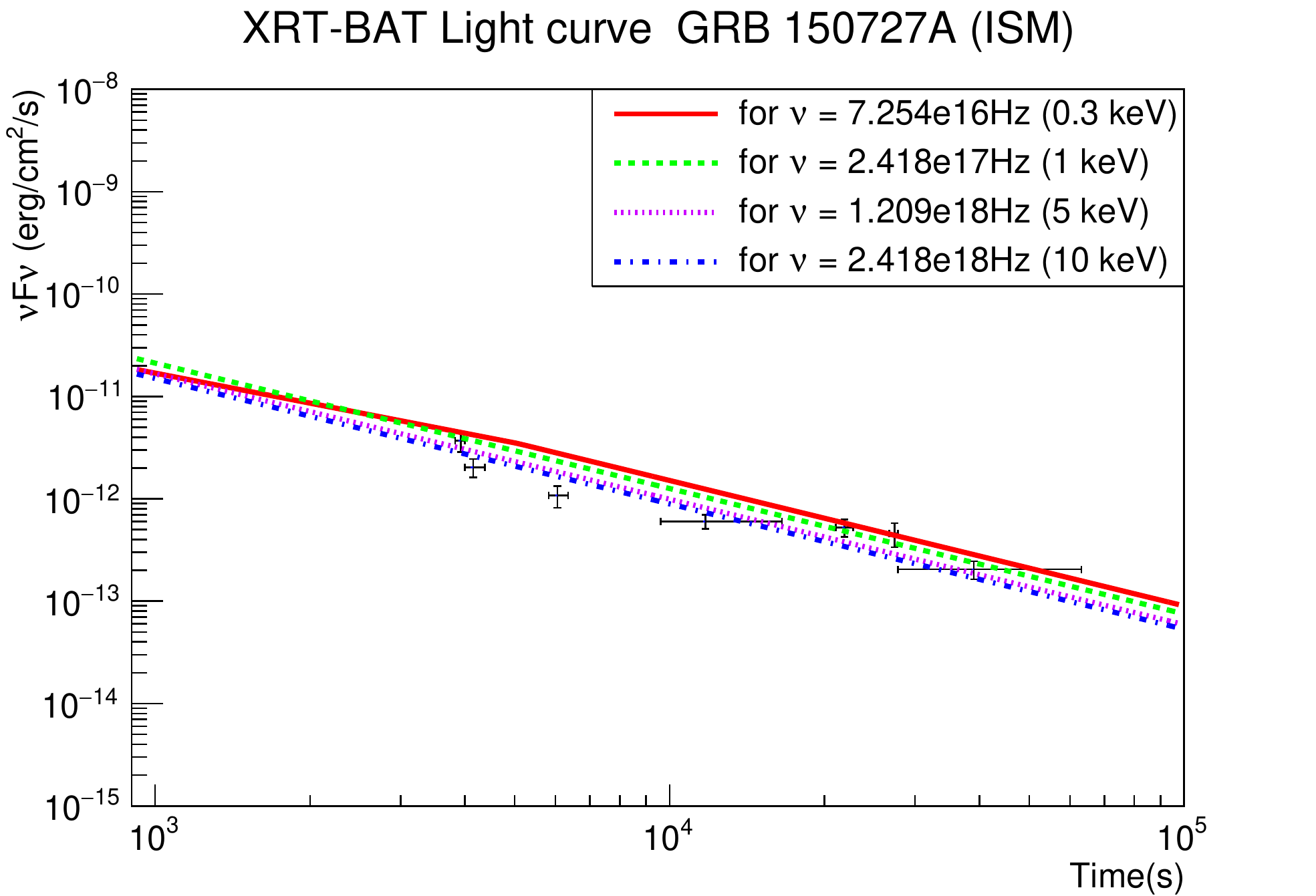}
\includegraphics[trim =  0 21 0 10, width=0.85\columnwidth]{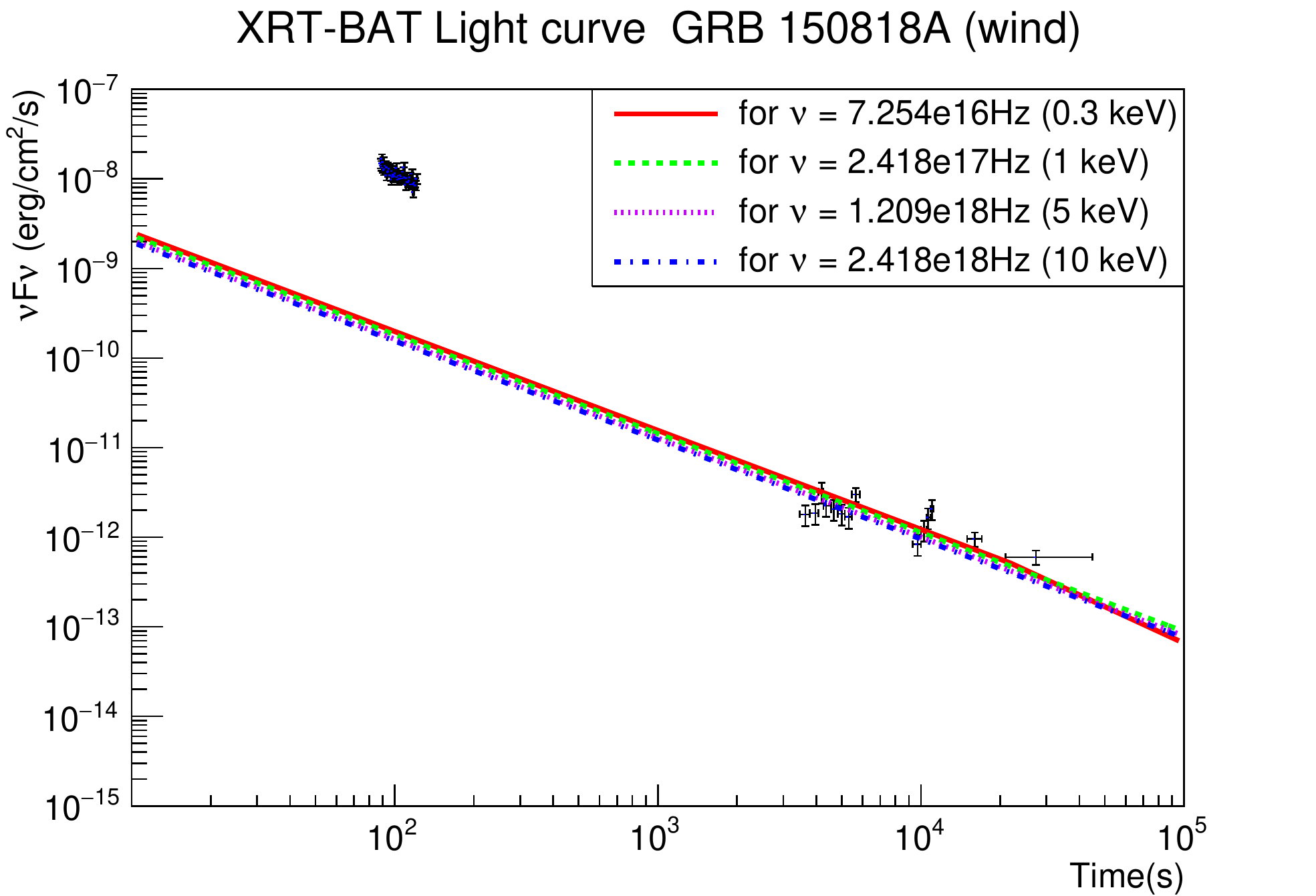}
\vspace{0.5 cm}
\includegraphics[trim =  0 21 0 10, width=0.85\columnwidth]{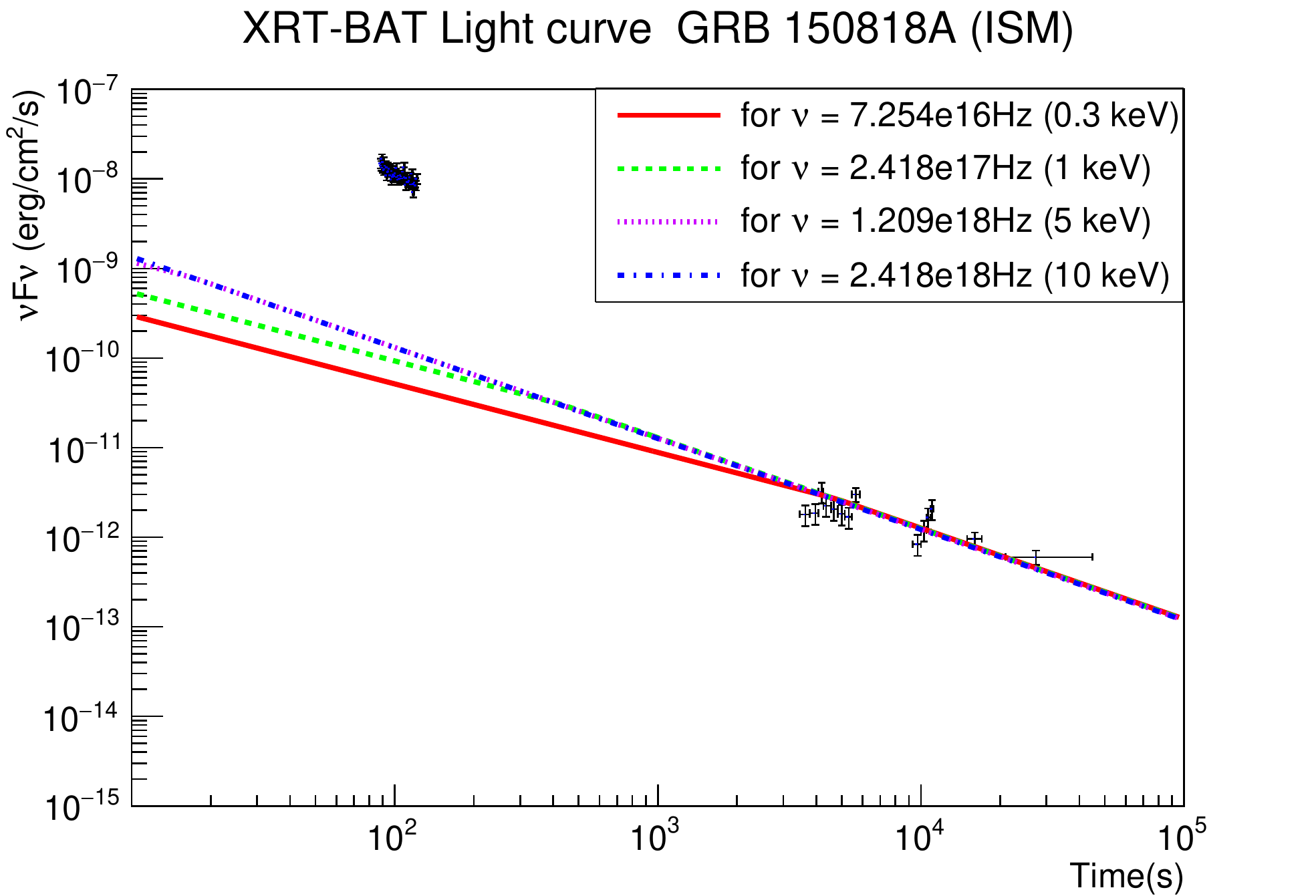}
\caption{\label{fig6lc} {Same as Fig.~\ref{fig1lc} but for GRB~130702A, GRB~130831A, GRB~150727A and GRB~150818A.}}
\end{figure*}

\begin{figure*}[th!]
\includegraphics[trim =  0 21 0 10, width=0.85\columnwidth]{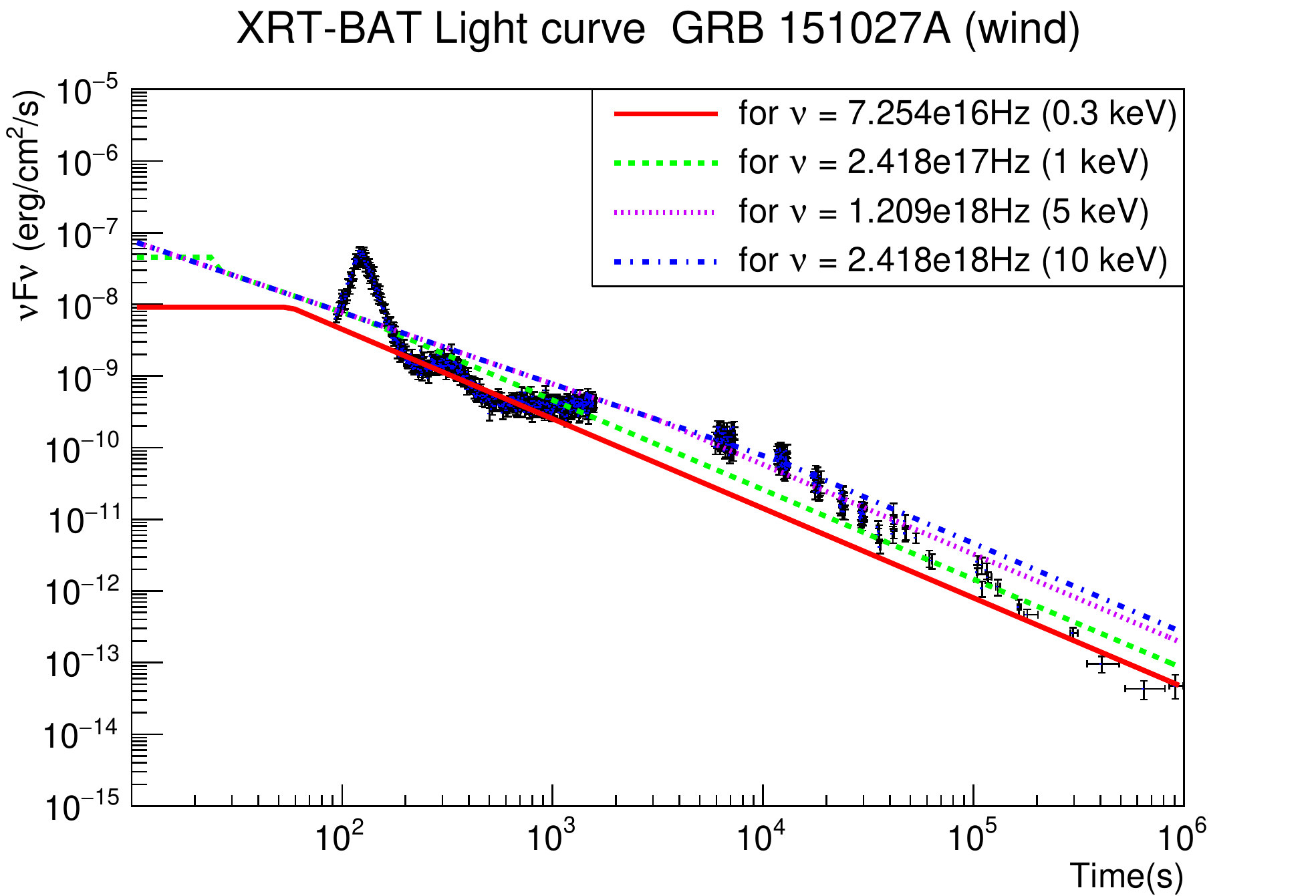}
\vspace{0.5 cm}
\includegraphics[trim =  0 21 0 10, width=0.85\columnwidth]{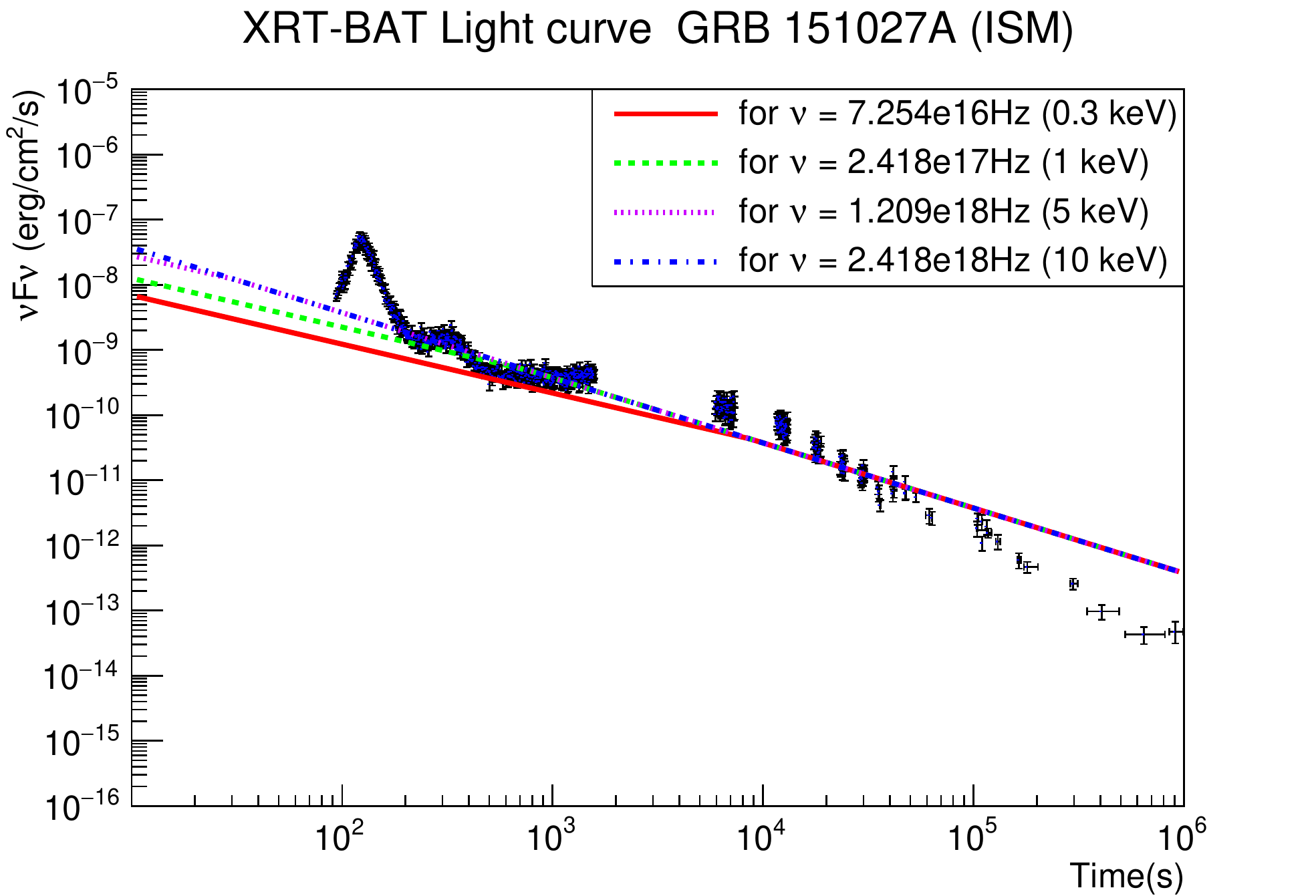}
\includegraphics[trim =  0 21 0 10, width=0.85\columnwidth]{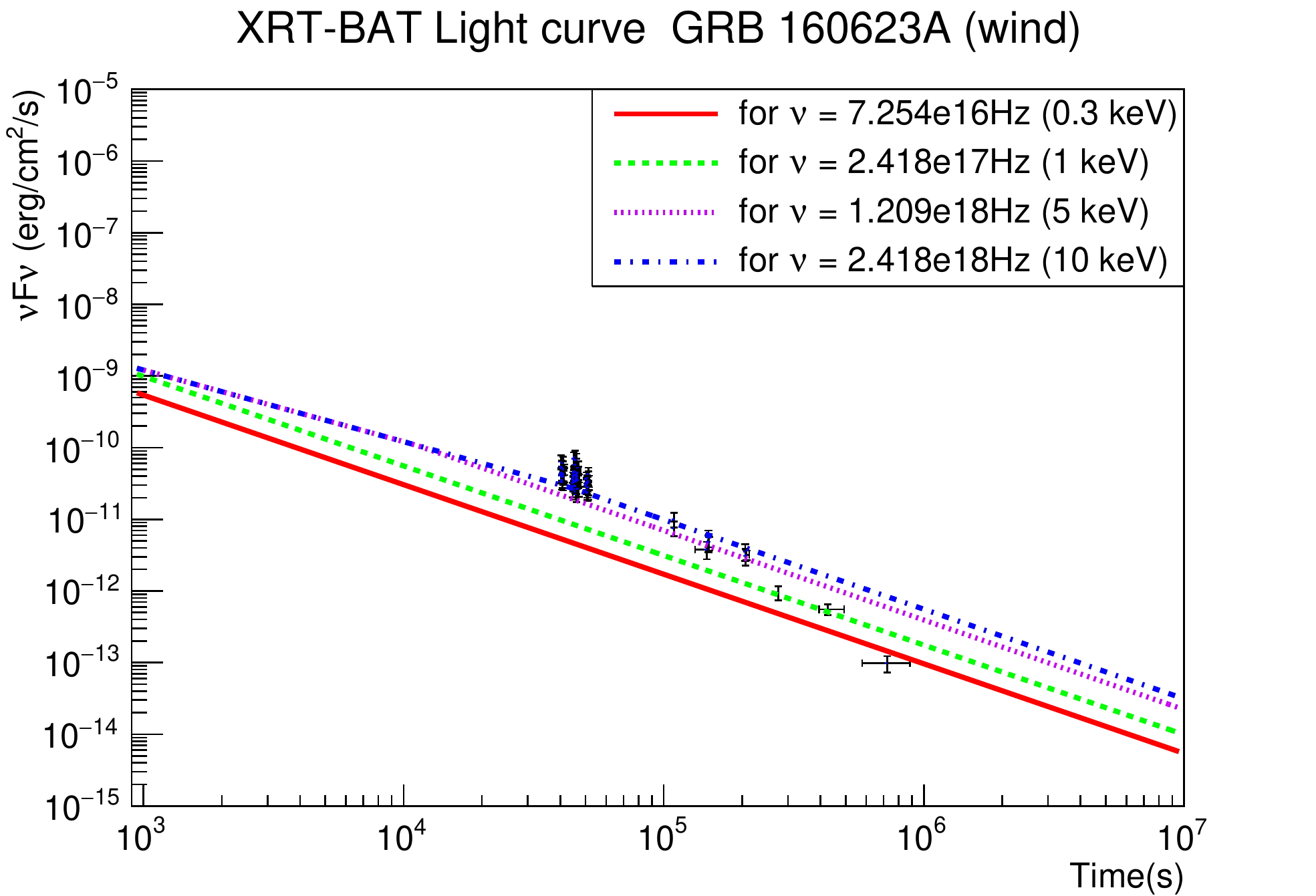}
\vspace{0.5 cm} 
\includegraphics[trim =  0 21 0 10, width=0.85\columnwidth]{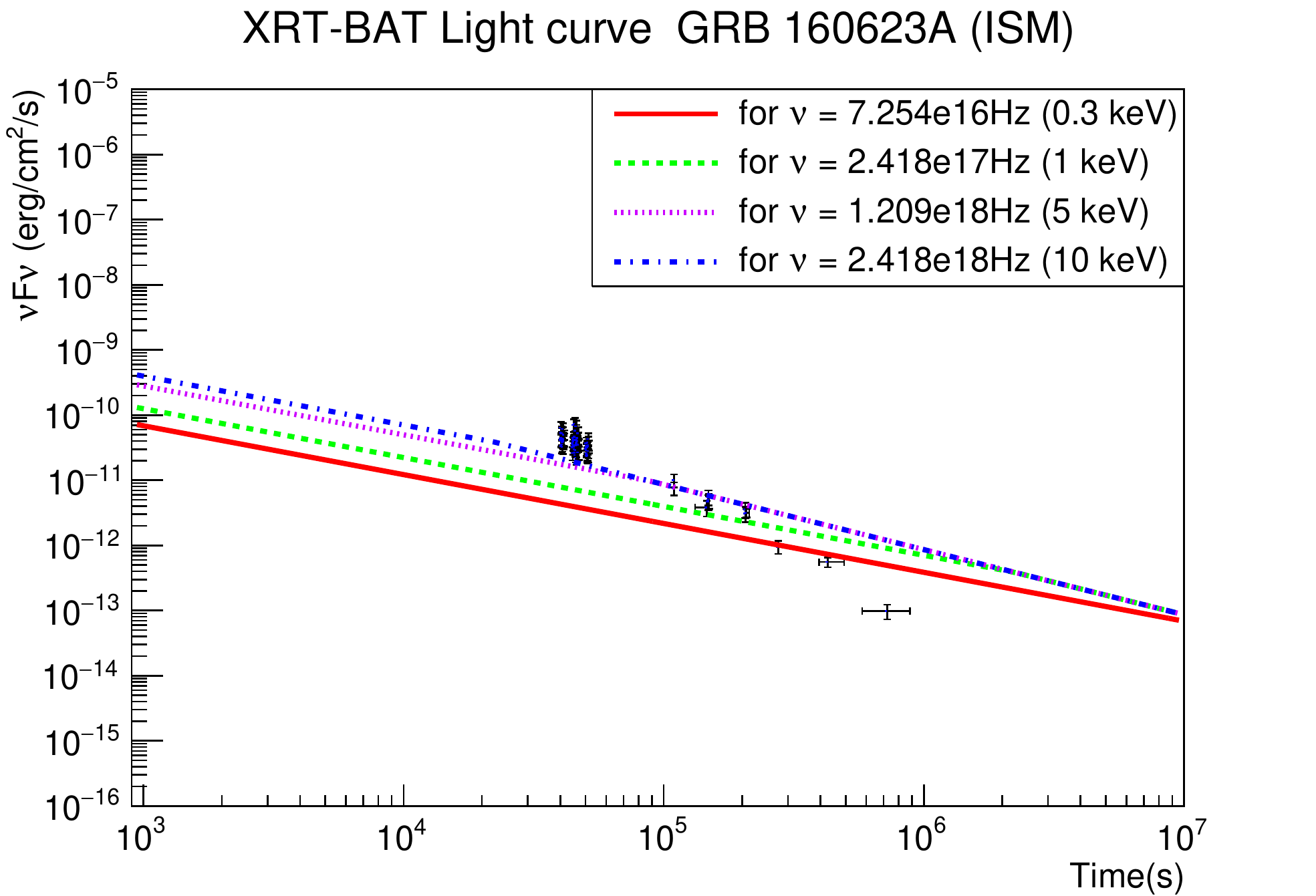}
\includegraphics[trim =  0 21 0 10, width=0.85\columnwidth]{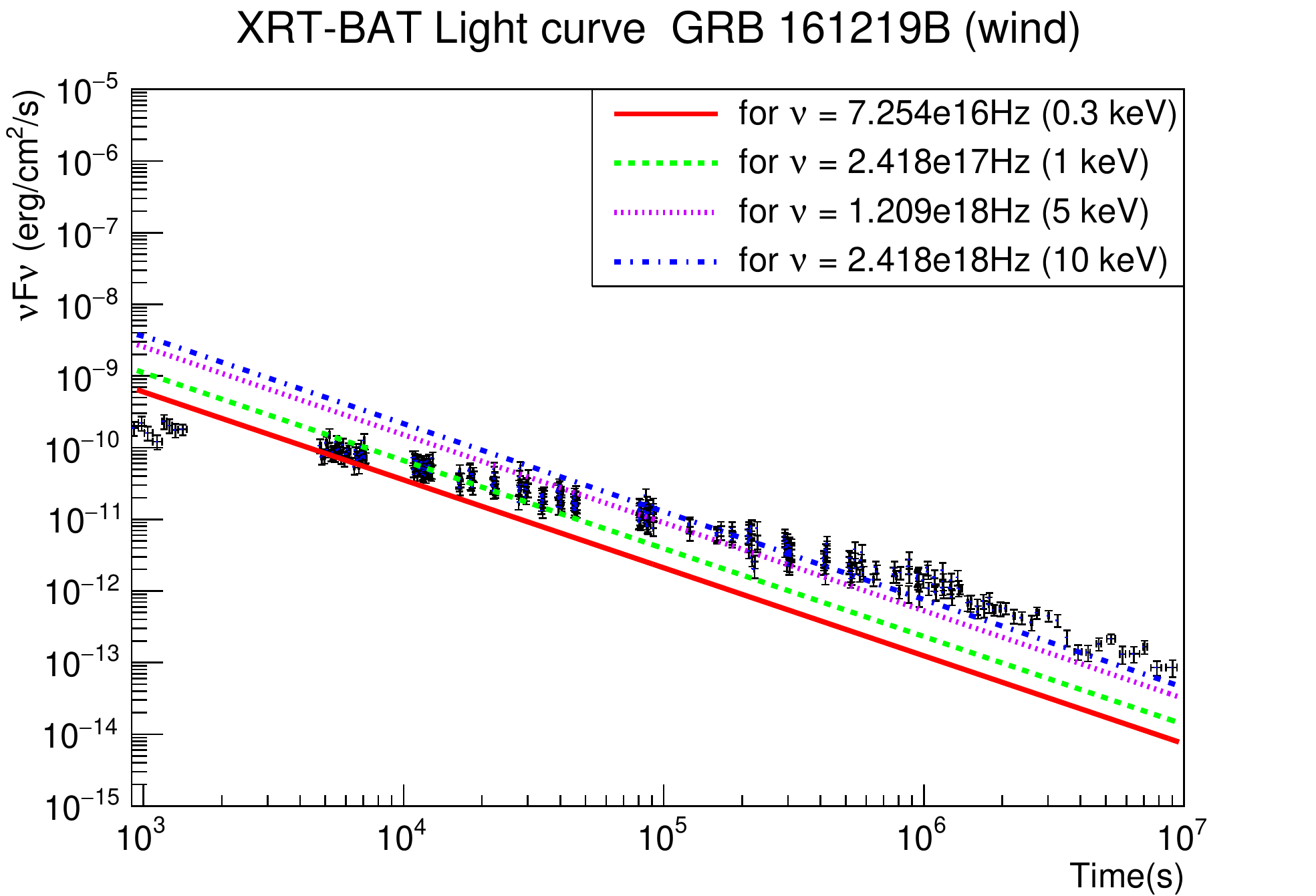}
\vspace{0.5 cm} 
\includegraphics[trim =  0 21 0 10, width=0.85\columnwidth]{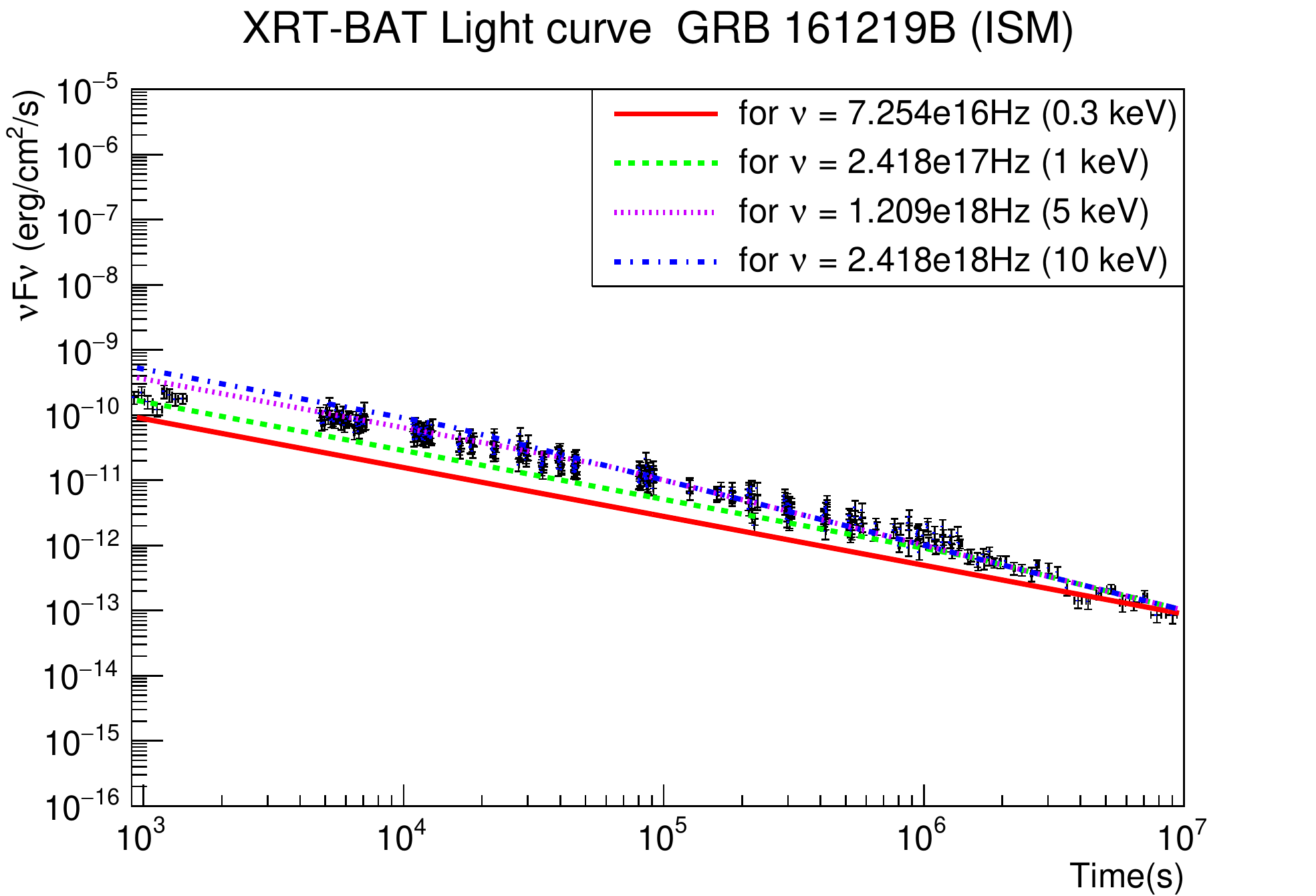}
\caption{\label{fig5lc} Same as Fig.~\ref{fig1lc} but for GRB~151027A, GRB~160623A and GRB~161219B.}
\end{figure*}

%%%%%%%%%%%%%%%%%%%%%%%%%%%%%%%%%%%%
\label{model-result}
We have built spectral energy distributions (SED) of all 23 GRBs in our sample for as many time intervals as possible with available flux data in various frequencies (see Tables \ref{tab1}, \ref{tab2} and \ref{tab3}) after the duration of the prompt emission phase defined as $T_{90}$.  These are shown in Figs.~\ref{fig1sed}-\ref{fig5sed}.  The left and right panels for each GRB represent the same afterglow data but the fits are for the wind and ISM models, respectively.  In Figs.~\ref{fig1lc}-\ref{fig5lc} we plot light curves at different frequencies for our sample of 23 GRBs. Again, the left and right panels for each GRB represent the same afterglow data but the fits are for the wind and ISM models, respectively.

%%%%%%%%%%%%%%%%%%%%%%%%%%%%%%%%%%
We have fitted synchrotron flux model in equations~(\ref{fastcool}) and (\ref{slowcool}), for both the ISM and wind environments, to the broadband SEDs of all 23 GRBs in our sample.  These model fits are represented by lines in Figs.~\ref{fig1sed}-\ref{fig5sed}.  Note that our numerical code automatically changes from the fast- to slow-cooling spectrum based on parameter values of the model and time.  In general we could not make a strong distinction between the ISM and wind models, although in case of GRBs 060218, 130702A, 130831A and 130427A a wind model is preferred while in case of GRBs 051109B, 051117B, 061021, 111225A and 151027A an ISM model is preferred. Our model fits are also constrained by the light curves at different frequencies, and are shown as lines in Figs.~\ref{fig1lc}-\ref{fig5lc}.  Note that the breaks in a number of light curves model are due to a transition from the fast- to slow-cooling spectrum.  However in a number of cases we could not find good fit for either ISM or wind model to the available data, which we assume because of the discrepancies involve with the simple forward shock GRB afterglow models and is inadequate to capture all physics.  Additional complications such as reverse shock emission \cite{Piran:2004ba}, refreshed shock emission \cite{Rees:1997nx} or off-axis emission \cite{vanEerten:2010zh} may be in play for observed afterglow data.  However, those scenarios also require additional model parameters.

The model parameters extracted from the fits in the ISM environment are reported in Table~\ref{tab4param} and for the wind environment in Table~\ref{tab5param}.  Given the large number of free parameters we tried to keep $n_0$ and $A_*$ close to their nominal values and vary other parameters. Similarly, we have kept the electron index $p$ close to $\sim 2$ as expected from Fermi shock acceleration.  In case of a wind environment the kinetic energy of the GRBs varied from $\sim 3.4\times 10^{49}$~erg for GRB 051109B to $\sim 1.1\times 10^{55}$~erg for GRB 130427A, which is one of the most energetic GRBs ever detected \cite{maseli}.  The ranges of microphysical parameters we obtained are $\epsilon_e \sim 10^{-2}$-$10^{-3}$ and $\epsilon_b \sim 10^{-2}$-$10^{-4}$ for the wind model, which are typical.  In case of an ISM environment, the kinetic energies are quite similar to the wind case, in the range $\sim 3.4\times 10^{49}$-$10^{55}$~erg.  The ranges of microphysical parameters in the ISM case are $\epsilon_e \sim 10^{-2}$-$5\times 10^{-4}$ and $\epsilon_b \sim 10^{-2}$-$10^{-4}$, which, again, are typical.  When possible (e.g., for GRB 130427A), we compared our model-fit parameters with other published results and found reasonable consistency.

We use the fit parameters, both for the wind and ISM models, obtained from modeling data to compute neutrino flux from the 23 nearby GRBs next.

\begin{table*}
\centering
\setlength\tabcolsep{0.3cm}
\caption{Model parameters obtained from synchrotron modeling in the wind environment}
\label{tab4param}
\begin{tabular}{crrrrrrrr}\hline \hline
   GRB   & $z$ & $T_{90} (s) $ & $E_{iso} (erg)$ & $E_{kin}(10^{55}erg)$ & $\epsilon_e$ & $\epsilon_b$ & $p$ &  $A*$\\ \hline \\
  050803&0.422&85$\pm$ 10&2.45$\times10^{51}$&$2.45\times10^{-3}$&$2\times10^{-2}$&$1\times10^{-3}$&2.0&0.1\\
  050826&0.297&35$\pm$ 8&3.39$\times10^{50}$&$3.39\times10^{-4}$&$1\times10^{-2}$&$5\times10^{-3}$&2.12&0.1\\
  051109B&0.08&15$\pm$1&3.4$\times10^{48}$&$3.4\times10^{-6}$&$2\times10^{-2}$&$1\times10^{-3}$&2.0&0.1\\   
  051117B&0.481&8$\pm$1&2.77$\times10^{51}$&$2.77\times10^{-3}$&$2\times10^{-3}$&$3\times10^{-4}$&2.1&0.1\\ 
  060218 &0.033&$~$2100&$ 1.9\times10^{49}$&$1.9\times10^{-5}$&$1\times10^{-2}$&$6.8\times10^{-2}$&2.27&0.1\\
  060512&0.443&8.6$\pm$ 2&1.99$\times10^{50}$&$1.99\times10^{-4}$&$5\times10^{-2}$&$1\times10^{-2}$&2.18&0.1\\
  060614 & 0.125&102$\pm$5&$ 8.4\times10^{50}$&$8.4\times10^{-4}$&$3\times10^{-3}$&$2\times10^{-3}$&2.1&0.1\\
  061021&0.346&46$\pm$1&4.06$\times10^{51}$&$4.06\times10^{-3}$&$6\times10^{-3}$&$4\times10^{-3}$&2.0&0.1\\
  091127&0.49&7.1$\pm$0.2&1.60$\times10^{52}$&$1.6\times10^{-2}$&$3\times10^{-2}$&$5\times10^{-3}$&2.0&0.1\\
  090417B&0.345& $>$ 260 & $6.98\times10^{50}$ & $7.98\times10^{-3}$& $4\times10^{-4}$ & $2\times10^{-3}$&1.5&0.01\\
 100316D&0.059&$>$1300&9.81$\times10^{48}$&$9.8\times10^{-6}$&$8\times10^{-2}$&$1\times10^{-3}$&2.25&0.1\\
  101225A& 0.40&  1088$\pm$20 &$1.28\times10^{50}$& $9.21\times10^{-3}$&$8\times10^{-3}$&$1\times10^{-2}$&1.7&0.01\\
  111225A&0.297&106.8$\pm$26.7&2.88$\times10^{50}$&$2.88\times10^{-4}$&$1\times10^{-2}$&$1.45\times10^{-2}$&2.18&0.1\\
  120422A&0.28&5.35$\pm$1.4&1.28$\times10^{51}$&$1.28\times10^{-3}$&$3.5\times10^{-2}$&$1\times10^{-4}$&2.4&1.0\\
  120714B&0.398&159 $\pm$ 34&4.51$\times10^{51}$&$4.5\times10^{-3}$&$1\times10^{-2}$&$1.2\times10^{-3}$&2.3&0.1\\
  130427A&0.34&162.83$\pm$1.36&8.5$\times10^{53}$&1.1&$9\times10^{-3}$&$1\times10^{-4}$&2.0&1.0\\
  130702A&0.145&59$\pm$1&7.8$\times10^{50}$&$7\times10^{-4}$&$2\times10^{-2}$&$5\times10^{-2}$&1.9&0.1\\
  130831A&0.479&32.5 $\pm$ 2.5&4.56$\times10^{51}$&$8\times10^{-3}$&$9\times10^{-3}$&$1\times10^{-3}$&2.0&1.0\\
  150727A& 0.313&  88$\pm$13 &$9.21\times10^{50}$& $9.1\times10^{-4}$&$3.3\times10^{-2}$&$3\times10^{-3}$&2.3&1.0\\
  150818A&0.282 &123.3$\pm$31.3&1$\times10^{51}$&$1\times10^{-3}$&$6\times10^{-3}$&$1\times10^{-2}$& 2.14 &0.1\\     
  151027A& 0.38&  129.69$\pm$5.5&$4\times10^{52}$& $6.42\times10^{-3}$&$5\times10^{-2}$&$3\times10^{-2}$&2.0&0.01\\
  160623A& 0.367&  13.5$\pm$0.6 &$2.26\times10^{53}$& $2.26\times10^{-1}$&$2\times10^{-3}$&$7\times10^{-3}$&2.0&0.1\\
  161219B& 0.148&  6.94$\pm$0.79 &$1.6\times10^{52}$& $1.6\times10^{-2}$& $7.5\times10^{-3}$&$1\times10^{-3}$&2.0&0.1\\    
\hline

\end{tabular}
\label{tab:accuracy1}
\end{table*}

\begin{table*} 
\centering
\setlength\tabcolsep{0.3cm}
\caption{Model parameters obtained from synchrotron modeling in the ISM environment}
\label{tab5param}
\begin{tabular}{crrrrrrrrrr}\hline \hline
  ISM GRB &  $z$ & $T_{90} (s)$ & $E_{iso} (erg)$ & $E_{kin}(10^{55}erg)$ & $\epsilon_e$ & $\epsilon_b$ & $p$ & $n_{0}$ \\\hline \\
  050803&0.422&85$\pm$10&$2.45\times10^{51}$&$2.45\times10^{-3}$&$7\times10^{-3}$&$1\times10^{-3}$&2.0&1.0\\
  050826&0.297&35$\pm$8&$3.39\times10^{50}$&$3.39\times10^{-4}$&$4.8\times10^{-3}$&$1\times10^{-2}$&2.14&1.0\\
  051109B&0.08&15$\pm$1&$3.46\times10^{48}$&$3.4\times10^{-6}$&$2\times10^{-2}$&$3\times10^{-2}$&2.1&1.0\\
  051117B&0.481&8$\pm$1&$2.77\times10^{51}$&$2.77\times10^{-3}$&$5\times10^{-4}$&$1\times10^{-3}$&2.0&1.0\\
  060218 &0.033&$~$2100&$ 1.9\times10^{49}$&$1.9\times10^{-5}$&$1\times10^{-3}$&$8\times10^{-3}$&2.08&1.0\\
  060512&0.443&8.6$\pm$ 2&$1.99\times10^{50}$&$1.99\times10^{-4}$&$2\times10^{-2}$&$3\times10^{-2}$&2.18&1.0\\
  060614 & 0.125&102$\pm$5&$ 8.4\times10^{50}$&$8.4\times10^{-4}$&$2\times10^{-3}$&$2\times10^{-3}$&2.18&1.0\\
  061021&0.346&46$\pm$1&$4.06\times10^{51}$&$4.06\times10^{-3}$&$4\times10^{-3}$&$3\times10^{-3}$&2.0&0.1\\
  090417B&0.345& $>$ 260 & $6.98\times10^{50}$ & $5.5\times10^{-3}$ & $1\times10^{-4}$ & $1\times10^{-3}$&1.5&0.01\\  
  091127&0.49&7.1$\pm$0.2&$1.60\times10^{52}$&$1.6\times10^{-2}$&$1.2\times10^{-2}$&$3.8\times10^{-3}$&2.0&0.1\\
 100316D&0.059&$>$1300&$9.81\times10^{48}$&$9.8\times10^{-6}$&$8\times10^{-2}$&$6.5\times10^{-2}$&2.25&0.1\\
 101225A& 0.40&  1088$\pm$20 &$1.28\times10^{50}$& $1\times10^{-3}$&$4\times10^{-3}$&$2\times10^{-2}$&1.7&0.1\\ 111225A&0.297&106.8$\pm$26.7&$2.88\times10^{50}$&$2.88\times10^{-4}$&$4.5\times10^{-3}$&$1.6\times10^{-2}$&2.19&0.1\\
  120422A&0.28&5.35$\pm$1.4&$1.28\times10^{51}$&$1.28\times10^{-3}$&$1\times10^{-3}$&$1.3\times10^{-2}$&2.12&1.0\\
  120714B&0.398&159 $\pm$ 34&$4.51\times10^{51}$&$4.5\times10^{-3}$&$2\times10^{-3}$&$2\times10^{-3}$&2.28&1.0\\
  130427A&0.34&162.83$\pm$1.36&$8.5\times10^{53}$&$8.5\times10^{-1}$&$1.2\times10^{-3}$&$1\times10^{-4}$&2.0&1.0\\
  130702A&0.145&59$\pm$1&$7.8\times10^{50}$&$7\times10^{-4}$&$5\times10^{-3}$&$8\times10^{-2}$&1.87&0.1\\
  130831A&0.479&32.5 $\pm$ 2.5&$4.56\times10^{51}$&$8\times10^{-3}$&$3\times10^{-3}$&$4\times10^{-3}$ &2.0&1.0\\ 
  150727A& 0.313&  88$\pm$13 &$9.21\times10^{50}$& $8\times10^{-4}$&$2.3\times10^{-2}$&$2\times10^{-2}$&2.4&1.0\\
  150818A&0.282&123.3$\pm$ 31.3&$1\times10^{51}$&$1\times10^{-3}$&$1\times10^{-3}$&$1\times10^{-2}$&2.02&1\\  
  151027A& 0.38&  129.69$\pm$5.5&$4\times10^{52}$& $6.42\times10^{-3}$&$8.5\times10^{-3}$&$2\times10^{-2}$&2.0&0.1\\
  160623A& 0.367&  13.5$\pm$0.6 &$2.26\times10^{53}$& $2.26\times10^{-1}$&$5\times10^{-4}$&$2\times10^{-3}$&2.0&0.01\\
  161219B& 0.148&  6.94$\pm$0.79 &$1.6\times10^{52}$& $1.6\times10^{-2}$& $9\times10^{-4}$&$2.7\times10^{-4}$&2.0&1.0\\
 \hline
\end{tabular}
\label{tab:accuracy2}
\end{table*}

%%%%%%%Sec.IV%%%%%%%%%%%%
\section{Afterglow neutrino flux calculation}
%%%%%%%%%%%%%%%%%%%%%%
\label{neutrino-flux}
Long-duration GRBs are one of the candidate sources of UHECRs and the $p\gamma$ interaction of these cosmic rays with synchrotron radiated photons can produce UHE neutrinos~\cite{Dermer:2000yd, Razzaque:2013dsa}. Here we have calculated neutrino flux, following Ref.~\cite{Razzaque:2013dsa}, from the 23 long-duration nearby GRBs within $z=0.5$ in our sample.  We have used the same parameters obtained from modeling the synchrotron afterglow of these GRBs to calculate $p\gamma$ interaction efficiencies and spectra of CR protons.  The resulting neutrino flux typically peaks in the energy range of $10^{15}$-$10^{18}$~eV. 

The shock-accelerated UHE protons interact with synchrotron emission during the afterglow through photo-meson interaction (production of pions and kaons), subsequently producing UHE neutrinos.  Here we have calculated the neutrino flux from pion and muon decays, where pions are produced via $\Delta^+$ resonance from $p\gamma$ interaction, $p\gamma \rightarrow \Delta^+ \rightarrow n\pi^+$ or $p\pi^0$ and $\pi^+ \rightarrow  \mu^+ + \nu_{\mu} \rightarrow  e^+ +\nu_e +\nu_{\mu }+ \bar \nu_\mu.$  The neutrino flux from the $p\gamma$ interactions depends on the proper density of synchrotron photons, the flux of cosmic ray protons from the GRB blast wave and the optical depth for the interaction. 
The proper density of the synchrotron photons, $n'_\gamma(E')$, relevant for $p\gamma$ opacity calculation, in the co-moving frame of the GRB blast wave depends on the bulk Lorentz factor $\Gamma (t)$ and the radius $R(t)$ of the GRB blast wave. The spectrum of these photons with energy $E' = h\nu(1+z)/\Gamma$ can be expressed~\cite{Piran:2004ba, Razzaque:2013dsa} for the fast-cooling synchrotron spectrum in equation~(\ref{fastcool}) as,
\begin{equation}
 \begin{array} {ll}
n'_\gamma(E') & =\frac{2d_l^2(1+z)F_{\nu, {\rm max}}}{R^2c\Gamma E'_c}  \\
& \times \left  \{ \begin{array} {ll}
\left(\frac{E'}{E'_a}\right) \left(\frac{E'_a}{E'_c}\right)^{-\frac{2}{3}}; &   E'<E'_a \\
\left(\frac{E'}{E'_c}\right)^{- \frac{2}{3}}; &   E'_a \le E' < E'_c \\                                         
\left(\frac{E'}{E'_c}\right)^{- \frac{3}{2}}; &   E'_c \le E' < E'_m \\
\left(\frac{E'_m}{E'_c}\right)^{- \frac{3}{2}} \left(\frac{E'}{E'_m}\right)^{-\frac{p}{2} -1} e^{-\frac{E'}{E_s'}}; & E' \ge E'_m \\ 
\end{array} \right.
\end{array}
\label{pgfast}
\end{equation}
and for the slow-cooling spectrum in equation~(\ref{slowcool}) as,
\begin{equation}
\begin{array} {ll}
n'_\gamma(E') & = \frac{2d_l^2(1+z)F_{\nu, {\rm max}}}{R^2c\Gamma E'_m} \\
& \times \left \{ \begin{array} {ll}
\left(\frac{E'}{E'_a}\right) \left(\frac{E'_a}{E'_m}\right)^{- \frac{2}{3}};  &   E'<E'_a \\
\left(\frac{E'}{E'_m}\right)^{- \frac{2}{3}};   &  E'_a \le E' < E'_m \\                                         
\left(\frac{E'}{E'_m}\right)^{-\frac{p+1}{2}};  &  E'_m \le E' < E'_c \\
\left(\frac{E'_c}{E'_m}\right)^{-\frac{p+1}{2}} \left(\frac{E'}{E'_c}\right)^{-\frac{p}{2} -1} 
e^{-\frac{E'}{E_s'}};  &  E' \ge E'_c \\ 
\end{array} \right.
\end{array}
\label{pgslow}
\end{equation}
Following Ref.~\cite{Razzaque:2014ola} we write the $p\gamma$ opacity as a function of the proton energy $E_p$ in an observer's frame.  In case of a fast-cooling synchrotron spectrum the $p\gamma$ opacity, using equation~(\ref{pgfast}), is given by
\ba
\tau_{p\gamma} (E_p) &=& \tau_{p\g} (E_{p,l}) \cr
&\times &
\begin{cases} 
\left( \frac{E_p}{E_{p,l}} \right)^{\frac{p}{2}} 
\,;\, E_p \le E_{p,l} \cr
\left( \frac{E_p}{E_{p,l}} \right)^{\frac{1}{2}} 
\,;\, E_{p,l} < E_p < E_{p,h} \cr
\left( \frac{E_{p,h}}{E_{p,l}} \right)^{\frac{1}{2}} 
\,;\, E_p \ge E_{p,h}
\end{cases}
\label{opacity_fast}
\ea
and in the case of a slow-cooling spectrum, using equation~(\ref{pgslow}), it is
\ba
\tau_{p\gamma} (E_p) &=& \tau_{p\g} (E_{p,l}) \cr
&\times &
\begin{cases} 
\left( \frac{E_{p,h}}{E_{p,l}} \right)^{\frac{p-1}{2}} 
\left( \frac{E_p}{E_{p,h}} \right)^{\frac{p}{2}} 
\,;\, E_p \le E_{p,h} \cr 
\left( \frac{E_p}{E_{p,l}} \right)^{\frac{p-1}{2}} 
\,;\, E_{p,h} < E_p < E_{p,l} \cr 
3  \,;\, E_p \ge E_{p,l}
\end{cases}
\label{opacity_slow}
\ea
Here $E_{p,l}$ is the minimum energy corresponding to the break energy $h\nu_m$ in the synchrotron spectrum, $E_{p,h}$ is the energy corresponding to the break energy $h\nu_c$ and $E_{p,s}$ is the highest energy by which the protons are accelerated in the forward shock.  

In correlation with the synchrotron break frequencies, the above-mentioned proton energies also evolve with time. In particular, in the case of a wind environment these energies are~\cite{Razzaque:2013dsa}, 
\ba
E_{p,l} &=& 9.1\times10^{7}(1+z)^{-2} \nonumber \\
&& \times \epsilon_{b,0.1}^{-1/2}\epsilon_{e,0.1}^{-2}A_*^{-1/2}t_2 \,\rm{GeV}
\ea
\ba
E_{p,h} &=& 3.2\times10^{12}\epsilon_{b,0.1}^{3/2}A_*^{3/2}t_2^{-1} \,\rm{GeV}
\ea
\ba
E_{p,s} &=& 6\times10^{9}(1+z)^{-5/4} \nonumber \\
&& \times \epsilon_{b,0.1}^{1/2}\phi_1^{-1}A_{*}^{-1/4}t_2^{1/4}E_{55}^{3/4} \,\rm{GeV},
\label{wind_Eps}
\ea
and in the case of a constant density ISM environment,
\ba
E_{p,l} &=& 1.3\times10^{8} (1+z)^{-7/4}\epsilon_{b,0.1}^{-1/2}\epsilon_{e,0.1}^{-2}n_0^{-1/4} \nonumber \\
&& \times E_{55}^{-1/4}t_2^{3/4}\, \rm{GeV} \\
E_{p,h} &=& 1.0\times10^{12}(1+z)^{-3/4}\epsilon_{b,0.1}^{3/2}n_0^{3/4} \nonumber \\
 && \times E_{55}^{3/4}t_2^{-1/4}\, \rm{GeV} \\
E_{p,s} &=& 2.3\times 10^{10}(1+z)^{-7/8}\epsilon_{b,0.1}^{1/2}\phi_1^{-1} \nonumber \\
&& \times n_0^{1/8}t_2^{-1/8} E_{55}^{3/8}\,\rm{GeV}.
\label{ism_Eps}
\ea
Here $t_2 = (t/100\,{\rm s})$ and $t$ is the time after the GRB prompt emission.  The $p\gamma$ opacity at the reference energy $E_{p,l}$ in equations~(\ref{opacity_fast}) and (\ref{opacity_slow}) are given by
\begin{equation}
\tau_{p\gamma}(E_{p,l})= 6.0(1+z)^{1/2}\epsilon_{b,0.1}^{1/2}A_*^{2}t_2^{-1/2}E_{55}^{-1/2}
\end{equation}
in the wind environment and
\begin{equation}
\tau_{p\gamma}(E_{p,l}) = 0.7(1+z)^{-1/2}\epsilon_{b,0.1}^{1/2}n_0\,t_2^{1/2}E_{55}^{1/2}
\end{equation}
in the ISM environment, respectively.

Finally we calculate the neutrino flux of each flavor, following Refs.~\cite{Razzaque:2013dsa, Razzaque:2014ola}.  For example, for $\pi^+ \to \nu_\mu$ flux,
\ba
J_{\nu_\mu} (E_\nu) = \int_0^1 \frac{dx}{x} \frac{\Theta(1-r_\pi - x)}{1-r_\pi} J_\pi \left( \frac{E_\nu}{x} \right)\,,
\label{nu_flux}
\ea
where $x=E_\nu/E_\pi$, $r_\pi = m_\mu^2/m_\pi^2$ and $\Theta$ is a Heaviside step function.  The intermediate charged pion flux is
\ba
J_\pi (E_\pi) &\approx & \frac{1}{2\langle x\rangle} J_p \left( \frac{E_\pi}{\langle x\rangle} \right) \nonumber \\
&& \times {\rm min} \left\{ \tau_{p\g} \left(\frac{E_\pi(1+z)}{\langle x\rangle \Gamma} \right), 3 \right\}.
\ea
Here $\langle x\rangle \approx 0.2$ is the average fraction of a proton energy transferred to pion in a $p\gamma$ interaction.  The proton flux $J_p$ (if it could escape freely from the GRB blast wave) in equation~(\ref{nu_flux}) is given by
\ba 
E_p^2 J_p(E_p) &=& 1.2\times 10^{-8} (1+z)^{1/2} \xi_1^{-1} \eps_p A_*^{1/2} 
\nonumber \\ && \times
E_{55}^{1/2} t_2^{-1/2} d_{l,28}^{-2}  ~{\rm GeV~cm}^{-2}~{\rm s}^{-1}
\label{Jp_ad_w}
\ea
in the wind and
\ba
E_p^2 J_p(E_p) &=& 4.8\times 10^{-9} (1+z)^{1/4} \xi_1^{-1} \eps_p n_0^{1/4} 
\nonumber \\ && \times
E_{55}^{3/4} t_2^{-1/4} d_{l,28}^{-2} ~{\rm GeV~cm}^{-2}~{\rm  s}^{-1}
\label{Jp_ad_i}
\ea
in the ISM environment, respectively~\cite{Razzaque:2013dsa}.  Here $\xi_1 \approx 1$ is a spectral correction factor for a $\propto E_p^{-2}$ assumed proton spectrum and $\eps_p \lesssim 1$ is the fraction of the blast wave kinetic energy carried by the shock-accelerated protons.

\begin{figure*}[th!]
\includegraphics[trim =  0 21 0 10, width=0.8\columnwidth]{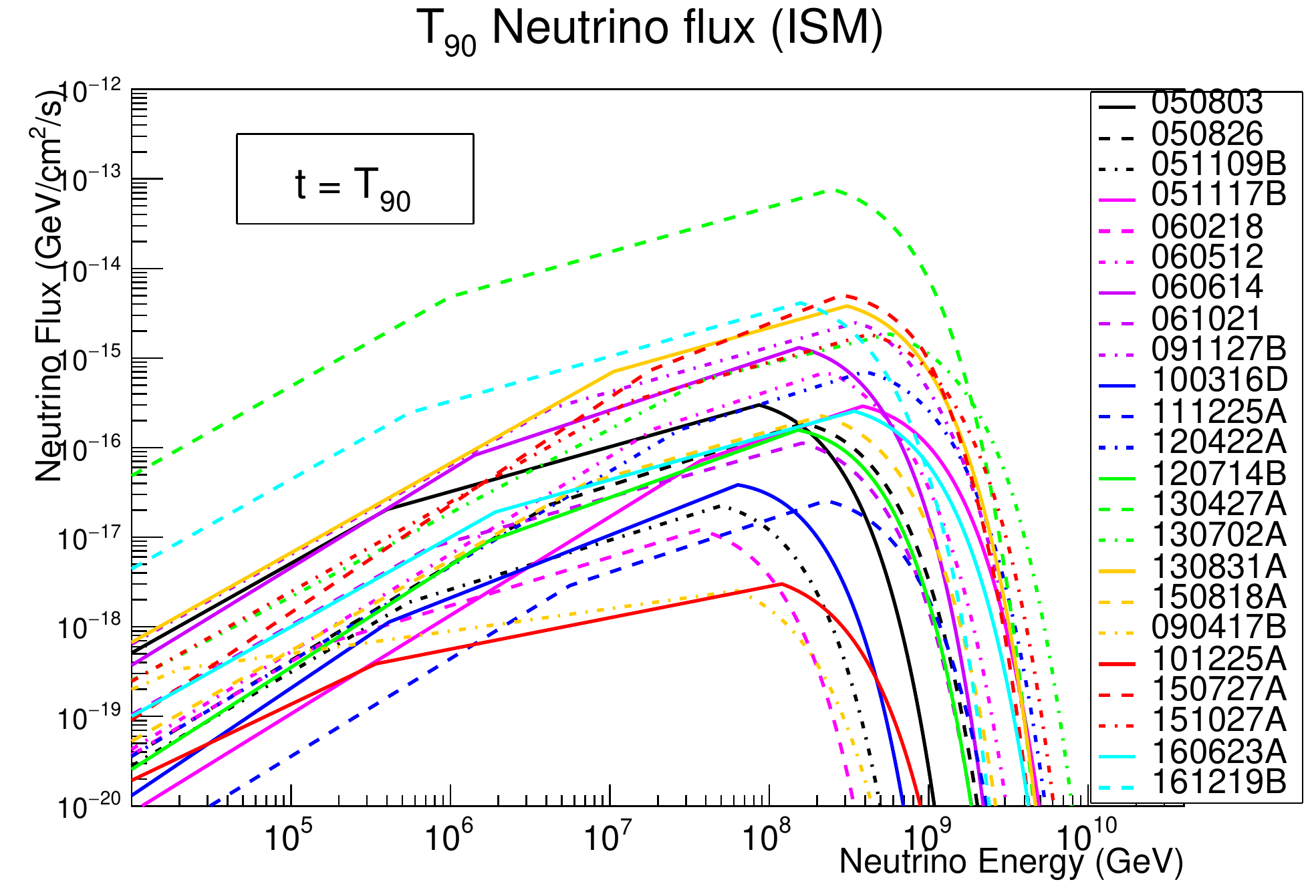}
\vspace{0.5cm}
\includegraphics[trim =  0 21 0 10, width=0.8\columnwidth]{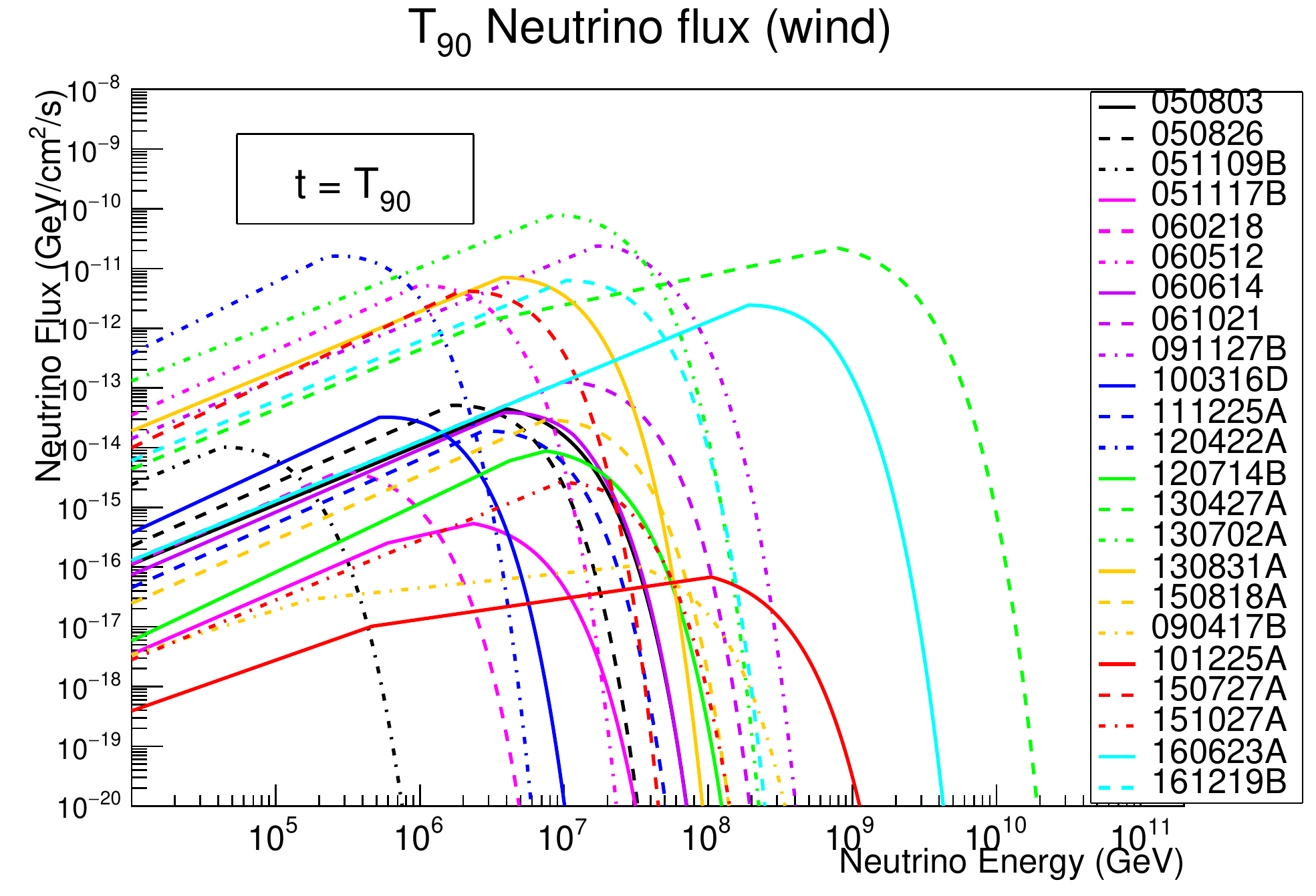}
\includegraphics[trim =  0 21 0 10, width=0.8\columnwidth]{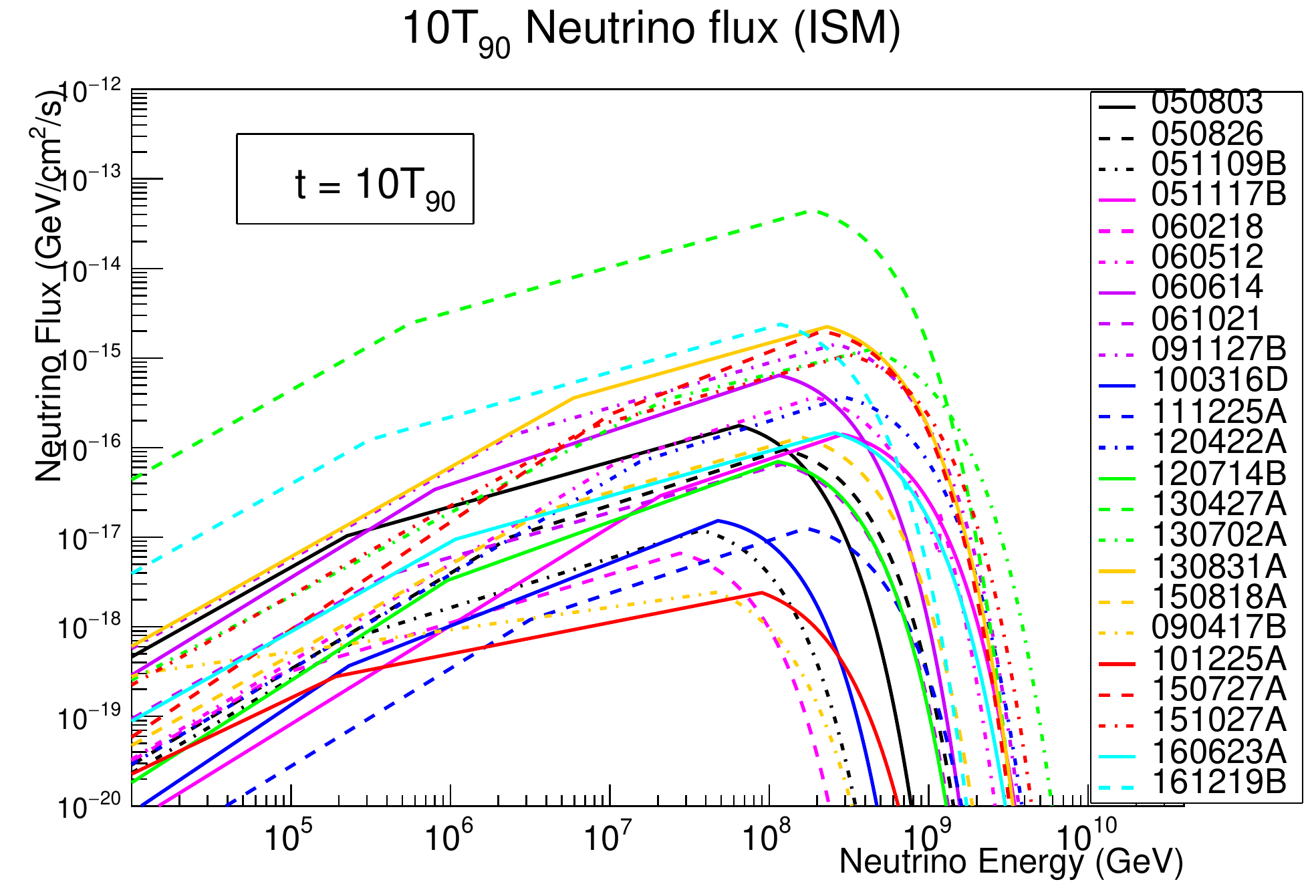}
\vspace{0.5cm}
\includegraphics[trim =  0 21 0 10, width=0.8\columnwidth]{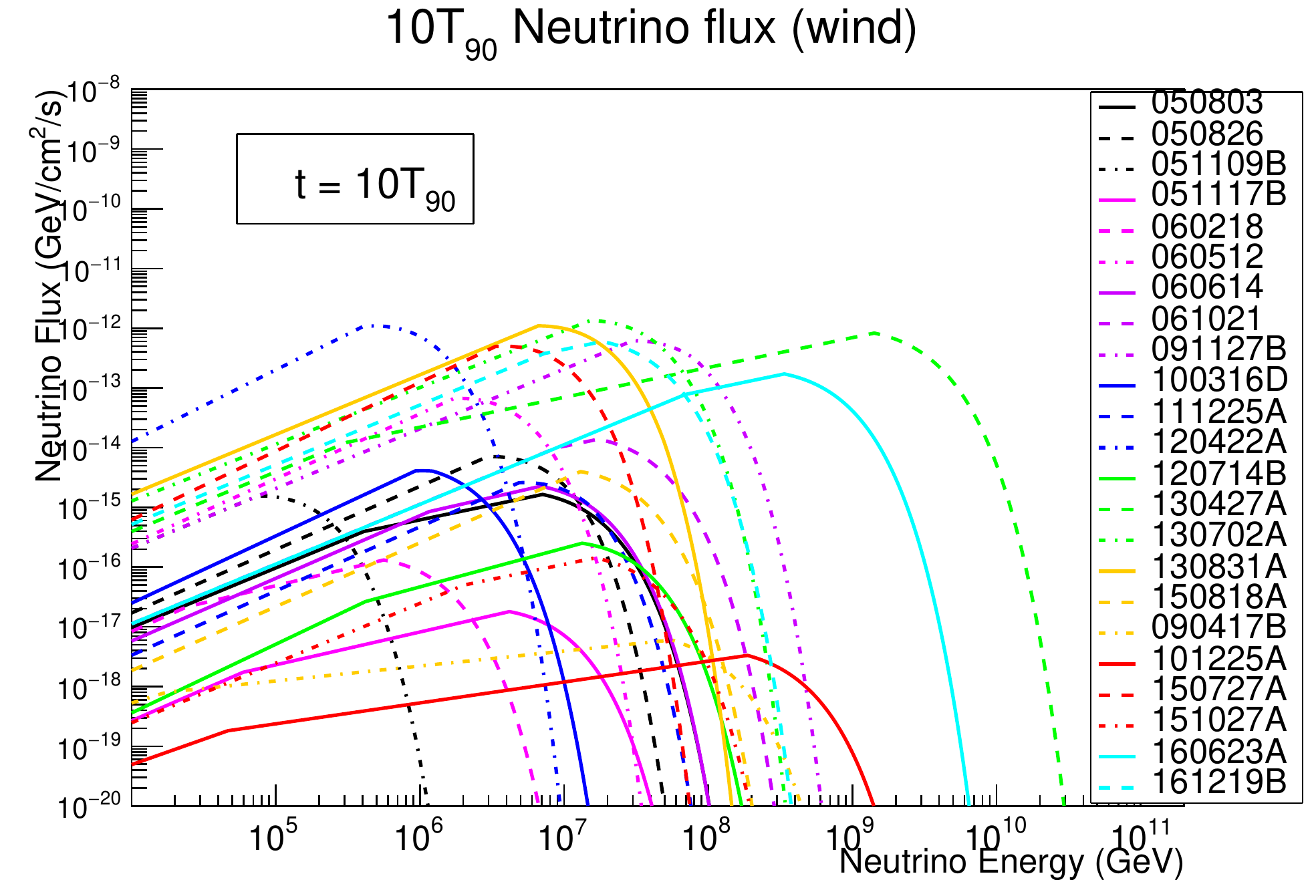}
\includegraphics[trim =  0 21 0 10, width=0.8\columnwidth]{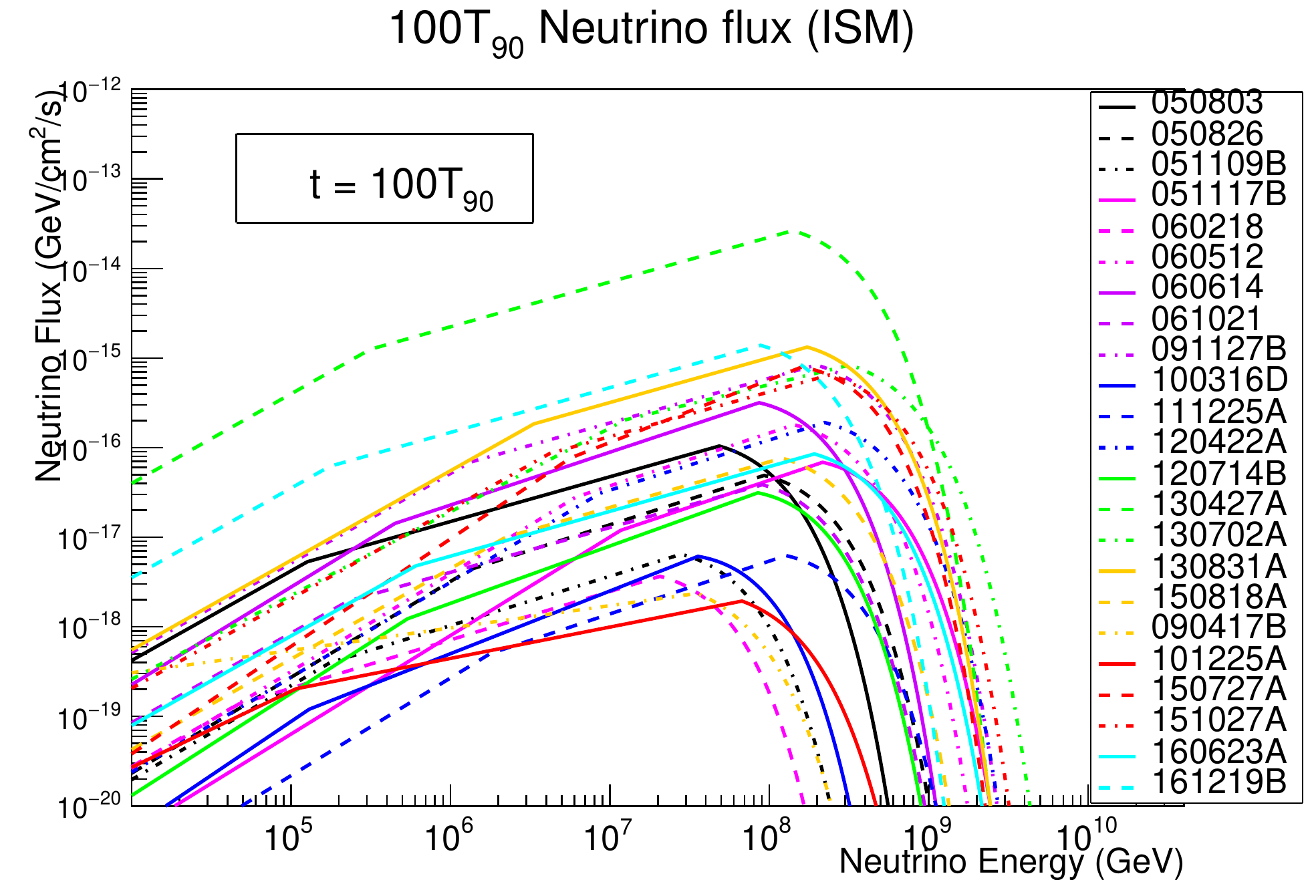}
\vspace{0.5cm}
\includegraphics[trim =  0 21 0 10, width=0.8\columnwidth]{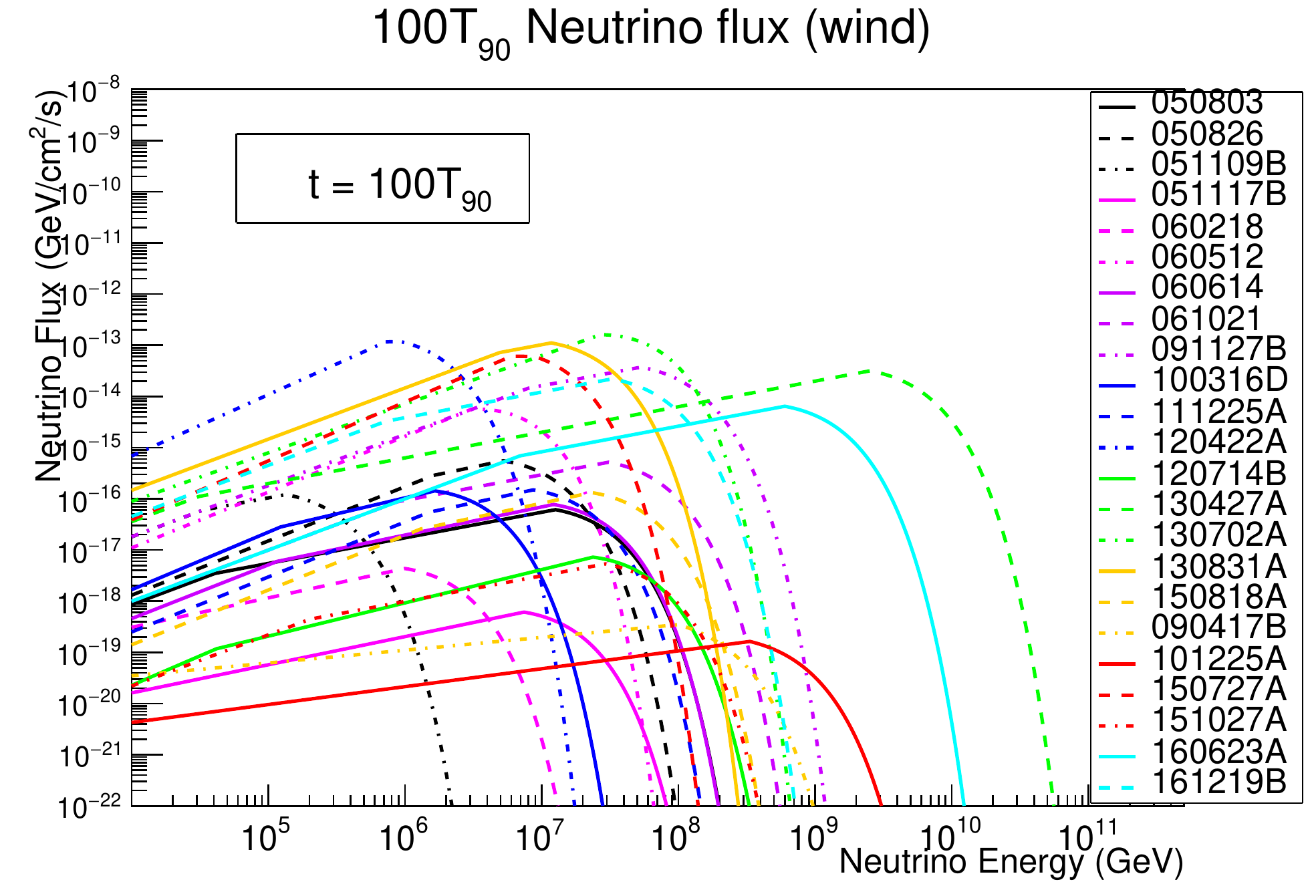}
\includegraphics[trim =  0 21 0 10, width=0.8\columnwidth]{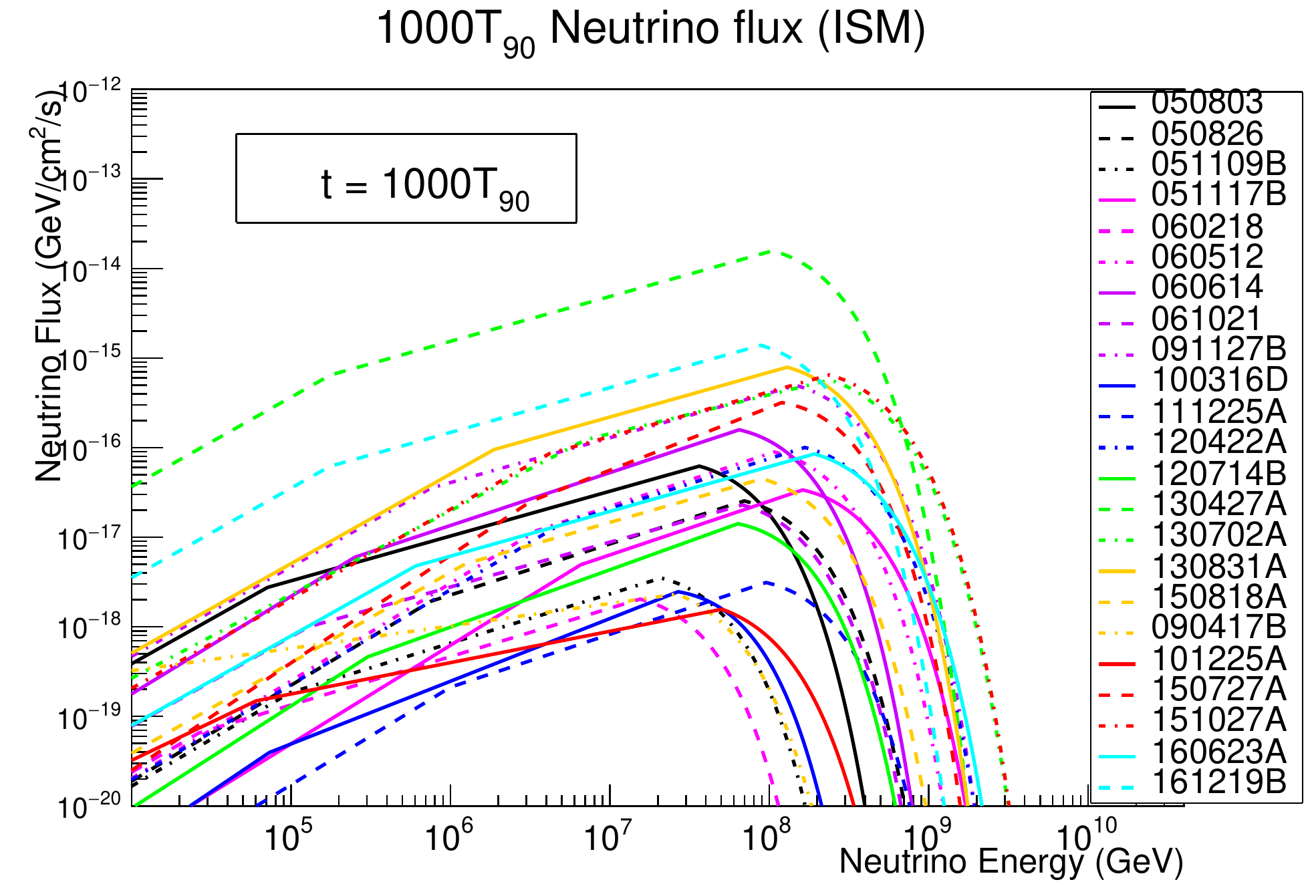}
\vspace{0.5cm}
\includegraphics[trim =  0 21 0 10, width=0.8\columnwidth]{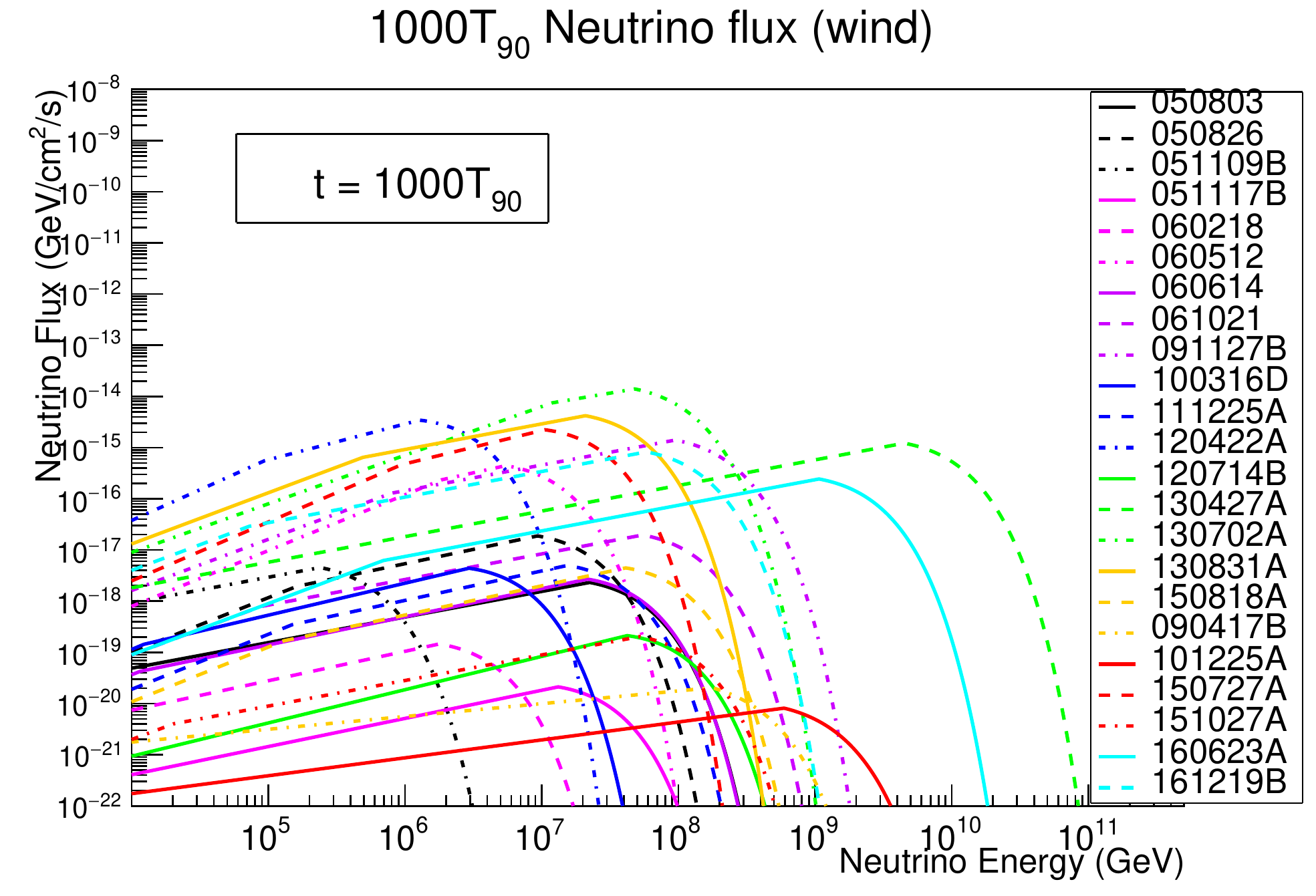}
\caption{\label{neu_flux} Neutrino flux for 23 nearby long GRBs in the ISM (left panels) and wind (right panels) environment.  Fluxes are calculated, from the top to bottom panels, at times $T_{90}$, $10\times T_{90}$, $100\times T_{90}$  and $1000\times T_{90}$, where $T_{90}$ is the duration of the GRB prompt phase, assumed approximately the same as the blast wave deceleration time scale.}
\end{figure*}

The neutrino fluxes for individual GRBs are plotted in Fig.~\ref{neu_flux} at the blast wave deceleration time $T_{90}$, assumed approximately the same as the duration of the prompt emission (top two panels).  Also plotted, from top to bottom, fluxes at $10\times T_{90}$,  $100\times T_{90}$ and $1000\times T_{90}$.  The left and right panels correspond to fluxes from a blast wave evolving in the ISM and wind environment, respectively.  Note that the neutrino flux evolves differently from a blast wave in an ISM and wind environment, but as expected, in both cases the flux decreases with time.  A difference between the evolution of flux in the ISM and wind is that the energy at which the flux peaks decreases (increases) with time for ISM (wind).  This can be understood from the time-dependence of $E_{p,s}$ in equations~(\ref{wind_Eps}) and (\ref{ism_Eps}).

In Fig.~\ref{neu_flux} the neutrino flux from GRB 130427A at $z=0.34$ dominates at the highest energies, in both environment.  This is understandably due to exceptional energy release from this GRB.  Other GRBs with expected high neutrino flux are GRBs 130831A ($z=0.479$), 130702A ($z=0.145$) and 091127B ($z=0.49$).

%%%%%%%%%%%%%%%%%%%%%%%
\section{UHE GRB neutrino flux detectability}
%%%%%%%%%%%%%%%%%%%%%%%
\label{detection}
IceCube Neutrino Observatory at the south pole detected cosmic neutrinos in the $\gtrsim 20$~TeV to $\approx 2$~PeV range~\cite{aartsen2014observation}.  A proposed upgrade of IceCube, called IceCube Gen-2~\cite{icecube_gen2}, and planned future experiments, such as the Askary'an Radio Array~\cite{Allison:2011wk}  and ARIANNA~\cite{Barwick:2006tg}, will increase sensitivity to UHE neutrino fluxes.  The IceCube Gen-2, the high energy extension of the IceCube experiment aims at improving the sensitivity for the detection of neutrinos with few hundreds of TeV and energies beyond~\cite{Aartsen:2015dkp}. 

The KM3NeT-Astroparticle Research with Cosmics in the Abyss (ARCA) is an upcoming deep-sea research observatory, currently under construction, in the Mediterranean Sea.  This neutrino telescope will have a volume of at least one cubic kilometer~\cite{Adrian-Martinez:2016fdl}.  KM3NeT-ARCA will provide much needed sensitivity to UHE cosmic neutrinos in the northern hemisphere.  Note that since the earth becomes opaque to neutrinos at $\gtrsim 1$ PeV~\cite{Gandhi:1995tf}, both IceCube and KM3NeT-ARCA are sensitive mostly to neutrinos arriving above the horizon at UHE.

Although the primary goal of the surface detector of the Pierre Auger Cosmic Ray Observatory, which is located in the province of Mendoza, Argentina at an altitude of 1400 m above the sea level, is to detect UHECRs, it can also detect UHE neutrinos~\cite{Aab:2015kma}.  The Surface Detector array of the Pierre Auger Observatory can detect neutrinos with energies at $\sim 1$~EeV and above. 

We considered the above three UHE neutrino detectors to explore detectability of modeled UHE neutrino flux from the 23 GRBs in our sample.  Neutrino events of a given flavor from individual GRB flux $J_\nu$ in equation~(\ref{nu_flux}) can be calculated as, 
\begin{equation}
N_{\nu} = \int_{E_{\nu, {\rm min}}}^{E_{\nu, {\rm max}}} \int_{T_{90}}^{t_{\rm max}} A_{\rm eff} (E_\nu) J_\nu (E_\nu) dE_{\nu}dt.
\label{nu_events}
\end{equation}
Here we take the neutrino flavor-dependent effective area $A_{\rm eff}$ for the detectors from Refs.~\cite{icecube_gen2, Aab:2015kma, Adrian-Martinez:2016fdl}.  The neutrino energy range of different detectors are $(E_{\nu, {\rm min}}, E_{\nu, {\rm max}}) = (10^6, 10^9)$~GeV for IceCube Gen-2 and $(2\times10^{2}, 10^{8})$~GeV for KM3NeT-ARCA.  For the PAO this energy range is$(2\times10^{7},  3\times10^{11})$~GeV for Earth-skimming tau neutrinos and higher for shallow zenith angles.  The maximum time we use for calculation is $t_{\rm max} = 100T_{90}$.  The flux decreases considerably at later time.
 
\begin{figure*}[th!]
\vskip -1cm
\includegraphics[width=39pc]{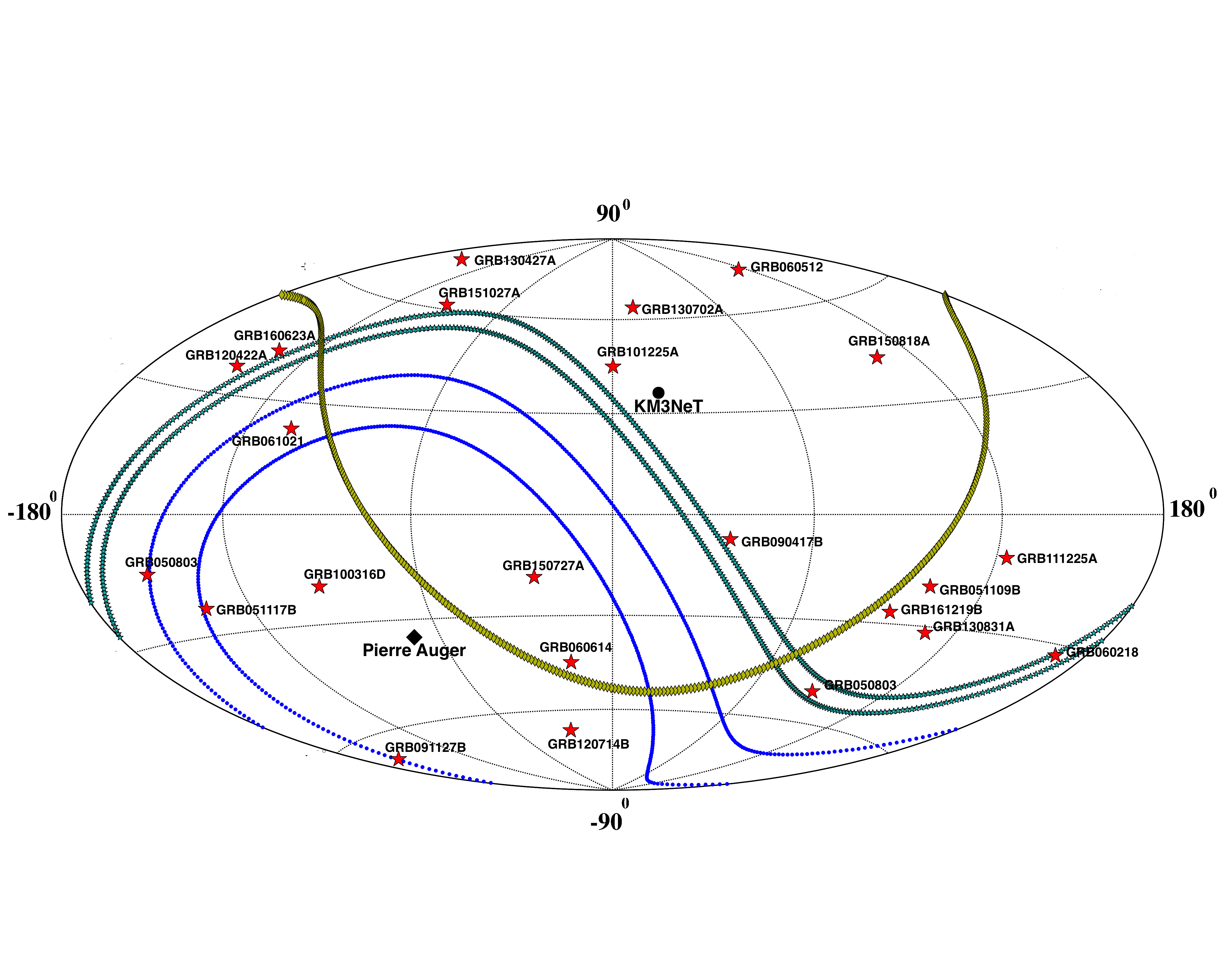}
\vskip -1cm
\caption{\label{skymap}The skymap for the 23 GRBs in the Galactic coordinates. The zenith-angle bins $60^{\circ}-75^{\circ}$ and $90^{\circ}-95^{\circ}$ for the PAO are shown with blue dotted line and cyan starred line (color rings), respectively.  The skymap with zenith angle $90^{\circ}$ with respect to the KM3NeT-ARCA is shown with yellow line (squares).  The position of the PAO and KM3NeT-ARCA are shown with black diamonds and circles respectively.}
\end{figure*}

As mentioned earlier, IceCube-Gen2 and KM3NeT-ARCA are mostly sensitive to UHE neutrino flux in the northern and southern hemisphere, respectively.  Specifically within the zenith angle of $90^{\circ}$ there are 11 GRBs in case of IceCube-Gen2 (GRB 050826, GRB 051117, GRB 060614, GRB 061021, GRB 090417, GRB 091127, GRB100316, GRB 101225, GRB 120714, GRB 160623, GRB 161219); whereas for KM3NeT-ARCA there are 9 GRBs falling within $90^{\circ}$ zenith angle (GRB 060512, GRB 060614, GRB 090417B, GRB 101125A, GRB 130427A, GRB 130702A, GRB 150727A, GRB 151027A and GRB 150818A), as shown in Fig.~\ref{skymap}.

The PAO is mostly sensitive to detect neutrinos coming from the horizon, the so-called Earth-skimming (zenith angle $90^{\circ}$-$95^{\circ}$) events and has published exposure for zenith angle bins $60^{\circ}$-$75^{\circ}$, $75^{\circ}$-$90^{\circ}$ and $90^{\circ}$-$95^{\circ}$~\cite{Aab:2015kma} . The Earth-skimming exposure is for tau neutrinos only.  The GRBs (see Fig.~\ref{skymap}) within the corresponding zenith angle bins has been taken for analysis.  Within  $60^{\circ}$-$75^{\circ}$ zenith angle bin we find 3 GRBs (GRB 091127B, GRB 061021 and GRB 050826), and within the zenith bin $90^{\circ}$-$95^{\circ}$ there are only 2 GRBs (GRB 050803 and GRB 060218), whereas the $75^{\circ}$-$90^{\circ}$ bin does not contain any GRBs.  We have converted the exposure for each zenith angle bin given in~\cite{Aab:2015kma} to effective areas, taking into account that the analysis is from 1 January 2004 to 20 June 2013.

\begin{figure*}[th!]
\includegraphics[trim =  0 21 0 10, width=0.8\columnwidth]{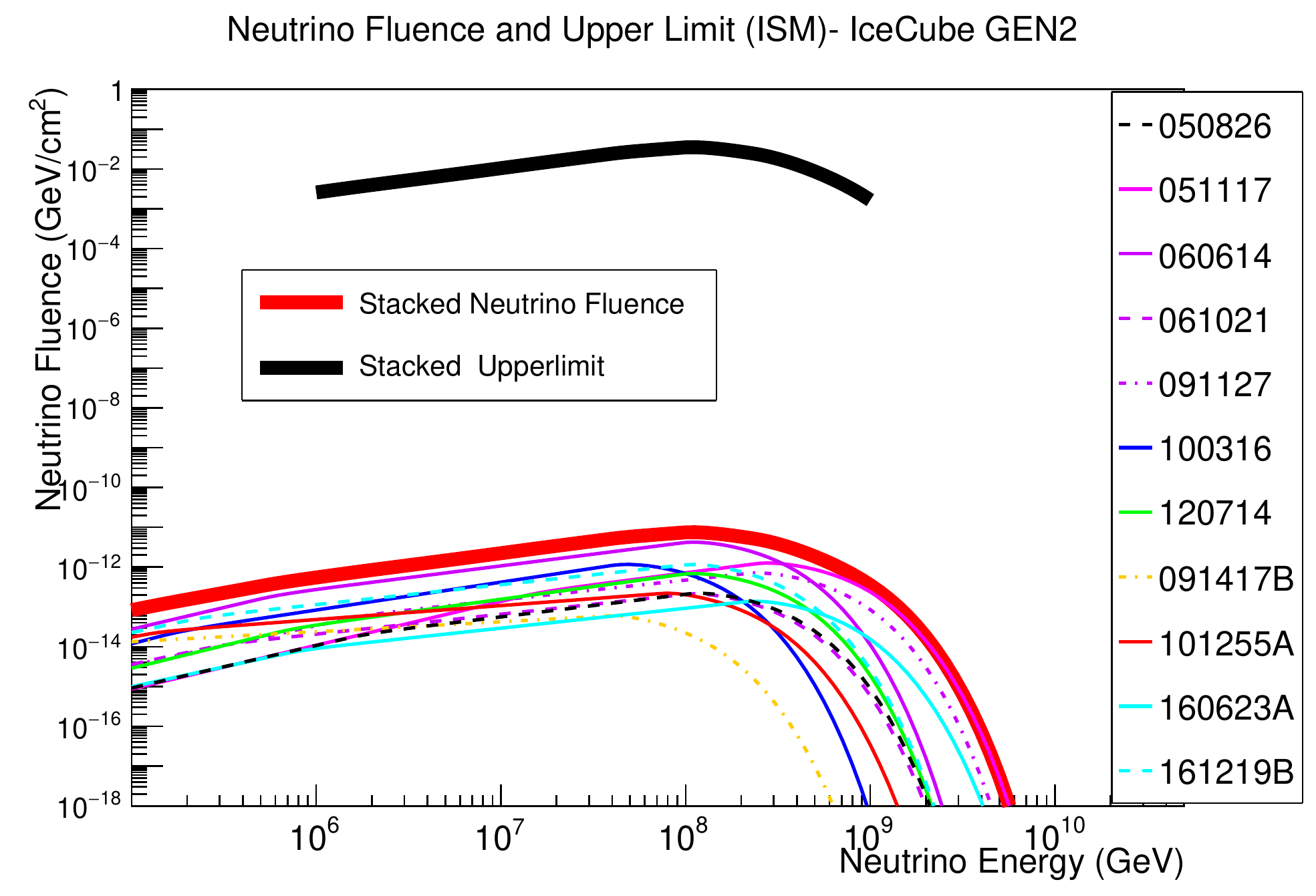}
\vspace{0.5cm}
\includegraphics[trim =  0 21 0 10, width=0.8\columnwidth]{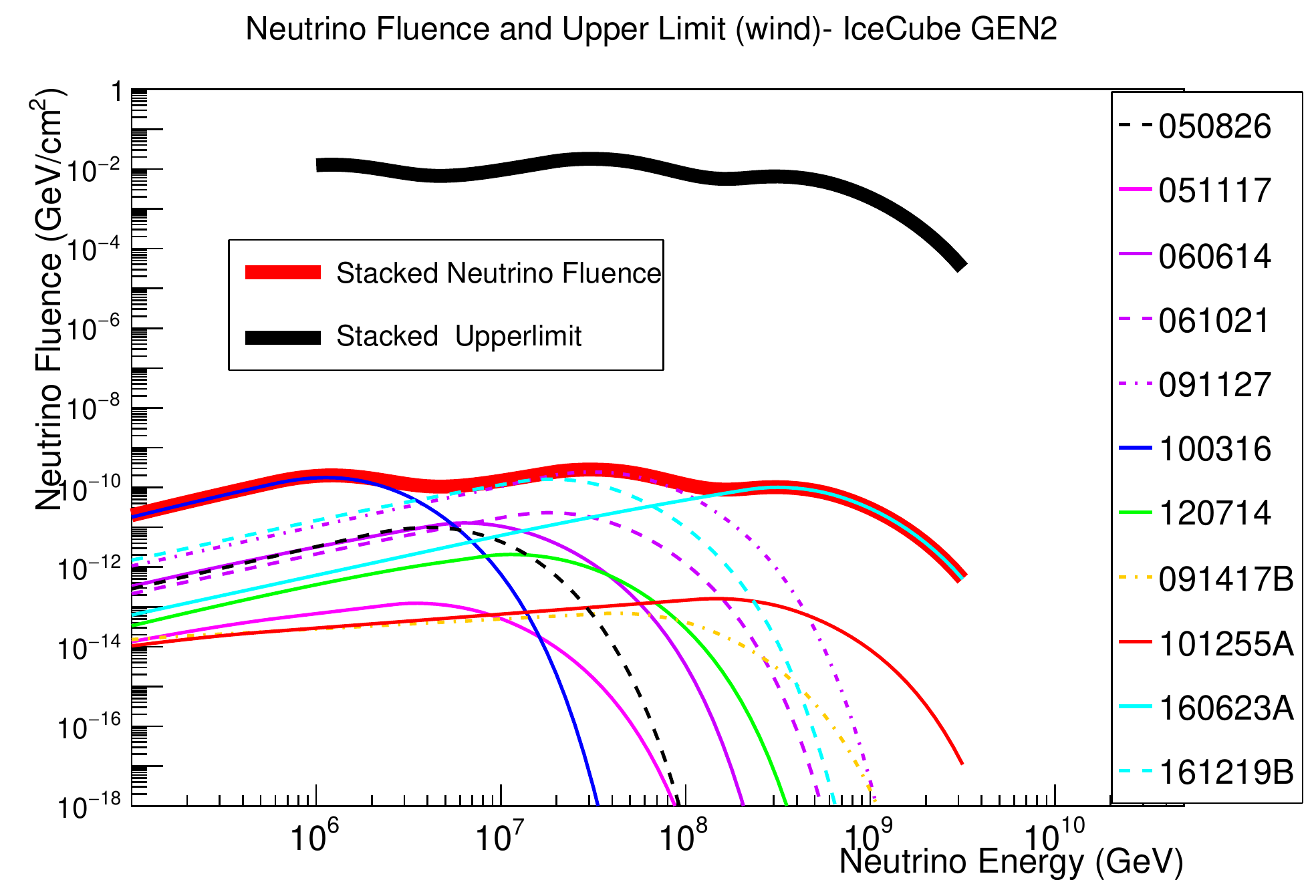}
\includegraphics[trim =  0 21 0 10, width=0.8\columnwidth]{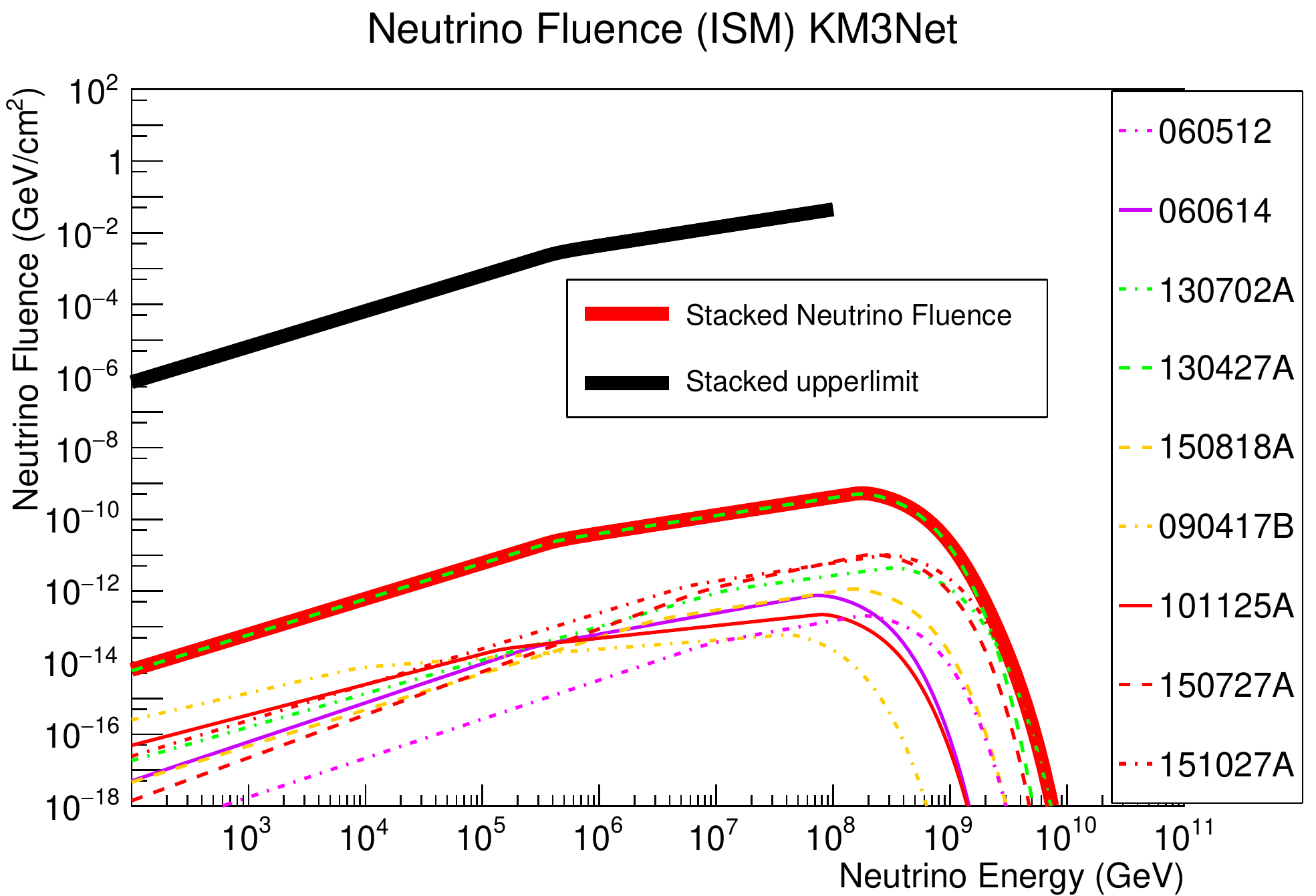}
\vspace{0.5cm}
\includegraphics[trim =  0 21 0 10, width=0.8\columnwidth]{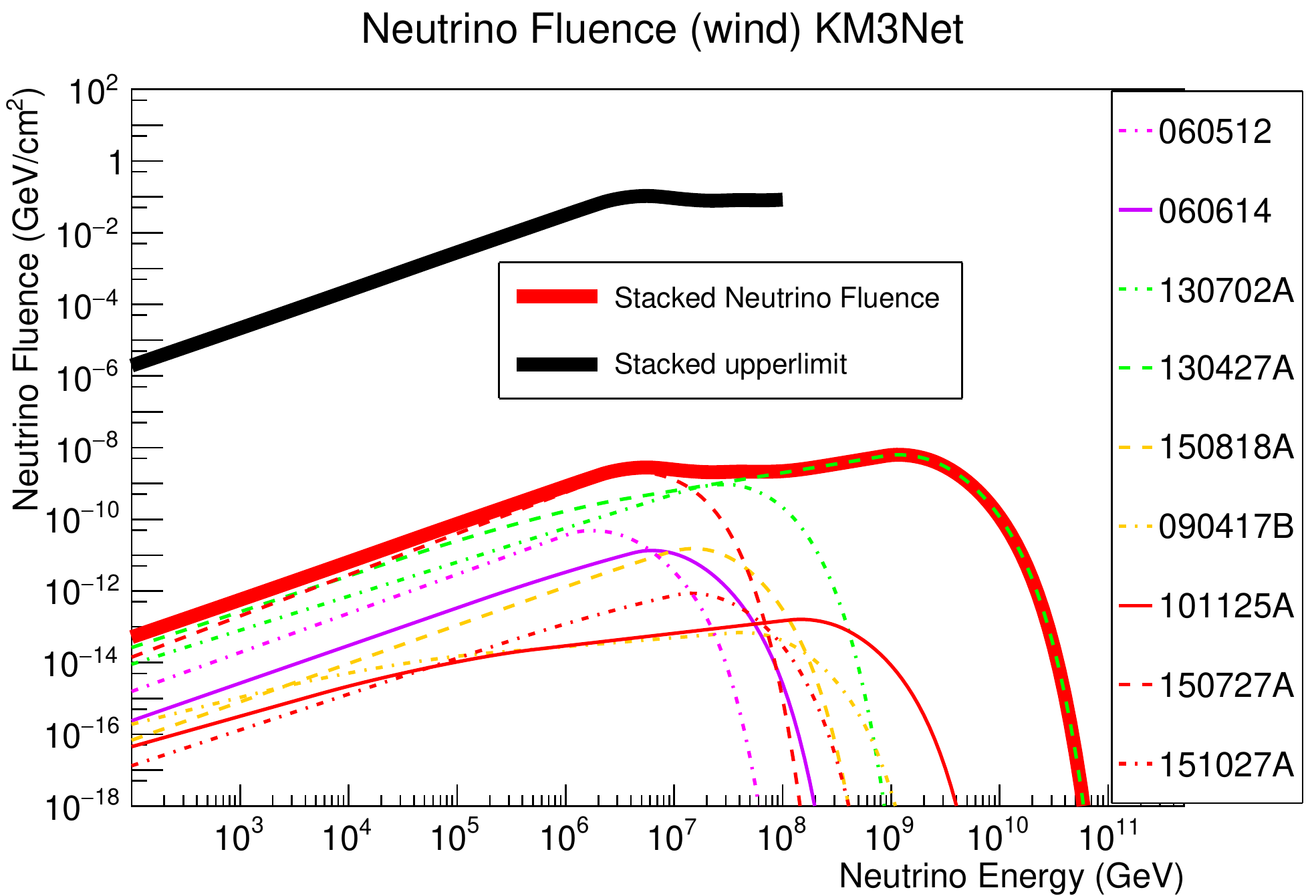}
\includegraphics[trim =  0 21 0 10, width=0.8\columnwidth]{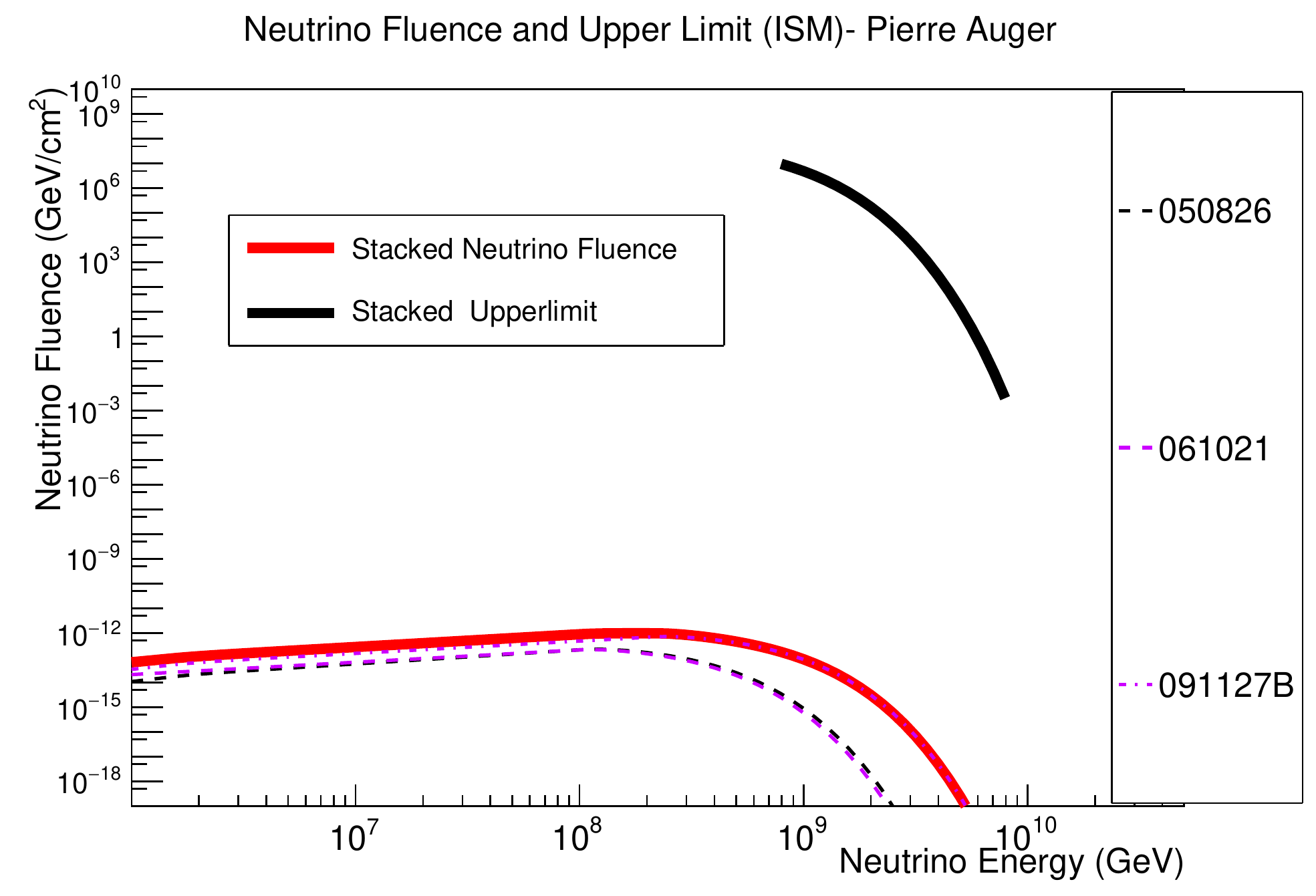}
\vspace{0.5cm}
\includegraphics[trim =  0 21 0 10, width=0.8\columnwidth]{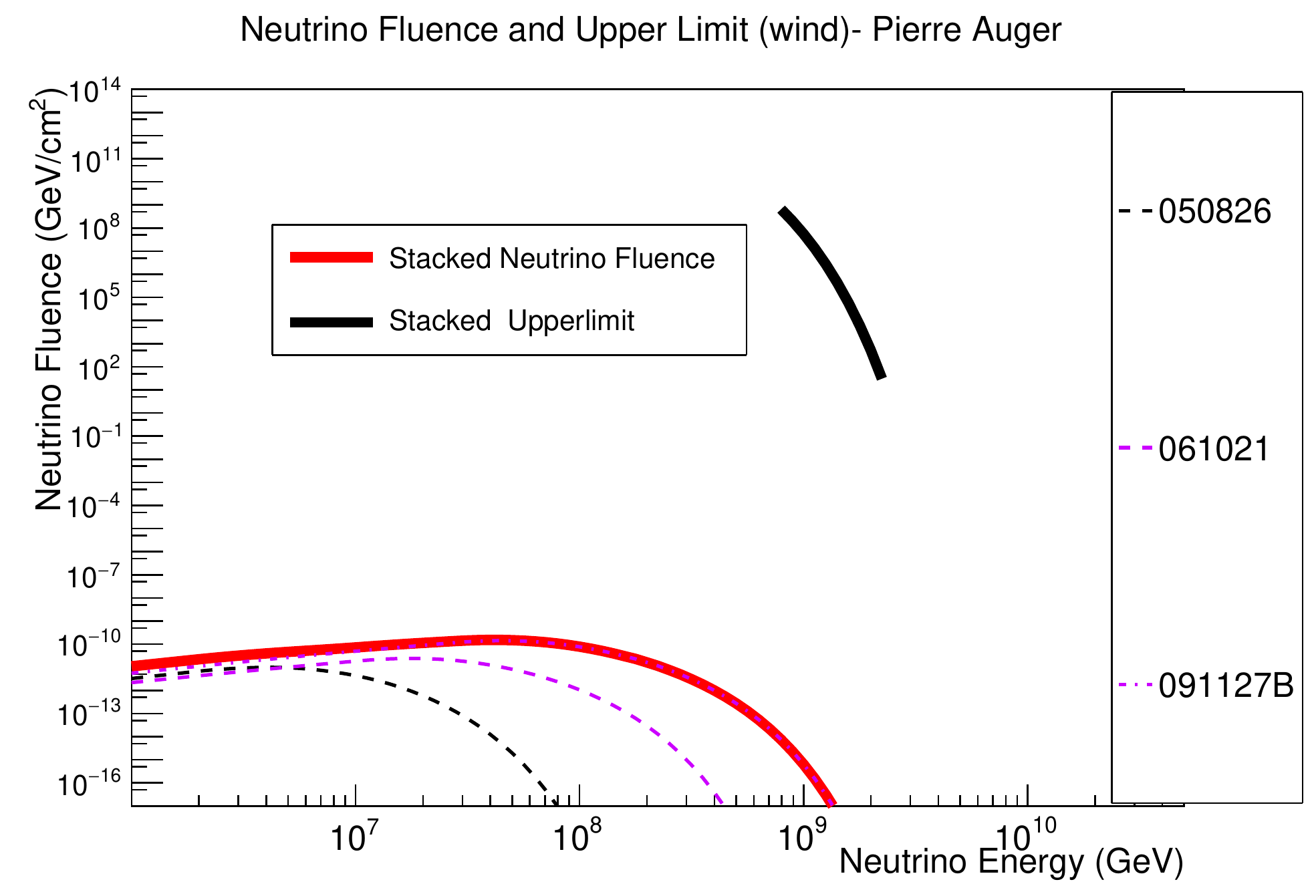}
\includegraphics[trim =  0 21 0 10, width=0.8\columnwidth]{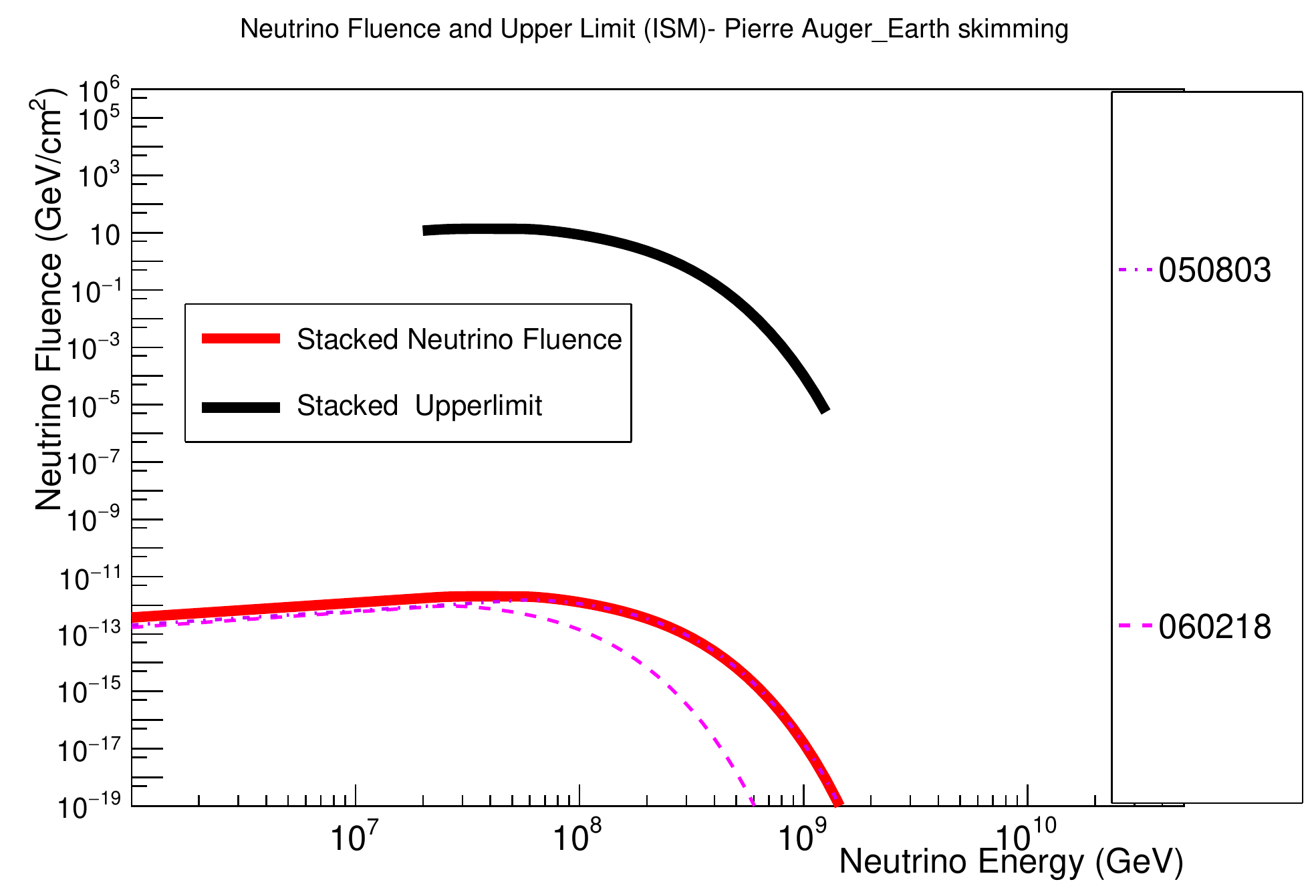}
\vspace{0.5cm}
\includegraphics[trim =  0 21 0 10, width=0.8\columnwidth]{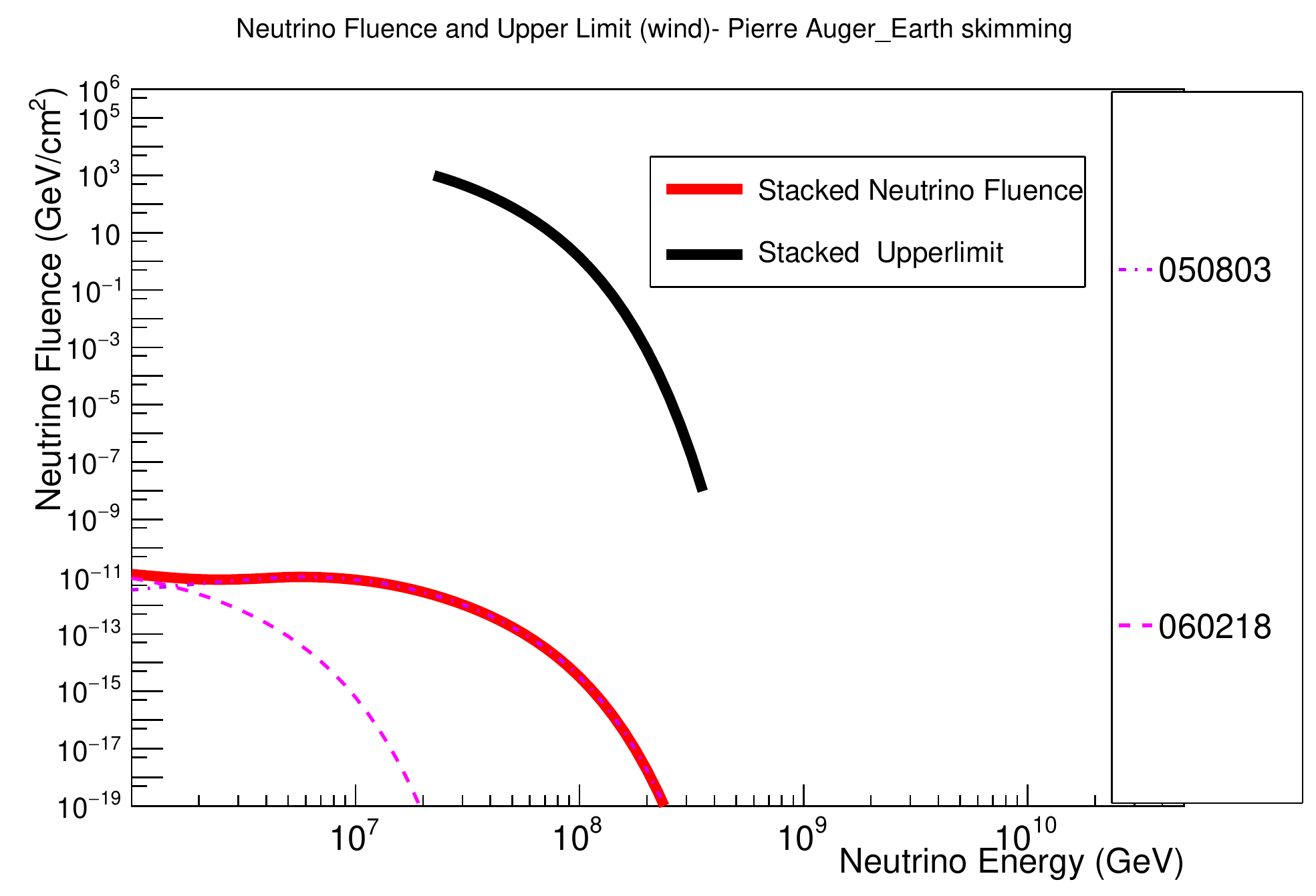}
\caption{\label{flu} Individual neutrino fluence calculated from GRB afterglows, the stacked fluence and the upper limits for the stacked fluence in the ISM and wind environment for IceCube Gen-2, Pierre Auger Observatory and KM3NeT-ARCA are shown in different panels.  Details are given in the main text. 
}
\end{figure*}

We have calculated UHE neutrino fluence of 23 selected GRBs in our sample, integrating flux for $t=T_{90}$-$100T_{90}$, as shown in Fig.~\ref{flu}.  Different panels of Fig.~\ref{flu} shows fluence of GRBs visible to each of the IceCube Gen-2, KM3NeT-ARCA (above horizon) and for different PAO cases, both for the ISM and wind environment of the GRB afterglow.   The first and second row plots are for IceCube Gen-2 and KM3Net-ARCA, respectively, whereas the third and fourth row plots are for the PAO. We have also plotted the stacked fluence from individual GRB fluence in each case, shown as thick lines in Fig.~\ref{flu}.

In all the above cases we found that a detection is not possible, i.e., $N_\nu \ll 1$ in equation~(\ref{nu_events}).  Therefore we have calculated the $90\%$ upper limit on the integrated stacked fluence for all the neutrino detectors.  These upper limits are shown as the black solid lines in Fig.~\ref{flu} for individual neutrino telescopes in case of non-detection of any neutrino events.   Note that zero background has been assumed for this calculation, which is expected at UHE in the short time scales that we have considered.  These upper limits can be compared with those in Ref.~\cite{Razzaque:2006qa}, where neutrino afterglow models in Refs.~\cite{Waxman:1999ai, Dai+01} were used.

\section{Results and discussion}
\label{discussion}
Ultra-high energy neutrinos can carry astrophysical information from large distances directly to us, therefore can identify the sources of ultrahigh-energy cosmic rays.  Gamma-ray bursts have been considered as the sources of UHECRs, being the most powerful electromagnetic explosions in the universe.  We have considered a model where UHECRs are accelerated in the blast wave of the GRB, expanding in a circumburst medium, which is also responsible for the observed afterglow radiation in the gamma-ray to radio wavelengths.  UHE neutrinos are produced by the $p\gamma$ interactions of UHECRs with afterglow photons.

We selected a sample of 23 long duration GRBs with known redshift below $z=0.5$.  Multi-wavelength afterglow data are available for these GRBs and we have fitted spectral energy distribution at different time intervals and light curves at different wavelengths using afterglow synchrotron model of a blast wave evolving in a constant density ISM and in a varying-density wind medium.  These fits allowed us to obtain the parameters of the afterglow model, thus characteristics of the target photons for $p\gamma$ interactions.   Next we have calculated expected neutrino flux from the GRBs in our sample based on the blast wave properties.  To our knowledge, this is the first time such a realistic neutrino flux calculation has been performed using broadband electromagnetic data from GRB afterglows.  The results from our investigation are listed below.
\begin{itemize}

\item  Our synchrotron afterglow model fit to multi-wavelength data includes broadband SED at different times and light curves at different frequencies.  A wind environment  for the GRB blast wave evolution is preferred for GRBs 060218, 130702A, 130831A and 130427A.  A constant density ISM is preferred for GRBs 051109B, 051117B, 061021, 111225A and 151027A.  A clear preference could not be found for other GRBs.  For a number of GRBs neither wind nor ISM model could fit data satisfactorily.  Additional emission mechanism may be required in such cases.  

\item  Fits to afterglow data give GRB blast wave kinetic energy $E_{kin} \sim 3\times 10^{49}$-$10^{55}$~erg, both in case of a wind environment and an ISM environment.  The values of microphysical parameters we found, $\epsilon_e \sim 10^{-3}$-$10^{-2}$ and $\epsilon_b \sim 10^{-4}$-$10^{-2}$ are typical. 

\item  We have calculated neutrino flux from the 23 GRBs using $p\gamma$ interaction efficiency calculated from target synchrotron afterglow photons from above fits.  GRB 130427A ($z=0.34$) dominates the flux, which is one of the brightest GRBs detected to-date.  GRB 130831A ($z=0.479$), GRB 130702A ($z=0.145$) and GRB 091127B ($z=0.49$) also had high neutrino flux according to our calculation.  Unless a significant fraction of the observed electromagnetic afterglow emission comes from a different mechanism than the forward shock discussed here, our neutrino flux calculation will not be affected by considering more complicated afterglow scenarios.   

\item  We found that ultrahigh-energy neutrinos from the nearby 23 long-duration GRBs in our sample cannot be detected by the currently operating and upcoming neutrino detectors: IceCube Gen-2, KM3NeT-ARCA and the Pierre Auger Observatory.  This in turn implies low efficiency of neutrino production in the blast wave of these GRBs.

\item  In case of non-detection of neutrinos from the GRBs in our sample, we have calculated upper limits on the stacked fluence from relevant GRBs for individual neutrino detectors.  In general these upper limits are orders of magnitude higher than the neutrino fluence.

\end{itemize}

Our calculations are useful to estimate sensitivity of new generation of neutrino telescopes for detecting ultrahigh-energy ($>$ 1 PeV) neutrinos  from the afterglows of long-durations GRBs and explore connection between ultrahigh-energy cosmic rays and gamma-ray bursts.

\acknowledgments
This work was supported in part by the National Research Foundation (South Africa) Grant No.\ 87823
(CPRR) to S.R.  J.K.T also acknowledges receiving PhD bursary from the National Research Foundation.

\end{document}